\documentclass[11pt,twoside]{article}

\usepackage{amsmath,amssymb,mathrsfs,bigdelim,bm,bold-extra}
\usepackage{newtxtext,newtxmath}
\usepackage{multirow,dcolumn,filecontents,stackengine,empheq,cases,braket}
\usepackage{booktabs,arydshln,dashrule}
\usepackage[hang,flushmargin]{footmisc}
\usepackage{graphicx,color}
\usepackage[dvipsnames]{xcolor}
\usepackage{tokcycle}
\usepackage[T1]{fontenc}
\usepackage{fancyhdr,pdflscape,afterpage,capt-of,graphbox,soul}
\usepackage[makeroom]{cancel}
\usepackage[normalem]{ulem}
\usepackage{mwe,here,fancybox,tikz,fourier-orns,pgfornament,psvectorian}
\usepackage{setspace,sectsty,framed,mdframed,comment,tocloft,enumitem,titlesec,titletoc,endnotes}

\def\hltcol{cyan}
\setulcolor{\hltcol}
\definecolor{TB}{RGB}{10,186,181}
\definecolor{tb}{rgb}{0.24,0.43,0.91}
\definecolor{shadecolor}{RGB}{10,186,181}

\usepackage[sort&compress,numbers,square]{natbib}
\renewcommand*{\cite}[1]{\textcolor{tb}{\citep{#1}}}

\usepackage[hyphens]{xurl}
\usepackage[pdfencoding=auto,linkbordercolor=CornflowerBlue,citecolor=tb,linkcolor=tb,pdfpagelabels=false]{hyperref}
\hypersetup{colorlinks=true,allbordercolors=CornflowerBlue}
\usepackage[font={small},labelsep=space]{caption}
\usepackage{subcaption}
\DeclareCaptionLabelFormat{vertf}{\textbf{{#1} {#2}\!}}
\DeclareCaptionLabelFormat{vertt}{\textbf{{#1} {#2}~|}}
\newcommand{\q}[1]{`#1'}

\makeatletter
\newcommand*{\myfnsymbol}[1]{\ensuremath{%
\ifcase#1 \or \ast \or \dagger \or \spadesuit \or \diamondsuit \or \clubsuit \or \heartsuit \else \@ctrerr \fi}}
\makeatother

\def\ai{Artificial Intelligence}
\def\quantum{Quantum Science}
\def\bio{Biotechnology}
\def\nano{Nanotechnology}
\def\agri{Agricultural Engineering}
\def\particle{Particle Physics}
\def\aerospace{Aerospace Engineering}
\def\nuclear{Nuclear Engineering}
\def\marine{Marine Engineering}
\def\neuro{Neuroscience}
\def\condensed{Condensed Matter Physics}
\def\envi{Environmental Engineering}
\def\earth{Earth Science}
\def\astro{Astronomy}
\def\math{Pure Mathematics}

\newlength\replength
\newcommand\ruleht{-4pt}
\newcommand\rulewidth{6.5pt}
\newcommand\sphl[1]{\sbox0{#1}\ooalign{\makebox[0pt][l]{%
  \smash{\color{\rulecolor}\rule[\ruleht]{\wd0}{\rulewidth}}}\cr#1}}
\newcommand\hlt[1]{%
\resettokcycle
  \Characterdirective{\addcytoks{\nobreak\hspace{0pt minus .6pt}%
    \sphl{##1}}}%
  \Spacedirective{\addcytoks{\nobreak\hspace{0pt minus .6pt}%
    \discretionary{}{}{\sphl{\ }}}}%
\cytoks{}%
\def\rulecolor{pink!60}%
\tokcyclexpress{#1}%
\the\cytoks%
}

\newcommand{\guide}[2]{%
\vspace{1em}
\noindent
{\large\textcolor{NavyBlue}{\underline{\textbf{#1}}}}
\begin{itemize}
#2
\end{itemize}
}

\newcommand{\cdend}[5]{%
    \begin{tabular}{c}
    \begin{minipage}{0.03\hsize}
\begin{center}
    \rotatebox{0}{\hspace{1.0cm}\textcolor{black}{[\textsf{#4}]}}
\end{center}
\end{minipage}
\begin{minipage}{0.97\hsize}
\begin{center}
\includegraphics[align=c, scale=0.47, vmargin=0mm]{fig/cdend_inst_#1_#2.pdf}\\[-3em]
\end{center}
    \end{minipage}
    \end{tabular}
\quad\\[-2em]
\dotfill 
\quad\\[-2em]
    \begin{tabular}{c}
    \begin{minipage}{0.03\hsize}
\begin{center}
    \rotatebox{0}{\hspace{1.0cm}\textcolor{black}{[\textsf{#5}]}}
\end{center}
\end{minipage}
\begin{minipage}{0.97\hsize}
\begin{center}
\includegraphics[align=c, scale=0.47, vmargin=0mm]{fig/cdend_inst_#1_#3.pdf}\\[-3em]
\end{center}
    \end{minipage}
    \end{tabular}
}

\begin{document}

\quad\vspace{-1.4cm}
\begin{flushright}
August 2023~~[v1]
\end{flushright}

\vspace{1.0cm}

\begin{center}
\fontsize{15pt}{16pt}\selectfont\bfseries
\textit{Atlas of Science Collaboration}, 1971--2020
\end{center}

\renewcommand*{\thefootnote}{\textcolor{black}{\myfnsymbol{\value{footnote}}}}

\vspace*{0.8cm}
\centerline{%
{Keisuke Okamura}\,\footnote{\,{\tt okamura@ifi.u-tokyo.ac.jp}}${}^{,}$%
\footnote{\,\href{https://orcid.org/0000-0002-0988-6392}{\tt \textcolor{black}{orcid.org/0000-0002-0988-6392}}}${}^{;\,1,\,2}$}

\vspace*{0.6cm}
{\small\centerline{\textit{
${}^{1}$Institute for Future Initiatives (IFI), The University of Tokyo,}}
\centerline{\textit{
7-3-1 Hongo, Bunkyo-ku, Tokyo 113-0033, Japan.}}
\vspace*{3mm}\centerline{\textit{
${}^{2}$SciREX Center, National Graduate Institute for Policy Studies (GRIPS),}}
\centerline{\textit{
7-22-1 Roppongi, Minato-ku, Tokyo 106-8677, Japan.}}}

\vspace{2.0cm}
\noindent\textbf{Abstract.}
\quad
The evolving landscape of interinstitutional collaborative research across 15 natural science disciplines is explored using the open data sourced from OpenAlex. 
This extensive exploration spans the years from 1971 to 2020, facilitating a thorough investigation of leading scientific output producers and their collaborative relationships based on coauthorships.
The findings are visually presented on world maps and other diagrams, offering a clear and insightful portrayal of notable variations in both national and international collaboration patterns across various fields and time periods. 
These visual representations serve as valuable resources for science policymakers, diplomats and institutional researchers, providing them with a comprehensive overview of global collaboration and aiding their intuitive grasp of the evolving nature of these partnerships over time.

\vspace{0.8cm}
\noindent\textbf{Keywords.}
\quad
International\,/\,interinstitutional research collaboration | OpenAlex | Open Bibliometrics

\vfill

\thispagestyle{empty}
\setcounter{footnote}{0}
\setcounter{figure}{0}
\setcounter{table}{-1}
\setcounter{equation}{0}

\setlength{\skip\footins}{10mm}
\setlength{\footnotesep}{4mm}

\vspace{-1.6cm}

\newpage

\setlength{\skip\footins}{10mm}
\setlength{\footnotesep}{4mm}

\renewcommand{\contentsname}{Table of Contents}
\setlength\cftaftertoctitleskip{3em}
\dottedcontents{section}[2em]{}{2.0em}{0.5pc}

\thispagestyle{empty}
\begin{center}
\vspace{0.7cm}
\begin{spacing}{1.6}
\tableofcontents
\end{spacing}

\vfill
\begin{mdframed}[linecolor=TB]
\begin{tabular}{l}
\begin{minipage}{0.09\hsize}
\vspace{-0.7em}
\hspace{-2.5mm}\includegraphics[width=1.4cm,clip]{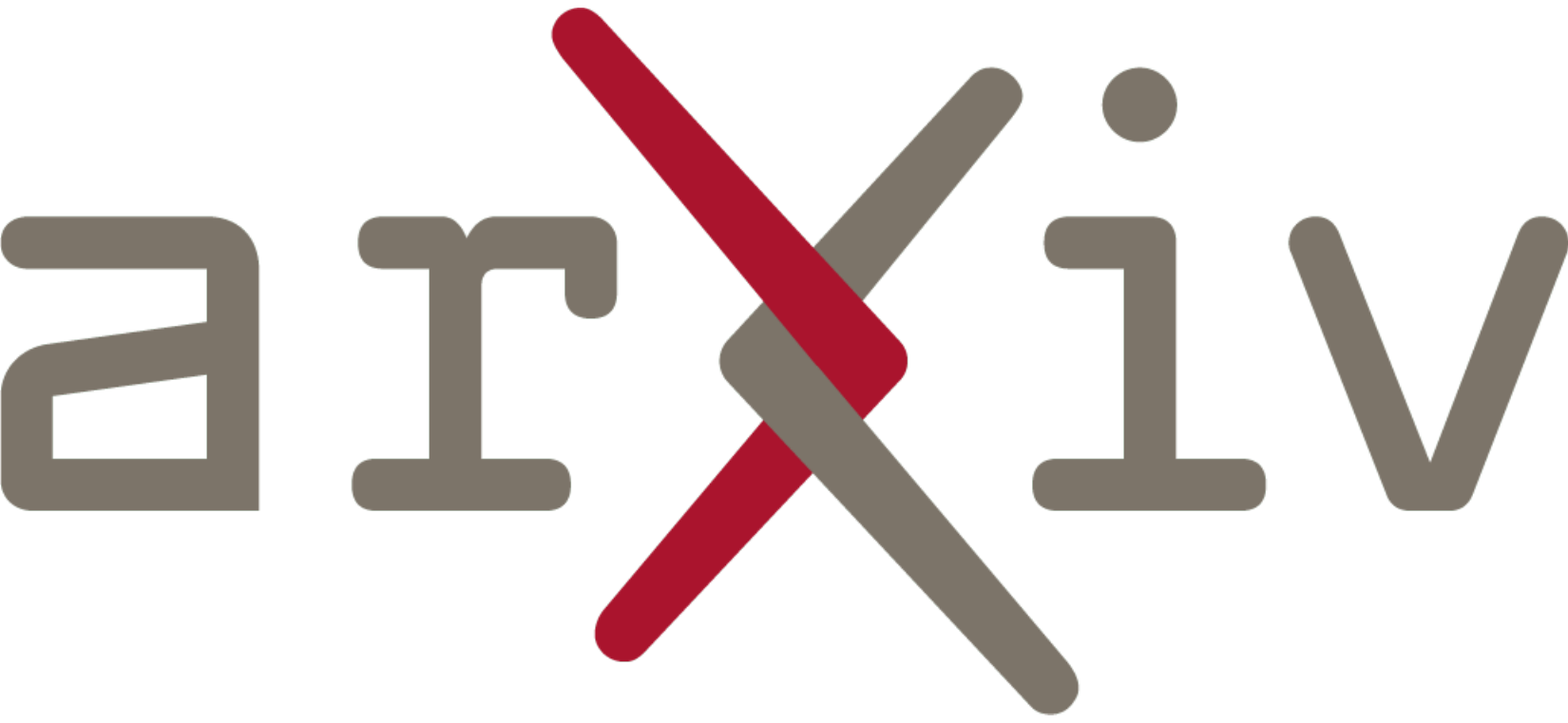}\\[-1.8em]

\hspace{-3.5mm}\includegraphics[width=1.6cm,clip]{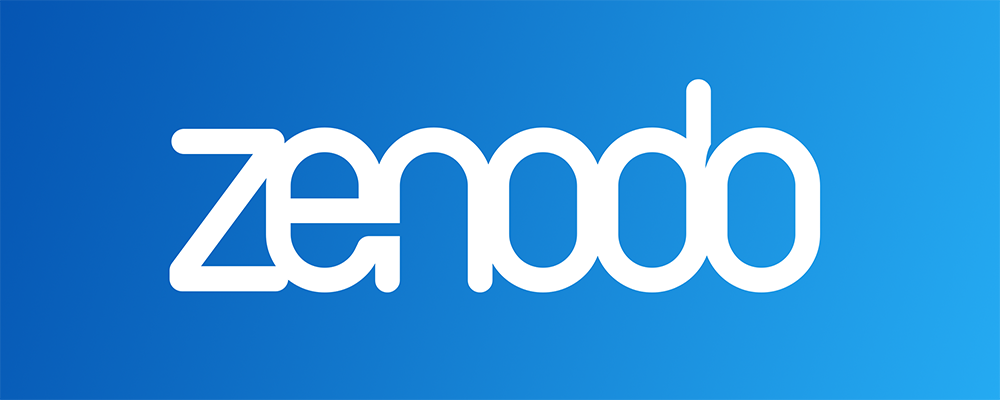}
\end{minipage}
\begin{minipage}{0.88\hsize}
\setstretch{0.9}
\noindent
\textcolor{TB}{\footnotesize\textsf{%
This booklet presents a preliminary set of illustrative diagrams and tables on international and interinstitutional research at a global scale, posted on arXiv and Zenodo (DOI:\href{https://doi.org/10.5281/zenodo.8266166}{10.5281/zenodo.8266166}).
The source data is retrieved from OpenAlex (\href{https://openalex.org}{https://openalex.org}) as of August 2023.}}
\end{minipage}
\end{tabular}
\end{mdframed}
\end{center}

\pagestyle{fancy}
\let\oldheadrule\headrule
\let\oldfootrule\footrule
\renewcommand{\headrule}{\color[rgb]{0.04, 0.73, 0.71}\oldheadrule}
\renewcommand{\footrule}{\color[rgb]{0.04, 0.73, 0.71}\oldfootrule}
\renewcommand\fbox{\fcolorbox{white}{TB}}
\renewcommand{\headrulewidth}{0pt}
\renewcommand{\footrulewidth}{1pt}
\fancyhead[O]{\parbox{\textwidth}{\setlength{\FrameSep}{\outerbordwidth}
    \begin{shaded}
\setlength{\fboxsep}{0pt}\makebox[\textwidth][r]{\setlength{\fboxsep}{1pt}\fcolorbox{TB}{TB!70}{\textbf{{\mbox{~}\makebox[\textwidth-0.49em][r]{\textcolor{white}{~~\large\textsc{\leftmark}~~}}\vphantom{q\^{M}}}}}}
    \end{shaded}
  }\vspace{-5pt}}
\fancyhead[E]{\parbox{\textwidth}{\setlength{\FrameSep}{\outerbordwidth}
    \begin{shaded}
\setlength{\fboxsep}{0pt}\makebox[\textwidth][l]{\setlength{\fboxsep}{1pt}\fcolorbox{TB}{TB!70}{\textbf{{\mbox{~}\makebox[\textwidth-0.49em][l]{\textcolor{white}{~~\large\textsc{\leftmark}~~}}\vphantom{q\^{M}}}}}}
    \end{shaded}
  }\vspace{-5pt}}
\fancyfoot[O]{\color[rgb]{0.04, 0.73, 0.71}{\raisebox{3pt}[8pt][0pt]{\tiny\textit{Atlas of Science Collaboration} (v1; August 2023) | {https://doi.org/10.5281/zenodo.8266166}}}\hfill\raisebox{3pt}[8pt][0pt]{\fbox{\textcolor{white}{\footnotesize\textbf{\textsf{\thepage}}}}}}
\fancyfoot[E]{\color[rgb]{0.04, 0.73, 0.71}{\raisebox{3pt}[8pt][0pt]{\fbox{\textcolor{white}{\footnotesize\textbf{\textsf{\thepage}}}}}\hfill\raisebox{3pt}[8pt][0pt]{\tiny\textit{Atlas of Science Collaboration} (v1; August 2023) | {https://doi.org/10.5281/zenodo.8266166}}}}
\fancyfoot[C]{}

\thispagestyle{empty}

\afterpage{\clearpage%
\pagenumbering{roman}
\setcounter{page}{3}

\quad\vspace{1em}
\markboth{\textsc{User Guide and Notes}}{}
\quad\vspace{-0.1cm}
\begin{center}
{\Large\bfseries\textit{Atlas of Science Collaboration}: User Guide and Notes}
\end{center}

\addcontentsline{toc}{section}{User Guide and Notes}

\vspace{0.7cm}

\guide{Intended Readership}{%

\item This booklet, entitled \textit{\q{Atlas of Science Collaboration}}, aims to offer a broad overview of international and interinstitutional research collaboration, shedding light on its present status and evolution on a global scale. 
While it might not delve into intricate scholarly or academic data analysis, it remains a valuable resource for those seeking a general understanding of the collaborative relationships that have been established between research institutions in the world of science.

\item The intended readership includes science and technology (S\&T) policymakers and diplomats, government research and development (R\&D) agencies, international organisations, S\&T think tanks, as well as institutional research divisions of universities or R\&D institutions.
}

\guide{Data Source}{%

\item The \textit{Atlas of Science Collaboration} is based on data retrieved from \href{https://docs.openalex.org/}{OpenAlex} \cite{Priem22},\endnote{https://docs.openalex.org/} a free and open (the CC0 license) catalogue of the world's scholarly papers, researchers, journals and institutions. 
Launched in January 2022, OpenAlex replaced \href{https://www.microsoft.com/en-us/research/project/microsoft-academic-graph/}{Microsoft Academic Graph (MAG)},\endnote{https://www.microsoft.com/en-us/research/project/microsoft-academic-graph/} which retired at the beginning of 2022.

\item OpenAlex collects information on scientific publications, including journal articles, non-journal articles, preprints, conference papers, books and datasets---hereafter collectively referred to as \q{works}---from various platforms such as \href{https://www.crossref.org/}{Crossref},\endnote{https://www.crossref.org/} \href{https://orcid.org/}{ORCID},\endnote{https://orcid.org/} \href{https://ror.org/}{ROR},\endnote{https://ror.org/} \href{https://pubmed.ncbi.nlm.nih.gov}{PubMed},\endnote{https://pubmed.ncbi.nlm.nih.gov} preprint servers like \href{https://arxiv.org/}{arXiv},\endnote{https://arxiv.org/} and institutional or disciplinary repositories like \href{https://zenodo.org/}{Zenodo}.\endnote{https://zenodo.org/}
For comparison with other scholarly data sources such as \href{https://www.scopus.com/}{Scopus},\endnote{https://www.scopus.com/} \href{https://clarivate.com/products/scientific-and-academic-research/research-discovery-and-workflow-solutions/webofscience-platform/}{Web of Science}\endnote{https://clarivate.com/products/scientific-and-academic-research/research-discovery-and-workflow-solutions/webofscience-platform/} and \href{https://www.dimensions.ai/}{Dimensions},\endnote{https://www.dimensions.ai/} please refer to \href{https://openalex.org/about\#comparison}{OpenAlex's website}.\endnote{https://openalex.org/about\#comparison}

\item OpenAlex offers extensive coverage of meta-information across a diverse spectrum of works, encompassing not only journal publications but also non-journal works, non-English works and contributions from the Global South. 
This attribute proves beneficial by providing a more precise augmentation of the extent of R\&D activities, along with their associated scholarly outputs. 
This is especially crucial in fields where journals are not the predominant channel for disseminating research outcomes.
Furthermore, OpenAlex effectively captures outputs in the preprint format, which might persist for varying durations, spanning from months to years or even indefinitely, without necessarily transitioning into journal publications.

\item The present edition (August 2023) of the \textit{Atlas of Science Collaboration} was compiled using data obtained via the \href{https://docs.openalex.org/how-to-use-the-api/api-overview}{OpenAlex API}\endnote{https://docs.openalex.org/how-to-use-the-api/api-overview} during the period from the 12th to the 15th of August 2023.
It is essential to note that OpenAlex is an ongoing project, continuously updating its data and improving its system.
Consequently, the visualisations in this booklet may not provide the most comprehensive view or accurate data.
Expect more accurate results when acquiring data in the future as OpenAlex undergoes further upgrades.
Revised editions of the \textit{Atlas of Science Collaboration} may be made available on \href{https://zenodo.org/}{Zenodo} or other open platforms beyond this release.
}

\guide{R\&D Disciplines}{%
\item In this current edition, the primary focus centres around the level-1 \q{concepts} listed in Table \ref{tab:disciplines} sourced from the OpenAlex classification, as previously explored in Ref.~\cite{Okamura23}.
Each level-1 concept is accompanied by \q{related concepts}, which can offer a finer or broader delineation compared to the level-1 concept.
Using this characteristic, an enhanced notion of R\&D discipline is constructed by including all associated subconcepts of level 2 or higher for each of the 15 level-1 concepts. 
For instance, the defined discipline of \q{Artificial Intelligence} includes OpenAlex's level-2 concepts of \q{\href{https://explore.openalex.org/concepts/C50644808}{Artificial Neural Network}} and \q{\href{https://explore.openalex.org/concepts/C108583219}{Deep Learning}}, but not the level-0 concepts of \q{\href{https://explore.openalex.org/concepts/C41008148}{Computer Science}} or \q{\href{https://explore.openalex.org/concepts/C33923547}{Mathematics}}.
}

\begin{table}[t]
\centering
\vspace{-0.3cm}
\caption{\textbf{R\&D disciplines and the associated OpenAlex concept IDs.}}
\vspace{-0.3cm}
\label{tab:disciplines}
{\footnotesize
{\renewcommand{\arraystretch}{1.05}
\begin{tabular}{rl@{\hspace{1.2cm}}l}\\[-2mm] \toprule[1pt] \\[-1.4em]
\multicolumn{2}{l}{R\&D discipline} & {Key OpenAlex ID} \\[-0.25em] \midrule[0.3pt]
{1.} & \!\!\ai 				& ~~\href{https://explore.openalex.org/concepts/C154945302}{C154945302}	\\
{2.} & \!\!\quantum 		& ~~\href{https://explore.openalex.org/concepts/C62520636}{C62520636}	\\
{3.} & \!\!\bio 			& ~~\href{https://explore.openalex.org/concepts/C150903083}{C150903083}	\\
{4.} & \!\!\nano 			& ~~\href{https://explore.openalex.org/concepts/C171250308}{C171250308}	\\
{5.} & \!\!\agri 			& ~~\href{https://explore.openalex.org/concepts/C88463610}{C88463610}	\\
{6.} & \!\!\particle 		& ~~\href{https://explore.openalex.org/concepts/C109214941}{C109214941}	\\
{7.} & \!\!\aerospace 		& ~~\href{https://explore.openalex.org/concepts/C146978453}{C146978453}	\\
{8.} & \!\!\nuclear 		& ~~\href{https://explore.openalex.org/concepts/C116915560}{C116915560}	\\
{9.} & \!\!\marine 			& ~~\href{https://explore.openalex.org/concepts/C199104240}{C199104240}	\\
{10.} & \!\!\neuro 			& ~~\href{https://explore.openalex.org/concepts/C169760540}{C169760540}	\\
{11.} & \!\!\condensed 		& ~~\href{https://explore.openalex.org/concepts/C26873012}{C26873012}	\\
{12.} & \!\!\envi			& ~~\href{https://explore.openalex.org/concepts/C87717796}{C87717796}	\\
{13.} & \!\!\earth 			& ~~\href{https://explore.openalex.org/concepts/C1965285}{C1965285}		\\
{14.} & \!\!\astro 			& ~~\href{https://explore.openalex.org/concepts/C1276947}{C1276947} 	\\
{15.} & \!\!\math 			& ~~\href{https://explore.openalex.org/concepts/C202444582}{C202444582} \\[0mm] \bottomrule[1pt] \\
\end{tabular}}
}
\end{table}

\guide{Analysis and Visualisation}{%

\item First, the \textbf{\textit{World Map of Science Collaboration}} divides the period from 1971 to 2020 into four intervals: 1971--1990, 1991--2000, 2001--2010 and 2011--2020.
For each period and discipline, bubbles represent the top 199 research institutions in terms of work production.
Additionally, for the top 50 research institutions, their locations are connected on the world map using great circle curves (the shortest route between them) to illustrate bilateral coauthorship relationships.
Coauthorship relationships with fewer than five coauthored papers are not displayed.
The background world map utilises the {\texttt{world}} data from the \href{https://CRAN.R-project.org/package=maps}{\texttt{maps}} package\endnote{https://CRAN.R-project.org/package=maps} in R.
The connection visualisation between two research institutions leverages the \texttt{gcIntermediate()} function from the \href{https://CRAN.R-project.org/package=geosphere}{\texttt{geosphere}} package\endnote{https://CRAN.R-project.org/package=geosphere} in R.
The sizes of the bubbles are proportional to the volume of work and can be compared across the different period panels.

\item Second, the \textbf{\textit{Top 30 Productive Institutions on the World Map}} displays the leading 30 institutions in terms of work production on the {World Map} for each discipline and the three respective periods: 1991--2000, 2001--2010 and 2011--2020.
The background world map employs the {\texttt{world}} data from the \href{https://CRAN.R-project.org/package=maps}{\texttt{maps}} package in R along with the \href{https://CRAN.R-project.org/package=ggplot2}{\texttt{ggplot2}} package\endnote{https://CRAN.R-project.org/package=ggplot2} in R. 
The sizes of the bubbles are proportional to the volume of work, standardised within each period panel, and cannot be compared across panels.

\item Third, the \textbf{\textit{Interregional Collaboration Matrix Diagram}} exhibits a half-matrix diagram at the country level for each discipline and the three respective periods: 1991--2000, 2001--2010 and 2011--2020.
It counts the number of bilateral coauthorship relationships represented on the {World Map}.
Each bubble's size (area) displayed in the matrix cell is proportional to the number of bilateral coauthorship relationships.
This edition particularly focuses on five pivotal parties: the US, China, EU27, the UK and Japan. 
These parties were specifically selected due to their substantial contributions to work production across all scientific fields from 1971 to 2020. 
These choices also align with the nations acclaimed as the \q{Big 5 science nations} (the US, China, Germany, the UK and Japan) in the \href{https://www.nature.com/articles/d41586-022-00569-7}{\textit{Nature Index}} \cite{NatureIndex22}.
Please note that the {Matrix Diagram} only takes into account the top 50 institutions in terms of work production for each period and discipline.
Therefore, if a cell shows zero (as small dots), it does not necessarily imply the absence of coauthorship relationships for the corresponding bilateral pair.

\item Forth, the \textbf{\textit{Interinstitutional Collaboration Dendrogram}} elucidates the development and evolution of interinstitutional research collaboration clusters spanning the last five decades. 
This is accomplished through hierarchical clustering analysis of institutions, considering the top 50 institutions in terms of work production across the four periods: 1971--1990, 1991--2000, 2001--2010 and 2011--2020.

\begin{itemize}[leftmargin=1em]
\item[--] The method used for hierarchical clustering analysis is the same as developed in Ref.~\cite{Okamura23}.
The distance between institutions X and Y is defined as the number of works with nationalities from both X and Y divided by the total number of works with nationalities from at least one of X and Y, subtracted from 1.
Hierarchical clustering analysis was performed on the distance matrix using the {\texttt{hclust}} function implemented in R with the {\texttt{ward.D2}} option (i.e.~the original Ward’s method) specified.

\item[--] The method of dendrogram visualisation is primarily derived from an example detailed on the \href{https://cran.r-project.org/web/packages/dendextend/vignettes/dendextend.html}{\texttt{dendextend} website}.\endnote{https://cran.r-project.org/web/packages/dendextend/vignettes/dendextend.html} 
Circular dendrograms were created using the \href{https://CRAN.R-project.org/package=dendextend}{\texttt{dendextend}}\endnote{https://CRAN.R-project.org/package=dendextend} and \href{https://CRAN.R-project.org/package=circlize}{\texttt{circlize}}\endnote{https://CRAN.R-project.org/package=circlize} packages in R. 
As one moves inward from the outer edge of the circle towards its centre, institutions or clusters of institutions that are in closer proximity to each other merge earlier.

\item[--] To indicate the country where the institutions are located, the country names are included at the beginning of the terms of research institutions, using the two-letter \href{https://en.wikipedia.org/wiki/ISO_3166-1_alpha-2}{ISO 3166-1 alpha-2} code.\endnote{https://en.wikipedia.org/wiki/ISO\_3166-1\_alpha-2}
The accompanied circularised bar graphs represent the number of works for the institutions involved.
\href{https://ror.org}{ROR}s are used as the canonical identifiers of the research institutions.
Readers of this booklet in PDF format can click on the ROR-based URL (\q{{https://ror.org/...}}) in the diagrams to view the corresponding ROR webpage from their browser.
\end{itemize}

\item Additionally, for each discipline and the respective periods of 1991--2000, 2001--2010 and 2011--2020, the top 100 institutions in terms of work production are displayed in tabular format, showing their respective country codes and production volumes.
If multiple research institutions have equal production volumes during each period, they are organised alphabetically by country codes and then by organisation names. 
Even if distinct rankings are shown, they lack significance and are treated as ties.

}

\guide{Important Notes}{%

\item It is worth reiterating that \hlt{the data from OpenAlex used to compile the \textit{Atlas of Science Collaboration}, even when incorporating bibliometric data related to past works, lacks consistent finality}.
As of the data acquisition for this version (August 2023), OpenAlex encompassed information regarding approximately 240 million works, with an additional influx of about 50,000 new data entries related to works being added daily.
Furthermore, for a substantial portion of these works, information regarding the corresponding institution to which the authors belong remains unknown.
As a result, \hlt{should the same analyses as those embedded within this booklet be replicated in the future,} although the qualitative extent of change remains uncertain, \hlt{it is undeniable that quantitatively distinct data will be acquired}.
Nonetheless, for individuals seeking an understanding of the global scope and evolution of international and interinstitutional collaborative research, the potential availability of this booklet or an enhanced, continuously updated evidence base holds inherent value. 

\item Further, it is worth reiterating that the term `works' encompasses a wide variety of scholarly publications. 
\hlt{The analyses conducted in the compilation of this booklet do not take into consideration whether these works are peer-reviewed articles or not, nor do they encompass considerations of their prominence, impact or quality}.
It is emphasised that \hlt{the primary intent behind the visualisations in this booklet is to quantitatively capture the momentum of scholarly knowledge production outputs from diverse research institutions, and to identify how productive institutions collaborate internationally and interinstitutionally}.
\hlt{Caution must be exercised, with acknowledgment that relying solely on the quantity of scholarly output produced by institutions falls short in encompassing discussions about their research potential, contributions to academia, or their relative superiority or inferiority}.
Further, it is recommended to consider the limitations discussed in Ref.~\cite{Okamura23} when using this booklet.
}

\guide{Miscellaneous}{%

\item It is important to note that some research institutions may encounter difficulties in accurately assessing the actual production volume at the institutional level within each analysis period due to challenges related to name disambiguation and the influence of historical organisational changes in bibliometric databases.

\item For the Interinstitutional Collaboration Dendrograms and the rankings of the top 100 productive institutions, entities like universities and R\&D institutions are primarily identified using the nomenclature employed in OpenAlex.
However, certain portions have been presented through abbreviations or acronyms, both for illustrative purposes and to effectively accommodate limited space. 
For instance, \q{University of} is abbreviated as \q{U.}, \q{Institution} and \q{Institute} as \q{Inst}, \q{National Laboratory} as \q{NL}, and \q{Science} and \q{Technology} as \q{Sci} and \q{Tech}, correspondingly, among others.
Should readers possess more fitting suggestions for abbreviations specific to particular organisations, or any other ideas aimed at enhancing the content of this booklet, the author would greatly appreciate their input.
}

{\small
\vspace{2em}
\noindent
\hrulefill{\quad List of URLs\quad }\hrulefill{}
\renewcommand*{\notesname}{}
\let\enotesize\scriptsize
\vspace{-4.5em}
\theendnotes
\markboth{\textsc{User Guide and Notes}}{}
}
}

\renewcommand{\sectionmark}[1]{\markboth{\thesection.~~#1}{}}
\newlength{\outerbordwidth}
\setlength{\outerbordwidth}{2pt}


\afterpage{\clearpage%
\pagenumbering{arabic}
}

%

\titleformat{\section}{\sc\centering\LARGE\bfseries}{\textsc{\thesection}.\!\!}{1em}{}

\afterpage{\clearpage%
\markboth{\textbf \textsc{\ai}}{}
\thispagestyle{empty}
\quad
\vspace{2cm}
\begin{center}
\pgfornament[width=0.5*\textwidth,symmetry=h]{89}\\[2em]
\section{\ai}
\vspace{1em}
\pgfornament[width=0.5*\textwidth]{89}
\end{center}
}

\afterpage{\clearpage%

\begin{figure}[!tp]
\centering
\vspace{-1em}
{\large \textbf{\textrm{{World Map of \textcolor{violet}{\textit{\ai}} Collaboration}}}~|~1971--2020}\\
\vspace{0.3cm}
\includegraphics[align=c, scale=0.054, trim={9.5cm 0 9.5cm 0},clip]{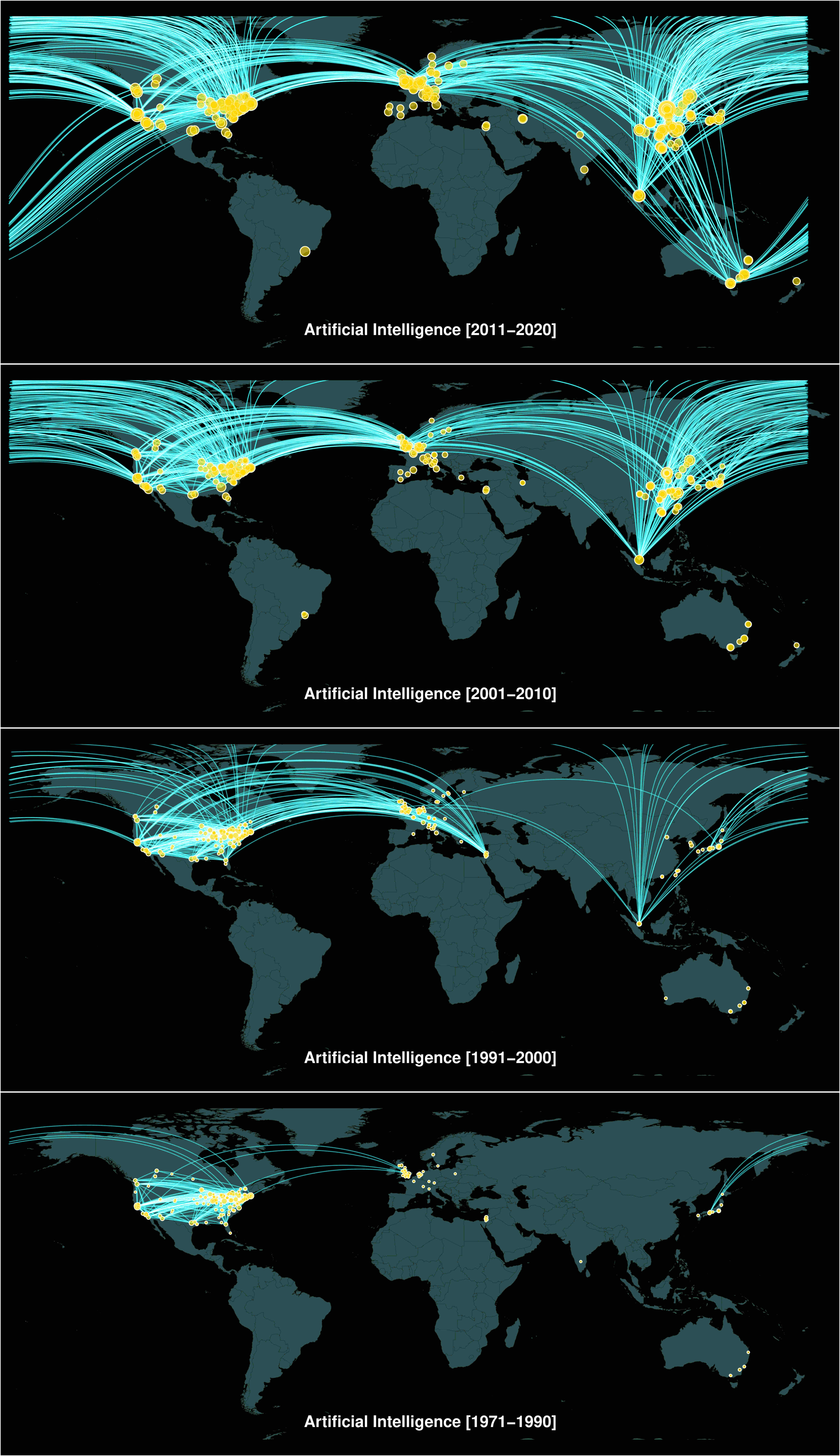}
\caption[{\ai}]{\textbf{(a)~|~The World Map of \textcolor{violet}{\textit{\ai}} Collaboration.}
The bubbles represent the top 199 institutions in terms of work production, with their sizes proportional to the work volume. 
The connecting lines depict coauthorship relationships among the top 50 institutions.}
\label{fig:wmap_ai}
\end{figure}
}
\afterpage{\clearpage%
\begin{figure}[!tp]\ContinuedFloat
\centering
\vspace{-1em}
{\large \textbf{Top 30 Productive Institutions on the World Map: \textcolor{violet}{\textit{\ai}}}~|~1991--2020}\\
\vspace{-0em}
\hspace*{-3em}                                                           
\includegraphics[align=c, scale=0.83]{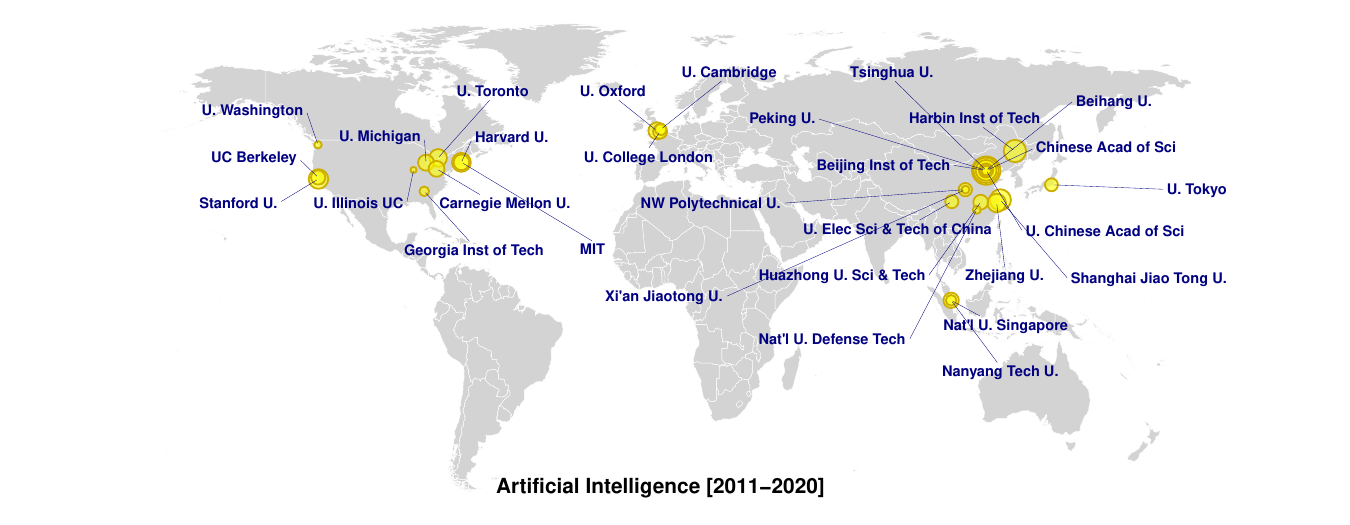}\\[-0.5em]
\quad\\[-1em]
\dotfill 
\quad\\[-0em]
\hspace*{-3em}                                                           
\includegraphics[align=c, scale=0.83]{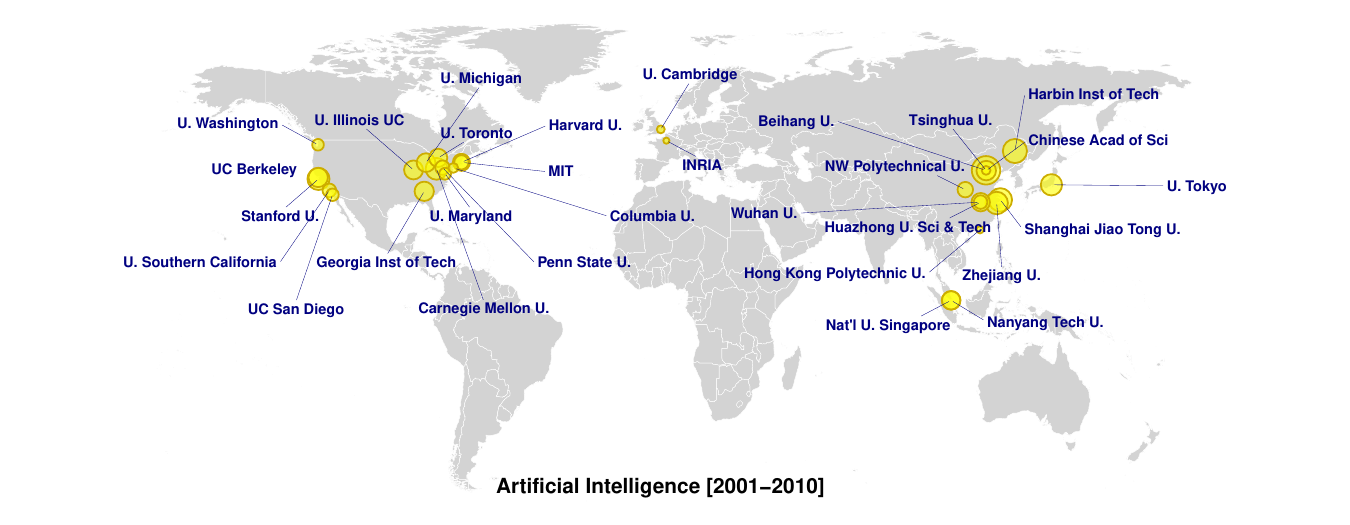}\\[-0.5em]
\quad\\[-1em]
\dotfill 
\quad\\[-0em]
\hspace*{-3em}                                                           
\includegraphics[align=c, scale=0.83]{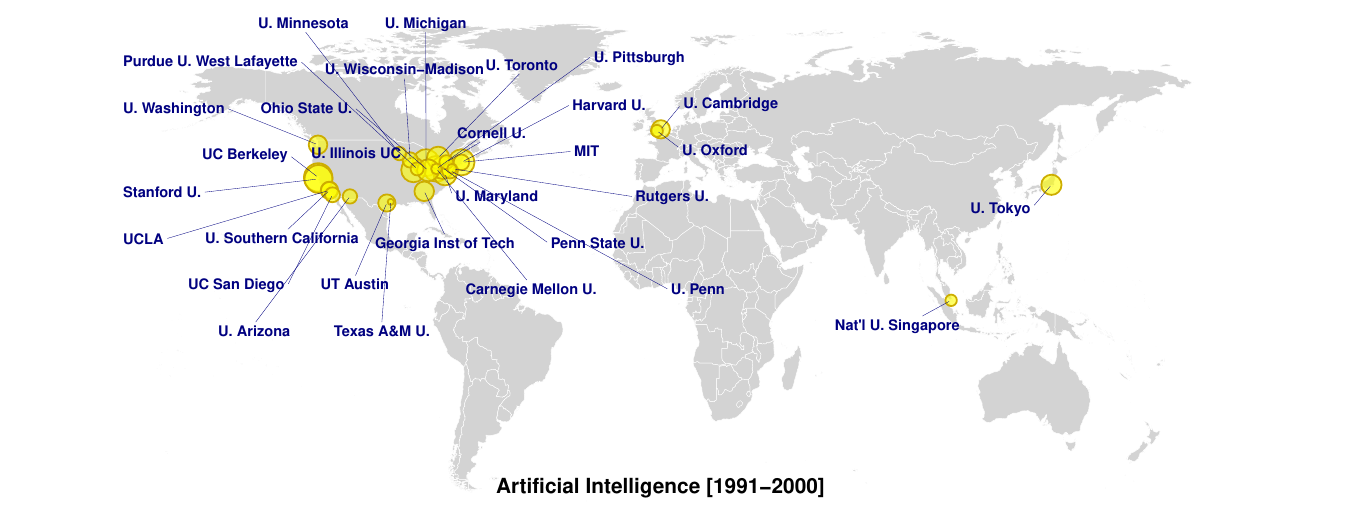}\\[-0.8em]
\caption[{\ai}]{\textbf{(b)~|~The top 30 productive institutions on the World Map: \textcolor{violet}{\textit{\ai}}.}
The bubbles represent the top 30 institutions in terms of work production, with their sizes proportional to the work volume.}
\label{fig:wmap_topinst_ai}
\end{figure}
}
\afterpage{\clearpage%
\begin{figure}[!tp]\ContinuedFloat
\centering
\vspace{-1em}
{\large \textbf{\textrm{{Interregional \textcolor{violet}{\textit{\ai}} Collaboration}}}~|~1991--2020}\\
\vspace{0.5em}
\hspace{-5em}\includegraphics[align=c, scale=1.7, vmargin=0mm]{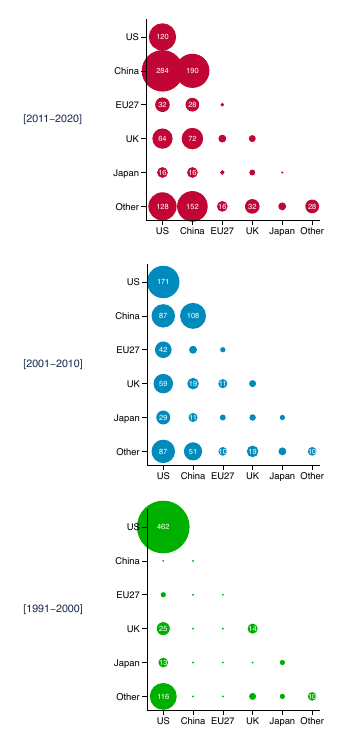}
\vspace{-1em}
\caption[]{\textbf{(c)~|~The Interregional \textcolor{violet}{\textit{\ai}} Collaboration Matrix Diagram.}
The bubble size represents the number of coauthorship relationships for the top 50 institutions in terms of work production. 
If the number is equal to or greater than 10, it is displayed inside the bubble.}
\label{fig:halfmat_ai}
\end{figure}
}
\afterpage{\clearpage%
\begin{figure}[!tp]\ContinuedFloat
\centering
\vspace{-1em}
{\large \textbf{\textrm{Interinstitutional \textcolor{violet}{\textit{\ai}} Collaboration}}~|~2001--2020\quad {\footnotesize \emph{(continued to next page)}}}\\
\cdend{ai}{2010}{2000}{2011--2020}{2001--2010}\\[-1.5em]
\caption[]{\textbf{(d)~|~The Interinstitutional \textcolor{violet}{\textit{\ai}} Collaboration Dendrogram.}
The top 50 institutions in terms of work production, indicated by the circularised bar graphs, are displayed.
}
\label{fig:cdend1_ai}
\end{figure}
}
\afterpage{\clearpage%
\begin{figure}[!tp]\ContinuedFloat
\centering
\vspace{-1em}
{\large \textbf{\textrm{Interinstitutional \textcolor{violet}{\textit{\ai}} Collaboration}}~|~1971--2000\quad {\footnotesize \emph{(continued from previous page)}}}\\
\cdend{ai}{1990}{1980}{1991--2000}{1971--1990}\\[-1.5em]
\caption[]{\textbf{(d)~|~The Interinstitutional \textcolor{violet}{\textit{\ai}} Collaboration Dendrogram.} \emph{(Cont.)}\hfill~}
\label{fig:cdend2_ai}
\end{figure}
}
\afterpage{\clearpage%
\begin{landscape}
\begin{table}[!t]
\vspace{-0.5em}
\caption{\textbf{The top 100 productive institutions: \textcolor{violet}{\textit{\ai}}.}}
\label{tab:r1_ai}
\vspace{2em}
\centering
{\tiny
{\renewcommand{\arraystretch}{1.2}
\begin{tabular}{rp{5cm}lr@{\hspace{4em}}p{5cm}lr@{\hspace{4em}}p{5cm}lr}\\[-5em] \toprule[1pt] \\[-1.4em]    
 & {\scriptsize \textbf{1991--2000}} & \multicolumn{2}{c}{No.~Works} & {\scriptsize \textbf{2001--2010}} & \multicolumn{2}{c}{No.~Works} & {\scriptsize \textbf{2011--2020}} & \multicolumn{2}{r}{No.~Works} \\[-0.2em] \cmidrule[0.5pt](lr{4em}){2-4} \cmidrule[0.5pt](l{-0em}r{4em}){5-7} \cmidrule[0.5pt](l{-0em}r{1em}){8-10}
\textit{1} & University of California, Berkeley & US & 4,734 & Tsinghua University & CN & 16,391 & Chinese Academy of Sciences & CN & 36,084 \\  
\textit{2} & Stanford University & US & 4,607 & Harbin Institute of Technology & CN & 12,701 & Tsinghua University & CN & 29,571 \\  
\textit{3} & University of Michigan{\textendash}Ann Arbor & US & 4,348 & Shanghai Jiao Tong University & CN & 12,218 & Harbin Institute of Technology & CN & 24,414 \\  
\textit{4} & Massachusetts Institute of Technology & US & 4,210 & Zhejiang University & CN & 11,593 & Shanghai Jiao Tong University & CN & 22,843 \\  
\textit{5} & Carnegie Mellon University & US & 4,129 & Stanford University & US & 11,390 & Beihang University & CN & 22,449 \\  
\textit{6} & University of Illinois Urbana-Champaign & US & 3,901 & Carnegie Mellon University & US & 11,186 & Stanford University & US & 20,870 \\  
\textit{7} & University of Maryland, College Park & US & 3,661 & The University of Tokyo & JP & 10,888 & Harvard University & US & 20,113 \\  
\textit{8} & Pennsylvania State University & US & 3,541 & Chinese Academy of Sciences & CN & 9,980 & Zhejiang University & CN & 19,766 \\  
\textit{9} & University of Toronto & CA & 3,520 & University of California, Berkeley & US & 9,898 & University of Toronto & CA & 18,885 \\  
\textit{10} & The Ohio State University & US & 3,450 & Georgia Institute of Technology & US & 9,882 & University of Oxford & GB & 18,412 \\  
\hdashline          
\textit{11} & The University of Tokyo & JP & 3,208 & University of Illinois Urbana-Champaign & US & 9,729 & University of Michigan{\textendash}Ann Arbor & US & 17,543 \\  
\textit{12} & Georgia Institute of Technology & US & 3,122 & University of Michigan{\textendash}Ann Arbor & US & 9,618 & Carnegie Mellon University & US & 17,449 \\  
\textit{13} & University of Washington & US & 3,006 & Nanyang Technological University & SG & 9,602 & Massachusetts Institute of Technology & US & 17,309 \\  
\textit{14} & University of Cambridge & GB & 2,982 & Huazhong University of Science and Technology & CN & 9,482 & Nanyang Technological University & SG & 17,125 \\  
\textit{15} & University of California, Los Angeles & US & 2,876 & National University of Singapore & SG & 9,464 & Huazhong University of Science and Technology & CN & 16,712 \\  
\textit{16} & The University of Texas at Austin & US & 2,866 & University of Toronto & CA & 9,415 & Beijing Institute of Technology & CN & 16,383 \\  
\textit{17} & University of Southern California & US & 2,833 & Massachusetts Institute of Technology & US & 8,975 & University of California, Berkeley & US & 16,366 \\  
\textit{18} & University of Pennsylvania & US & 2,752 & Northwestern Polytechnical University & CN & 8,417 & University College London & GB & 16,243 \\  
\textit{19} & Harvard University & US & 2,727 & Pennsylvania State University & US & 8,129 & Northwestern Polytechnical University & CN & 15,893 \\  
\textit{20} & University of Wisconsin{\textendash}Madison & US & 2,697 & Harvard University & US & 8,087 & University of Electronic Science and Technology of China & CN & 15,805 \\  
\hdashline          
\textit{21} & University of California, San Diego & US & 2,671 & University of Southern California & US & 7,923 & The University of Tokyo & JP & 15,665 \\  
\textit{22} & University of Arizona & US & 2,631 & Wuhan University & CN & 7,919 & Peking University & CN & 15,180 \\  
\textit{23} & University of Minnesota & US & 2,570 & University of California, San Diego & US & 7,678 & University of Cambridge & GB & 15,013 \\  
\textit{24} & Cornell University & US & 2,551 & University of Maryland, College Park & US & 7,563 & National University of Singapore & SG & 14,757 \\  
\textit{25} & Purdue University West Lafayette & US & 2,547 & University of Washington & US & 7,524 & University of Chinese Academy of Sciences & CN & 14,486 \\  
\textit{26} & University of Oxford & GB & 2,491 & Columbia University & US & 7,274 & Georgia Institute of Technology & US & 14,466 \\  
\textit{27} & National University of Singapore & SG & 2,468 & Beihang University & CN & 7,173 & National University of Defense Technology & CN & 14,185 \\  
\textit{28} & University of Pittsburgh & US & 2,444 & University of Cambridge & GB & 7,110 & Xi'an Jiaotong University & CN & 14,163 \\  
\textit{29} & Rutgers, The State University of New Jersey & US & 2,410 & Hong Kong Polytechnic University & CN & 7,086 & University of Washington & US & 14,110 \\  
\textit{30} & Texas A\&M University & US & 2,357 & \scalebox{0.9}[1]{French Institute for Research in Computer Science and Automation} & FR & 7,060 & University of Illinois Urbana-Champaign & US & 14,064 \\  
\hdashline          
\textit{31} & Columbia University & US & 2,352 & The Ohio State University & US & 6,900 & Imperial College London & GB & 13,928 \\  
\textit{32} & McGill University & CA & 2,350 & University of British Columbia & CA & 6,863 & Southeast University & CN & 13,744 \\  
\textit{33} & AT\&T (United States) & US & 2,297 & University of California, Los Angeles & US & 6,837 & University of California, San Diego & US & 13,720 \\  
\textit{34} & Yale University & US & 2,259 & The University of Texas at Austin & US & 6,834 & ETH Zurich & CH & 13,581 \\  
\textit{35} & Boston University & US & 2,224 & Xi'an Jiaotong University & CN & 6,829 & Wuhan University & CN & 13,376 \\  
\textit{36} & University of Manchester & GB & 2,221 & University College London & GB & 6,822 & KU Leuven & BE & 13,307 \\  
\textit{37} & University of British Columbia & CA & 2,196 & Delft University of Technology & NL & 6,713 & University of Pennsylvania & US & 12,942 \\  
\textit{38} & Virginia Tech & US & 2,176 & Osaka University & JP & 6,656 & UNSW Sydney & AU & 12,771 \\  
\textit{39} & Johns Hopkins University & US & 2,152 & University of Pennsylvania & US & 6,618 & University of Southern California & US & 12,756 \\  
\textit{40} & Technion {\textendash} Israel Institute of Technology & IL & 2,117 & University of Oxford & GB & 6,608 & Columbia University & US & 12,605 \\  
\hdashline          
\textit{41} & Osaka University & JP & 2,072 & University of Minnesota & US & 6,477 & University of California, Los Angeles & US & 12,603 \\  
\textit{42} & Delft University of Technology & NL & 2,044 & Tianjin University & CN & 6,351 & Johns Hopkins University & US & 12,558 \\  
\textit{43} & Arizona State University & US & 2,042 & KU Leuven & BE & 6,340 & Technical University of Munich & DE & 12,400 \\  
\textit{44} & University College London & GB & 2,028 & Northeastern University & CN & 6,313 & Beijing University of Posts and Telecommunications & CN & 12,382 \\  
\textit{45} & University of Florida & US & 2,011 & University of Manchester & GB & 6,310 & Nanjing University of Aeronautics and Astronautics & CN & 12,362 \\  
\textit{46} & Tel Aviv University & IL & 1,999 & Tokyo Institute of Technology & JP & 6,287 & Tianjin University & CN & 12,358 \\  
\textit{47} & University of Edinburgh & GB & 1,997 & National University of Defense Technology & CN & 6,232 & University of British Columbia & CA & 12,152 \\  
\textit{48} & Princeton University & US & 1,970 & University of Alberta & CA & 6,174 & University of Sydney & AU & 12,117 \\  
\textit{49} & University of Sheffield & GB & 1,954 & South China University of Technology & CN & 6,133 & University of Melbourne & AU & 12,091 \\  
\textit{50} & Kyoto University & JP & 1,935 & Chinese University of Hong Kong & CN & 6,045 & Pennsylvania State University & US & 12,031 \\  
          
 \\[-1.4em]
\hdashline \\[-1em]
\multicolumn{10}{r}{\scriptsize \emph{(continued to next page)}}
\end{tabular}}
}
\end{table}
\end{landscape}
}
\afterpage{\clearpage%
\begin{landscape}
\begin{table}[!t]\ContinuedFloat
\vspace{-3.3em}
\caption{\textbf{The top 100 productive institutions: \textcolor{violet}{\textit{\ai}}.} \emph{(Cont.)}}
\label{tab:r2_ai}
\vspace{2em}
{\tiny
{\renewcommand{\arraystretch}{1.2}
\begin{tabular}{rp{5cm}lr@{\hspace{4em}}p{5cm}lr@{\hspace{4em}}p{5cm}lr}\\[-5em] \toprule[1pt] \\[-1.4em]    
 & {\scriptsize \textbf{1991--2000}} & \multicolumn{2}{c}{No.\ Works} & {\scriptsize \textbf{2001--2010}} & \multicolumn{2}{c}{No.\ Works} & {\scriptsize \textbf{2011--2020}} & \multicolumn{2}{r}{No.\ Works} \\[-0.2em] \cmidrule[0.5pt](lr{4em}){2-4} \cmidrule[0.5pt](l{-0em}r{4em}){5-7} \cmidrule[0.5pt](l{-0em}r{1em}){8-10}
\textit{51} & \'{E}cole Polytechnique F\'{e}d\'{e}rale de Lausanne & CH & 1,934 & Beijing Institute of Technology & CN & 6,012 & Xidian University & CN & 11,994 \\  
\textit{52} & California Institute of Technology & US & 1,889 & Imperial College London & GB & 5,915 & Delft University of Technology & NL & 11,634 \\  
\textit{53} & KU Leuven & BE & 1,866 & McGill University & CA & 5,818 & The University of Texas at Austin & US & 11,558 \\  
\textit{54} & Tokyo Institute of Technology & JP & 1,840 & \'{E}cole Polytechnique F\'{e}d\'{e}rale de Lausanne & CH & 5,804 & McGill University & CA & 11,382 \\  
\textit{55} & Korea Advanced Institute of Science and Technology & KR & 1,819 & University of Waterloo & CA & 5,778 & New York University & US & 11,334 \\  
\textit{56} & University of Amsterdam & NL & 1,807 & Cornell University & US & 5,773 & South China University of Technology & CN & 11,254 \\  
\textit{57} & Sapienza University of Rome & IT & 1,797 & University of Melbourne & AU & 5,757 & Universidade de S\~{a}o Paulo & BR & 11,242 \\  
\textit{58} & University of Chicago & US & 1,791 & Kyoto University & JP & 5,748 & University of Science and Technology of China & CN & 11,227 \\  
\textit{59} & University of Iowa & US & 1,786 & Texas A\&M University & US & 5,739 & North China Electric Power University & CN & 10,848 \\  
\textit{60} & University of Waterloo & CA & 1,782 & University of Florida & US & 5,680 & Chinese University of Hong Kong & CN & 10,812 \\  
\hdashline          
\textit{61} & New York University & US & 1,745 & UNSW Sydney & AU & 5,657 & King's College London & GB & 10,706 \\  
\textit{62} & Imperial College London & GB & 1,740 & Southeast University & CN & 5,653 & Tongji University & CN & 10,626 \\  
\textit{63} & \scalebox{0.9}[1]{French Institute for Research in Computer Science and Automation} & FR & 1,735 & Peking University & CN & 5,636 & Dalian University of Technology & CN & 10,552 \\  
\textit{64} & University of California, Davis & US & 1,715 & Johns Hopkins University & US & 5,602 & Hong Kong Polytechnic University & CN & 10,526 \\  
\textit{65} & Duke University & US & 1,712 & University of Wisconsin{\textendash}Madison & US & 5,572 & University of Maryland, College Park & US & 10,505 \\  
\textit{66} & University of California, Santa Barbara & US & 1,704 & City University of Hong Kong & CN & 5,556 & The Ohio State University & US & 10,473 \\  
\textit{67} & Jet Propulsion Laboratory & US & 1,691 & Seoul National University & KR & 5,545 & Northeastern University & CN & 10,443 \\  
\textit{68} & University of Sydney & AU & 1,690 & University of Pittsburgh & US & 5,506 & University of Edinburgh & GB & 10,441 \\  
\textit{69} & University of California, Irvine & US & 1,683 & North China Electric Power University & CN & 5,466 & Boston University & US & 10,436 \\  
\textit{70} & Michigan State University & US & 1,611 & Arizona State University & US & 5,365 & University of Wisconsin{\textendash}Madison & US & 10,359 \\  
\hdashline          
\textit{71} & University of North Carolina at Chapel Hill & US & 1,601 & Nanjing University of Aeronautics and Astronautics & CN & 5,233 & Arizona State University & US & 10,245 \\  
\textit{72} & Iowa State University & US & 1,595 & Virginia Tech & US & 5,230 & Chongqing University & CN & 10,113 \\  
\textit{73} & UNSW Sydney & AU & 1,592 & Chongqing University & CN & 5,192 & University of Waterloo & CA & 10,026 \\  
\textit{74} & Nanyang Technological University & SG & 1,588 & University of Sydney & AU & 5,138 & Beijing Jiaotong University & CN & 10,022 \\  
\textit{75} & State University of New York & US & 1,567 & Purdue University West Lafayette & US & 5,123 & Seoul National University & KR & 10,013 \\  
\textit{76} & University of Virginia & US & 1,566 & Xidian University & CN & 5,101 & Nanjing University of Science and Technology & CN & 10,000 \\  
\textit{77} & University of Melbourne & AU & 1,554 & Boston University & US & 5,076 & University of Minnesota & US & 9,834 \\  
\textit{78} & University of Wales & GB & 1,553 & National Taiwan University & TW & 5,075 & University of Alberta & CA & 9,803 \\  
\textit{79} & University of Alberta & CA & 1,542 & Xiaomi (China) & CN & 5,069 & Yale University & US & 9,801 \\  
\textit{80} & Northwestern University & US & 1,526 & Universitat Polit\`{e}cnica de Catalunya & ES & 5,041 & \'{E}cole Polytechnique F\'{e}d\'{e}rale de Lausanne & CH & 9,707 \\  
\hdashline          
\textit{81} & Ames Research Center & US & 1,496 & Yale University & US & 5,016 & Duke University & US & 9,627 \\  
\textit{82} & Tsinghua University & CN & 1,479 & Central South University & CN & 4,970 & University of Manchester & GB & 9,598 \\  
\textit{83} & North Carolina State University & US & 1,461 & University of Electronic Science and Technology of China & CN & 4,939 & University of Pittsburgh & US & 9,522 \\  
\textit{84} & Tohoku University & JP & 1,440 & National Yang Ming Chiao Tung University & TW & 4,913 & Shandong University & CN & 9,383 \\  
\textit{85} & Case Western Reserve University & US & 1,433 & University of Edinburgh & GB & 4,908 & University of Tehran & IR & 9,342 \\  
\textit{86} & University of Southampton & GB & 1,423 & Korea Advanced Institute of Science and Technology & KR & 4,864 & Korea Advanced Institute of Science and Technology & KR & 9,286 \\  
\textit{87} & Washington University in St. Louis & US & 1,408 & University of Arizona & US & 4,861 & University of Amsterdam & NL & 9,279 \\  
\textit{88} & Langley Research Center & US & 1,389 & New York University & US & 4,825 & Monash University & AU & 9,244 \\  
\textit{89} & National Yang Ming Chiao Tung University & TW & 1,385 & Universidade de S\~{a}o Paulo & BR & 4,808 & Google (United States) & US & 9,108 \\  
\textit{90} & University of Massachusetts Amherst & US & 1,383 & University of Southampton & GB & 4,808 & Cornell University & US & 9,068 \\  
\hdashline          
\textit{91} & Australian National University & AU & 1,375 & Rutgers, The State University of New Jersey & US & 4,792 & Sapienza University of Rome & IT & 9,067 \\  
\textit{92} & Nagoya University & JP & 1,375 & Tohoku University & JP & 4,738 & University of Queensland & AU & 9,019 \\  
\textit{93} & University of Queensland & AU & 1,371 & University of Sheffield & GB & 4,725 & Politecnico di Milano & IT & 9,013 \\  
\textit{94} & University of Colorado Boulder & US & 1,361 & Jilin University & CN & 4,715 & University of Florida & US & 8,917 \\  
\textit{95} & University of Illinois at Chicago & US & 1,360 & Dalian University of Technology & CN & 4,697 & Central South University & CN & 8,774 \\  
\textit{96} & Seoul National University & KR & 1,332 & Michigan State University & US & 4,693 & Texas A\&M University & US & 8,704 \\  
\textit{97} & University of Rochester & US & 1,331 & Beijing Jiaotong University & CN & 4,681 & Inserm & FR & 8,650 \\  
\textit{98} & Rensselaer Polytechnic Institute & US & 1,324 & University of Science and Technology of China & CN & 4,669 & \scalebox{0.9}[1]{French Institute for Research in Computer Science and Automation} & FR & 8,553 \\  
\textit{99} & University of Leeds & GB & 1,318 & Duke University & US & 4,665 & Jilin University & CN & 8,505 \\  
\textit{100} & Hitachi (Japan) & JP & 1,309 & Princeton University & US & 4,660 & Osaka University & JP & 8,426 \\  
          
 \\[-1.4em]
\bottomrule
\end{tabular}}
}
\end{table}
\end{landscape}
}
%

\titleformat{\section}{\sc\centering\LARGE\bfseries}{\textsc{\thesection}.\!\!}{1em}{}

\afterpage{\clearpage%
\markboth{\textbf \textsc{\quantum}}{}
\thispagestyle{empty}
\quad
\vspace{2cm}
\begin{center}
\pgfornament[width=0.5*\textwidth,symmetry=h]{89}\\[2em]
\section{\quantum}
\vspace{1em}
\pgfornament[width=0.5*\textwidth]{89}
\end{center}
}

\afterpage{\clearpage%

\begin{figure}[!tp]
\centering
\vspace{-1em}
{\large \textbf{\textrm{{World Map of \textcolor{violet}{\textit{\quantum}} Collaboration}}}~|~1971--2020}\\
\vspace{0.3cm}
\includegraphics[align=c, scale=0.054, trim={9.5cm 0 9.5cm 0},clip]{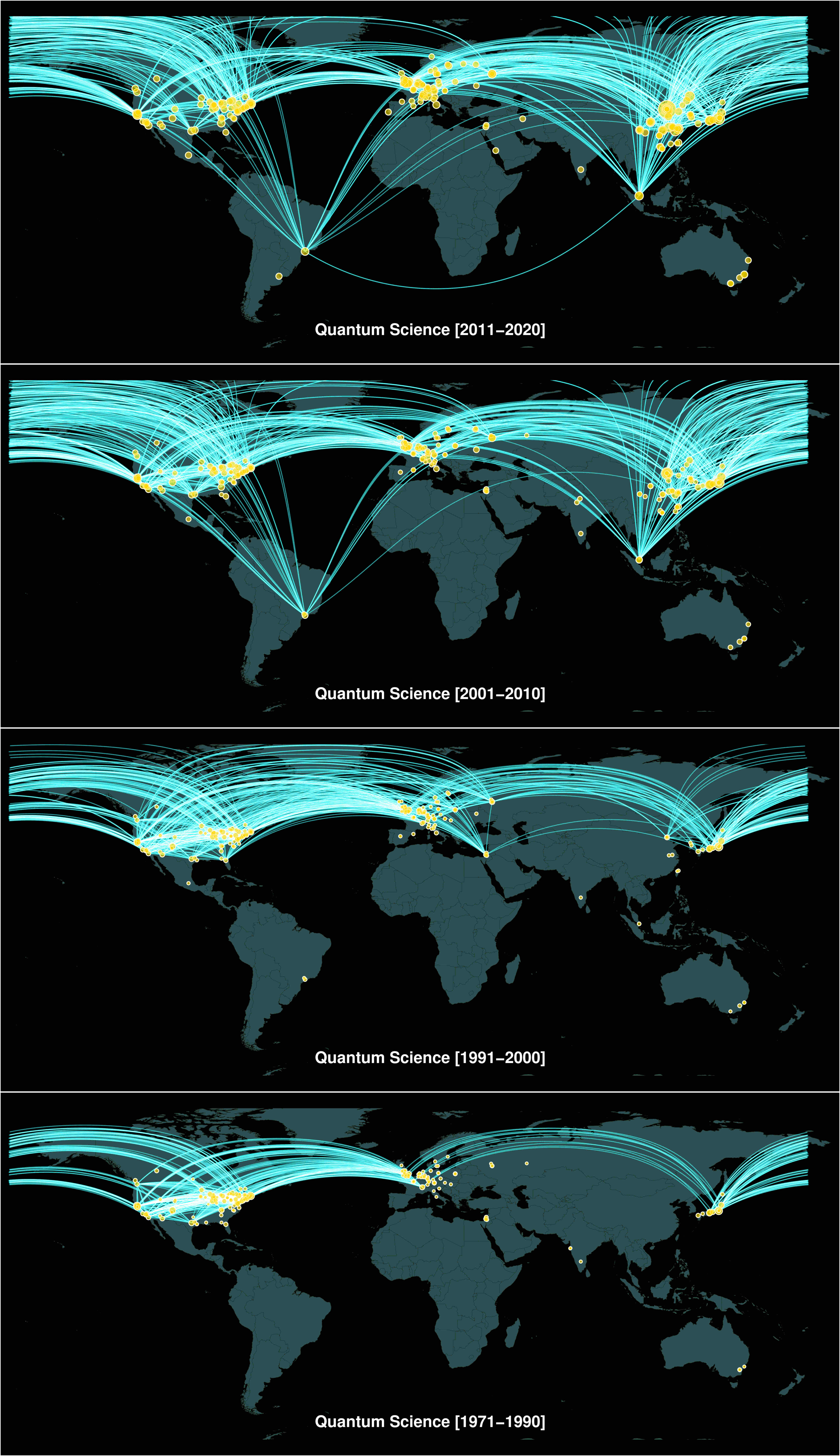}
\caption[{\quantum}]{\textbf{(a)~|~The World Map of \textcolor{violet}{\textit{\quantum}} Collaboration.}
The bubbles represent the top 199 institutions in terms of work production, with their sizes proportional to the work volume. 
The connecting lines depict coauthorship relationships among the top 50 institutions.}
\label{fig:wmap_quantum}
\end{figure}
}
\afterpage{\clearpage%
\begin{figure}[!tp]\ContinuedFloat
\centering
\vspace{-1em}
{\large \textbf{Top 30 Productive Institutions on the World Map: \textcolor{violet}{\textit{\quantum}}}~|~1991--2020}\\
\vspace{-0em}
\hspace*{-3em}                                                           
\includegraphics[align=c, scale=0.83]{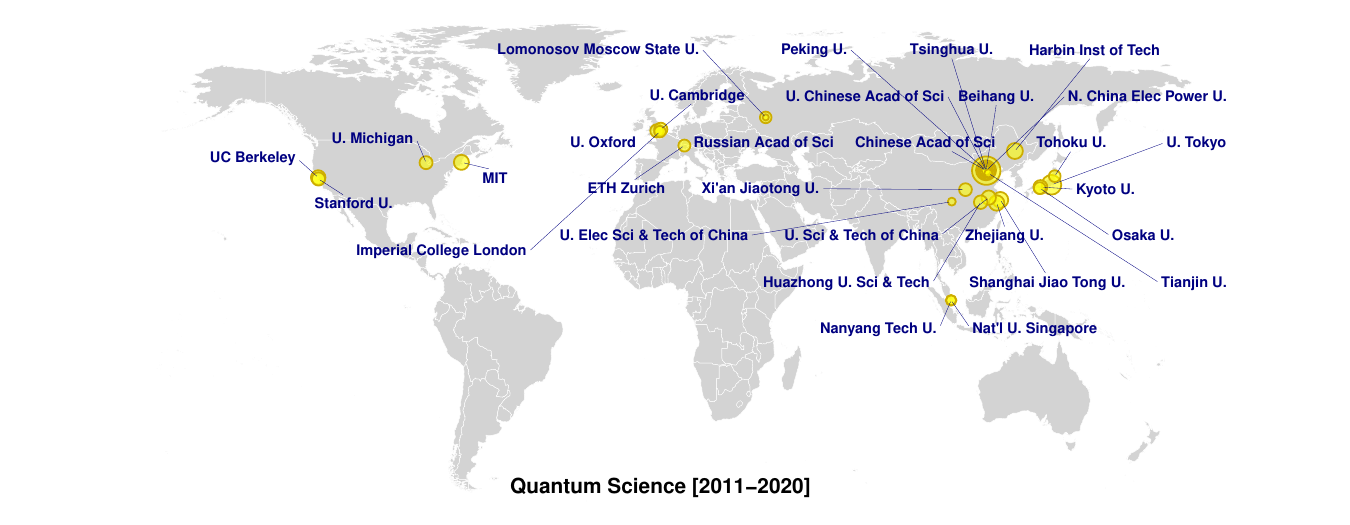}\\[-0.5em]
\quad\\[-1em]
\dotfill 
\quad\\[-0em]
\hspace*{-3em}                                                           
\includegraphics[align=c, scale=0.83]{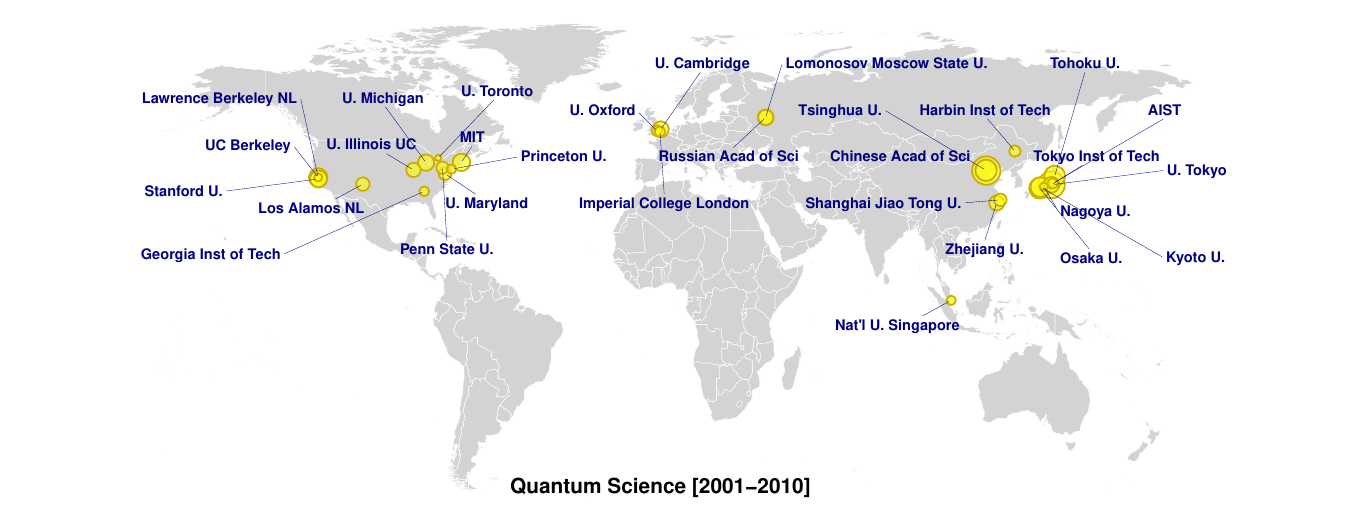}\\[-0.5em]
\quad\\[-1em]
\dotfill 
\quad\\[-0em]
\hspace*{-3em}                                                           
\includegraphics[align=c, scale=0.83]{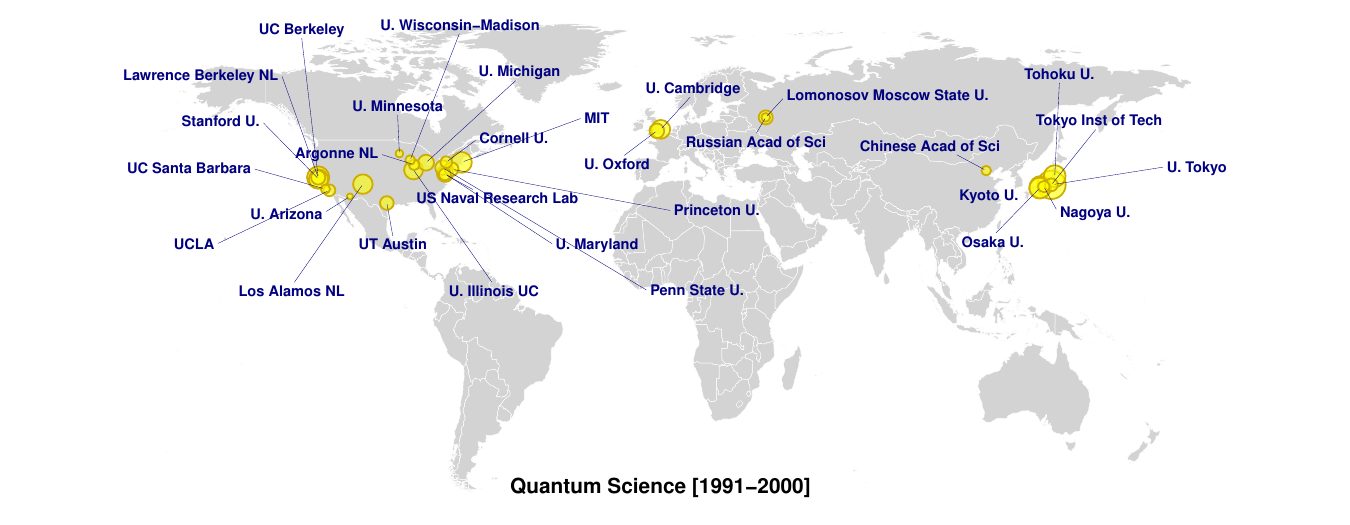}\\[-0.8em]
\caption[{\quantum}]{\textbf{(b)~|~The top 30 productive institutions on the World Map: \textcolor{violet}{\textit{\quantum}}.}
The bubbles represent the top 30 institutions in terms of work production, with their sizes proportional to the work volume.}
\label{fig:wmap_topinst_quantum}
\end{figure}
}
\afterpage{\clearpage%
\begin{figure}[!tp]\ContinuedFloat
\centering
\vspace{-1em}
{\large \textbf{\textrm{{Interregional \textcolor{violet}{\textit{\quantum}} Collaboration}}}~|~1991--2020}\\
\vspace{0.5em}
\hspace{-5em}\includegraphics[align=c, scale=1.7, vmargin=0mm]{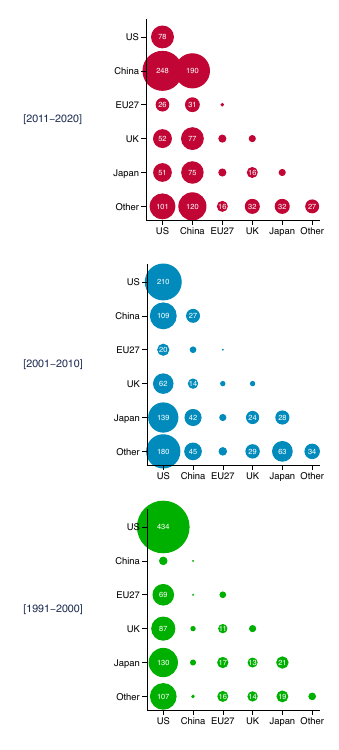}
\vspace{-1em}
\caption[]{\textbf{(c)~|~The Interregional \textcolor{violet}{\textit{\quantum}} Collaboration Matrix Diagram.}
The bubble size represents the number of coauthorship relationships for the top 50 institutions in terms of work production. 
If the number is equal to or greater than 10, it is displayed inside the bubble.}
\label{fig:halfmat_quantum}
\end{figure}
}
\afterpage{\clearpage%
\begin{figure}[!tp]\ContinuedFloat
\centering
\vspace{-1em}
{\large \textbf{\textrm{Interinstitutional \textcolor{violet}{\textit{\quantum}} Collaboration}}~|~2001--2020\quad {\footnotesize \emph{(continued to next page)}}}\\
\cdend{quantum}{2010}{2000}{2011--2020}{2001--2010}\\[-1.5em]
\caption[]{\textbf{(d)~|~The Interinstitutional \textcolor{violet}{\textit{\quantum}} Collaboration Dendrogram.}
The top 50 institutions in terms of work production, indicated by the circularised bar graphs, are displayed.
}
\label{fig:cdend1_quantum}
\end{figure}
}
\afterpage{\clearpage%
\begin{figure}[!tp]\ContinuedFloat
\centering
\vspace{-1em}
{\large \textbf{\textrm{Interinstitutional \textcolor{violet}{\textit{\quantum}} Collaboration}}~|~1971--2000\quad {\footnotesize \emph{(continued from previous page)}}}\\
\cdend{quantum}{1990}{1980}{1991--2000}{1971--1990}\\[-1.5em]
\caption[]{\textbf{(d)~|~The Interinstitutional \textcolor{violet}{\textit{\quantum}} Collaboration Dendrogram.} \emph{(Cont.)}\hfill~}
\label{fig:cdend2_quantum}
\end{figure}
}
\afterpage{\clearpage%
\begin{landscape}
\begin{table}[!t]
\vspace{-0.5em}
\caption{\textbf{The top 100 productive institutions: \textcolor{violet}{\textit{\quantum}}.}}
\label{tab:r1_quantum}
\vspace{2em}
\centering
{\tiny
{\renewcommand{\arraystretch}{1.2}
\begin{tabular}{rp{5cm}lr@{\hspace{4em}}p{5cm}lr@{\hspace{4em}}p{5cm}lr}\\[-5em] \toprule[1pt] \\[-1.4em]    
 & {\scriptsize \textbf{1991--2000}} & \multicolumn{2}{c}{No.~Works} & {\scriptsize \textbf{2001--2010}} & \multicolumn{2}{c}{No.~Works} & {\scriptsize \textbf{2011--2020}} & \multicolumn{2}{r}{No.~Works} \\[-0.2em] \cmidrule[0.5pt](lr{4em}){2-4} \cmidrule[0.5pt](l{-0em}r{4em}){5-7} \cmidrule[0.5pt](l{-0em}r{1em}){8-10}
\textit{1} & The University of Tokyo & JP & 13,688 & Chinese Academy of Sciences & CN & 27,417 & Chinese Academy of Sciences & CN & 66,552 \\  
\textit{2} & University of California, Berkeley & US & 8,849 & The University of Tokyo & JP & 24,020 & Tsinghua University & CN & 29,227 \\  
\textit{3} & Kyoto University & JP & 8,592 & Tsinghua University & CN & 16,092 & The University of Tokyo & JP & 28,241 \\  
\textit{4} & Tohoku University & JP & 8,554 & Osaka University & JP & 15,442 & University of Chinese Academy of Sciences & CN & 22,786 \\  
\textit{5} & Osaka University & JP & 8,106 & Tohoku University & JP & 15,412 & Harbin Institute of Technology & CN & 20,551 \\  
\textit{6} & Massachusetts Institute of Technology & US & 8,058 & Kyoto University & JP & 14,430 & Shanghai Jiao Tong University & CN & 20,248 \\  
\textit{7} & Los Alamos National Laboratory & US & 7,420 & University of California, Berkeley & US & 13,128 & Zhejiang University & CN & 19,643 \\  
\textit{8} & Stanford University & US & 7,237 & Massachusetts Institute of Technology & US & 12,402 & Massachusetts Institute of Technology & US & 19,161 \\  
\textit{9} & University of Illinois Urbana-Champaign & US & 7,151 & Stanford University & US & 12,086 & University of California, Berkeley & US & 18,428 \\  
\textit{10} & University of Cambridge & GB & 7,115 & University of Michigan{\textendash}Ann Arbor & US & 11,562 & University of Science and Technology of China & CN & 18,236 \\  
\hdashline          
\textit{11} & University of Maryland, College Park & US & 6,191 & University of Cambridge & GB & 11,360 & Peking University & CN & 18,082 \\  
\textit{12} & University of Michigan{\textendash}Ann Arbor & US & 6,092 & Lomonosov Moscow State University & RU & 11,219 & Kyoto University & JP & 17,655 \\  
\textit{13} & University of Oxford & GB & 5,744 & Tokyo Institute of Technology & JP & 11,207 & Huazhong University of Science and Technology & CN & 17,257 \\  
\textit{14} & Pennsylvania State University & US & 5,734 & Russian Academy of Sciences & RU & 10,990 & University of Oxford & GB & 16,993 \\  
\textit{15} & Lawrence Berkeley National Laboratory & US & 5,657 & Zhejiang University & CN & 10,839 & University of Cambridge & GB & 16,936 \\  
\textit{16} & Lomonosov Moscow State University & RU & 5,620 & University of Illinois Urbana-Champaign & US & 10,513 & Stanford University & US & 16,898 \\  
\textit{17} & Princeton University & US & 5,571 & Los Alamos National Laboratory & US & 10,079 & University of Michigan{\textendash}Ann Arbor & US & 16,695 \\  
\textit{18} & The University of Texas at Austin & US & 5,515 & University of Maryland, College Park & US & 9,709 & Xi'an Jiaotong University & CN & 16,502 \\  
\textit{19} & Tokyo Institute of Technology & JP & 5,338 & Shanghai Jiao Tong University & CN & 9,701 & Beihang University & CN & 15,995 \\  
\textit{20} & United States Naval Research Laboratory & US & 5,206 & Pennsylvania State University & US & 9,602 & ETH Zurich & CH & 15,783 \\  
\hdashline          
\textit{21} & University of California, Los Angeles & US & 5,178 & Harbin Institute of Technology & CN & 9,289 & Osaka University & JP & 15,647 \\  
\textit{22} & Nagoya University & JP & 5,117 & University of Oxford & GB & 9,169 & Tohoku University & JP & 15,546 \\  
\textit{23} & Cornell University & US & 5,102 & Imperial College London & GB & 9,070 & Lomonosov Moscow State University & RU & 15,460 \\  
\textit{24} & Argonne National Laboratory & US & 5,046 & \scalebox{0.82}[1]{National Institute of Advanced Industrial Science and Technology (AIST)} & JP & 9,012 & Imperial College London & GB & 15,419 \\  
\textit{25} & University of Wisconsin{\textendash}Madison & US & 4,911 & Georgia Institute of Technology & US & 8,910 & Nanyang Technological University & SG & 14,764 \\  
\textit{26} & Chinese Academy of Sciences & CN & 4,899 & Princeton University & US & 8,828 & National University of Singapore & SG & 14,434 \\  
\textit{27} & University of California, Santa Barbara & US & 4,876 & National University of Singapore & SG & 8,821 & North China Electric Power University & CN & 14,088 \\  
\textit{28} & Russian Academy of Sciences & RU & 4,831 & Nagoya University & JP & 8,662 & Tianjin University & CN & 13,702 \\  
\textit{29} & University of Minnesota & US & 4,775 & Lawrence Berkeley National Laboratory & US & 8,641 & University of Electronic Science and Technology of China & CN & 13,632 \\  
\textit{30} & University of Arizona & US & 4,744 & University of Toronto & CA & 8,538 & Russian Academy of Sciences & RU & 13,428 \\  
\hdashline          
\textit{31} & California Institute of Technology & US & 4,675 & California Institute of Technology & US & 8,365 & Southeast University & CN & 13,397 \\  
\textit{32} & The Ohio State University & US & 4,671 & The University of Texas at Austin & US & 8,241 & Nanjing University & CN & 13,123 \\  
\textit{33} & National Institute of Standards and Technology & US & 4,647 & University of California, Los Angeles & US & 8,116 & University of Toronto & CA & 12,679 \\  
\textit{34} & AT\&T (United States) & US & 4,609 & Peking University & CN & 8,112 & University of Illinois Urbana-Champaign & US & 12,653 \\  
\textit{35} & University of Paris-Sud & FR & 4,588 & ETH Zurich & CH & 7,925 & The University of Texas at Austin & US & 12,512 \\  
\textit{36} & Lawrence Livermore National Laboratory & US & 4,581 & University of Science and Technology of China & CN & 7,856 & University of Maryland, College Park & US & 12,456 \\  
\textit{37} & Imperial College London & GB & 4,550 & University of California, San Diego & US & 7,793 & Beijing Institute of Technology & CN & 12,373 \\  
\textit{38} & Oak Ridge National Laboratory & US & 4,495 & Huazhong University of Science and Technology & CN & 7,782 & Georgia Institute of Technology & US & 12,358 \\  
\textit{39} & European Organization for Nuclear Research & CH & 4,480 & Texas A\&M University & US & 7,735 & \'{E}cole Polytechnique F\'{e}d\'{e}rale de Lausanne & CH & 12,284 \\  
\textit{40} & Polish Academy of Sciences & PL & 4,400 & Cornell University & US & 7,728 & University College London & GB & 12,190 \\  
\hdashline          
\textit{41} & University of California, San Diego & US & 4,357 & Kyushu University & JP & 7,717 & Harvard University & US & 12,153 \\  
\textit{42} & University of Toronto & CA & 4,345 & \'{E}cole Polytechnique F\'{e}d\'{e}rale de Lausanne & CH & 7,712 & Universidade de S\~{a}o Paulo & BR & 12,125 \\  
\textit{43} & Technical University of Munich & DE & 4,132 & Nanyang Technological University & SG & 7,558 & Jilin University & CN & 12,023 \\  
\textit{44} & Kyushu University & JP & 4,078 & Seoul National University & KR & 7,556 & Shandong University & CN & 11,989 \\  
\textit{45} & Harvard University & US & 4,071 & University of Wisconsin{\textendash}Madison & US & 7,542 & Pennsylvania State University & US & 11,860 \\  
\textit{46} & University of Florida & US & 4,058 & Max Planck Society & DE & 7,454 & Lawrence Berkeley National Laboratory & US & 11,778 \\  
\textit{47} & Institute of Physics & PL & 4,057 & National Taiwan University & TW & 7,291 & Princeton University & US & 11,657 \\  
\textit{48} & University of Washington & US & 4,051 & The Ohio State University & US & 7,256 & University of California, Los Angeles & US & 11,652 \\  
\textit{49} & Tel Aviv University & IL & 3,959 & Harvard University & US & 7,102 & Technical University of Munich & DE & 11,533 \\  
\textit{50} & University of Manchester & GB & 3,931 & Universidade de S\~{a}o Paulo & BR & 7,085 & Karlsruhe Institute of Technology & DE & 11,524 \\  
          
 \\[-1.4em]
\hdashline \\[-1em]
\multicolumn{10}{r}{\scriptsize \emph{(continued to next page)}}
\end{tabular}}
}
\end{table}
\end{landscape}
}
\afterpage{\clearpage%
\begin{landscape}
\begin{table}[!t]\ContinuedFloat
\vspace{-3.3em}
\caption{\textbf{The top 100 productive institutions: \textcolor{violet}{\textit{\quantum}}.} \emph{(Cont.)}}
\label{tab:r2_quantum}
\vspace{2em}
{\tiny
{\renewcommand{\arraystretch}{1.2}
\begin{tabular}{rp{5cm}lr@{\hspace{4em}}p{5cm}lr@{\hspace{4em}}p{5cm}lr}\\[-5em] \toprule[1pt] \\[-1.4em]    
 & {\scriptsize \textbf{1991--2000}} & \multicolumn{2}{c}{No.\ Works} & {\scriptsize \textbf{2001--2010}} & \multicolumn{2}{c}{No.\ Works} & {\scriptsize \textbf{2011--2020}} & \multicolumn{2}{r}{No.\ Works} \\[-0.2em] \cmidrule[0.5pt](lr{4em}){2-4} \cmidrule[0.5pt](l{-0em}r{4em}){5-7} \cmidrule[0.5pt](l{-0em}r{1em}){8-10}
\textit{51} & Rutgers, The State University of New Jersey & US & 3,791 & University of California, Santa Barbara & US & 7,084 & Dalian University of Technology & CN & 11,507 \\  
\textit{52} & Texas A\&M University & US & 3,716 & University of Florida & US & 6,985 & Northwestern Polytechnical University & CN & 11,362 \\  
\textit{53} & \'{E}cole Polytechnique F\'{e}d\'{e}rale de Lausanne & CH & 3,604 & KU Leuven & BE & 6,976 & University of California, San Diego & US & 11,167 \\  
\textit{54} & Forschungszentrum J\"{u}lich & DE & 3,549 & Japan Science and Technology Agency (JST) & JP & 6,958 & Nagoya University & JP & 10,976 \\  
\textit{55} & Max Planck Society & DE & 3,535 & University of Paris-Sud & FR & 6,936 & Tokyo Institute of Technology & JP & 10,818 \\  
\textit{56} & Brookhaven National Laboratory & US & 3,504 & Nanjing University & CN & 6,893 & KU Leuven & BE & 10,774 \\  
\textit{57} & Goddard Space Flight Center & US & 3,499 & University of Washington & US & 6,886 & South China University of Technology & CN & 10,709 \\  
\textit{58} & Columbia University & US & 3,463 & University of Manchester & GB & 6,870 & Sapienza University of Rome & IT & 10,574 \\  
\textit{59} & Tokyo University of Science & JP & 3,455 & University of Minnesota & US & 6,870 & Seoul National University & KR & 10,545 \\  
\textit{60} & Iowa State University & US & 3,442 & National Academy of Sciences of Ukraine & UA & 6,864 & Texas A\&M University & US & 10,520 \\  
\hdashline          
\textit{61} & Purdue University West Lafayette & US & 3,432 & Delft University of Technology & NL & 6,839 & University of Manchester & GB & 10,437 \\  
\textit{62} & University of Tsukuba & JP & 3,427 & Petersburg Nuclear Physics Institute & RU & 6,640 & Delft University of Technology & NL & 10,398 \\  
\textit{63} & Sapienza University of Rome & IT & 3,424 & Polish Academy of Sciences & PL & 6,622 & Chongqing University & CN & 10,360 \\  
\textit{64} & Hokkaido University & JP & 3,408 & University of Arizona & US & 6,574 & Wuhan University & CN & 10,331 \\  
\textit{65} & University of Chicago & US & 3,392 & Argonne National Laboratory & US & 6,490 & University of Washington & US & 10,220 \\  
\textit{66} & University of Colorado Boulder & US & 3,369 & Sapienza University of Rome & IT & 6,418 & UNSW Sydney & AU & 10,131 \\  
\textit{67} & Technion {\textendash} Israel Institute of Technology & IL & 3,368 & Columbia University & US & 6,352 & Los Alamos National Laboratory & US & 10,102 \\  
\textit{68} & University of Pennsylvania & US & 3,335 & National Institute of Standards and Technology & US & 6,351 & Fudan University & CN & 10,069 \\  
\textit{69} & Hitachi (Japan) & JP & 3,316 & Hokkaido University & JP & 6,136 & University of Wisconsin{\textendash}Madison & US & 10,015 \\  
\textit{70} & ETH Zurich & CH & 3,294 & University College London & GB & 6,124 & Columbia University & US & 9,879 \\  
\hdashline          
\textit{71} & Georgia Institute of Technology & US & 3,225 & Technical University of Munich & DE & 6,106 & Royal Institute of Technology & SE & 9,775 \\  
\textit{72} & KU Leuven & BE & 3,206 & Tianjin University & CN & 5,975 & The Ohio State University & US & 9,706 \\  
\textit{73} & University of British Columbia & CA & 3,176 & Xi'an Jiaotong University & CN & 5,961 & Tongji University & CN & 9,686 \\  
\textit{74} & Chalmers University of Technology & SE & 3,157 & United States Naval Research Laboratory & US & 5,932 & California Institute of Technology & US & 9,681 \\  
\textit{75} & Johns Hopkins University & US & 3,156 & Oak Ridge National Laboratory & US & 5,896 & Nanjing University of Aeronautics and Astronautics & CN & 9,609 \\  
\textit{76} & Delft University of Technology & NL & 3,113 & University of California, Davis & US & 5,885 & National Taiwan University & TW & 9,576 \\  
\textit{77} & Northwestern University & US & 3,103 & Goddard Space Flight Center & US & 5,847 & University of British Columbia & CA & 9,420 \\  
\textit{78} & Yale University & US & 3,093 & University of British Columbia & CA & 5,841 & University of Paris-Saclay & FR & 9,387 \\  
\textit{79} & University of California, Davis & US & 3,079 & Institute of Physics & PL & 5,803 & University of Minnesota & US & 9,365 \\  
\textit{80} & University of Southern California & US & 3,054 & Purdue University West Lafayette & US & 5,758 & University of Waterloo & CA & 9,294 \\  
\hdashline          
\textit{81} & State University of New York & US & 3,048 & University of Southampton & GB & 5,749 & University of Sydney & AU & 9,202 \\  
\textit{82} & Australian National University & AU & 3,027 & Jilin University & CN & 5,672 & Sichuan University & CN & 9,108 \\  
\textit{83} & University of Stuttgart & DE & 3,004 & Tel Aviv University & IL & 5,670 & Politecnico di Milano & IT & 9,093 \\  
\textit{84} & McGill University & CA & 2,984 & National Yang Ming Chiao Tung University & TW & 5,634 & University of Southampton & GB & 8,975 \\  
\textit{85} & University of Rochester & US & 2,983 & University of Bologna & IT & 5,619 & Central South University & CN & 8,908 \\  
\textit{86} & Universit\'{e} Paris Cit\'{e} & FR & 2,933 & Tokyo University of Science & JP & 5,603 & Oak Ridge National Laboratory & US & 8,891 \\  
\textit{87} & Uppsala University & SE & 2,897 & Wuhan University & CN & 5,568 & Purdue University West Lafayette & US & 8,867 \\  
\textit{88} & Arizona State University & US & 2,885 & University of Alberta & CA & 5,563 & University of Colorado Boulder & US & 8,851 \\  
\textit{89} & Universidade de S\~{a}o Paulo & BR & 2,882 & Rutgers, The State University of New Jersey & US & 5,516 & Cornell University & US & 8,786 \\  
\textit{90} & University of Amsterdam & NL & 2,872 & University of Waterloo & CA & 5,504 & Argonne National Laboratory & US & 8,775 \\  
\hdashline          
\textit{91} & Boston University & US & 2,848 & Royal Institute of Technology & SE & 5,450 & Kyushu University & JP & 8,716 \\  
\textit{92} & Jet Propulsion Laboratory & US & 2,843 & Australian National University & AU & 5,445 & National University of Defense Technology & CN & 8,709 \\  
\textit{93} & North Carolina State University & US & 2,826 & University of Tsukuba & JP & 5,430 & Technical University of Denmark & DK & 8,705 \\  
\textit{94} & Stony Brook University & US & 2,814 & Dalian University of Technology & CN & 5,413 & Nanjing University of Science and Technology & CN & 8,626 \\  
\textit{95} & University College London & GB & 2,805 & Shandong University & CN & 5,410 & \scalebox{0.82}[1]{National Institute of Advanced Industrial Science and Technology (AIST)} & JP & 8,461 \\  
\textit{96} & Ruhr University Bochum & DE & 2,788 & Arizona State University & US & 5,384 & RWTH Aachen University & DE & 8,437 \\  
\textit{97} & University of Southampton & GB & 2,781 & Lawrence Livermore National Laboratory & US & 5,384 & University of Alberta & CA & 8,379 \\  
\textit{98} & Hiroshima University & JP & 2,778 & Korea Advanced Institute of Science and Technology & KR & 5,343 & McGill University & CA & 8,376 \\  
\textit{99} & Weizmann Institute of Science & IL & 2,774 & North China Electric Power University & CN & 5,332 & Universidad Nacional Aut\'{o}noma de M\'{e}xico & MX & 8,374 \\  
\textit{100} & Hebrew University of Jerusalem & IL & 2,748 & Iowa State University & US & 5,322 & University of Paris-Sud & FR & 8,299 \\  
          
 \\[-1.4em]
\bottomrule
\end{tabular}}
}
\end{table}
\end{landscape}
}
%

\titleformat{\section}{\sc\centering\LARGE\bfseries}{\textsc{\thesection}.\!\!}{1em}{}

\afterpage{\clearpage%
\markboth{\textbf \textsc{\bio}}{}
\thispagestyle{empty}
\quad
\vspace{2cm}
\begin{center}
\pgfornament[width=0.5*\textwidth,symmetry=h]{89}\\[2em]
\section{\bio}
\vspace{1em}
\pgfornament[width=0.5*\textwidth]{89}
\end{center}
}

\afterpage{\clearpage%

\begin{figure}[!tp]
\centering
\vspace{-1em}
{\large \textbf{\textrm{{World Map of \textcolor{violet}{\textit{\bio}} Collaboration}}}~|~1971--2020}\\
\vspace{0.3cm}
\includegraphics[align=c, scale=0.054, trim={9.5cm 0 9.5cm 0},clip]{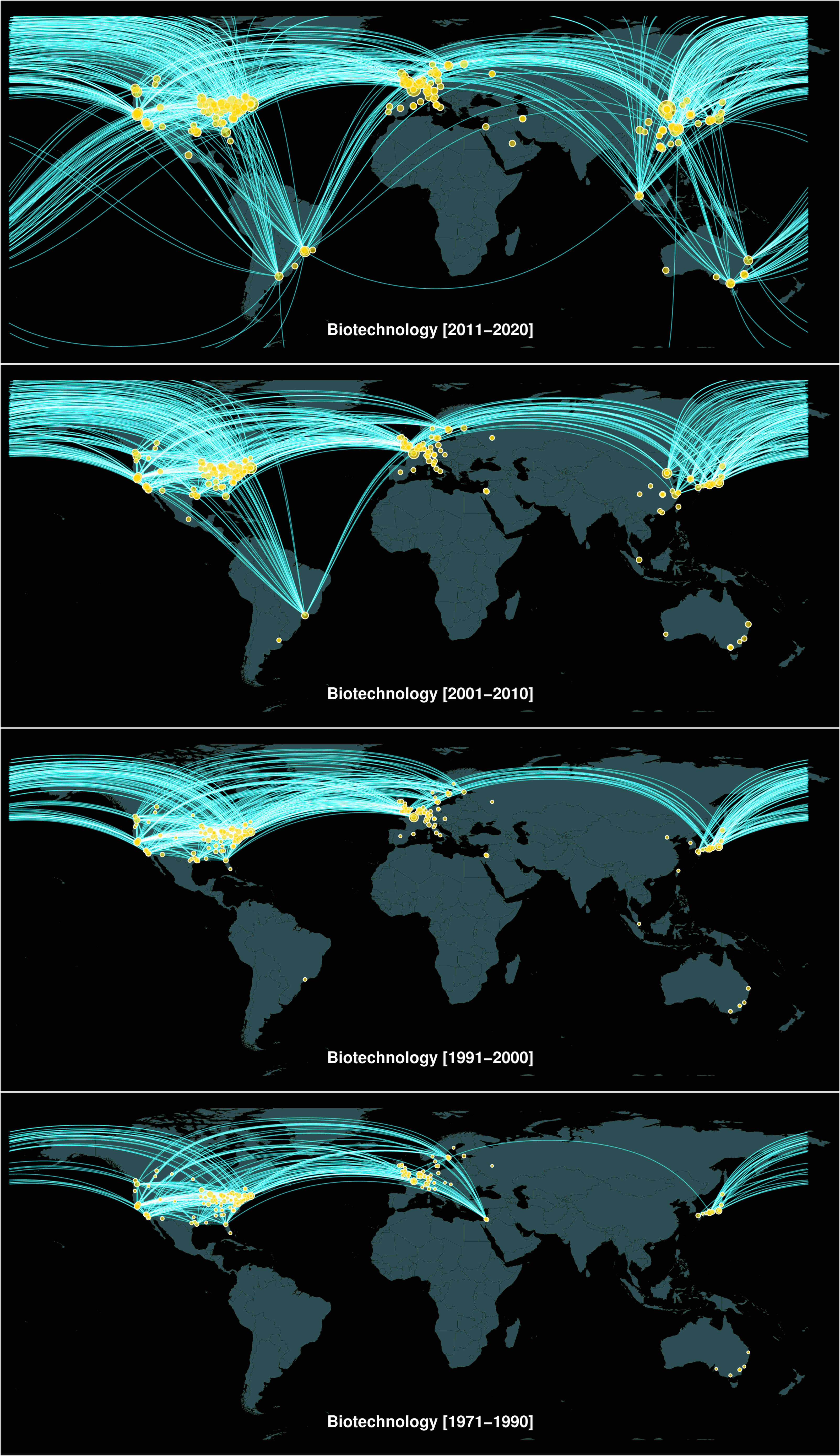}
\caption[{\bio}]{\textbf{(a)~|~The World Map of \textcolor{violet}{\textit{\bio}} Collaboration.}
The bubbles represent the top 199 institutions in terms of work production, with their sizes proportional to the work volume. 
The connecting lines depict coauthorship relationships among the top 50 institutions.}
\label{fig:wmap_bio}
\end{figure}
}
\afterpage{\clearpage%
\begin{figure}[!tp]\ContinuedFloat
\centering
\vspace{-1em}
{\large \textbf{Top 30 Productive Institutions on the World Map: \textcolor{violet}{\textit{\bio}}}~|~1991--2020}\\
\vspace{-0em}
\hspace*{-3em}                                                           
\includegraphics[align=c, scale=0.83]{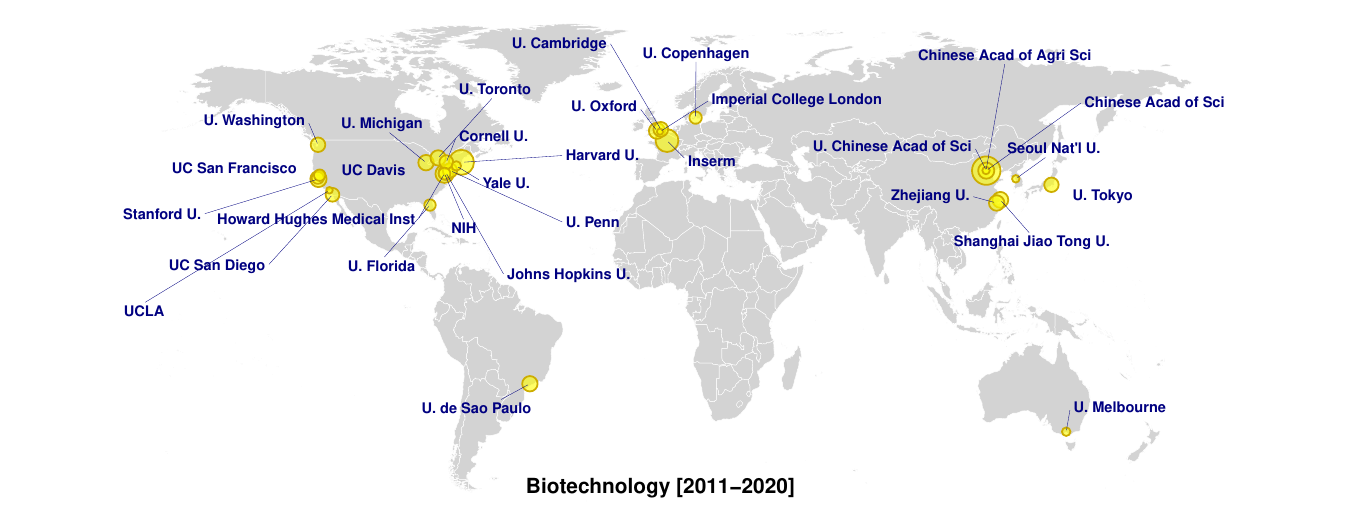}\\[-0.5em]
\quad\\[-1em]
\dotfill 
\quad\\[-0em]
\hspace*{-3em}                                                           
\includegraphics[align=c, scale=0.83]{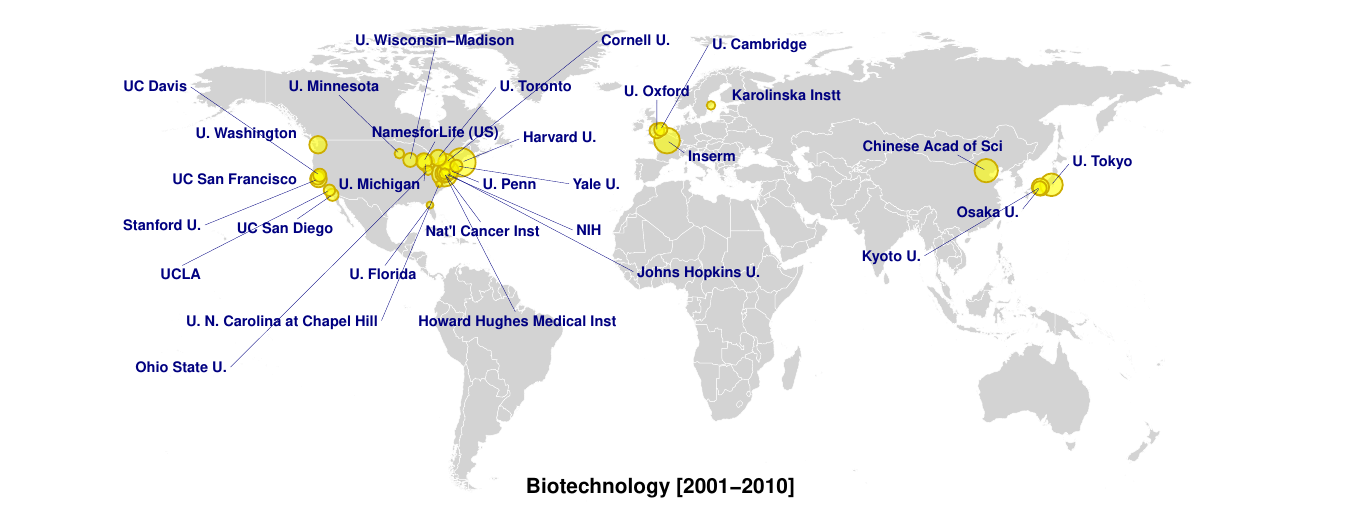}\\[-0.5em]
\quad\\[-1em]
\dotfill 
\quad\\[-0em]
\hspace*{-3em}                                                           
\includegraphics[align=c, scale=0.83]{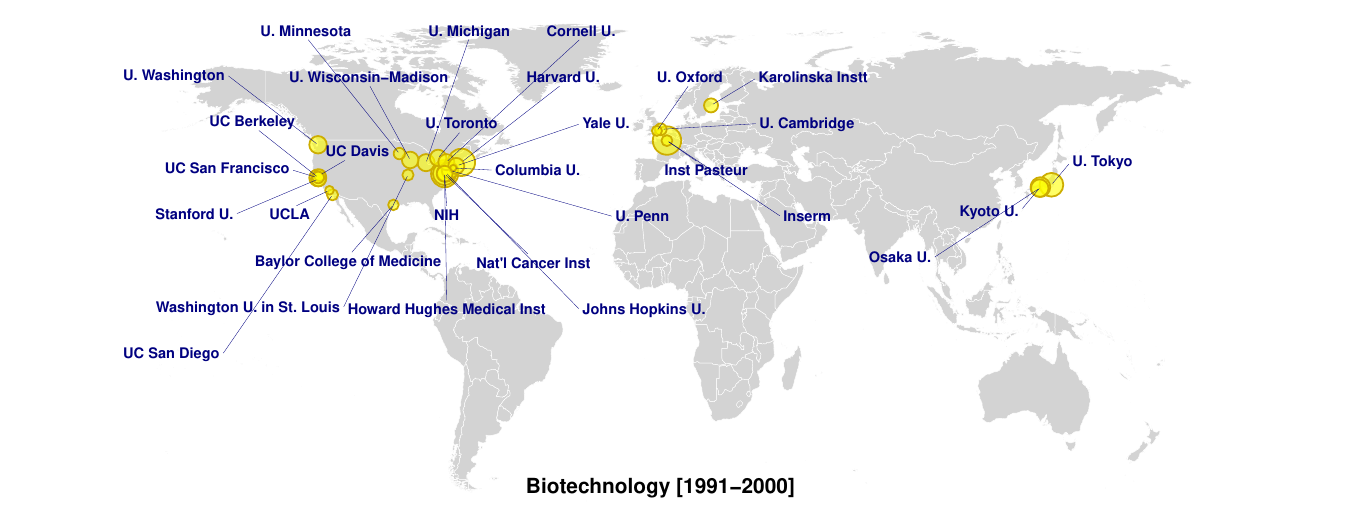}\\[-0.8em]
\caption[{\bio}]{\textbf{(b)~|~The top 30 productive institutions on the World Map: \textcolor{violet}{\textit{\bio}}.}
The bubbles represent the top 30 institutions in terms of work production, with their sizes proportional to the work volume.}
\label{fig:wmap_topinst_bio}
\end{figure}
}
\afterpage{\clearpage%
\begin{figure}[!tp]\ContinuedFloat
\centering
\vspace{-1em}
{\large \textbf{\textrm{{Interregional \textcolor{violet}{\textit{\bio}} Collaboration}}}~|~1991--2020}\\
\vspace{0.5em}
\hspace{-5em}\includegraphics[align=c, scale=1.7, vmargin=0mm]{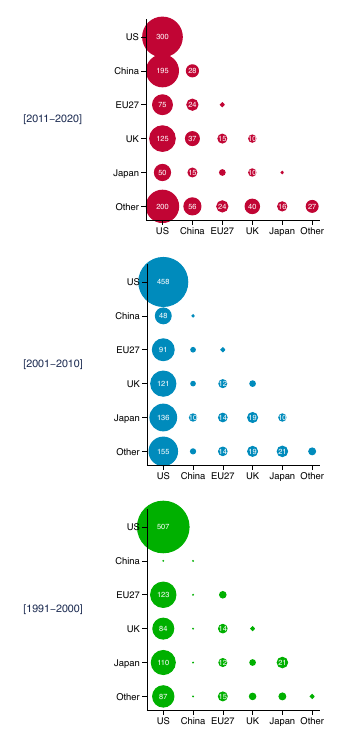}
\vspace{-1em}
\caption[]{\textbf{(c)~|~The Interregional \textcolor{violet}{\textit{\bio}} Collaboration Matrix Diagram.}
The bubble size represents the number of coauthorship relationships for the top 50 institutions in terms of work production. 
If the number is equal to or greater than 10, it is displayed inside the bubble.}
\label{fig:halfmat_bio}
\end{figure}
}
\afterpage{\clearpage%
\begin{figure}[!tp]\ContinuedFloat
\centering
\vspace{-1em}
{\large \textbf{\textrm{Interinstitutional \textcolor{violet}{\textit{\bio}} Collaboration}}~|~2001--2020\quad {\footnotesize \emph{(continued to next page)}}}\\
\cdend{bio}{2010}{2000}{2011--2020}{2001--2010}\\[-1.5em]
\caption[]{\textbf{(d)~|~The Interinstitutional \textcolor{violet}{\textit{\bio}} Collaboration Dendrogram.}
The top 50 institutions in terms of work production, indicated by the circularised bar graphs, are displayed.
}
\label{fig:cdend1_bio}
\end{figure}
}
\afterpage{\clearpage%
\begin{figure}[!tp]\ContinuedFloat
\centering
\vspace{-1em}
{\large \textbf{\textrm{Interinstitutional \textcolor{violet}{\textit{\bio}} Collaboration}}~|~1971--2000\quad {\footnotesize \emph{(continued from previous page)}}}\\
\cdend{bio}{1990}{1980}{1991--2000}{1971--1990}\\[-1.5em]
\caption[]{\textbf{(d)~|~The Interinstitutional \textcolor{violet}{\textit{\bio}} Collaboration Dendrogram.} \emph{(Cont.)}\hfill~}
\label{fig:cdend2_bio}
\end{figure}
}
\afterpage{\clearpage%
\begin{landscape}
\begin{table}[!t]
\vspace{-0.5em}
\caption{\textbf{The top 100 productive institutions: \textcolor{violet}{\textit{\bio}}.}}
\label{tab:r1_bio}
\vspace{2em}
\centering
{\tiny
{\renewcommand{\arraystretch}{1.2}
\begin{tabular}{rp{5cm}lr@{\hspace{4em}}p{5cm}lr@{\hspace{4em}}p{5cm}lr}\\[-5em] \toprule[1pt] \\[-1.4em]    
 & {\scriptsize \textbf{1991--2000}} & \multicolumn{2}{c}{No.~Works} & {\scriptsize \textbf{2001--2010}} & \multicolumn{2}{c}{No.~Works} & {\scriptsize \textbf{2011--2020}} & \multicolumn{2}{r}{No.~Works} \\[-0.2em] \cmidrule[0.5pt](lr{4em}){2-4} \cmidrule[0.5pt](l{-0em}r{4em}){5-7} \cmidrule[0.5pt](l{-0em}r{1em}){8-10}
\textit{1} & Inserm & FR & 15,837 & Harvard University & US & 26,878 & Chinese Academy of Sciences & CN & 61,036 \\  
\textit{2} & Harvard University & US & 14,595 & Inserm & FR & 22,131 & Harvard University & US & 45,412 \\  
\textit{3} & National Institutes of Health & US & 14,350 & National Institutes of Health & US & 21,905 & Inserm & FR & 37,370 \\  
\textit{4} & The University of Tokyo & JP & 11,083 & Chinese Academy of Sciences & CN & 18,722 & National Institutes of Health & US & 26,148 \\  
\textit{5} & Howard Hughes Medical Institute & US & 10,541 & The University of Tokyo & JP & 17,775 & Stanford University & US & 22,565 \\  
\textit{6} & Kyoto University & JP & 8,381 & Howard Hughes Medical Institute & US & 13,378 & University of Chinese Academy of Sciences & CN & 22,495 \\  
\textit{7} & Johns Hopkins University & US & 7,576 & Johns Hopkins University & US & 13,091 & University of Oxford & GB & 22,314 \\  
\textit{8} & Osaka University & JP & 7,376 & University of Washington & US & 12,896 & Shanghai Jiao Tong University & CN & 22,090 \\  
\textit{9} & University of Washington & US & 7,334 & University of Pennsylvania & US & 12,889 & University of Michigan{\textendash}Ann Arbor & US & 21,508 \\  
\textit{10} & University of Michigan{\textendash}Ann Arbor & US & 7,184 & Kyoto University & JP & 12,802 & Zhejiang University & CN & 21,451 \\  
\hdashline          
\textit{11} & University of California, San Francisco & US & 7,166 & University of Michigan{\textendash}Ann Arbor & US & 12,491 & Universidade de S\~{a}o Paulo & BR & 21,371 \\  
\textit{12} & University of Wisconsin{\textendash}Madison & US & 6,999 & Stanford University & US & 12,417 & University of Toronto & CA & 21,242 \\  
\textit{13} & University of Toronto & CA & 6,931 & University of Toronto & CA & 12,101 & University of Cambridge & GB & 20,538 \\  
\textit{14} & University of Pennsylvania & US & 6,917 & NamesforLife (United States) & US & 11,926 & Johns Hopkins University & US & 20,333 \\  
\textit{15} & Yale University & US & 6,775 & Cornell University & US & 11,868 & University of Washington & US & 20,315 \\  
\textit{16} & National Cancer Institute & US & 6,719 & University of Oxford & GB & 11,309 & The University of Tokyo & JP & 20,230 \\  
\textit{17} & Cornell University & US & 6,576 & University of Wisconsin{\textendash}Madison & US & 10,904 & University of Pennsylvania & US & 20,154 \\  
\textit{18} & Stanford University & US & 6,414 & Osaka University & JP & 10,897 & University of California, San Diego & US & 19,442 \\  
\textit{19} & Karolinska Institutet & SE & 6,158 & University of California, San Francisco & US & 10,778 & University of Copenhagen & DK & 18,132 \\  
\textit{20} & University of Cambridge & GB & 5,643 & University of Cambridge & GB & 10,602 & University of California, San Francisco & US & 17,984 \\  
\hdashline          
\textit{21} & University of California, Davis & US & 5,613 & University of California, Davis & US & 10,558 & Howard Hughes Medical Institute & US & 17,725 \\  
\textit{22} & University of California, Berkeley & US & 5,612 & Yale University & US & 10,263 & University of Florida & US & 17,488 \\  
\textit{23} & University of Minnesota & US & 5,581 & University of California, San Diego & US & 10,229 & University of California, Davis & US & 17,468 \\  
\textit{24} & University of California, San Diego & US & 5,554 & University of California, Los Angeles & US & 10,004 & Cornell University & US & 17,388 \\  
\textit{25} & Washington University in St. Louis & US & 5,552 & National Cancer Institute & US & 9,891 & Yale University & US & 16,410 \\  
\textit{26} & Institut Pasteur & FR & 5,506 & University of Minnesota & US & 9,586 & University of Melbourne & AU & 16,054 \\  
\textit{27} & Baylor College of Medicine & US & 5,453 & The Ohio State University & US & 9,548 & Seoul National University & KR & 15,825 \\  
\textit{28} & University of Oxford & GB & 5,394 & Karolinska Institutet & SE & 9,410 & Chinese Academy of Agricultural Sciences & CN & 15,821 \\  
\textit{29} & University of California, Los Angeles & US & 5,245 & University of Florida & US & 9,268 & University of California, Los Angeles & US & 15,725 \\  
\textit{30} & Columbia University & US & 5,142 & University of North Carolina at Chapel Hill & US & 9,250 & Imperial College London & GB & 15,714 \\  
\hdashline          
\textit{31} & Kyushu University & JP & 5,007 & University of California, Berkeley & US & 9,186 & University of Queensland & AU & 15,554 \\  
\textit{32} & University of North Carolina at Chapel Hill & US & 4,921 & Zhejiang University & CN & 8,953 & The Ohio State University & US & 15,547 \\  
\textit{33} & McGill University & CA & 4,890 & Universidade de S\~{a}o Paulo & BR & 8,886 & University of Minnesota & US & 15,462 \\  
\textit{34} & Hokkaido University & JP & 4,870 & Seoul National University & KR & 8,886 & University of North Carolina at Chapel Hill & US & 15,410 \\  
\textit{35} & The Ohio State University & US & 4,792 & Washington University in St. Louis & US & 8,777 & Fudan University & CN & 15,388 \\  
\textit{36} & Nagoya University & JP & 4,788 & Columbia University & US & 8,607 & The University of Texas MD Anderson Cancer Center & US & 15,355 \\  
\textit{37} & University of Florida & US & 4,572 & University College London & GB & 8,340 & University of Wisconsin{\textendash}Madison & US & 15,328 \\  
\textit{38} & Lund University & SE & 4,521 & The University of Texas MD Anderson Cancer Center & US & 8,273 & Kyoto University & JP & 15,233 \\  
\textit{39} & Agricultural Research Service & US & 4,485 & University of British Columbia & CA & 8,213 & University of British Columbia & CA & 15,043 \\  
\textit{40} & University of Illinois Urbana-Champaign & US & 4,303 & Baylor College of Medicine & US & 8,163 & University College London & GB & 14,793 \\  
\hdashline          
\textit{41} & Massachusetts General Hospital & US & 4,302 & Imperial College London & GB & 8,090 & Massachusetts General Hospital & US & 14,696 \\  
\textit{42} & University College London & GB & 4,272 & University of Pittsburgh & US & 8,072 & Karolinska Institutet & SE & 14,445 \\  
\textit{43} & University of British Columbia & CA & 4,242 & University of Illinois Urbana-Champaign & US & 7,856 & Boston University & US & 14,229 \\  
\textit{44} & University of Chicago & US & 4,214 & McGill University & CA & 7,689 & University of Pittsburgh & US & 13,705 \\  
\textit{45} & Tohoku University & JP & 4,182 & Massachusetts General Hospital & US & 7,614 & University of California, Berkeley & US & 13,696 \\  
\textit{46} & University of Pittsburgh & US & 4,068 & Hokkaido University & JP & 7,572 & National University of Singapore & SG & 13,628 \\  
\textit{47} & Pennsylvania State University & US & 4,015 & Agricultural Research Service & US & 7,541 & Peking University & CN & 13,619 \\  
\textit{48} & University of Iowa & US & 3,977 & Boston University & US & 7,535 & Nanjing Medical University & CN & 13,403 \\  
\textit{49} & Rutgers, The State University of New Jersey & US & 3,940 & Kyushu University & JP & 7,322 & University of Edinburgh & GB & 13,297 \\  
\textit{50} & University of Helsinki & FI & 3,917 & KU Leuven & BE & 7,274 & Washington University in St. Louis & US & 13,189 \\  
          
 \\[-1.4em]
\hdashline \\[-1em]
\multicolumn{10}{r}{\scriptsize \emph{(continued to next page)}}
\end{tabular}}
}
\end{table}
\end{landscape}
}
\afterpage{\clearpage%
\begin{landscape}
\begin{table}[!t]\ContinuedFloat
\vspace{-3.3em}
\caption{\textbf{The top 100 productive institutions: \textcolor{violet}{\textit{\bio}}.} \emph{(Cont.)}}
\label{tab:r2_bio}
\vspace{2em}
{\tiny
{\renewcommand{\arraystretch}{1.2}
\begin{tabular}{rp{5cm}lr@{\hspace{4em}}p{5cm}lr@{\hspace{4em}}p{5cm}lr}\\[-5em] \toprule[1pt] \\[-1.4em]    
 & {\scriptsize \textbf{1991--2000}} & \multicolumn{2}{c}{No.\ Works} & {\scriptsize \textbf{2001--2010}} & \multicolumn{2}{c}{No.\ Works} & {\scriptsize \textbf{2011--2020}} & \multicolumn{2}{r}{No.\ Works} \\[-0.2em] \cmidrule[0.5pt](lr{4em}){2-4} \cmidrule[0.5pt](l{-0em}r{4em}){5-7} \cmidrule[0.5pt](l{-0em}r{1em}){8-10}
\textit{51} & Case Western Reserve University & US & 3,851 & Pennsylvania State University & US & 7,213 & Consejo Nacional de Investigaciones Cient\'{i}ficas y T\'{e}cnicas & AR & 13,062 \\  
\textit{52} & State University of New York & US & 3,850 & Lund University & SE & 7,178 & KU Leuven & BE & 13,002 \\  
\textit{53} & Johns Hopkins Medicine & US & 3,840 & University of Helsinki & FI & 7,147 & Brigham and Women's Hospital & US & 12,959 \\  
\textit{54} & Massachusetts Institute of Technology & US & 3,826 & Tohoku University & JP & 7,109 & Huazhong University of Science and Technology & CN & 12,845 \\  
\textit{55} & The University of Texas MD Anderson Cancer Center & US & 3,819 & GTx (United States) & US & 7,061 & China Agricultural University & CN & 12,786 \\  
\textit{56} & Duke Medical Center & US & 3,762 & Brigham and Women's Hospital & US & 7,044 & Nanjing Agricultural University & CN & 12,762 \\  
\textit{57} & The University of Texas Southwestern Medical Center & US & 3,702 & Emory University & US & 6,804 & Sun Yat-sen University & CN & 12,720 \\  
\textit{58} & University of Milan & IT & 3,670 & University of Edinburgh & GB & 6,768 & Baylor College of Medicine & US & 12,438 \\  
\textit{59} & Leiden University & NL & 3,646 & University of Chicago & US & 6,759 & Columbia University & US & 12,412 \\  
\textit{60} & Texas A\&M University & US & 3,635 & Huazhong University of Science and Technology & CN & 6,744 & University of Sydney & AU & 12,380 \\  
\hdashline          
\textit{61} & KU Leuven & BE & 3,596 & University of Copenhagen & DK & 6,737 & McGill University & CA & 12,359 \\  
\textit{62} & University of Alberta & CA & 3,566 & Nagoya University & JP & 6,706 & Massachusetts Institute of Technology & US & 12,349 \\  
\textit{63} & University of Arizona & US & 3,565 & University of Queensland & AU & 6,532 & University of Illinois Urbana-Champaign & US & 12,336 \\  
\textit{64} & Michigan State University & US & 3,538 & Massachusetts Institute of Technology & US & 6,484 & Emory University & US & 12,206 \\  
\textit{65} & University of Amsterdam & NL & 3,531 & Rutgers, The State University of New Jersey & US & 6,482 & Shandong University & CN & 11,967 \\  
\textit{66} & Boston University & US & 3,526 & University of Southern California & US & 6,469 & Osaka University & JP & 11,596 \\  
\textit{67} & Hebrew University of Jerusalem & IL & 3,434 & University of Melbourne & AU & 6,346 & Agricultural Research Service & US & 11,583 \\  
\textit{68} & Scripps Research Institute & US & 3,423 & University of Milan & IT & 6,247 & University of Helsinki & FI & 11,580 \\  
\textit{69} & University of Utah & US & 3,398 & University of Utah & US & 6,239 & Wageningen University \& Research & NL & 11,525 \\  
\textit{70} & University of Edinburgh & GB & 3,382 & Vanderbilt University & US & 6,205 & Tsinghua University & CN & 11,421 \\  
\hdashline          
\textit{71} & University of Alabama at Birmingham & US & 3,358 & National University of Singapore & SG & 6,183 & Iowa State University & US & 11,343 \\  
\textit{72} & Vanderbilt University & US & 3,328 & New York University & US & 6,133 & Duke University & US & 11,335 \\  
\textit{73} & New York University & US & 3,324 & University of Manchester & GB & 6,101 & Sichuan University & CN & 11,321 \\  
\textit{74} & University of Southern California & US & 3,290 & Scripps Research Institute & US & 6,095 & King's College London & GB & 11,297 \\  
\textit{75} & Utrecht University & NL & 3,266 & Duke University & US & 6,069 & University of Milan & IT & 11,249 \\  
\textit{76} & Brigham and Women's Hospital & US & 3,262 & University of Alberta & CA & 5,978 & National Cancer Institute & US & 11,233 \\  
\textit{77} & University of Zurich & CH & 3,082 & Michigan State University & US & 5,969 & Pennsylvania State University & US & 10,999 \\  
\textit{78} & Albert Einstein College of Medicine & US & 3,078 & Charit\'{e} {\textendash} Universit\"{a}tsmedizin Berlin & DE & 5,930 & Michigan State University & US & 10,977 \\  
\textit{79} & Duke University Hospital & US & 3,052 & Texas A\&M University & US & 5,906 & University of Southern California & US & 10,969 \\  
\textit{80} & Tokyo University of Science & JP & 3,025 & University of Iowa & US & 5,906 & Rutgers, The State University of New Jersey & US & 10,927 \\  
\hdashline          
\textit{81} & Emory University & US & 3,014 & Uppsala University & SE & 5,850 & University of Chicago & US & 10,904 \\  
\textit{82} & Ludwig-Maximilians-Universit\"{a}t M\"{u}nchen & DE & 3,009 & Case Western Reserve University & US & 5,820 & Ghent University & BE & 10,873 \\  
\textit{83} & Uppsala University & SE & 2,962 & University of Alabama at Birmingham & US & 5,806 & Huazhong Agricultural University & CN & 10,756 \\  
\textit{84} & Sapienza University of Rome & IT & 2,954 & Johns Hopkins Medicine & US & 5,791 & University of Zurich & CH & 10,228 \\  
\textit{85} & Purdue University West Lafayette & US & 2,917 & Ludwig-Maximilians-Universit\"{a}t M\"{u}nchen & DE & 5,789 & University of Manchester & GB & 10,209 \\  
\textit{86} & Wayne State University & US & 2,876 & University of Arizona & US & 5,767 & ETH Zurich & CH & 10,178 \\  
\textit{87} & University of W\"{u}rzburg & DE & 2,847 & Northwestern University & US & 5,752 & Technical University of Munich & DE & 10,159 \\  
\textit{88} & Rockefeller University & US & 2,842 & Wageningen University \& Research & NL & 5,737 & Johns Hopkins Medicine & US & 10,119 \\  
\textit{89} & Freie Universit\"{a}t Berlin & DE & 2,831 & Institut Pasteur & FR & 5,724 & Northwestern University & US & 10,108 \\  
\textit{90} & University of Georgia & US & 2,827 & Shanghai Jiao Tong University & CN & 5,693 & Monash University & AU & 9,973 \\  
\hdashline          
\textit{91} & Iowa State University & US & 2,824 & Duke Medical Center & US & 5,631 & University of Alberta & CA & 9,970 \\  
\textit{92} & Hiroshima University & JP & 2,817 & Ghent University & BE & 5,596 & Jilin University & CN & 9,761 \\  
\textit{93} & University of California, Irvine & US & 2,817 & The University of Texas Southwestern Medical Center & US & 5,577 & University of Utah & US & 9,758 \\  
\textit{94} & University of Queensland & AU & 2,809 & King's College London & GB & 5,445 & North West Agriculture and Forestry University & CN & 9,755 \\  
\textit{95} & University of Manchester & GB & 2,808 & University of Sydney & AU & 5,404 & S\~{a}o Paulo State University & BR & 9,691 \\  
\textit{96} & University of Virginia & US & 2,779 & Iowa State University & US & 5,343 & Dana-Farber Cancer Institute & US & 9,671 \\  
\textit{97} & Northwestern University & US & 2,746 & University of Zurich & CH & 5,305 & Texas A\&M University & US & 9,628 \\  
\textit{98} & Radboud University Nijmegen & NL & 2,689 & Fudan University & CN & 5,281 & Icahn School of Medicine at Mount Sinai & US & 9,607 \\  
\textit{99} & Tel Aviv University & IL & 2,687 & Peking University & CN & 5,194 & Universidad Nacional Aut\'{o}noma de M\'{e}xico & MX & 9,573 \\  
\textit{100} & University of Copenhagen & DK & 2,671 & University of California, Irvine & US & 5,164 & Capital Medical University & CN & 9,549 \\  
          
 \\[-1.4em]
\bottomrule
\end{tabular}}
}
\end{table}
\end{landscape}
}
%

\titleformat{\section}{\sc\centering\LARGE\bfseries}{\textsc{\thesection}.\!\!}{1em}{}

\afterpage{\clearpage%
\markboth{\textbf \textsc{\nano}}{}
\thispagestyle{empty}
\quad
\vspace{2cm}
\begin{center}
\pgfornament[width=0.5*\textwidth,symmetry=h]{89}\\[2em]
\section{\nano}
\vspace{1em}
\pgfornament[width=0.5*\textwidth]{89}
\end{center}
}

\afterpage{\clearpage%

\begin{figure}[!tp]
\centering
\vspace{-1em}
{\large \textbf{\textrm{{World Map of \textcolor{violet}{\textit{\nano}} Collaboration}}}~|~1971--2020}\\
\vspace{0.3cm}
\includegraphics[align=c, scale=0.054, trim={9.5cm 0 9.5cm 0},clip]{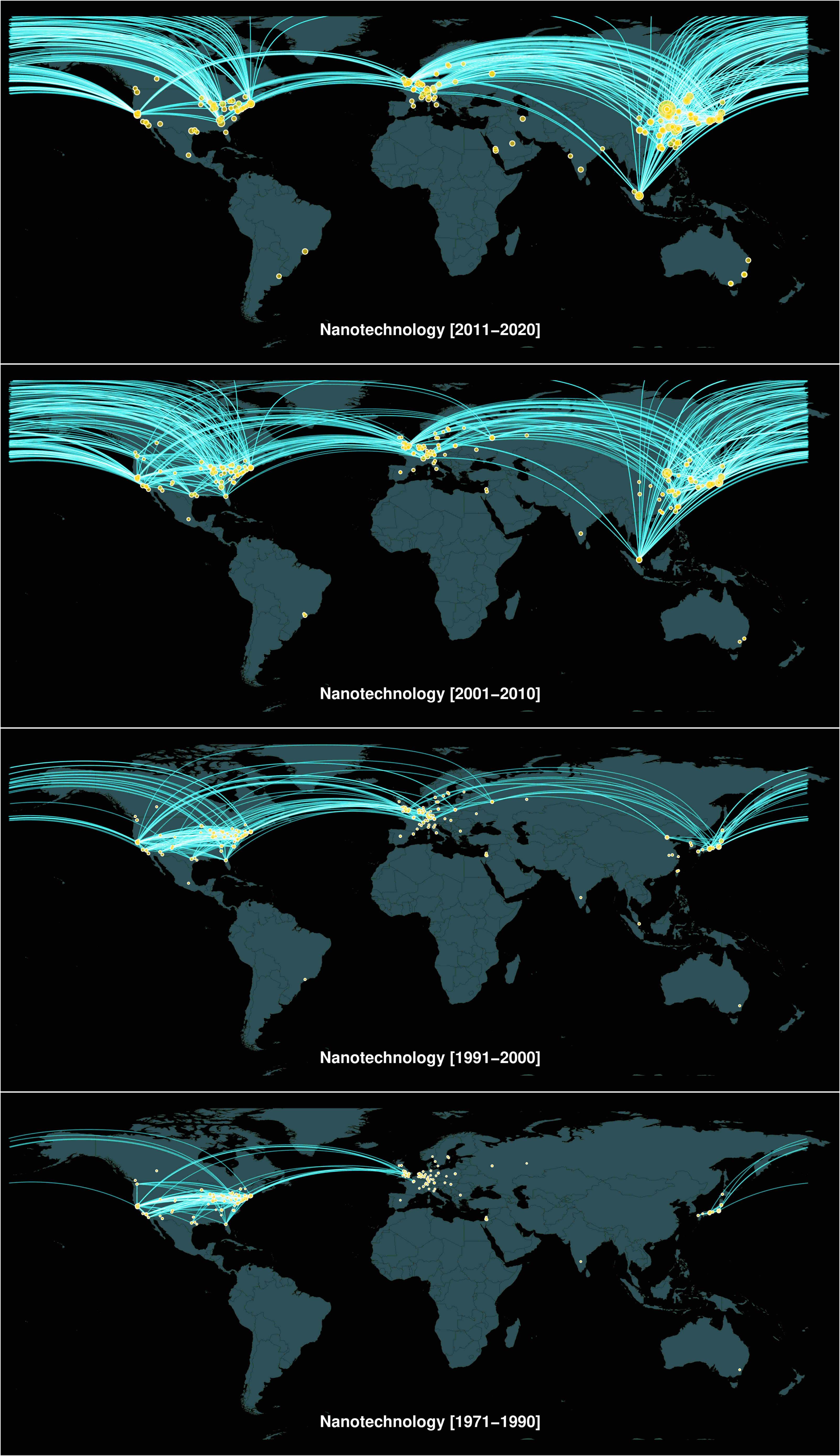}
\caption[{\nano}]{\textbf{(a)~|~The World Map of \textcolor{violet}{\textit{\nano}} Collaboration.}
The bubbles represent the top 199 institutions in terms of work production, with their sizes proportional to the work volume. 
The connecting lines depict coauthorship relationships among the top 50 institutions.}
\label{fig:wmap_nano}
\end{figure}
}
\afterpage{\clearpage%
\begin{figure}[!tp]\ContinuedFloat
\centering
\vspace{-1em}
{\large \textbf{Top 30 Productive Institutions on the World Map: \textcolor{violet}{\textit{\nano}}}~|~1991--2020}\\
\vspace{-0em}
\hspace*{-3em}                                                           
\includegraphics[align=c, scale=0.83]{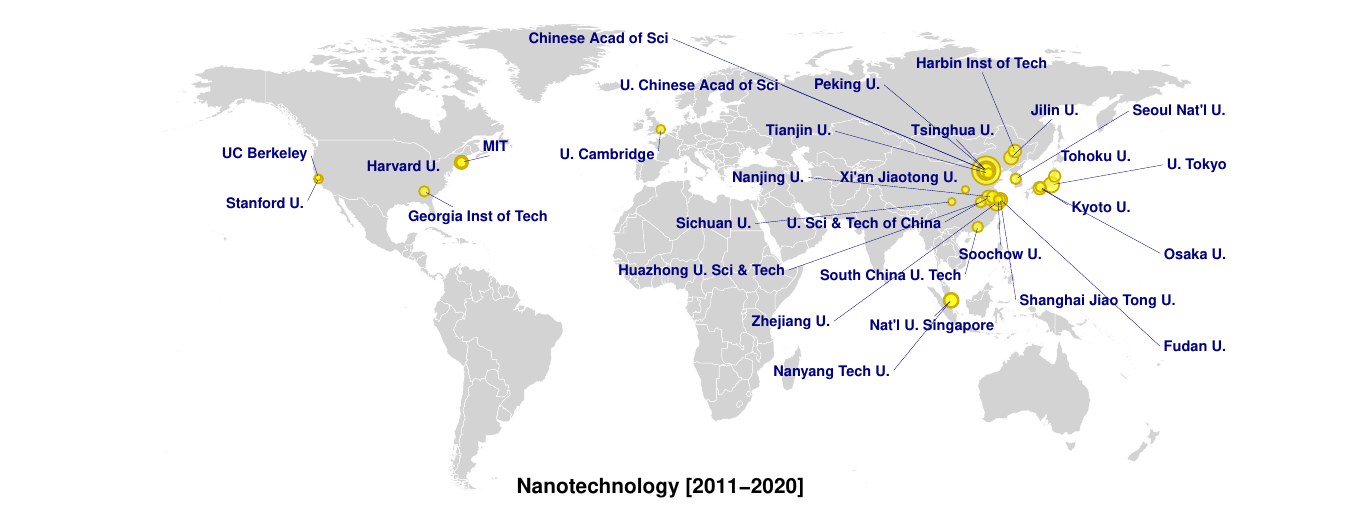}\\[-0.5em]
\quad\\[-1em]
\dotfill 
\quad\\[-0em]
\hspace*{-3em}                                                           
\includegraphics[align=c, scale=0.83]{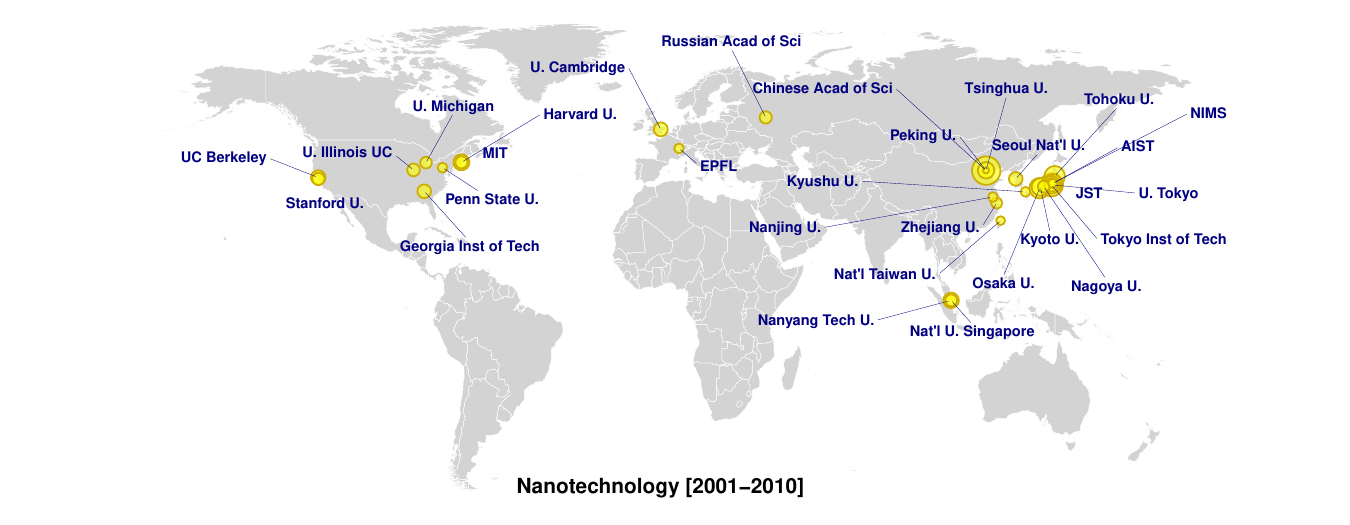}\\[-0.5em]
\quad\\[-1em]
\dotfill 
\quad\\[-0em]
\hspace*{-3em}                                                           
\includegraphics[align=c, scale=0.83]{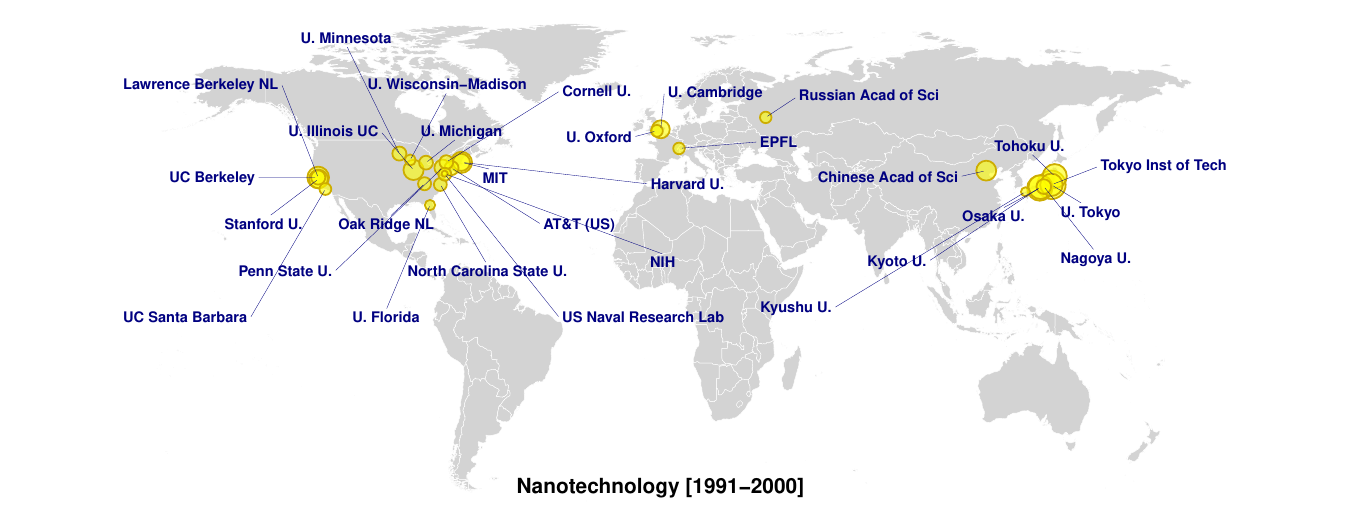}\\[-0.8em]
\caption[{\nano}]{\textbf{(b)~|~The top 30 productive institutions on the World Map: \textcolor{violet}{\textit{\nano}}.}
The bubbles represent the top 30 institutions in terms of work production, with their sizes proportional to the work volume.}
\label{fig:wmap_topinst_nano}
\end{figure}
}
\afterpage{\clearpage%
\begin{figure}[!tp]\ContinuedFloat
\centering
\vspace{-1em}
{\large \textbf{\textrm{{Interregional \textcolor{violet}{\textit{\nano}} Collaboration}}}~|~1991--2020}\\
\vspace{0.5em}
\hspace{-5em}\includegraphics[align=c, scale=1.7, vmargin=0mm]{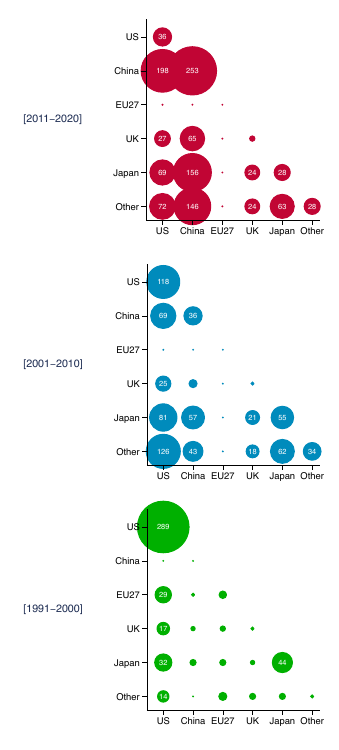}
\vspace{-1em}
\caption[]{\textbf{(c)~|~The Interregional \textcolor{violet}{\textit{\nano}} Collaboration Matrix Diagram.}
The bubble size represents the number of coauthorship relationships for the top 50 institutions in terms of work production. 
If the number is equal to or greater than 10, it is displayed inside the bubble.}
\label{fig:halfmat_nano}
\end{figure}
}
\afterpage{\clearpage%
\begin{figure}[!tp]\ContinuedFloat
\centering
\vspace{-1em}
{\large \textbf{\textrm{Interinstitutional \textcolor{violet}{\textit{\nano}} Collaboration}}~|~2001--2020\quad {\footnotesize \emph{(continued to next page)}}}\\
\cdend{nano}{2010}{2000}{2011--2020}{2001--2010}\\[-1.5em]
\caption[]{\textbf{(d)~|~The Interinstitutional \textcolor{violet}{\textit{\nano}} Collaboration Dendrogram.}
The top 50 institutions in terms of work production, indicated by the circularised bar graphs, are displayed.
}
\label{fig:cdend1_nano}
\end{figure}
}
\afterpage{\clearpage%
\begin{figure}[!tp]\ContinuedFloat
\centering
\vspace{-1em}
{\large \textbf{\textrm{Interinstitutional \textcolor{violet}{\textit{\nano}} Collaboration}}~|~1971--2000\quad {\footnotesize \emph{(continued from previous page)}}}\\
\cdend{nano}{1990}{1980}{1991--2000}{1971--1990}\\[-1.5em]
\caption[]{\textbf{(d)~|~The Interinstitutional \textcolor{violet}{\textit{\nano}} Collaboration Dendrogram.} \emph{(Cont.)}\hfill~}
\label{fig:cdend2_nano}
\end{figure}
}
\afterpage{\clearpage%
\begin{landscape}
\begin{table}[!t]
\vspace{-0.5em}
\caption{\textbf{The top 100 productive institutions: \textcolor{violet}{\textit{\nano}}.}}
\label{tab:r1_nano}
\vspace{2em}
\centering
{\tiny
{\renewcommand{\arraystretch}{1.2}
\begin{tabular}{rp{5cm}lr@{\hspace{4em}}p{5cm}lr@{\hspace{4em}}p{5cm}lr}\\[-5em] \toprule[1pt] \\[-1.4em]    
 & {\scriptsize \textbf{1991--2000}} & \multicolumn{2}{c}{No.~Works} & {\scriptsize \textbf{2001--2010}} & \multicolumn{2}{c}{No.~Works} & {\scriptsize \textbf{2011--2020}} & \multicolumn{2}{r}{No.~Works} \\[-0.2em] \cmidrule[0.5pt](lr{4em}){2-4} \cmidrule[0.5pt](l{-0em}r{4em}){5-7} \cmidrule[0.5pt](l{-0em}r{1em}){8-10}
\textit{1} & The University of Tokyo & JP & 5,507 & Chinese Academy of Sciences & CN & 21,443 & Chinese Academy of Sciences & CN & 67,495 \\  
\textit{2} & Osaka University & JP & 4,365 & The University of Tokyo & JP & 13,354 & University of Chinese Academy of Sciences & CN & 21,935 \\  
\textit{3} & Tohoku University & JP & 4,076 & Osaka University & JP & 10,393 & Tsinghua University & CN & 19,028 \\  
\textit{4} & Kyoto University & JP & 4,042 & Tohoku University & JP & 10,212 & The University of Tokyo & JP & 15,433 \\  
\textit{5} & Tokyo Institute of Technology & JP & 3,666 & \scalebox{0.82}[1]{National Institute of Advanced Industrial Science and Technology (AIST)} & JP & 9,197 & Zhejiang University & CN & 15,151 \\  
\textit{6} & University of California, Berkeley & US & 3,474 & Tokyo Institute of Technology & JP & 8,332 & Nanyang Technological University & SG & 14,901 \\  
\textit{7} & Massachusetts Institute of Technology & US & 3,372 & Kyoto University & JP & 8,150 & University of Science and Technology of China & CN & 14,084 \\  
\textit{8} & University of Illinois Urbana-Champaign & US & 3,170 & Tsinghua University & CN & 7,507 & Peking University & CN & 13,798 \\  
\textit{9} & Chinese Academy of Sciences & CN & 3,125 & Massachusetts Institute of Technology & US & 7,233 & Jilin University & CN & 13,753 \\  
\textit{10} & Stanford University & US & 2,909 & National University of Singapore & SG & 6,965 & Shanghai Jiao Tong University & CN & 13,569 \\  
\hdashline          
\textit{11} & University of Cambridge & GB & 2,811 & University of California, Berkeley & US & 6,544 & National University of Singapore & SG & 13,161 \\  
\textit{12} & Harvard University & US & 2,688 & University of Cambridge & GB & 6,302 & Massachusetts Institute of Technology & US & 12,896 \\  
\textit{13} & Pennsylvania State University & US & 2,565 & Georgia Institute of Technology & US & 6,219 & Harbin Institute of Technology & CN & 12,319 \\  
\textit{14} & Nagoya University & JP & 2,401 & Seoul National University & KR & 6,193 & Osaka University & JP & 12,148 \\  
\textit{15} & Lawrence Berkeley National Laboratory & US & 2,362 & Japan Science and Technology Agency (JST) & JP & 6,126 & Nanjing University & CN & 11,670 \\  
\textit{16} & AT\&T (United States) & US & 2,342 & University of Illinois Urbana-Champaign & US & 5,886 & Tohoku University & JP & 11,186 \\  
\textit{17} & University of Minnesota & US & 2,338 & Stanford University & US & 5,794 & South China University of Technology & CN & 10,714 \\  
\textit{18} & University of Michigan{\textendash}Ann Arbor & US & 2,319 & Russian Academy of Sciences & RU & 5,752 & Seoul National University & KR & 10,670 \\  
\textit{19} & Cornell University & US & 2,293 & Harvard University & US & 5,717 & Harvard University & US & 10,600 \\  
\textit{20} & Oak Ridge National Laboratory & US & 2,257 & Nanyang Technological University & SG & 5,647 & Huazhong University of Science and Technology & CN & 10,546 \\  
\hdashline          
\textit{21} & North Carolina State University & US & 2,246 & Nagoya University & JP & 5,640 & Tianjin University & CN & 10,496 \\  
\textit{22} & United States Naval Research Laboratory & US & 2,244 & University of Michigan{\textendash}Ann Arbor & US & 5,594 & Georgia Institute of Technology & US & 10,479 \\  
\textit{23} & University of Oxford & GB & 2,171 & Zhejiang University & CN & 5,512 & Kyoto University & JP & 10,269 \\  
\textit{24} & University of California, Santa Barbara & US & 2,170 & Pennsylvania State University & US & 5,224 & Fudan University & CN & 10,166 \\  
\textit{25} & Russian Academy of Sciences & RU & 2,150 & Kyushu University & JP & 5,215 & Stanford University & US & 9,831 \\  
\textit{26} & University of Wisconsin{\textendash}Madison & US & 2,127 & Nanjing University & CN & 5,191 & University of Cambridge & GB & 9,826 \\  
\textit{27} & \'{E}cole Polytechnique F\'{e}d\'{e}rale de Lausanne & CH & 2,121 & National Institute for Materials Science (NIMS) & JP & 5,138 & Soochow University & CN & 9,688 \\  
\textit{28} & University of Florida & US & 2,091 & \'{E}cole Polytechnique F\'{e}d\'{e}rale de Lausanne & CH & 5,125 & Sichuan University & CN & 9,436 \\  
\textit{29} & Kyushu University & JP & 2,043 & National Taiwan University & TW & 5,078 & Xi'an Jiaotong University & CN & 9,409 \\  
\textit{30} & National Institutes of Health & US & 2,000 & Peking University & CN & 4,921 & University of California, Berkeley & US & 9,320 \\  
\hdashline          
\textit{31} & The University of Texas at Austin & US & 1,992 & Lawrence Berkeley National Laboratory & US & 4,761 & Shandong University & CN & 9,317 \\  
\textit{32} & University of California, Los Angeles & US & 1,917 & Jilin University & CN & 4,731 & Korea Advanced Institute of Science and Technology & KR & 8,953 \\  
\textit{33} & National Institute of Standards and Technology & US & 1,912 & Shanghai Jiao Tong University & CN & 4,576 & ETH Zurich & CH & 8,716 \\  
\textit{34} & Hitachi (Japan) & JP & 1,906 & Korea Advanced Institute of Science and Technology & KR & 4,454 & \scalebox{0.82}[1]{National Institute of Advanced Industrial Science and Technology (AIST)} & JP & 8,685 \\  
\textit{35} & Polish Academy of Sciences & PL & 1,906 & University of Science and Technology of China & CN & 4,390 & National Taiwan University & TW & 8,552 \\  
\textit{36} & Technical University of Munich & DE & 1,892 & University of Oxford & GB & 4,385 & National Institute for Materials Science (NIMS) & JP & 8,514 \\  
\textit{37} & Tokyo University of Science & JP & 1,857 & The University of Texas at Austin & US & 4,363 & \'{E}cole Polytechnique F\'{e}d\'{e}rale de Lausanne & CH & 8,309 \\  
\textit{38} & California Institute of Technology & US & 1,806 & Lomonosov Moscow State University & RU & 4,344 & Tokyo Institute of Technology & JP & 8,287 \\  
\textit{39} & University of Toronto & CA & 1,789 & National Cheng Kung University & TW & 4,300 & Beijing National Laboratory for Molecular Sciences & CN & 8,202 \\  
\textit{40} & Northwestern University & US & 1,789 & Cornell University & US & 4,183 & University of Michigan{\textendash}Ann Arbor & US & 8,122 \\  
\hdashline          
\textit{41} & Hokkaido University & JP & 1,784 & University of California, Santa Barbara & US & 4,153 & Sungkyunkwan University & KR & 8,043 \\  
\textit{42} & Lomonosov Moscow State University & RU & 1,743 & National Yang Ming Chiao Tung University & TW & 4,115 & University of Oxford & GB & 7,934 \\  
\textit{43} & Sandia National Laboratories & US & 1,738 & Northwestern University & US & 4,113 & Imperial College London & GB & 7,910 \\  
\textit{44} & University of Paris-Sud & FR & 1,737 & National Tsing Hua University & TW & 4,109 & Central South University & CN & 7,824 \\  
\textit{45} & Los Alamos National Laboratory & US & 1,715 & University of Toronto & CA & 4,098 & University Surgical Associates & US & 7,720 \\  
\textit{46} & Argonne National Laboratory & US & 1,704 & Hokkaido University & JP & 3,959 & Lawrence Berkeley National Laboratory & US & 7,669 \\  
\textit{47} & Max Planck Society & DE & 1,696 & University of Washington & US & 3,946 & Dalian University of Technology & CN & 7,653 \\  
\textit{48} & The Ohio State University & US & 1,683 & University of Florida & US & 3,903 & Nankai University & CN & 7,632 \\  
\textit{49} & Johns Hopkins University & US & 1,682 & University of California, Los Angeles & US & 3,856 & University of Illinois Urbana-Champaign & US & 7,521 \\  
\textit{50} & Forschungszentrum J\"{u}lich & DE & 1,671 & ETH Zurich & CH & 3,760 & Nagoya University & JP & 7,483 \\  
          
 \\[-1.4em]
\hdashline \\[-1em]
\multicolumn{10}{r}{\scriptsize \emph{(continued to next page)}}
\end{tabular}}
}
\end{table}
\end{landscape}
}
\afterpage{\clearpage%
\begin{landscape}
\begin{table}[!t]\ContinuedFloat
\vspace{-3.3em}
\caption{\textbf{The top 100 productive institutions: \textcolor{violet}{\textit{\nano}}.} \emph{(Cont.)}}
\label{tab:r2_nano}
\vspace{2em}
{\tiny
{\renewcommand{\arraystretch}{1.2}
\begin{tabular}{rp{5cm}lr@{\hspace{4em}}p{5cm}lr@{\hspace{4em}}p{5cm}lr}\\[-5em] \toprule[1pt] \\[-1.4em]    
 & {\scriptsize \textbf{1991--2000}} & \multicolumn{2}{c}{No.\ Works} & {\scriptsize \textbf{2001--2010}} & \multicolumn{2}{c}{No.\ Works} & {\scriptsize \textbf{2011--2020}} & \multicolumn{2}{r}{No.\ Works} \\[-0.2em] \cmidrule[0.5pt](lr{4em}){2-4} \cmidrule[0.5pt](l{-0em}r{4em}){5-7} \cmidrule[0.5pt](l{-0em}r{1em}){8-10}
\textit{51} & Arizona State University & US & 1,666 & Fudan University & CN & 3,749 & Sun Yat-sen University & CN & 7,477 \\  
\textit{52} & University of Washington & US & 1,646 & Harbin Institute of Technology & CN & 3,748 & Russian Academy of Sciences & RU & 7,465 \\  
\textit{53} & University of Pennsylvania & US & 1,632 & Oak Ridge National Laboratory & US & 3,735 & Southeast University & CN & 7,393 \\  
\textit{54} & University of California, San Diego & US & 1,621 & University of Minnesota & US & 3,631 & The University of Texas at Austin & US & 7,382 \\  
\textit{55} & University of Arizona & US & 1,618 & Yonsei University & KR & 3,625 & Beihang University & CN & 7,367 \\  
\textit{56} & IBM Research - Thomas J. Watson Research Center & US & 1,574 & Purdue University West Lafayette & US & 3,613 & Karlsruhe Institute of Technology & DE & 7,361 \\  
\textit{57} & Imperial College London & GB & 1,540 & National Institute of Standards and Technology & US & 3,601 & Xiamen University & CN & 7,318 \\  
\textit{58} & Purdue University West Lafayette & US & 1,533 & University of Wisconsin{\textendash}Madison & US & 3,588 & Lomonosov Moscow State University & RU & 7,243 \\  
\textit{59} & Technical University of Berlin & DE & 1,521 & Polish Academy of Sciences & PL & 3,571 & Wuhan University & CN & 7,204 \\  
\textit{60} & Inserm & FR & 1,517 & Carnegie Mellon University & US & 3,563 & Beijing Institute of Technology & CN & 7,189 \\  
\hdashline          
\textit{61} & Yale University & US & 1,512 & Hanyang University & KR & 3,545 & Yonsei University & KR & 7,095 \\  
\textit{62} & University of Utah & US & 1,506 & The Ohio State University & US & 3,541 & Korea University & KR & 7,079 \\  
\textit{63} & Rutgers, The State University of New Jersey & US & 1,474 & Imperial College London & GB & 3,512 & University of Toronto & CA & 6,820 \\  
\textit{64} & KU Leuven & BE & 1,456 & Tokyo University of Science & JP & 3,490 & Kyushu University & JP & 6,803 \\  
\textit{65} & University of Tsukuba & JP & 1,433 & Shandong University & CN & 3,479 & Hanyang University & KR & 6,764 \\  
\textit{66} & Carnegie Mellon University & US & 1,425 & KU Leuven & BE & 3,439 & Northwestern University & US & 6,758 \\  
\textit{67} & Uppsala University & SE & 1,412 & Samsung (South Korea) & KR & 3,434 & Beijing University of Chemical Technology & CN & 6,753 \\  
\textit{68} & Georgia Institute of Technology & US & 1,409 & University of Tsukuba & JP & 3,348 & Pennsylvania State University & US & 6,672 \\  
\textit{69} & University of North Carolina at Chapel Hill & US & 1,406 & Arizona State University & US & 3,340 & University of Electronic Science and Technology of China & CN & 6,618 \\  
\textit{70} & Columbia University & US & 1,399 & University of California, San Diego & US & 3,302 & Northwestern Polytechnical University & CN & 6,515 \\  
\hdashline          
\textit{71} & National University of Singapore & SG & 1,386 & University of Pennsylvania & US & 3,240 & Hunan University & CN & 6,513 \\  
\textit{72} & Korea Advanced Institute of Science and Technology & KR & 1,384 & National Institutes of Health & US & 3,223 & University of Science and Technology Beijing & CN & 6,506 \\  
\textit{73} & University of Maryland, College Park & US & 1,383 & North Carolina State University & US & 3,222 & Shanghai University & CN & 6,415 \\  
\textit{74} & Max Planck Institute for Solid State Research & DE & 1,367 & Technical University of Munich & DE & 3,213 & UNSW Sydney & AU & 6,399 \\  
\textit{75} & RIKEN & JP & 1,342 & California Institute of Technology & US & 3,203 & University College London & GB & 6,362 \\  
\textit{76} & Princeton University & US & 1,335 & Universidade de S\~{a}o Paulo & BR & 3,153 & National Tsing Hua University & TW & 6,300 \\  
\textit{77} & Lund University & SE & 1,294 & Argonne National Laboratory & US & 3,148 & Universidade de S\~{a}o Paulo & BR & 6,252 \\  
\textit{78} & Texas A\&M University & US & 1,289 & Max Planck Society & DE & 3,119 & Changchun Institute of Applied Chemistry & CN & 6,191 \\  
\textit{79} & Freie Universit\"{a}t Berlin & DE & 1,283 & Johns Hopkins University & US & 3,108 & University of California, Los Angeles & US & 6,121 \\  
\textit{80} & Lawrence Livermore National Laboratory & US & 1,281 & University of Maryland, College Park & US & 3,086 & East China University of Science and Technology & CN & 6,119 \\  
\hdashline          
\textit{81} & Chalmers University of Technology & SE & 1,271 & Pohang University of Science and Technology & KR & 3,039 & Argonne National Laboratory & US & 6,112 \\  
\textit{82} & University of Stuttgart & DE & 1,262 & Nankai University & CN & 3,022 & Oak Ridge National Laboratory & US & 6,103 \\  
\textit{83} & University of Manchester & GB & 1,250 & University of California, Davis & US & 2,960 & City University of Hong Kong & CN & 6,092 \\  
\textit{84} & ETH Zurich & CH & 1,242 & Texas A\&M University & US & 2,952 & National Yang Ming Chiao Tung University & TW & 6,018 \\  
\textit{85} & State University of New York & US & 1,236 & University of Paris-Sud & FR & 2,925 & University of Washington & US & 6,005 \\  
\textit{86} & Howard Hughes Medical Institute & US & 1,233 & Korea University & KR & 2,909 & University of California, San Diego & US & 5,996 \\  
\textit{87} & University of Erlangen-Nuremberg & DE & 1,229 & Delft University of Technology & NL & 2,902 & KU Leuven & BE & 5,982 \\  
\textit{88} & Hebrew University of Jerusalem & IL & 1,225 & University of Manchester & GB & 2,896 & Purdue University West Lafayette & US & 5,939 \\  
\textit{89} & Seoul National University & KR & 1,221 & Columbia University & US & 2,865 & Technical University of Munich & DE & 5,933 \\  
\textit{90} & Nanjing University & CN & 1,217 & Institute of Chemistry & CN & 2,856 & Chongqing University & CN & 5,915 \\  
\hdashline          
\textit{91} & Case Western Reserve University & US & 1,216 & Los Alamos National Laboratory & US & 2,840 & Pohang University of Science and Technology & KR & 5,888 \\  
\textit{92} & NEC (Japan) & JP & 1,213 & United States Naval Research Laboratory & US & 2,826 & Cornell University & US & 5,804 \\  
\textit{93} & Hiroshima University & JP & 1,211 & University of Southern California & US & 2,783 & University of Tehran & IR & 5,765 \\  
\textit{94} & University of W\"{u}rzburg & DE & 1,194 & Huazhong University of Science and Technology & CN & 2,775 & King Abdulaziz University & SA & 5,750 \\  
\textit{95} & Institute of Physics & PL & 1,187 & Sungkyunkwan University & KR & 2,738 & Italian Institute of Technology & IT & 5,741 \\  
\textit{96} & University of Southern California & US & 1,186 & Korea Institute of Science and Technology & KR & 2,729 & King Saud University & SA & 5,723 \\  
\textit{97} & IBM Research - Almaden & US & 1,183 & Eindhoven University of Technology & NL & 2,723 & Japan Science and Technology Agency (JST) & JP & 5,709 \\  
\textit{98} & University of Barcelona & ES & 1,179 & Indian Institute of Science Bangalore & IN & 2,706 & Korea Institute of Science and Technology & KR & 5,685 \\  
\textit{99} & Leiden University & NL & 1,179 & National Academy of Sciences of Ukraine & UA & 2,703 & Zhengzhou University & CN & 5,680 \\  
\textit{100} & University of California, Davis & US & 1,179 & University College London & GB & 2,692 & Wuhan University of Technology & CN & 5,657 \\  
          
 \\[-1.4em]
\bottomrule
\end{tabular}}
}
\end{table}
\end{landscape}
}
%

\titleformat{\section}{\sc\centering\LARGE\bfseries}{\textsc{\thesection}.\!\!}{1em}{}

\afterpage{\clearpage%
\markboth{\textbf \textsc{\agri}}{}
\thispagestyle{empty}
\quad
\vspace{2cm}
\begin{center}
\pgfornament[width=0.5*\textwidth,symmetry=h]{89}\\[2em]
\section{\agri}
\vspace{1em}
\pgfornament[width=0.5*\textwidth]{89}
\end{center}
}

\afterpage{\clearpage%

\begin{figure}[!tp]
\centering
\vspace{-1em}
{\large \textbf{\textrm{{World Map of \textcolor{violet}{\textit{\agri}} Collaboration}}}~|~1971--2020}\\
\vspace{0.3cm}
\includegraphics[align=c, scale=0.054, trim={9.5cm 0 9.5cm 0},clip]{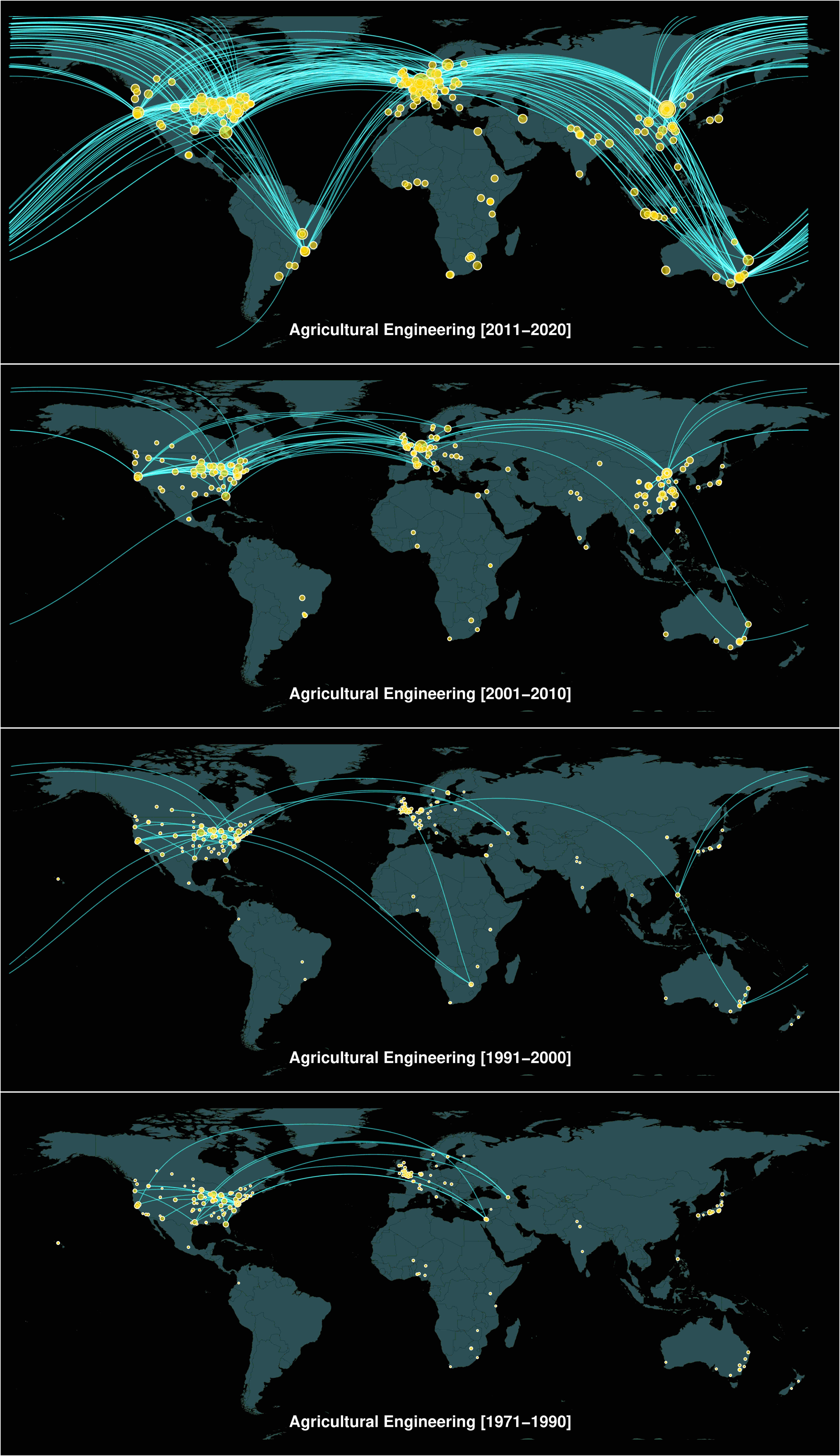}
\caption[{\agri}]{\textbf{(a)~|~The World Map of \textcolor{violet}{\textit{\agri}} Collaboration.}
The bubbles represent the top 199 institutions in terms of work production, with their sizes proportional to the work volume. 
The connecting lines depict coauthorship relationships among the top 50 institutions.}
\label{fig:wmap_agri}
\end{figure}
}
\afterpage{\clearpage%
\begin{figure}[!tp]\ContinuedFloat
\centering
\vspace{-1em}
{\large \textbf{Top 30 Productive Institutions on the World Map: \textcolor{violet}{\textit{\agri}}}~|~1991--2020}\\
\vspace{-0em}
\hspace*{-3em}                                                           
\includegraphics[align=c, scale=0.83]{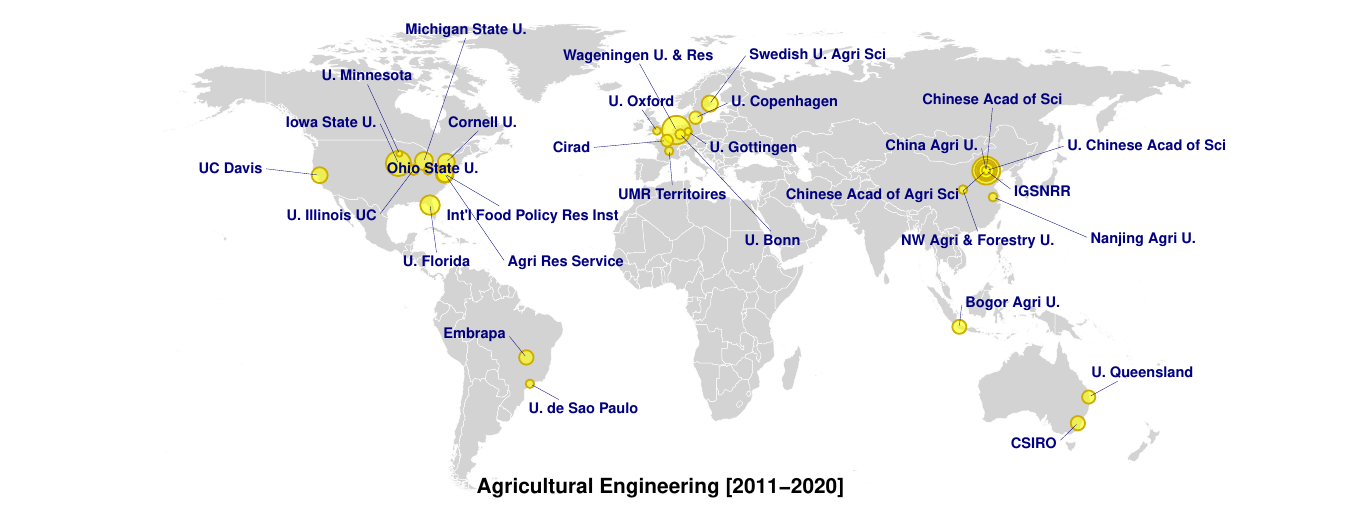}\\[-0.5em]
\quad\\[-1em]
\dotfill 
\quad\\[-0em]
\hspace*{-3em}                                                           
\includegraphics[align=c, scale=0.83]{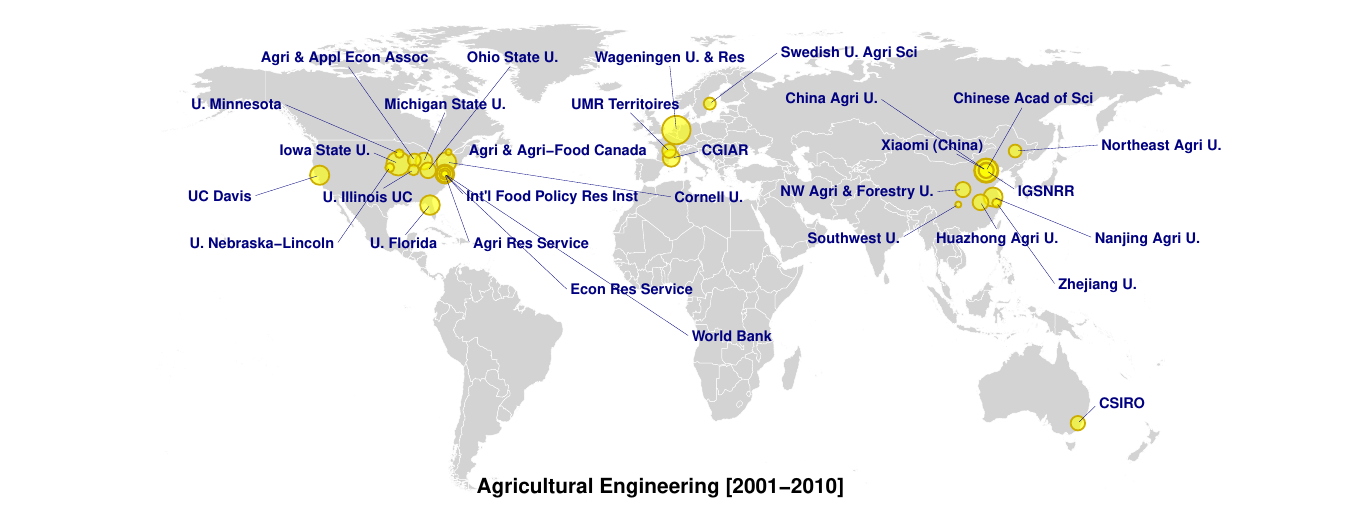}\\[-0.5em]
\quad\\[-1em]
\dotfill 
\quad\\[-0em]
\hspace*{-3em}                                                           
\includegraphics[align=c, scale=0.83]{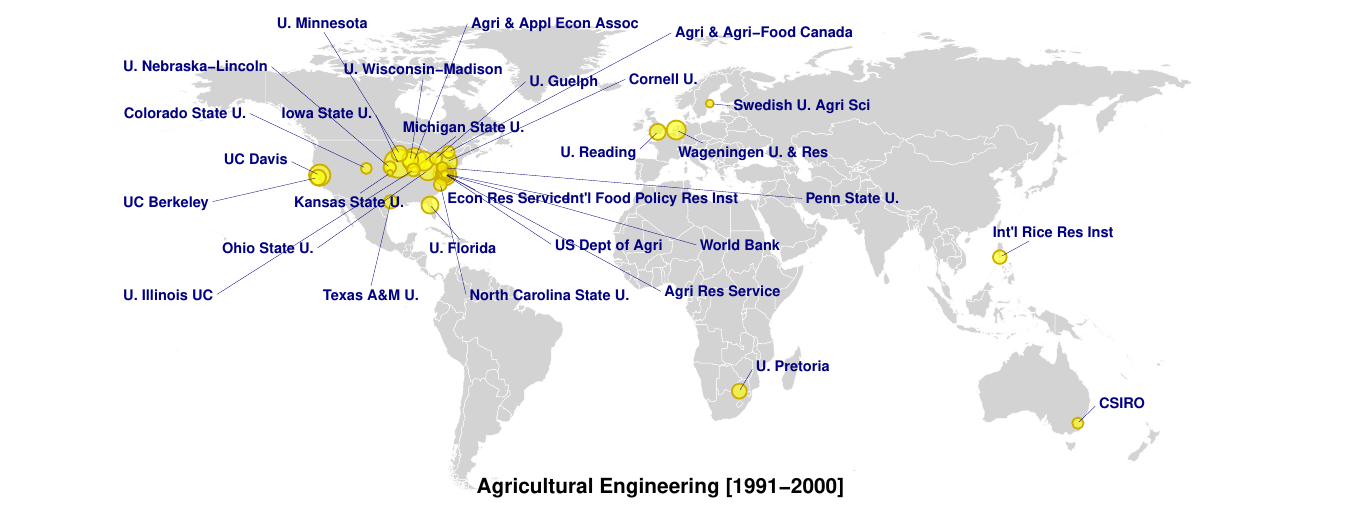}\\[-0.8em]
\caption[{\agri}]{\textbf{(b)~|~The top 30 productive institutions on the World Map: \textcolor{violet}{\textit{\agri}}.}
The bubbles represent the top 30 institutions in terms of work production, with their sizes proportional to the work volume.}
\label{fig:wmap_topinst_agri}
\end{figure}
}
\afterpage{\clearpage%
\begin{figure}[!tp]\ContinuedFloat
\centering
\vspace{-1em}
{\large \textbf{\textrm{{Interregional \textcolor{violet}{\textit{\agri}} Collaboration}}}~|~1991--2020}\\
\vspace{0.5em}
\hspace{-5em}\includegraphics[align=c, scale=1.7, vmargin=0mm]{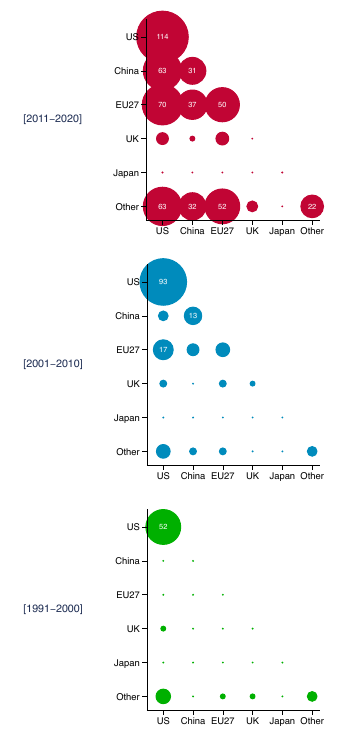}
\vspace{-1em}
\caption[]{\textbf{(c)~|~The Interregional \textcolor{violet}{\textit{\agri}} Collaboration Matrix Diagram.}
The bubble size represents the number of coauthorship relationships for the top 50 institutions in terms of work production. 
If the number is equal to or greater than 10, it is displayed inside the bubble.}
\label{fig:halfmat_agri}
\end{figure}
}
\afterpage{\clearpage%
\begin{figure}[!tp]\ContinuedFloat
\centering
\vspace{-1em}
{\large \textbf{\textrm{Interinstitutional \textcolor{violet}{\textit{\agri}} Collaboration}}~|~2001--2020\quad {\footnotesize \emph{(continued to next page)}}}\\
\cdend{agri}{2010}{2000}{2011--2020}{2001--2010}\\[-1.5em]
\caption[]{\textbf{(d)~|~The Interinstitutional \textcolor{violet}{\textit{\agri}} Collaboration Dendrogram.}
The top 50 institutions in terms of work production, indicated by the circularised bar graphs, are displayed.
}
\label{fig:cdend1_agri}
\end{figure}
}
\afterpage{\clearpage%
\begin{figure}[!tp]\ContinuedFloat
\centering
\vspace{-1em}
{\large \textbf{\textrm{Interinstitutional \textcolor{violet}{\textit{\agri}} Collaboration}}~|~1971--2000\quad {\footnotesize \emph{(continued from previous page)}}}\\
\cdend{agri}{1990}{1980}{1991--2000}{1971--1990}\\[-1.5em]
\caption[]{\textbf{(d)~|~The Interinstitutional \textcolor{violet}{\textit{\agri}} Collaboration Dendrogram.} \emph{(Cont.)}\hfill~}
\label{fig:cdend2_agri}
\end{figure}
}
\afterpage{\clearpage%
\begin{landscape}
\begin{table}[!t]
\vspace{-0.5em}
\caption{\textbf{The top 100 productive institutions: \textcolor{violet}{\textit{\agri}}.}}
\label{tab:r1_agri}
\vspace{2em}
\centering
{\tiny
{\renewcommand{\arraystretch}{1.2}
\begin{tabular}{rp{5cm}lr@{\hspace{4em}}p{5cm}lr@{\hspace{4em}}p{5cm}lr}\\[-5em] \toprule[1pt] \\[-1.4em]    
 & {\scriptsize \textbf{1991--2000}} & \multicolumn{2}{c}{No.~Works} & {\scriptsize \textbf{2001--2010}} & \multicolumn{2}{c}{No.~Works} & {\scriptsize \textbf{2011--2020}} & \multicolumn{2}{r}{No.~Works} \\[-0.2em] \cmidrule[0.5pt](lr{4em}){2-4} \cmidrule[0.5pt](l{-0em}r{4em}){5-7} \cmidrule[0.5pt](l{-0em}r{1em}){8-10}
\textit{1} & Iowa State University & US & 588 & Wageningen University \& Research & NL & 1,467 & Wageningen University \& Research & NL & 3,146 \\  
\textit{2} & Agricultural \& Applied Economics Association & US & 399 & Chinese Academy of Sciences & CN & 1,031 & Chinese Academy of Sciences & CN & 3,015 \\  
\textit{3} & Agricultural Research Service & US & 362 & Iowa State University & US & 1,003 & Iowa State University & US & 2,421 \\  
\textit{4} & University of California, Davis & US & 322 & China Agricultural University & CN & 910 & China Agricultural University & CN & 1,818 \\  
\textit{5} & Cornell University & US & 319 & University of Florida & US & 699 & University of Florida & US & 1,573 \\  
\textit{6} & The Ohio State University & US & 291 & Cornell University & US & 696 & Michigan State University & US & 1,435 \\  
\textit{7} & Michigan State University & US & 271 & University of California, Davis & US & 680 & University of Chinese Academy of Sciences & CN & 1,402 \\  
\textit{8} & Wageningen University \& Research & NL & 268 & Nanjing Agricultural University & CN & 672 & International Food Policy Research Institute & US & 1,394 \\  
\textit{9} & Economic Research Service & US & 244 & Agricultural Research Service & US & 662 & Swedish University of Agricultural Sciences & SE & 1,275 \\  
\textit{10} & University of Florida & US & 232 & CGIAR & FR & 582 & Cornell University & US & 1,269 \\  
\hdashline          
\textit{11} & University of Reading & GB & 217 & Michigan State University & US & 572 & Agricultural Research Service & US & 1,257 \\  
\textit{12} & University of Wisconsin{\textendash}Madison & US & 217 & World Bank & US & 540 & University of California, Davis & US & 1,250 \\  
\textit{13} & World Bank & US & 215 & Huazhong Agricultural University & CN & 538 & \scalebox{0.9}[1]{Institute of Geographic Sciences and Natural Resources Research} & CN & 1,213 \\  
\textit{14} & University of Minnesota & US & 212 & The Ohio State University & US & 533 & Brazilian Agricultural Research Corporation & BR & 1,153 \\  
\textit{15} & University of California, Berkeley & US & 210 & Xiaomi (China) & CN & 519 & \scalebox{0.9}[1]{Commonwealth Scientific and Industrial Research Organisation} & AU & 1,144 \\  
\textit{16} & University of Pretoria & ZA & 201 & North West Agriculture and Forestry University & CN & 516 & Bogor Agricultural University & ID & 1,132 \\  
\textit{17} & International Rice Research Institute & PH & 188 & \scalebox{0.9}[1]{Institute of Geographic Sciences and Natural Resources Research} & CN & 500 & University of Queensland & AU & 1,089 \\  
\textit{18} & Texas A\&M University & US & 186 & \scalebox{0.9}[1]{Commonwealth Scientific and Industrial Research Organisation} & AU & 493 & University of Copenhagen & DK & 1,060 \\  
\textit{19} & University of Illinois Urbana-Champaign & US & 181 & UMR Territoires & FR & 479 & \scalebox{0.7}[1]{Centre de Coop\'{e}ration Internationale en Recherche Agronomique pour le D\'{e}veloppement} & FR & 1,032 \\  
\textit{20} & North Carolina State University & US & 180 & Agricultural \& Applied Economics Association & US & 457 & University of Bonn & DE & 978 \\  
\hdashline          
\textit{21} & University of Nebraska{\textendash}Lincoln & US & 178 & Northeast Agricultural University & CN & 454 & University of Illinois Urbana-Champaign & US & 978 \\  
\textit{22} & University of Guelph & CA & 176 & Swedish University of Agricultural Sciences & SE & 439 & Nanjing Agricultural University & CN & 947 \\  
\textit{23} & International Food Policy Research Institute & US & 176 & International Food Policy Research Institute & US & 427 & North West Agriculture and Forestry University & CN & 947 \\  
\textit{24} & Agriculture and Agri-Food Canada & CA & 175 & University of Illinois Urbana-Champaign & US & 419 & Chinese Academy of Agricultural Sciences & CN & 946 \\  
\textit{25} & \scalebox{0.9}[1]{Commonwealth Scientific and Industrial Research Organisation} & AU & 169 & Zhejiang University & CN & 399 & The Ohio State University & US & 946 \\  
\textit{26} & Pennsylvania State University & US & 169 & University of Nebraska{\textendash}Lincoln & US & 398 & Universidade de S\~{a}o Paulo & BR & 942 \\  
\textit{27} & Colorado State University & US & 168 & University of Minnesota & US & 395 & UMR Territoires & FR & 941 \\  
\textit{28} & Swedish University of Agricultural Sciences & SE & 155 & Economic Research Service & US & 389 & University of Oxford & GB & 935 \\  
\textit{29} & United States Department of Agriculture & US & 154 & Agriculture and Agri-Food Canada & CA & 387 & University of G\"{o}ttingen & DE & 931 \\  
\textit{30} & Kansas State University & US & 153 & Southwest University & CN & 386 & University of Minnesota & US & 925 \\  
\hdashline          
\textit{31} & University of Arizona & US & 146 & Colorado State University & US & 376 & Aarhus University & DK & 903 \\  
\textit{32} & University of Queensland & AU & 142 & University of Wisconsin{\textendash}Madison & US & 375 & Pennsylvania State University & US & 903 \\  
\textit{33} & Washington State University & US & 142 & South China Agricultural University & CN & 369 & University of Wisconsin{\textendash}Madison & US & 870 \\  
\textit{34} & Virginia Tech & US & 139 & Australian National University & AU & 357 & University of Nebraska{\textendash}Lincoln & US & 836 \\  
\textit{35} & University of Oxford & GB & 138 & KU Leuven & BE & 356 & Beijing Normal University & CN & 822 \\  
\textit{36} & Oregon State University & US & 135 & University of Copenhagen & DK & 356 & Huazhong Agricultural University & CN & 822 \\  
\textit{37} & University of Maryland, College Park & US & 132 & University of California, Berkeley & US & 356 & Colorado State University & US & 809 \\  
\textit{38} & Purdue University West Lafayette & US & 121 & Food and Agriculture Organization of the United Nations & IT & 344 & Ghent University & BE & 805 \\  
\textit{39} & Public Economics & FR & 120 & University of Queensland & AU & 337 & Agriculture and Food & AU & 804 \\  
\textit{40} & Australian National University & AU & 119 & Pennsylvania State University & US & 337 & Gadjah Mada University & ID & 794 \\  
\hdashline          
\textit{41} & Kyoto University & JP & 114 & University of Reading & GB & 332 & S\~{a}o Paulo State University & BR & 778 \\  
\textit{42} & University of London & GB & 113 & University of Hohenheim & DE & 331 & University of Li\`{e}ge & BE & 761 \\  
\textit{43} & University of Georgia & GE & 112 & \scalebox{0.7}[1]{Centre de Coop\'{e}ration Internationale en Recherche Agronomique pour le D\'{e}veloppement} & FR & 330 & University of Tehran & IR & 761 \\  
\textit{44} & University of Kentucky & US & 112 & Kansas State University & US & 327 & Texas A\&M University & US & 755 \\  
\textit{45} & University of Georgia & US & 111 & Washington State University & US & 321 & University of California, Berkeley & US & 752 \\  
\textit{46} & University of Cambridge & GB & 110 & Hebei Agricultural University & CN & 320 & Australian National University & AU & 747 \\  
\textit{47} & University of East Anglia & GB & 110 & Ghent University & BE & 312 & University of Hohenheim & DE & 746 \\  
\textit{48} & \scalebox{0.9}[1]{National Research Institute for Agriculture, Food and Environment} & FR & 108 & Hunan Agricultural University & CN & 310 & Agro ParisTech & FR & 731 \\  
\textit{49} & University of Missouri & US & 108 & University of Guelph & CA & 309 & Agriculture and Agri-Food Canada & CA & 725 \\  
\textit{50} & University of Saskatchewan & CA & 104 & Institute of Soil and Water Conservation & CN & 308 & University of Guelph & CA & 717 \\  
          
 \\[-1.4em]
\hdashline \\[-1em]
\multicolumn{10}{r}{\scriptsize \emph{(continued to next page)}}
\end{tabular}}
}
\end{table}
\end{landscape}
}
\afterpage{\clearpage%
\begin{landscape}
\begin{table}[!t]\ContinuedFloat
\vspace{-3.3em}
\caption{\textbf{The top 100 productive institutions: \textcolor{violet}{\textit{\agri}}.} \emph{(Cont.)}}
\label{tab:r2_agri}
\vspace{2em}
{\tiny
{\renewcommand{\arraystretch}{1.2}
\begin{tabular}{rp{5cm}lr@{\hspace{4em}}p{5cm}lr@{\hspace{4em}}p{5cm}lr}\\[-5em] \toprule[1pt] \\[-1.4em]    
 & {\scriptsize \textbf{1991--2000}} & \multicolumn{2}{c}{No.\ Works} & {\scriptsize \textbf{2001--2010}} & \multicolumn{2}{c}{No.\ Works} & {\scriptsize \textbf{2011--2020}} & \multicolumn{2}{r}{No.\ Works} \\[-0.2em] \cmidrule[0.5pt](lr{4em}){2-4} \cmidrule[0.5pt](l{-0em}r{4em}){5-7} \cmidrule[0.5pt](l{-0em}r{1em}){8-10}
\textit{51} & Silsoe Research Institute & GB & 103 & University of Oxford & GB & 303 & University of Melbourne & AU & 715 \\  
\textit{52} & Newcastle University & GB & 102 & Shenyang Agricultural University & CN & 300 & University of KwaZulu-Natal & ZA & 708 \\  
\textit{53} & Kyushu University & JP & 99 & North Carolina State University & US & 299 & Food and Agriculture Organization of the United Nations & IT & 703 \\  
\textit{54} & Rothamsted Research & GB & 98 & Texas A\&M University & US & 298 & Indian Agricultural Research Institute & IN & 702 \\  
\textit{55} & Graduate School Experimental Plant Sciences & NL & 95 & Beijing Normal University & CN & 296 & Warsaw University of Life Sciences & PL & 699 \\  
\textit{56} & UMR Territoires & FR & 94 & University of G\"{o}ttingen & DE & 287 & University of British Columbia & CA & 698 \\  
\textit{57} & \scalebox{0.72}[1]{Centre d'Economie et de Sociologie Appliqu\'{e}es \`{a} l'Agriculture et aux Espaces Ruraux} & FR & 93 & Brazilian Agricultural Research Corporation & BR & 283 & North Carolina State University & US & 693 \\  
\textit{58} & The University of Tokyo & JP & 93 & Chinese Academy of Agricultural Sciences & CN & 283 & Washington State University & US & 693 \\  
\textit{59} & University College London & GB & 89 & University of Debrecen & HU & 282 & KU Leuven & BE & 691 \\  
\textit{60} & University of Bristol & GB & 89 & The University of Tokyo & JP & 276 & University of Western Australia & AU & 688 \\  
\hdashline          
\textit{61} & Agricultural Research Organization & IL & 89 & United States Department of Agriculture & US & 271 & Zhejiang University & CN & 683 \\  
\textit{62} & Hebrew University of Jerusalem & IL & 89 & China Tobacco & CN & 270 & University of Cambridge & GB & 681 \\  
\textit{63} & University of Edinburgh & GB & 88 & Fujian Agriculture and Forestry University & CN & 266 & Consejo Nacional de Investigaciones Cient\'{i}ficas y T\'{e}cnicas & AR & 673 \\  
\textit{64} & Institut National de la Recherche Agronomique du Niger & NE & 87 & University of Missouri & US & 263 & McGill University & CA & 671 \\  
\textit{65} & University of Alberta & CA & 86 & Oregon State University & US & 256 & Bangladesh Agricultural University & BD & 669 \\  
\textit{66} & Chinese Academy of Sciences & CN & 85 & Agricultural Research Center & EG & 252 & University of Pretoria & ZA & 652 \\  
\textit{67} & D\'{e}partement Environnement et Agronomie & FR & 84 & Wuhan University & CN & 247 & Agroecology & FR & 647 \\  
\textit{68} & Utah State University & US & 84 & Henan Agricultural University & CN & 246 & University of Sydney & AU & 629 \\  
\textit{69} & Rutgers, The State University of New Jersey & US & 83 & University of Bonn & DE & 245 & Kansas State University & US & 625 \\  
\textit{70} & KU Leuven & BE & 82 & Nanjing University & CN & 244 & The University of Tokyo & JP & 624 \\  
\hdashline          
\textit{71} & Auburn University & US & 82 & University of Cambridge & GB & 240 & Agricultural \& Applied Economics Association & US & 621 \\  
\textit{72} & Tokyo University of Agriculture & JP & 79 & University of Adelaide & AU & 237 & World Bank & US & 614 \\  
\textit{73} & World Agroforestry Centre & KE & 79 & Mississippi State University & US & 237 & China Tobacco & CN & 612 \\  
\textit{74} & University of Melbourne & AU & 78 & Renmin University of China & CN & 236 & Purdue University West Lafayette & US & 603 \\  
\textit{75} & University of New England & AU & 78 & University of Melbourne & AU & 232 & University of Maryland, College Park & US & 603 \\  
\textit{76} & University of Tsukuba & JP & 78 & \scalebox{0.9}[1]{Montpellier Interdisciplinary center on Sustainable Agri-food systems} & FR & 231 & Agricultural Research Center & EG & 598 \\  
\textit{77} & Environmental Protection Agency & US & 77 & Kyoto University & JP & 228 & United States Department of Agriculture & US & 580 \\  
\textit{78} & University of Adelaide & AU & 76 & University of Pretoria & ZA & 228 & South China Agricultural University & CN & 578 \\  
\textit{79} & Stanford University & US & 76 & Gansu Agricultural University & CN & 223 & Stanford University & US & 576 \\  
\textit{80} & University of British Columbia & CA & 75 & University of Chinese Academy of Sciences & CN & 223 & University of Natural Resources and Life Sciences, Vienna & AT & 575 \\  
\hdashline          
\textit{81} & University of Manchester & GB & 75 & University of Arizona & US & 221 & University of Missouri & US & 570 \\  
\textit{82} & \scalebox{0.9}[1]{International Center for Agricultural Research in the Dry Areas} & SY & 75 & Newcastle University & GB & 219 & ETH Zurich & CH & 568 \\  
\textit{83} & University of Western Australia & AU & 74 & Beijing Forestry University & CN & 218 & Peking University & CN & 568 \\  
\textit{84} & University of Sheffield & GB & 74 & Peking University & CN & 218 & University of Adelaide & AU & 566 \\  
\textit{85} & \scalebox{0.9}[1]{Agricultural Development Advisory Service (United Kingdom)} & GB & 73 & Rothamsted Research & GB & 217 & Columbia University & US & 562 \\  
\textit{86} & Making View (Norway) & NO & 73 & University of Western Australia & AU & 216 & Norwegian University of Life Sciences & NO & 549 \\  
\textit{87} & University of Wales & GB & 71 & Imperial College London & GB & 214 & Humboldt-Universit\"{a}t zu Berlin & DE & 546 \\  
\textit{88} & University of Tennessee at Knoxville & US & 71 & University of Sydney & AU & 209 & Arizona State University & US & 545 \\  
\textit{89} & University of Sussex & GB & 70 & Virginia Tech & US & 208 & Universiti Putra Malaysia & MY & 542 \\  
\textit{90} & Asian Institute of Technology & TH & 70 & Shihezi University & CN & 207 & University of Michigan{\textendash}Ann Arbor & US & 538 \\  
\hdashline          
\textit{91} & University of Zimbabwe & ZW & 70 & University of British Columbia & CA & 206 & Indian Council of Agricultural Research & IN & 533 \\  
\textit{92} & University of Sydney & AU & 69 & Yunnan Agricultural University & CN & 205 & University of Edinburgh & GB & 528 \\  
\textit{93} & University of Hohenheim & DE & 68 & University of Maryland, College Park & US & 205 & University of Ghana & GH & 528 \\  
\textit{94} & University of Leicester & GB & 68 & Lanzhou University & CN & 201 & Stellenbosch University & ZA & 528 \\  
\textit{95} & University of Birmingham & GB & 67 & University of Georgia & GE & 201 & University of Reading & GB & 524 \\  
\textit{96} & Oklahoma State University & US & 67 & McGill University & CA & 200 & Harvard University & US & 523 \\  
\textit{97} & Cranfield University & GB & 66 & Universidade de S\~{a}o Paulo & BR & 199 & Lund University & SE & 521 \\  
\textit{98} & GTx (United States) & US & 66 & Guizhou University & CN & 198 & University of Life Sciences in Lublin & PL & 518 \\  
\textit{99} & Natural Resources Conservation Service & US & 66 & Shanxi Agricultural University & CN & 198 & Oregon State University & US & 514 \\  
\textit{100} & University of Arkansas at Fayetteville & US & 66 & Aarhus University & DK & 195 & Universidade Federal de Vi\c{c}osa & BR & 511 \\  
          
 \\[-1.4em]
\bottomrule
\end{tabular}}
}
\end{table}
\end{landscape}
}
%

\titleformat{\section}{\sc\centering\LARGE\bfseries}{\textsc{\thesection}.\!\!}{1em}{}

\afterpage{\clearpage%
\markboth{\textbf \textsc{\particle}}{}
\thispagestyle{empty}
\quad
\vspace{2cm}
\begin{center}
\pgfornament[width=0.5*\textwidth,symmetry=h]{89}\\[2em]
\section{\particle}
\vspace{1em}
\pgfornament[width=0.5*\textwidth]{89}
\end{center}
}

\afterpage{\clearpage%

\begin{figure}[!tp]
\centering
\vspace{-1em}
{\large \textbf{\textrm{{World Map of \textcolor{violet}{\textit{\particle}} Collaboration}}}~|~1971--2020}\\
\vspace{0.3cm}
\includegraphics[align=c, scale=0.054, trim={9.5cm 0 9.5cm 0},clip]{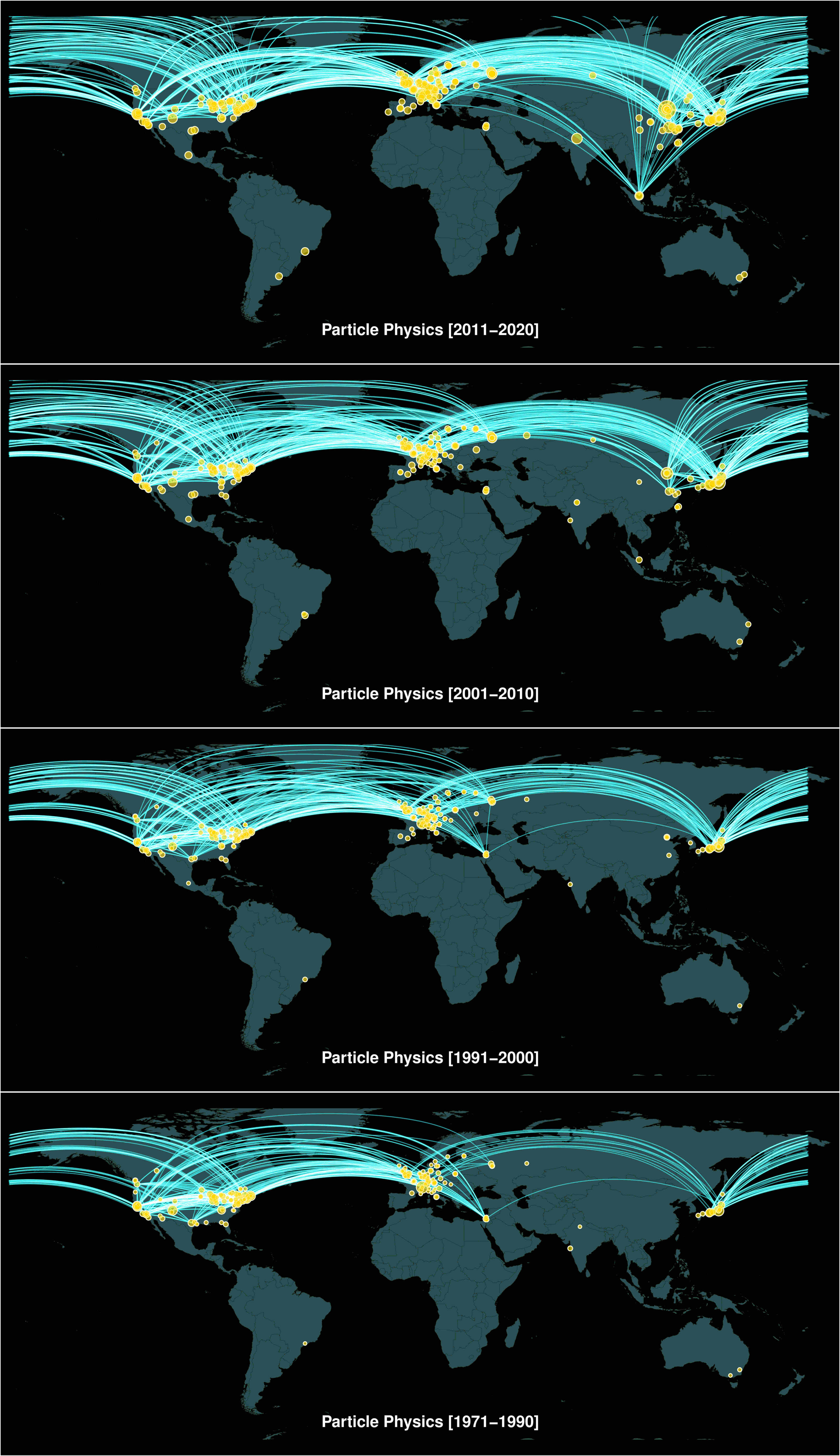}
\caption[{\particle}]{\textbf{(a)~|~The World Map of \textcolor{violet}{\textit{\particle}} Collaboration.}
The bubbles represent the top 199 institutions in terms of work production, with their sizes proportional to the work volume. 
The connecting lines depict coauthorship relationships among the top 50 institutions.}
\label{fig:wmap_particle}
\end{figure}
}
\afterpage{\clearpage%
\begin{figure}[!tp]\ContinuedFloat
\centering
\vspace{-1em}
{\large \textbf{Top 30 Productive Institutions on the World Map: \textcolor{violet}{\textit{\particle}}}~|~1991--2020}\\
\vspace{-0em}
\hspace*{-3em}                                                           
\includegraphics[align=c, scale=0.83]{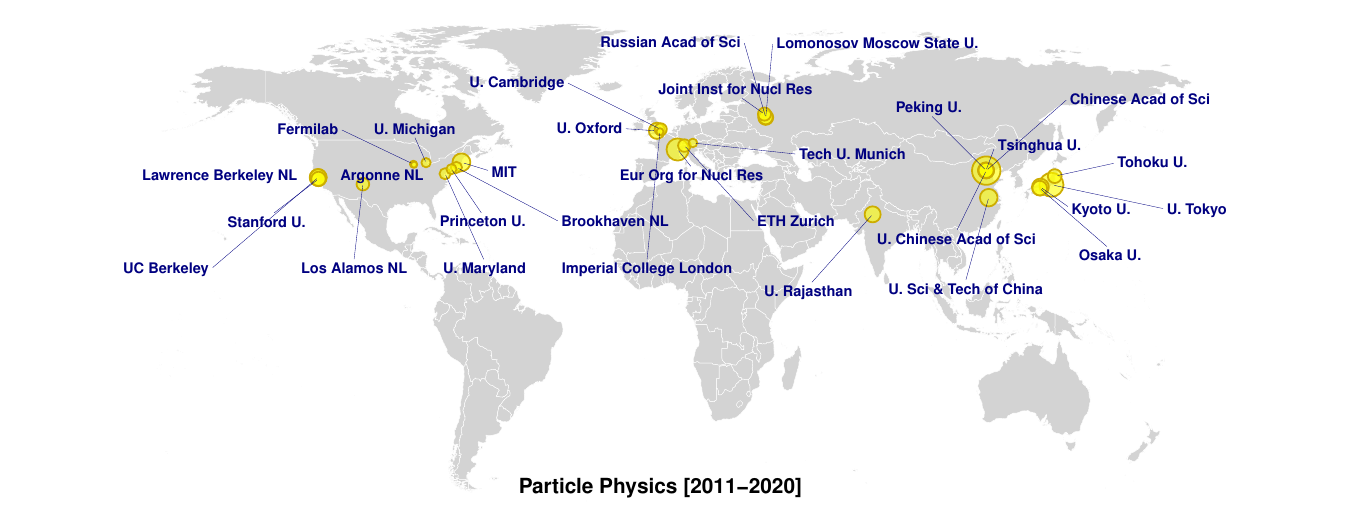}\\[-0.5em]
\quad\\[-1em]
\dotfill 
\quad\\[-0em]
\hspace*{-3em}                                                           
\includegraphics[align=c, scale=0.83]{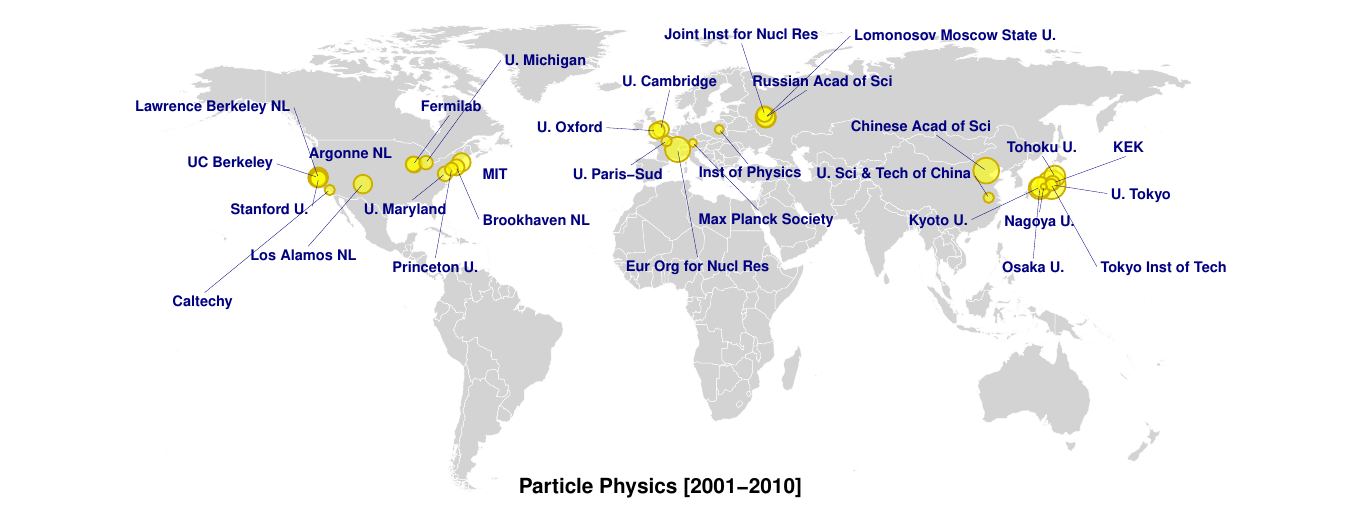}\\[-0.5em]
\quad\\[-1em]
\dotfill 
\quad\\[-0em]
\hspace*{-3em}                                                           
\includegraphics[align=c, scale=0.83]{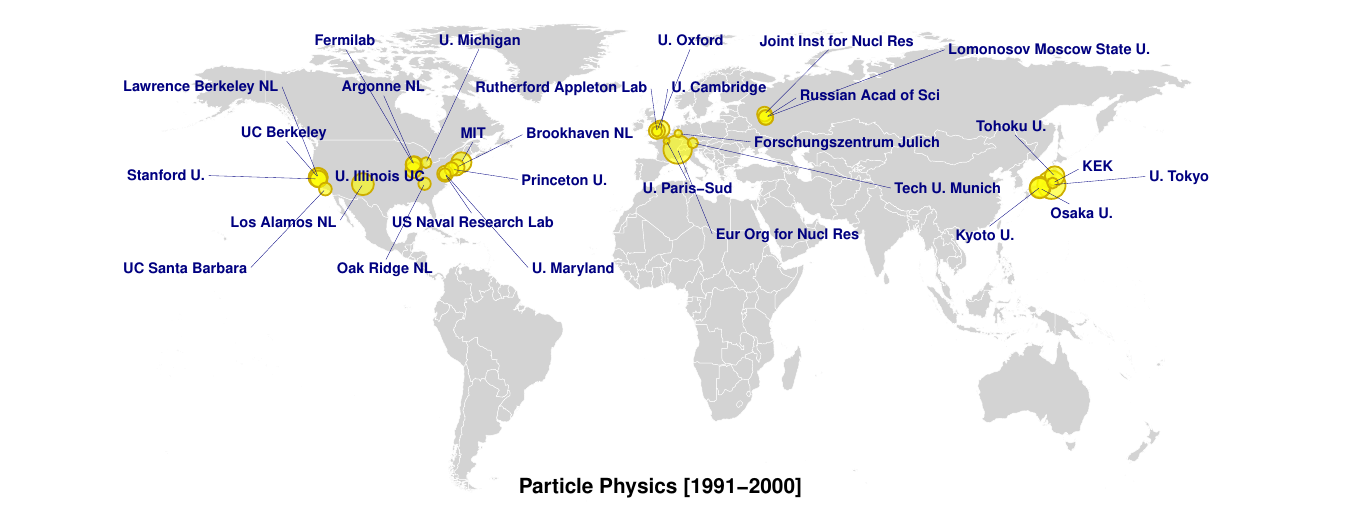}\\[-0.8em]
\caption[{\particle}]{\textbf{(b)~|~The top 30 productive institutions on the World Map: \textcolor{violet}{\textit{\particle}}.}
The bubbles represent the top 30 institutions in terms of work production, with their sizes proportional to the work volume.}
\label{fig:wmap_topinst_particle}
\end{figure}
}
\afterpage{\clearpage%
\begin{figure}[!tp]\ContinuedFloat
\centering
\vspace{-1em}
{\large \textbf{\textrm{{Interregional \textcolor{violet}{\textit{\particle}} Collaboration}}}~|~1991--2020}\\
\vspace{0.5em}
\hspace{-5em}\includegraphics[align=c, scale=1.7, vmargin=0mm]{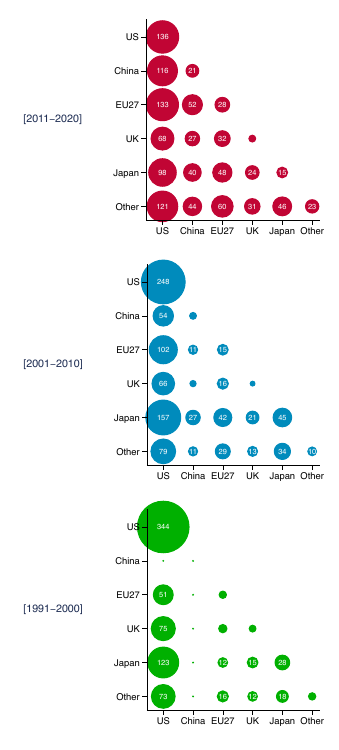}
\vspace{-1em}
\caption[]{\textbf{(c)~|~The Interregional \textcolor{violet}{\textit{\particle}} Collaboration Matrix Diagram.}
The bubble size represents the number of coauthorship relationships for the top 50 institutions in terms of work production. 
If the number is equal to or greater than 10, it is displayed inside the bubble.}
\label{fig:halfmat_particle}
\end{figure}
}
\afterpage{\clearpage%
\begin{figure}[!tp]\ContinuedFloat
\centering
\vspace{-1em}
{\large \textbf{\textrm{Interinstitutional \textcolor{violet}{\textit{\particle}} Collaboration}}~|~2001--2020\quad {\footnotesize \emph{(continued to next page)}}}\\
\cdend{particle}{2010}{2000}{2011--2020}{2001--2010}\\[-1.5em]
\caption[]{\textbf{(d)~|~The Interinstitutional \textcolor{violet}{\textit{\particle}} Collaboration Dendrogram.}
The top 50 institutions in terms of work production, indicated by the circularised bar graphs, are displayed.
}
\label{fig:cdend1_particle}
\end{figure}
}
\afterpage{\clearpage%
\begin{figure}[!tp]\ContinuedFloat
\centering
\vspace{-1em}
{\large \textbf{\textrm{Interinstitutional \textcolor{violet}{\textit{\particle}} Collaboration}}~|~1971--2000\quad {\footnotesize \emph{(continued from previous page)}}}\\
\cdend{particle}{1990}{1980}{1991--2000}{1971--1990}\\[-1.5em]
\caption[]{\textbf{(d)~|~The Interinstitutional \textcolor{violet}{\textit{\particle}} Collaboration Dendrogram.} \emph{(Cont.)}\hfill~}
\label{fig:cdend2_particle}
\end{figure}
}
\afterpage{\clearpage%
\begin{landscape}
\begin{table}[!t]
\vspace{-0.5em}
\caption{\textbf{The top 100 productive institutions: \textcolor{violet}{\textit{\particle}}.}}
\label{tab:r1_particle}
\vspace{2em}
\centering
{\tiny
{\renewcommand{\arraystretch}{1.2}
\begin{tabular}{rp{5cm}lr@{\hspace{4em}}p{5cm}lr@{\hspace{4em}}p{5cm}lr}\\[-5em] \toprule[1pt] \\[-1.4em]    
 & {\scriptsize \textbf{1991--2000}} & \multicolumn{2}{c}{No.~Works} & {\scriptsize \textbf{2001--2010}} & \multicolumn{2}{c}{No.~Works} & {\scriptsize \textbf{2011--2020}} & \multicolumn{2}{r}{No.~Works} \\[-0.2em] \cmidrule[0.5pt](lr{4em}){2-4} \cmidrule[0.5pt](l{-0em}r{4em}){5-7} \cmidrule[0.5pt](l{-0em}r{1em}){8-10}
\textit{1} & European Organization for Nuclear Research & CH & 5,367 & The University of Tokyo & JP & 7,970 & Chinese Academy of Sciences & CN & 14,559 \\  
\textit{2} & The University of Tokyo & JP & 5,308 & Chinese Academy of Sciences & CN & 6,555 & The University of Tokyo & JP & 9,871 \\  
\textit{3} & Los Alamos National Laboratory & US & 3,304 & European Organization for Nuclear Research & CH & 6,273 & European Organization for Nuclear Research & CH & 8,375 \\  
\textit{4} & Kyoto University & JP & 3,024 & Osaka University & JP & 4,925 & Massachusetts Institute of Technology & US & 5,919 \\  
\textit{5} & Osaka University & JP & 2,927 & Tohoku University & JP & 4,596 & University of Science and Technology of China & CN & 5,764 \\  
\textit{6} & Massachusetts Institute of Technology & US & 2,904 & Russian Academy of Sciences & RU & 4,390 & Lawrence Berkeley National Laboratory & US & 5,541 \\  
\textit{7} & Tohoku University & JP & 2,798 & Kyoto University & JP & 4,348 & Kyoto University & JP & 5,493 \\  
\textit{8} & Lawrence Berkeley National Laboratory & US & 2,723 & Lawrence Berkeley National Laboratory & US & 4,231 & University of California, Berkeley & US & 5,385 \\  
\textit{9} & University of California, Berkeley & US & 2,585 & Lomonosov Moscow State University & RU & 3,972 & University of Oxford & GB & 5,218 \\  
\textit{10} & University of Cambridge & GB & 2,491 & Los Alamos National Laboratory & US & 3,908 & Peking University & CN & 5,182 \\  
\hdashline          
\textit{11} & Argonne National Laboratory & US & 2,386 & Massachusetts Institute of Technology & US & 3,860 & University of Rajasthan & IN & 5,179 \\  
\textit{12} & Stanford University & US & 2,347 & University of California, Berkeley & US & 3,518 & University of Chinese Academy of Sciences & CN & 5,025 \\  
\textit{13} & Brookhaven National Laboratory & US & 2,290 & University of Cambridge & GB & 3,440 & Tsinghua University & CN & 4,953 \\  
\textit{14} & University of Illinois Urbana-Champaign & US & 2,281 & Stanford University & US & 3,405 & Lomonosov Moscow State University & RU & 4,845 \\  
\textit{15} & University of Oxford & GB & 2,273 & Brookhaven National Laboratory & US & 3,368 & Osaka University & JP & 4,779 \\  
\textit{16} & University of Maryland, College Park & US & 2,241 & Argonne National Laboratory & US & 3,318 & Stanford University & US & 4,693 \\  
\textit{17} & Fermilab & US & 2,154 & University of Oxford & GB & 3,314 & Russian Academy of Sciences & RU & 4,545 \\  
\textit{18} & Russian Academy of Sciences & RU & 2,132 & Joint Institute for Nuclear Research & RU & 3,262 & Tohoku University & JP & 4,532 \\  
\textit{19} & Joint Institute for Nuclear Research & RU & 2,116 & Fermilab & US & 3,226 & Joint Institute for Nuclear Research & RU & 4,354 \\  
\textit{20} & Lomonosov Moscow State University & RU & 2,095 & University of Maryland, College Park & US & 3,111 & Los Alamos National Laboratory & US & 4,254 \\  
\hdashline          
\textit{21} & Princeton University & US & 2,017 & High Energy Accelerator Research Organization (KEK) & JP & 3,072 & ETH Zurich & CH & 4,244 \\  
\textit{22} & University of California, Santa Barbara & US & 1,941 & University of Michigan{\textendash}Ann Arbor & US & 2,974 & University of Cambridge & GB & 4,154 \\  
\textit{23} & Oak Ridge National Laboratory & US & 1,917 & Princeton University & US & 2,897 & Brookhaven National Laboratory & US & 3,997 \\  
\textit{24} & United States Naval Research Laboratory & US & 1,874 & Tokyo Institute of Technology & JP & 2,775 & University of Maryland, College Park & US & 3,995 \\  
\textit{25} & High Energy Accelerator Research Organization (KEK) & JP & 1,825 & University of Science and Technology of China & CN & 2,676 & Princeton University & US & 3,854 \\  
\textit{26} & University of Michigan{\textendash}Ann Arbor & US & 1,821 & University of Paris-Sud & FR & 2,657 & University of Michigan{\textendash}Ann Arbor & US & 3,763 \\  
\textit{27} & Rutherford Appleton Laboratory & GB & 1,804 & California Institute of Technology & US & 2,657 & Technical University of Munich & DE & 3,713 \\  
\textit{28} & Technical University of Munich & DE & 1,800 & Institute of Physics & PL & 2,585 & Imperial College London & GB & 3,628 \\  
\textit{29} & Forschungszentrum J\"{u}lich & DE & 1,715 & Max Planck Society & DE & 2,560 & Fermilab & US & 3,622 \\  
\textit{30} & University of Paris-Sud & FR & 1,694 & Nagoya University & JP & 2,521 & Argonne National Laboratory & US & 3,612 \\  
\hdashline          
\textit{31} & Lawrence Livermore National Laboratory & US & 1,687 & Imperial College London & GB & 2,473 & University of Paris-Saclay & FR & 3,470 \\  
\textit{32} & Institute of Physics & PL & 1,660 & Technical University of Munich & DE & 2,470 & Karlsruhe Institute of Technology & DE & 3,412 \\  
\textit{33} & Nagoya University & JP & 1,647 & Tsinghua University & CN & 2,430 & Harvard University & US & 3,369 \\  
\textit{34} & California Institute of Technology & US & 1,611 & University of Illinois Urbana-Champaign & US & 2,405 & Nagoya University & JP & 3,347 \\  
\textit{35} & University of California, Los Angeles & US & 1,581 & University of California, Santa Barbara & US & 2,395 & Oak Ridge National Laboratory & US & 3,339 \\  
\textit{36} & Imperial College London & GB & 1,569 & Japan Science and Technology Agency (JST) & JP & 2,355 & National University of Singapore & SG & 3,335 \\  
\textit{37} & University of Washington & US & 1,542 & ETH Zurich & CH & 2,295 & California Institute of Technology & US & 3,313 \\  
\textit{38} & China Center of Advanced Science and Technology & CN & 1,538 & National Institute of Standards and Technology & US & 2,294 & University of Paris-Sud & FR & 3,260 \\  
\textit{39} & The University of Texas at Austin & US & 1,509 & Peking University & CN & 2,188 & Shanghai Jiao Tong University & CN & 3,231 \\  
\textit{40} & National Institute of Standards and Technology & US & 1,496 & \scalebox{0.82}[1]{National Institute of Advanced Industrial Science and Technology (AIST)} & JP & 2,173 & Nanjing University & CN & 3,006 \\  
\hdashline          
\textit{41} & Stony Brook University & US & 1,493 & Harvard University & US & 2,142 & National Institute of Standards and Technology & US & 2,990 \\  
\textit{42} & Pennsylvania State University & US & 1,474 & Oak Ridge National Laboratory & US & 2,142 & Moscow Engineering Physics Institute & RU & 2,988 \\  
\textit{43} & University of Tsukuba & JP & 1,466 & CEA Saclay & FR & 2,115 & High Energy Accelerator Research Organization (KEK) & JP & 2,964 \\  
\textit{44} & University of Wisconsin{\textendash}Madison & US & 1,450 & United States Naval Research Laboratory & US & 2,106 & SLAC National Accelerator Laboratory & US & 2,944 \\  
\textit{45} & Cornell University & US & 1,435 & University of Wisconsin{\textendash}Madison & US & 2,080 & University College London & GB & 2,911 \\  
\textit{46} & Tel Aviv University & IL & 1,425 & Forschungszentrum J\"{u}lich & DE & 2,077 & Forschungszentrum J\"{u}lich & DE & 2,896 \\  
\textit{47} & Max Planck Institute for Solid State Research & DE & 1,402 & University of Washington & US & 2,076 & University of California, Los Angeles & US & 2,892 \\  
\textit{48} & Tokyo Institute of Technology & JP & 1,397 & RIKEN & JP & 2,062 & \'{E}cole Polytechnique F\'{e}d\'{e}rale de Lausanne & CH & 2,884 \\  
\textit{49} & University of Chicago & US & 1,388 & University of California, Los Angeles & US & 2,061 & \scalebox{0.88}[1]{Institut National de Physique Nucl\'{e}aire et de Physique des Particules} & FR & 2,853 \\  
\textit{50} & Harvard University & US & 1,383 & Pennsylvania State University & US & 2,046 & Sapienza University of Rome & IT & 2,797 \\  
          
 \\[-1.4em]
\hdashline \\[-1em]
\multicolumn{10}{r}{\scriptsize \emph{(continued to next page)}}
\end{tabular}}
}
\end{table}
\end{landscape}
}
\afterpage{\clearpage%
\begin{landscape}
\begin{table}[!t]\ContinuedFloat
\vspace{-3.3em}
\caption{\textbf{The top 100 productive institutions: \textcolor{violet}{\textit{\particle}}.} \emph{(Cont.)}}
\label{tab:r2_particle}
\vspace{2em}
{\tiny
{\renewcommand{\arraystretch}{1.2}
\begin{tabular}{rp{5cm}lr@{\hspace{4em}}p{5cm}lr@{\hspace{4em}}p{5cm}lr}\\[-5em] \toprule[1pt] \\[-1.4em]    
 & {\scriptsize \textbf{1991--2000}} & \multicolumn{2}{c}{No.\ Works} & {\scriptsize \textbf{2001--2010}} & \multicolumn{2}{c}{No.\ Works} & {\scriptsize \textbf{2011--2020}} & \multicolumn{2}{r}{No.\ Works} \\[-0.2em] \cmidrule[0.5pt](lr{4em}){2-4} \cmidrule[0.5pt](l{-0em}r{4em}){5-7} \cmidrule[0.5pt](l{-0em}r{1em}){8-10}
\textit{51} & University of Minnesota & US & 1,382 & Cornell University & US & 2,031 & University of Chicago & US & 2,722 \\  
\textit{52} & \scalebox{0.9}[1]{P.N.\ Lebedev Physical Institute of the Russian Academy of Sciences} & RU & 1,373 & Physico-Technical Institute & RU & 2,010 & University of Illinois Urbana-Champaign & US & 2,706 \\  
\textit{53} & University of Arizona & US & 1,313 & Institute of Theoretical Physics & CN & 2,008 & University of California, Santa Barbara & US & 2,684 \\  
\textit{54} & The Ohio State University & US & 1,310 & Polish Academy of Sciences & PL & 1,989 & Universidade de S\~{a}o Paulo & BR & 2,679 \\  
\textit{55} & Johannes Gutenberg University Mainz & DE & 1,306 & University of Tsukuba & JP & 1,971 & Institute of Theoretical Physics & CN & 2,664 \\  
\textit{56} & AT\&T (United States) & US & 1,302 & National Academy of Sciences of Ukraine & UA & 1,964 & Quantum Group (United States) & US & 2,610 \\  
\textit{57} & University of California, San Diego & US & 1,284 & Lawrence Livermore National Laboratory & US & 1,963 & Moscow Institute of Physics and Technology & RU & 2,597 \\  
\textit{58} & ETH Zurich & CH & 1,278 & Rutherford Appleton Laboratory & GB & 1,953 & Johannes Gutenberg University Mainz & DE & 2,562 \\  
\textit{59} & Weizmann Institute of Science & IL & 1,253 & Karlsruhe Institute of Technology & DE & 1,941 & St Petersburg University & RU & 2,538 \\  
\textit{60} & Rutgers, The State University of New Jersey & US & 1,247 & Texas A\&M University & US & 1,918 & Tokyo Institute of Technology & JP & 2,521 \\  
\hdashline          
\textit{61} & State University of New York & US & 1,246 & Tokyo University of Science & JP & 1,892 & University of Valencia & ES & 2,519 \\  
\textit{62} & Polish Academy of Sciences & PL & 1,241 & Universidade de S\~{a}o Paulo & BR & 1,852 & University of Toronto & CA & 2,516 \\  
\textit{63} & Universit\'{e} Paris Cit\'{e} & FR & 1,231 & The University of Texas at Austin & US & 1,852 & The University of Texas at Austin & US & 2,486 \\  
\textit{64} & Tokyo University of Science & JP & 1,227 & Sapienza University of Rome & IT & 1,850 & Perimeter Institute & CA & 2,485 \\  
\textit{65} & Chinese Academy of Sciences & CN & 1,226 & Laboratoire de Physique Th\'{e}orique & FR & 1,834 & Heidelberg University & DE & 2,482 \\  
\textit{66} & Karlsruhe Institute of Technology & DE & 1,223 & University of Toronto & CA & 1,821 & \scalebox{0.9}[1]{P.N.\ Lebedev Physical Institute of the Russian Academy of Sciences} & RU & 2,446 \\  
\textit{67} & Texas A\&M University & US & 1,209 & \scalebox{0.9}[1]{P.N.\ Lebedev Physical Institute of the Russian Academy of Sciences} & RU & 1,818 & Texas A\&M University & US & 2,443 \\  
\textit{68} & University of Pennsylvania & US & 1,191 & Ruhr University Bochum & DE & 1,802 & University of Waterloo & CA & 2,441 \\  
\textit{69} & University of Rochester & US & 1,177 & The Ohio State University & US & 1,788 & Universidad Nacional Aut\'{o}noma de M\'{e}xico & MX & 2,434 \\  
\textit{70} & Columbia University & US & 1,170 & Columbia University & US & 1,785 & Pennsylvania State University & US & 2,434 \\  
\hdashline          
\textit{71} & University of Manchester & GB & 1,163 & University of Arizona & US & 1,766 & Kurchatov Institute & RU & 2,396 \\  
\textit{72} & Universit\"{a}t Hamburg & DE & 1,155 & National Taiwan University & TW & 1,733 & University of Wisconsin{\textendash}Madison & US & 2,379 \\  
\textit{73} & RIKEN & JP & 1,141 & University of California, San Diego & US & 1,726 & Institute of Physics & CN & 2,343 \\  
\textit{74} & Yale University & US & 1,141 & \'{E}cole Polytechnique F\'{e}d\'{e}rale de Lausanne & CH & 1,725 & University of Washington & US & 2,342 \\  
\textit{75} & Uppsala University & SE & 1,128 & Weizmann Institute of Science & IL & 1,710 & Max Planck Society & DE & 2,331 \\  
\textit{76} & \scalebox{0.88}[1]{Institut National de Physique Nucl\'{e}aire et de Physique des Particules} & FR & 1,124 & University of Chicago & US & 1,706 & Deutsches Elektronen-Synchrotron DESY & DE & 2,326 \\  
\textit{77} & Kurchatov Institute & RU & 1,123 & Seoul National University & KR & 1,680 & Zhejiang University & CN & 2,323 \\  
\textit{78} & Ruhr University Bochum & DE & 1,110 & Petersburg Nuclear Physics Institute & RU & 1,675 & National Taiwan University & TW & 2,287 \\  
\textit{79} & \scalebox{0.9}[1]{The Abdus Salam International Centre for Theoretical Physics (ICTP)} & IT & 1,108 & National University of Singapore & SG & 1,674 & University of California, San Diego & US & 2,286 \\  
\textit{80} & Lund University & SE & 1,100 & \scalebox{0.88}[1]{Institut National de Physique Nucl\'{e}aire et de Physique des Particules} & FR & 1,636 & Paul Scherrer Institute & CH & 2,281 \\  
\hdashline          
\textit{81} & Max Planck Society & DE & 1,096 & University of Valencia & ES & 1,628 & Ludwig-Maximilians-Universit\"{a}t M\"{u}nchen & DE & 2,275 \\  
\textit{82} & University of British Columbia & CA & 1,090 & Stony Brook University & US & 1,618 & University of Manchester & GB & 2,274 \\  
\textit{83} & Goethe University Frankfurt & DE & 1,089 & Johannes Gutenberg University Mainz & DE & 1,592 & University of Geneva & CH & 2,273 \\  
\textit{84} & Technion {\textendash} Israel Institute of Technology & IL & 1,080 & \'{E}cole Polytechnique & FR & 1,591 & Institute of Physics & PL & 2,242 \\  
\textit{85} & Florida State University & US & 1,067 & Institute for Theoretical and Experimental Physics & RU & 1,589 & University of Colorado Boulder & US & 2,232 \\  
\textit{86} & Sapienza University of Rome & IT & 1,055 & Institute of High Energy Physics & CN & 1,576 & Columbia University & US & 2,212 \\  
\textit{87} & Chalmers University of Technology & SE & 1,055 & Tel Aviv University & IL & 1,575 & Consejo Nacional de Investigaciones Cient\'{i}ficas y T\'{e}cnicas & AR & 2,208 \\  
\textit{88} & University of Florida & US & 1,054 & University of Florida & US & 1,568 & Institute of High Energy Physics & CN & 2,193 \\  
\textit{89} & Istituto Nazionale per la Fisica della Materia & IT & 1,050 & University of California, Davis & US & 1,562 & Goethe University Frankfurt & DE & 2,184 \\  
\textit{90} & \'{E}cole Polytechnique & FR & 1,045 & Kyushu University & JP & 1,556 & Universit\"{a}t Hamburg & DE & 2,168 \\  
\hdashline          
\textit{91} & Stanford Synchrotron Radiation Lightsource & US & 1,035 & Uppsala University & SE & 1,551 & Novosibirsk State University & RU & 2,165 \\  
\textit{92} & University of Valencia & ES & 1,000 & University of Bonn & DE & 1,543 & \'{E}cole Polytechnique & FR & 2,161 \\  
\textit{93} & Autonomous University of Madrid & ES & 999 & St Petersburg University & RU & 1,541 & Autonomous University of Madrid & ES & 2,141 \\  
\textit{94} & University of California, Davis & US & 993 & University of Minnesota & US & 1,538 & Australian National University & AU & 2,133 \\  
\textit{95} & Heidelberg University & DE & 992 & Universidad Nacional Aut\'{o}noma de M\'{e}xico & MX & 1,530 & Uppsala University & SE & 2,122 \\  
\textit{96} & Iowa State University & US & 990 & Yale University & US & 1,520 & The Ohio State University & US & 2,121 \\  
\textit{97} & Hiroshima University & JP & 986 & University College London & GB & 1,504 & Harbin Institute of Technology & CN & 2,110 \\  
\textit{98} & Hokkaido University & JP & 986 & Boston University & US & 1,499 & Nanyang Technological University & SG & 2,110 \\  
\textit{99} & Universidade de S\~{a}o Paulo & BR & 976 & Hokkaido University & JP & 1,472 & University of Southampton & GB & 2,106 \\  
\textit{100} & Humboldt-Universit\"{a}t zu Berlin & DE & 974 & Hiroshima University & JP & 1,471 & National Institute for Materials Science (NIMS) & JP & 2,106 \\  
          
 \\[-1.4em]
\bottomrule
\end{tabular}}
}
\end{table}
\end{landscape}
}
%

\titleformat{\section}{\sc\centering\LARGE\bfseries}{\textsc{\thesection}.\!\!}{1em}{}

\afterpage{\clearpage%
\markboth{\textbf \textsc{\aerospace}}{}
\thispagestyle{empty}
\quad
\vspace{2cm}
\begin{center}
\pgfornament[width=0.5*\textwidth,symmetry=h]{89}\\[2em]
\section{\aerospace}
\vspace{1em}
\pgfornament[width=0.5*\textwidth]{89}
\end{center}
}

\afterpage{\clearpage%

\begin{figure}[!tp]
\centering
\vspace{-1em}
{\large \textbf{\textrm{{World Map of \textcolor{violet}{\textit{\aerospace}} Collaboration}}}~|~1971--2020}\\
\vspace{0.3cm}
\includegraphics[align=c, scale=0.054, trim={9.5cm 0 9.5cm 0},clip]{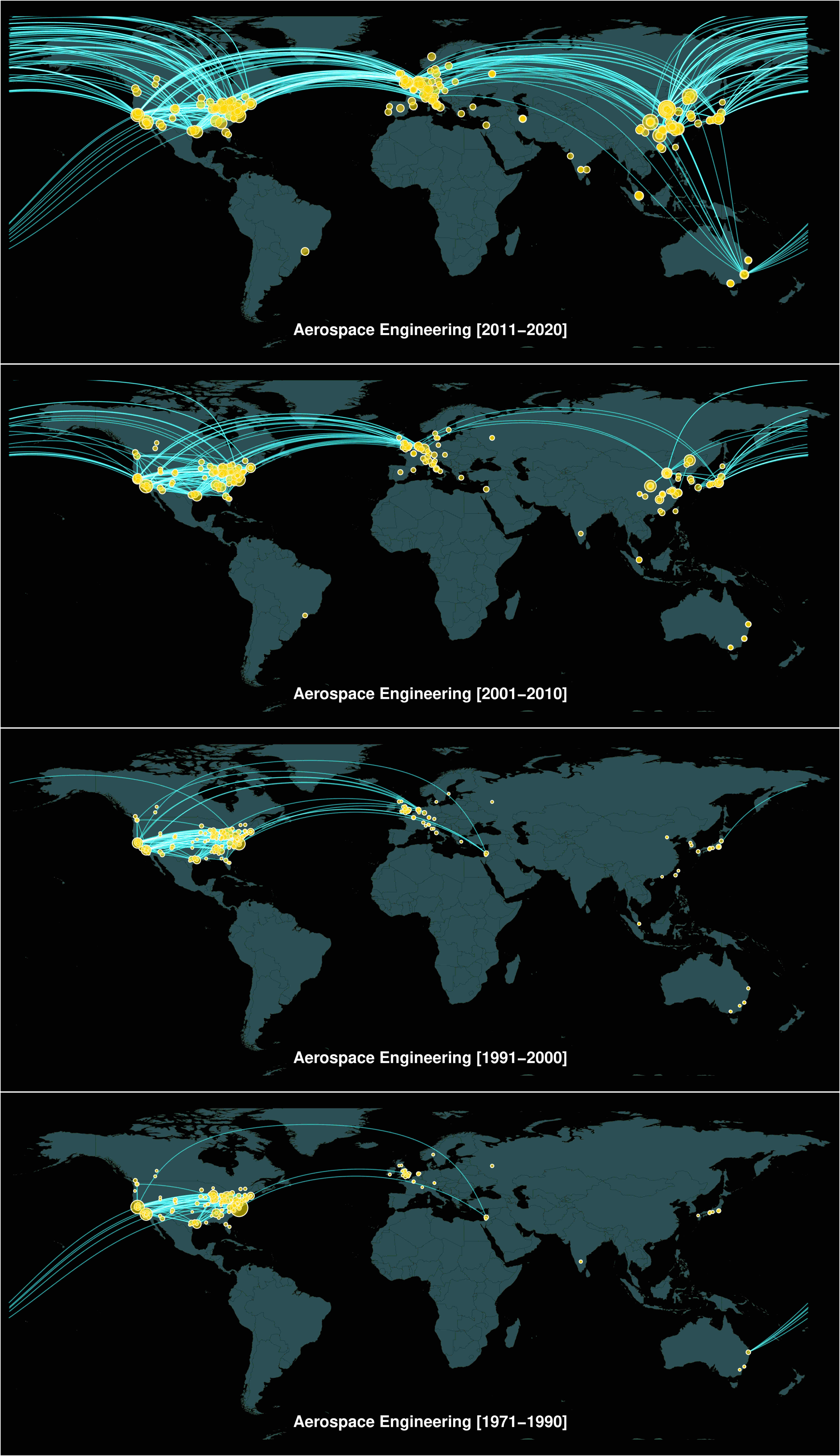}
\caption[{\aerospace}]{\textbf{(a)~|~The World Map of \textcolor{violet}{\textit{\aerospace}} Collaboration.}
The bubbles represent the top 199 institutions in terms of work production, with their sizes proportional to the work volume. 
The connecting lines depict coauthorship relationships among the top 50 institutions.}
\label{fig:wmap_aerospace}
\end{figure}
}
\afterpage{\clearpage%
\begin{figure}[!tp]\ContinuedFloat
\centering
\vspace{-1em}
{\large \textbf{Top 30 Productive Institutions on the World Map: \textcolor{violet}{\textit{\aerospace}}}~|~1991--2020}\\
\vspace{-0em}
\hspace*{-3em}                                                           
\includegraphics[align=c, scale=0.83]{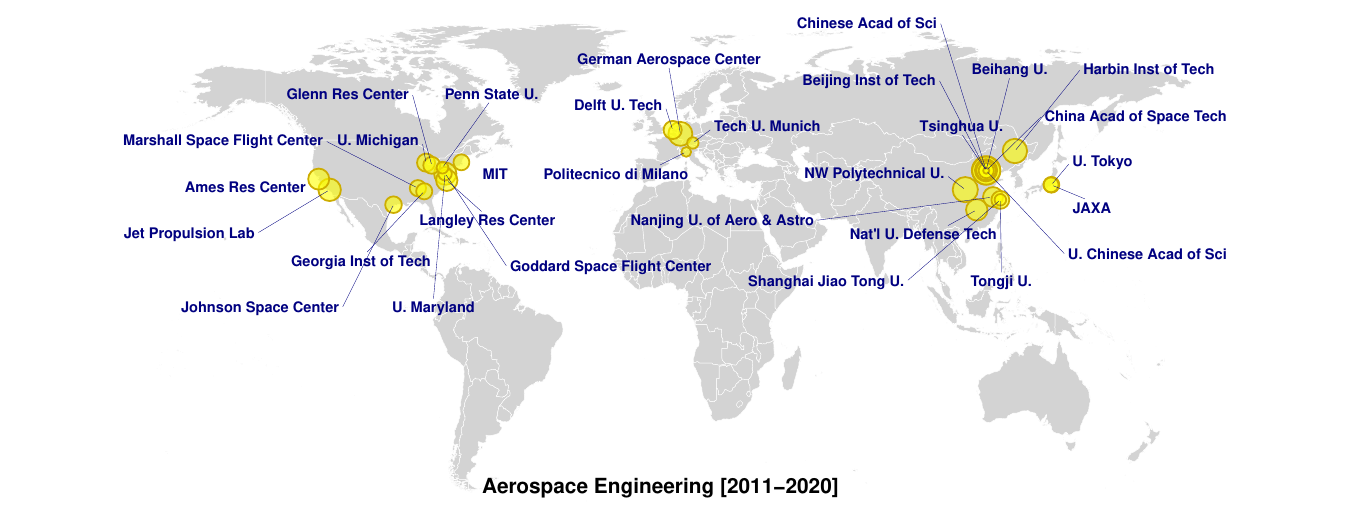}\\[-0.5em]
\quad\\[-1em]
\dotfill 
\quad\\[-0em]
\hspace*{-3em}                                                           
\includegraphics[align=c, scale=0.83]{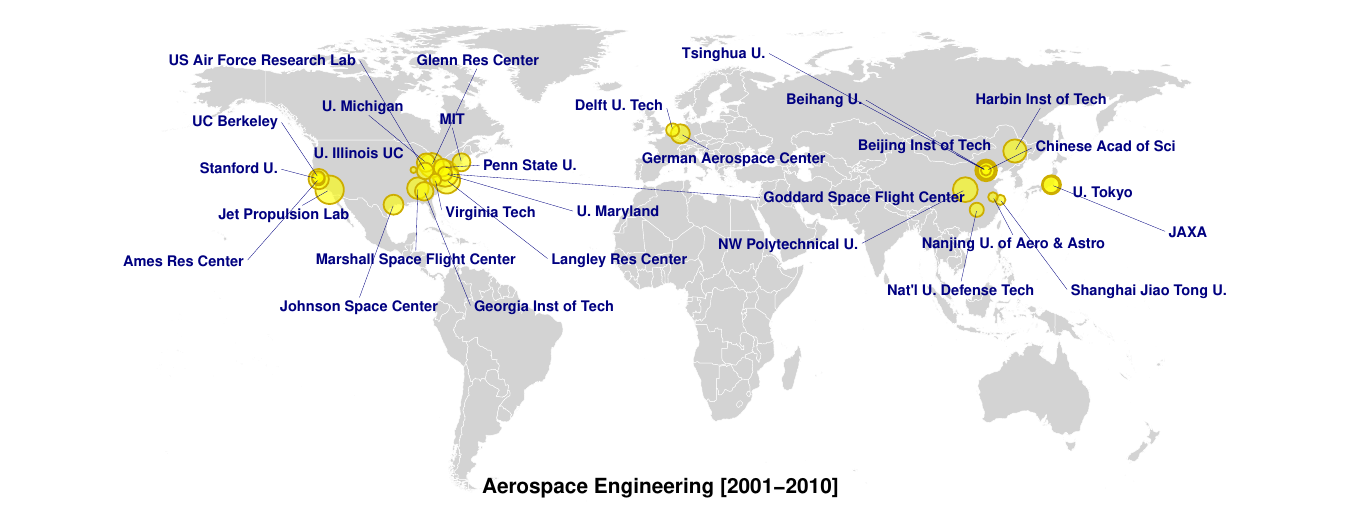}\\[-0.5em]
\quad\\[-1em]
\dotfill 
\quad\\[-0em]
\hspace*{-3em}                                                           
\includegraphics[align=c, scale=0.83]{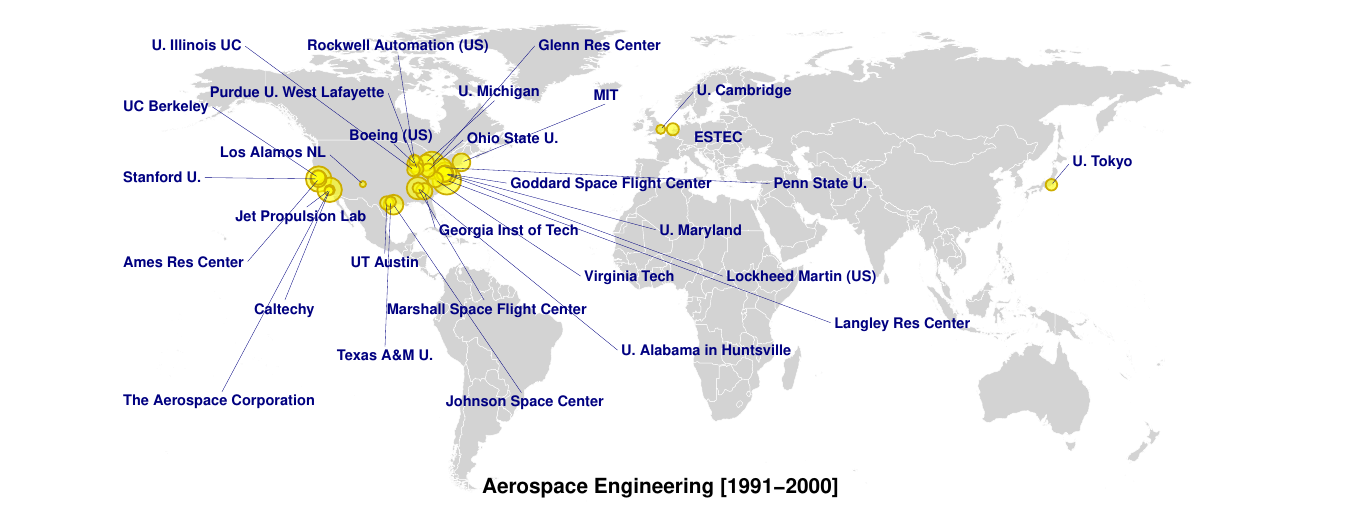}\\[-0.8em]
\caption[{\aerospace}]{\textbf{(b)~|~The top 30 productive institutions on the World Map: \textcolor{violet}{\textit{\aerospace}}.}
The bubbles represent the top 30 institutions in terms of work production, with their sizes proportional to the work volume.}
\label{fig:wmap_topinst_aerospace}
\end{figure}
}
\afterpage{\clearpage%
\begin{figure}[!tp]\ContinuedFloat
\centering
\vspace{-1em}
{\large \textbf{\textrm{{Interregional \textcolor{violet}{\textit{\aerospace}} Collaboration}}}~|~1991--2020}\\
\vspace{0.5em}
\hspace{-5em}\includegraphics[align=c, scale=1.7, vmargin=0mm]{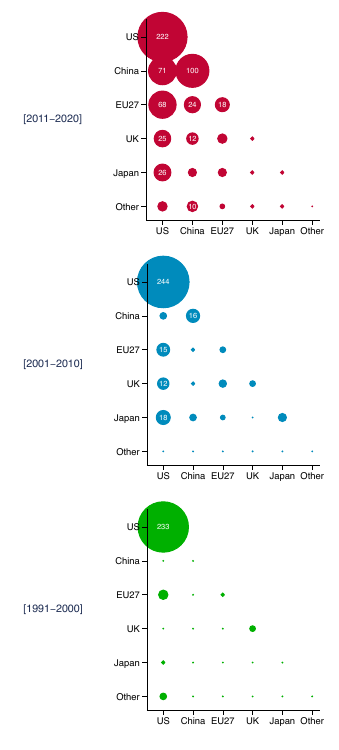}
\vspace{-1em}
\caption[]{\textbf{(c)~|~The Interregional \textcolor{violet}{\textit{\aerospace}} Collaboration Matrix Diagram.}
The bubble size represents the number of coauthorship relationships for the top 50 institutions in terms of work production. 
If the number is equal to or greater than 10, it is displayed inside the bubble.}
\label{fig:halfmat_aerospace}
\end{figure}
}
\afterpage{\clearpage%
\begin{figure}[!tp]\ContinuedFloat
\centering
\vspace{-1em}
{\large \textbf{\textrm{Interinstitutional \textcolor{violet}{\textit{\aerospace}} Collaboration}}~|~2001--2020\quad {\footnotesize \emph{(continued to next page)}}}\\
\cdend{aerospace}{2010}{2000}{2011--2020}{2001--2010}\\[-1.5em]
\caption[]{\textbf{(d)~|~The Interinstitutional \textcolor{violet}{\textit{\aerospace}} Collaboration Dendrogram.}
The top 50 institutions in terms of work production, indicated by the circularised bar graphs, are displayed.
}
\label{fig:cdend1_aerospace}
\end{figure}
}
\afterpage{\clearpage%
\begin{figure}[!tp]\ContinuedFloat
\centering
\vspace{-1em}
{\large \textbf{\textrm{Interinstitutional \textcolor{violet}{\textit{\aerospace}} Collaboration}}~|~1971--2000\quad {\footnotesize \emph{(continued from previous page)}}}\\
\cdend{aerospace}{1990}{1980}{1991--2000}{1971--1990}\\[-1.5em]
\caption[]{\textbf{(d)~|~The Interinstitutional \textcolor{violet}{\textit{\aerospace}} Collaboration Dendrogram.} \emph{(Cont.)}\hfill~}
\label{fig:cdend2_aerospace}
\end{figure}
}
\afterpage{\clearpage%
\begin{landscape}
\begin{table}[!t]
\vspace{-0.5em}
\caption{\textbf{The top 100 productive institutions: \textcolor{violet}{\textit{\aerospace}}.}}
\label{tab:r1_aerospace}
\vspace{2em}
\centering
{\tiny
{\renewcommand{\arraystretch}{1.2}
\begin{tabular}{rp{5cm}lr@{\hspace{4em}}p{5cm}lr@{\hspace{4em}}p{5cm}lr}\\[-5em] \toprule[1pt] \\[-1.4em]    
 & {\scriptsize \textbf{1991--2000}} & \multicolumn{2}{c}{No.~Works} & {\scriptsize \textbf{2001--2010}} & \multicolumn{2}{c}{No.~Works} & {\scriptsize \textbf{2011--2020}} & \multicolumn{2}{r}{No.~Works} \\[-0.2em] \cmidrule[0.5pt](lr{4em}){2-4} \cmidrule[0.5pt](l{-0em}r{4em}){5-7} \cmidrule[0.5pt](l{-0em}r{1em}){8-10}
\textit{1} & Langley Research Center & US & 3,016 & Jet Propulsion Laboratory & US & 3,616 & Beihang University & CN & 6,290 \\  
\textit{2} & Glenn Research Center & US & 2,569 & Langley Research Center & US & 3,136 & Chinese Academy of Sciences & CN & 6,104 \\  
\textit{3} & Ames Research Center & US & 2,169 & Goddard Space Flight Center & US & 3,066 & Northwestern Polytechnical University & CN & 4,888 \\  
\textit{4} & Jet Propulsion Laboratory & US & 1,927 & Northwestern Polytechnical University & CN & 2,743 & Harbin Institute of Technology & CN & 4,585 \\  
\textit{5} & Goddard Space Flight Center & US & 1,534 & Glenn Research Center & US & 2,679 & German Aerospace Center & DE & 4,447 \\  
\textit{6} & Marshall Space Flight Center & US & 1,487 & Harbin Institute of Technology & CN & 2,455 & Jet Propulsion Laboratory & US & 3,983 \\  
\textit{7} & Johnson Space Center & US & 1,107 & Beihang University & CN & 2,203 & Goddard Space Flight Center & US & 3,937 \\  
\textit{8} & Pennsylvania State University & US & 950 & Marshall Space Flight Center & US & 2,188 & Tsinghua University & CN & 3,898 \\  
\textit{9} & Massachusetts Institute of Technology & US & 904 & Ames Research Center & US & 2,012 & National University of Defense Technology & CN & 3,690 \\  
\textit{10} & University of Maryland, College Park & US & 797 & Johnson Space Center & US & 1,957 & Langley Research Center & US & 3,676 \\  
\hdashline          
\textit{11} & Georgia Institute of Technology & US & 771 & German Aerospace Center & DE & 1,856 & Ames Research Center & US & 3,578 \\  
\textit{12} & Stanford University & US & 753 & The University of Tokyo & JP & 1,851 & Nanjing University of Aeronautics and Astronautics & CN & 3,317 \\  
\textit{13} & University of Michigan{\textendash}Ann Arbor & US & 751 & Tsinghua University & CN & 1,799 & Beijing Institute of Technology & CN & 3,195 \\  
\textit{14} & Lockheed Martin (United States) & US & 723 & Georgia Institute of Technology & US & 1,794 & Delft University of Technology & NL & 2,954 \\  
\textit{15} & Boeing (United States) & US & 722 & University of Michigan{\textendash}Ann Arbor & US & 1,776 & University of Michigan{\textendash}Ann Arbor & US & 2,926 \\  
\textit{16} & Virginia Tech & US & 688 & Massachusetts Institute of Technology & US & 1,742 & Shanghai Jiao Tong University & CN & 2,903 \\  
\textit{17} & University of California, Berkeley & US & 640 & Pennsylvania State University & US & 1,701 & Glenn Research Center & US & 2,742 \\  
\textit{18} & The Ohio State University & US & 594 & Japan Aerospace Exploration Agency (JAXA) & JP & 1,610 & Johnson Space Center & US & 2,695 \\  
\textit{19} & The University of Texas at Austin & US & 587 & United States Air Force Research Laboratory & US & 1,609 & Marshall Space Flight Center & US & 2,631 \\  
\textit{20} & Purdue University West Lafayette & US & 551 & Chinese Academy of Sciences & CN & 1,601 & Georgia Institute of Technology & US & 2,617 \\  
\hdashline          
\textit{21} & University of Illinois Urbana-Champaign & US & 547 & National University of Defense Technology & CN & 1,428 & Massachusetts Institute of Technology & US & 2,584 \\  
\textit{22} & European Space Research and Technology Centre & NL & 529 & University of Maryland, College Park & US & 1,419 & The University of Tokyo & JP & 2,558 \\  
\textit{23} & The University of Tokyo & JP & 504 & Delft University of Technology & NL & 1,367 & University of Chinese Academy of Sciences & CN & 2,488 \\  
\textit{24} & University of Alabama in Huntsville & US & 496 & Stanford University & US & 1,345 & Japan Aerospace Exploration Agency (JAXA) & JP & 2,457 \\  
\textit{25} & Rockwell Automation (United States) & US & 494 & University of California, Berkeley & US & 1,337 & University of Maryland, College Park & US & 2,286 \\  
\textit{26} & Texas A\&M University & US & 485 & Virginia Tech & US & 1,298 & Pennsylvania State University & US & 2,144 \\  
\textit{27} & California Institute of Technology & US & 468 & Shanghai Jiao Tong University & CN & 1,248 & Tongji University & CN & 2,133 \\  
\textit{28} & The Aerospace Corporation & US & 449 & Beijing Institute of Technology & CN & 1,247 & Technical University of Munich & DE & 2,124 \\  
\textit{29} & University of Cambridge & GB & 447 & Nanjing University of Aeronautics and Astronautics & CN & 1,235 & Politecnico di Milano & IT & 2,027 \\  
\textit{30} & Los Alamos National Laboratory & US & 421 & University of Illinois Urbana-Champaign & US & 1,189 & China Academy of Space Technology & CN & 1,945 \\  
\hdashline          
\textit{31} & Lawrence Livermore National Laboratory & US & 418 & California Institute of Technology & US & 1,155 & Virginia Tech & US & 1,945 \\  
\textit{32} & Old Dominion University & US & 418 & European Space Research and Technology Centre & NL & 1,152 & Stanford University & US & 1,928 \\  
\textit{33} & Technion {\textendash} Israel Institute of Technology & IL & 415 & The Ohio State University & US & 1,104 & University of California, Berkeley & US & 1,918 \\  
\textit{34} & North Carolina State University & US & 413 & Tohoku University & JP & 1,064 & Zhejiang University & CN & 1,846 \\  
\textit{35} & Cornell University & US & 402 & Boeing (United States) & US & 1,035 & United States Air Force Research Laboratory & US & 1,845 \\  
\textit{36} & German Aerospace Center & DE & 398 & Texas A\&M University & US & 1,012 & Wuhan University & CN & 1,788 \\  
\textit{37} & Arizona State University & US & 395 & Lockheed Martin (United States) & US & 989 & University of Colorado Boulder & US & 1,745 \\  
\textit{38} & University of Arizona & US & 391 & Xiaomi (China) & CN & 985 & The Ohio State University & US & 1,740 \\  
\textit{39} & Princeton University & US & 378 & Tongji University & CN & 983 & University of Illinois Urbana-Champaign & US & 1,738 \\  
\textit{40} & University of Washington & US & 375 & Air Force Engineering University & CN & 974 & Harbin Engineering University & CN & 1,708 \\  
\hdashline          
\textit{41} & Armstrong Flight Research Center & US & 368 & The University of Texas at Austin & US & 974 & Purdue University West Lafayette & US & 1,705 \\  
\textit{42} & United States Naval Research Laboratory & US & 368 & Zhejiang University & CN & 924 & Imperial College London & GB & 1,699 \\  
\textit{43} & Wright-Patterson Air Force Base & US & 363 & University of Cambridge & GB & 893 & Texas A\&M University & US & 1,683 \\  
\textit{44} & Space Information Laboratories (United States) & US & 357 & University of Florida & US & 886 & University of Cambridge & GB & 1,672 \\  
\textit{45} & University of Florida & US & 356 & Kyushu University & JP & 865 & The University of Texas at Austin & US & 1,655 \\  
\textit{46} & University of Wisconsin{\textendash}Madison & US & 354 & Imperial College London & GB & 859 & Sapienza University of Rome & IT & 1,623 \\  
\textit{47} & University of California, Los Angeles & US & 352 & University of Southampton & GB & 833 & Xi'an Jiaotong University & CN & 1,612 \\  
\textit{48} & University of Colorado Boulder & US & 345 & University of Washington & US & 832 & European Space Research and Technology Centre & NL & 1,568 \\  
\textit{49} & Imperial College London & GB & 344 & Purdue University West Lafayette & US & 826 & UNSW Sydney & AU & 1,558 \\  
\textit{50} & University of Oxford & GB & 344 & University of Arizona & US & 824 & Polytechnic University of Turin & IT & 1,551 \\  
          
 \\[-1.4em]
\hdashline \\[-1em]
\multicolumn{10}{r}{\scriptsize \emph{(continued to next page)}}
\end{tabular}}
}
\end{table}
\end{landscape}
}
\afterpage{\clearpage%
\begin{landscape}
\begin{table}[!t]\ContinuedFloat
\vspace{-3.3em}
\caption{\textbf{The top 100 productive institutions: \textcolor{violet}{\textit{\aerospace}}.} \emph{(Cont.)}}
\label{tab:r2_aerospace}
\vspace{2em}
{\tiny
{\renewcommand{\arraystretch}{1.2}
\begin{tabular}{rp{5cm}lr@{\hspace{4em}}p{5cm}lr@{\hspace{4em}}p{5cm}lr}\\[-5em] \toprule[1pt] \\[-1.4em]    
 & {\scriptsize \textbf{1991--2000}} & \multicolumn{2}{c}{No.\ Works} & {\scriptsize \textbf{2001--2010}} & \multicolumn{2}{c}{No.\ Works} & {\scriptsize \textbf{2011--2020}} & \multicolumn{2}{r}{No.\ Works} \\[-0.2em] \cmidrule[0.5pt](lr{4em}){2-4} \cmidrule[0.5pt](l{-0em}r{4em}){5-7} \cmidrule[0.5pt](l{-0em}r{1em}){8-10}
\textit{51} & Johns Hopkins University & US & 343 & Wuhan University & CN & 821 & Nanjing University of Science and Technology & CN & 1,544 \\  
\textit{52} & Naval Postgraduate School & US & 343 & Sapienza University of Rome & IT & 787 & University of Oxford & GB & 1,536 \\  
\textit{53} & Office National d'\'{E}tudes et de Recherches A\'{e}rospatiales & FR & 339 & Seoul National University & KR & 781 & University of Toronto & CA & 1,510 \\  
\textit{54} & University of Minnesota & US & 335 & University of Colorado Boulder & US & 768 & Nanyang Technological University & SG & 1,510 \\  
\textit{55} & United States Army & US & 333 & Politecnico di Milano & IT & 759 & University of Southampton & GB & 1,478 \\  
\textit{56} & University of California, Davis & US & 332 & Cornell University & US & 753 & Tianjin University & CN & 1,454 \\  
\textit{57} & Institute of Space and Astronautical Science (ISAS) & JP & 328 & Cranfield University & GB & 731 & Iowa State University & US & 1,450 \\  
\textit{58} & University of Southampton & GB & 316 & Carnegie Mellon University & US & 730 & ETH Zurich & CH & 1,428 \\  
\textit{59} & Delft University of Technology & NL & 312 & Arizona State University & US & 729 & RWTH Aachen University & DE & 1,409 \\  
\textit{60} & Tohoku University & JP & 311 & Huazhong University of Science and Technology & CN & 726 & Peking University & CN & 1,399 \\  
\hdashline          
\textit{61} & Duke University & US & 305 & Kyoto University & JP & 721 & Huazhong University of Science and Technology & CN & 1,394 \\  
\textit{62} & University of Manchester & GB & 304 & University of Toronto & CA & 715 & University of Stuttgart & DE & 1,393 \\  
\textit{63} & United States Air Force Research Laboratory & US & 301 & Tokyo Institute of Technology & JP & 705 & University of Washington & US & 1,387 \\  
\textit{64} & University of British Columbia & CA & 294 & University of Wisconsin{\textendash}Madison & US & 704 & National University of Singapore & SG & 1,385 \\  
\textit{65} & University of Toronto & CA & 290 & North Carolina State University & US & 703 & Seoul National University & KR & 1,375 \\  
\textit{66} & Iowa State University & US & 287 & Nanyang Technological University & SG & 696 & Dalian University of Technology & CN & 1,350 \\  
\textit{67} & Kyoto University & JP & 286 & Harbin Engineering University & CN & 684 & Cranfield University & GB & 1,342 \\  
\textit{68} & Naval Surface Warfare Center & US & 286 & Hong Kong Polytechnic University & CN & 677 & Technical University of Denmark & DK & 1,334 \\  
\textit{69} & Rensselaer Polytechnic Institute & US & 286 & University of California, Los Angeles & US & 676 & California Institute of Technology & US & 1,333 \\  
\textit{70} & \scalebox{0.9}[1]{Science Applications International Corporation (United States)} & US & 286 & Technical University of Munich & DE & 672 & Jilin University & CN & 1,332 \\  
\hdashline          
\textit{71} & Kyushu University & JP & 271 & University of Oxford & GB & 670 & Tohoku University & JP & 1,324 \\  
\textit{72} & Cranfield University & GB & 269 & Osaka University & JP & 660 & Arizona State University & US & 1,320 \\  
\textit{73} & Sapienza University of Rome & IT & 269 & National University of Singapore & SG & 649 & University of Florida & US & 1,306 \\  
\textit{74} & Sandia National Laboratories & US & 267 & Princeton University & US & 649 & University of Electronic Science and Technology of China & CN & 1,265 \\  
\textit{75} & Carnegie Mellon University & US & 266 & Office National d'\'{E}tudes et de Recherches A\'{e}rospatiales & FR & 645 & Southeast University & CN & 1,255 \\  
\textit{76} & University of Stuttgart & DE & 265 & Xi'an Jiaotong University & CN & 644 & Norwegian University of Science and Technology & NO & 1,255 \\  
\textit{77} & Osaka University & JP & 265 & ETH Zurich & CH & 638 & Korea Advanced Institute of Science and Technology & KR & 1,249 \\  
\textit{78} & DEVCOM Army Research Laboratory & US & 262 & Polytechnic University of Turin & IT & 632 & KU Leuven & BE & 1,226 \\  
\textit{79} & Rutgers, The State University of New Jersey & US & 260 & University of British Columbia & CA & 631 & Karlsruhe Institute of Technology & DE & 1,224 \\  
\textit{80} & University of Cincinnati & US & 256 & University of Manchester & GB & 630 & University of Naples Federico II & IT & 1,214 \\  
\hdashline          
\textit{81} & University College London & GB & 249 & Technion {\textendash} Israel Institute of Technology & IL & 629 & Chalmers University of Technology & SE & 1,208 \\  
\textit{82} & National Aeronautics and Space Administration & US & 249 & University of Stuttgart & DE & 627 & University of Wisconsin{\textendash}Madison & US & 1,208 \\  
\textit{83} & University of Southern California & US & 242 & University of California, Davis & US & 627 & University of Technology Malaysia & MY & 1,202 \\  
\textit{84} & Johns Hopkins University Applied Physics Laboratory & US & 241 & University of Minnesota & US & 627 & Universidade de S\~{a}o Paulo & BR & 1,197 \\  
\textit{85} & Defense Systems (United States) & US & 240 & Boeing (Australia) & AU & 624 & University of Arizona & US & 1,195 \\  
\textit{86} & Loughborough University & GB & 231 & Jilin University & CN & 620 & Air Force Engineering University & CN & 1,194 \\  
\textit{87} & National Cheng Kung University & TW & 231 & Nagoya University & JP & 620 & Universitat Polit\`{e}cnica de Catalunya & ES & 1,188 \\  
\textit{88} & Daimler (Germany) & DE & 228 & Tianjin University & CN & 617 & Central South University & CN & 1,167 \\  
\textit{89} & University of Virginia & US & 228 & Korea Advanced Institute of Science and Technology & KR & 613 & Engineering (Italy) & IT & 1,153 \\  
\textit{90} & University of Surrey & GB & 227 & UNSW Sydney & AU & 605 & Universidad Politd\'{e}cnica de Madrid & ES & 1,148 \\  
\hdashline          
\textit{91} & Ford Motor Company (United States) & US & 227 & United States Naval Research Laboratory & US & 600 & Office National d'\'{E}tudes et de Recherches A\'{e}rospatiales & FR & 1,144 \\  
\textit{92} & University of Glasgow & GB & 225 & \'{E}cole Polytechnique F\'{e}d\'{e}rale de Lausanne & CH & 599 & Harvard University & US & 1,124 \\  
\textit{93} & Centre National d'\'{E}tudes Spatiales & FR & 224 & Peking University & CN & 598 & Hong Kong Polytechnic University & CN & 1,115 \\  
\textit{94} & University of California, San Diego & US & 223 & McGill University & CA & 597 & University of Science and Technology of China & CN & 1,105 \\  
\textit{95} & McGill University & CA & 221 & University of California, San Diego & US & 574 & China Aerodynamics Research and Development Center & CN & 1,102 \\  
\textit{96} & \'{E}cole Polytechnique F\'{e}d\'{e}rale de Lausanne & CH & 220 & Johns Hopkins University Applied Physics Laboratory & US & 572 & Southwest Jiaotong University & CN & 1,098 \\  
\textit{97} & University of Sheffield & GB & 220 & Nanjing University of Science and Technology & CN & 571 & Universit\'{e} de Toulouse & FR & 1,092 \\  
\textit{98} & University of Wales & GB & 219 & Institute of Space and Astronautical Science (ISAS) & JP & 571 & University of California, Los Angeles & US & 1,092 \\  
\textit{99} & University of Sydney & AU & 218 & University of Sheffield & GB & 564 & Wuhan University of Technology & CN & 1,087 \\  
\textit{100} & Tokyo Institute of Technology & JP & 212 & Chongqing University & CN & 563 & \'{E}cole Polytechnique F\'{e}d\'{e}rale de Lausanne & CH & 1,086 \\  
          
 \\[-1.4em]
\bottomrule
\end{tabular}}
}
\end{table}
\end{landscape}
}
%

\titleformat{\section}{\sc\centering\LARGE\bfseries}{\textsc{\thesection}.\!\!}{1em}{}

\afterpage{\clearpage%
\markboth{\textbf \textsc{\nuclear}}{}
\thispagestyle{empty}
\quad
\vspace{2cm}
\begin{center}
\pgfornament[width=0.5*\textwidth,symmetry=h]{89}\\[2em]
\section{\nuclear}
\vspace{1em}
\pgfornament[width=0.5*\textwidth]{89}
\end{center}
}

\afterpage{\clearpage%

\begin{figure}[!tp]
\centering
\vspace{-1em}
{\large \textbf{\textrm{{World Map of \textcolor{violet}{\textit{\nuclear}} Collaboration}}}~|~1971--2020}\\
\vspace{0.3cm}
\includegraphics[align=c, scale=0.054, trim={9.5cm 0 9.5cm 0},clip]{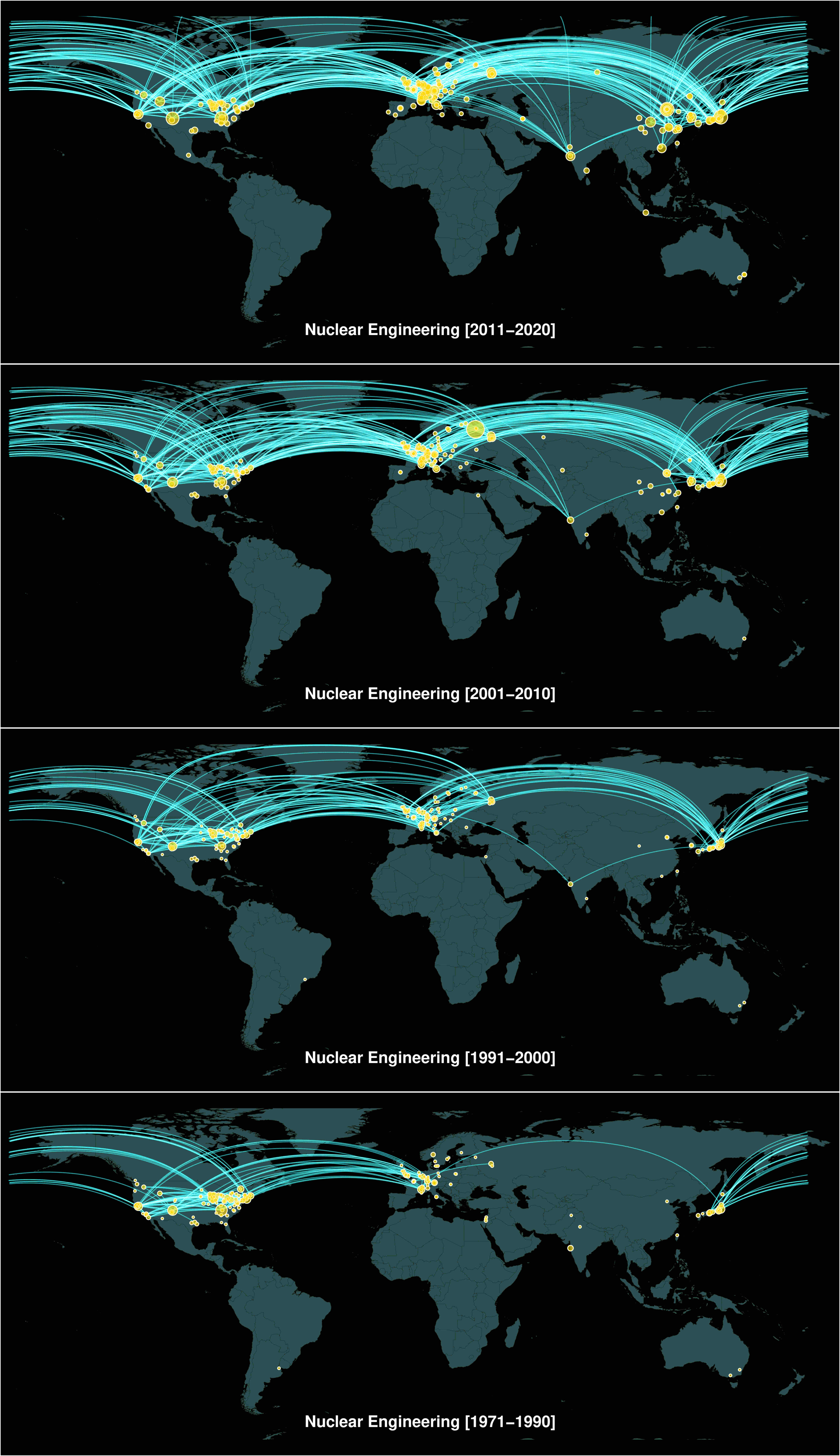}
\caption[{\nuclear}]{\textbf{(a)~|~The World Map of \textcolor{violet}{\textit{\nuclear}} Collaboration.}
The bubbles represent the top 199 institutions in terms of work production, with their sizes proportional to the work volume. 
The connecting lines depict coauthorship relationships among the top 50 institutions.}
\label{fig:wmap_nuclear}
\end{figure}
}
\afterpage{\clearpage%
\begin{figure}[!tp]\ContinuedFloat
\centering
\vspace{-1em}
{\large \textbf{Top 30 Productive Institutions on the World Map: \textcolor{violet}{\textit{\nuclear}}}~|~1991--2020}\\
\vspace{-0em}
\hspace*{-3em}                                                           
\includegraphics[align=c, scale=0.83]{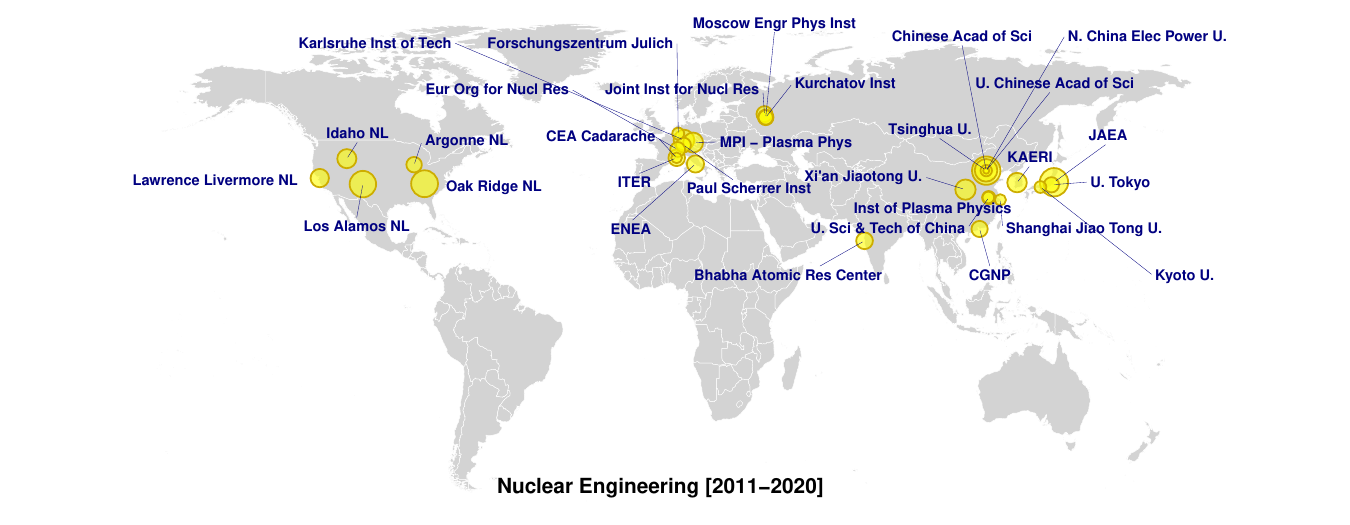}\\[-0.5em]
\quad\\[-1em]
\dotfill 
\quad\\[-0em]
\hspace*{-3em}                                                           
\includegraphics[align=c, scale=0.83]{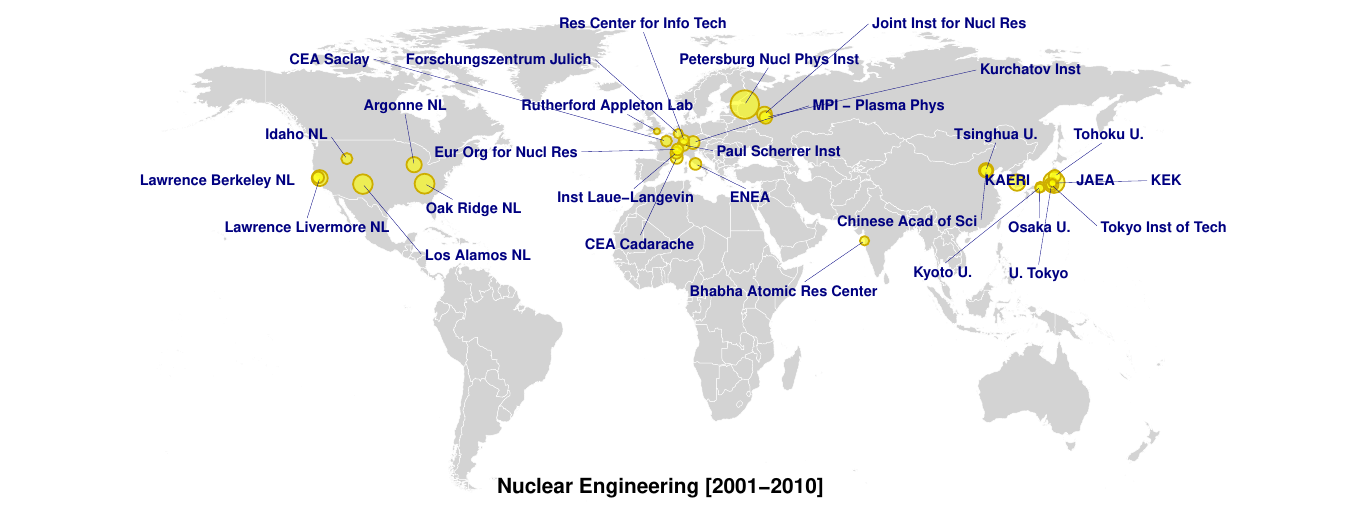}\\[-0.5em]
\quad\\[-1em]
\dotfill 
\quad\\[-0em]
\hspace*{-3em}                                                           
\includegraphics[align=c, scale=0.83]{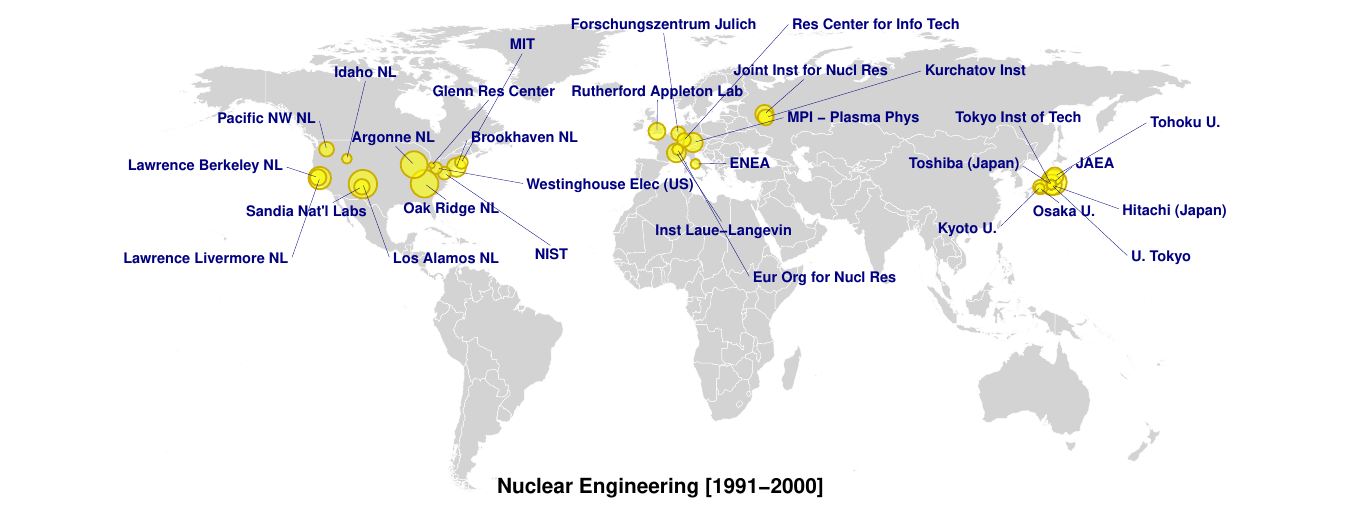}\\[-0.8em]
\caption[{\nuclear}]{\textbf{(b)~|~The top 30 productive institutions on the World Map: \textcolor{violet}{\textit{\nuclear}}.}
The bubbles represent the top 30 institutions in terms of work production, with their sizes proportional to the work volume.}
\label{fig:wmap_topinst_nuclear}
\end{figure}
}
\afterpage{\clearpage%
\begin{figure}[!tp]\ContinuedFloat
\centering
\vspace{-1em}
{\large \textbf{\textrm{{Interregional \textcolor{violet}{\textit{\nuclear}} Collaboration}}}~|~1991--2020}\\
\vspace{0.5em}
\hspace{-5em}\includegraphics[align=c, scale=1.7, vmargin=0mm]{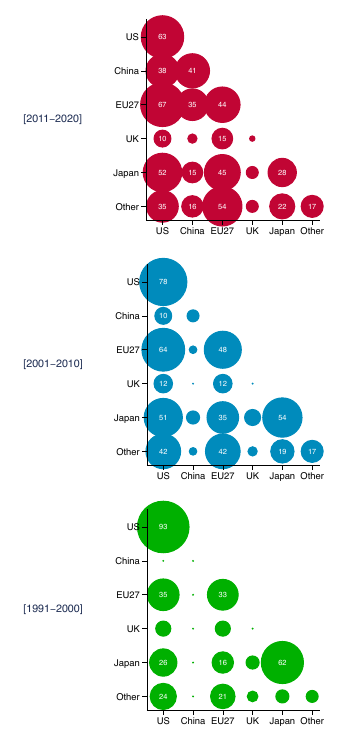}
\vspace{-1em}
\caption[]{\textbf{(c)~|~The Interregional \textcolor{violet}{\textit{\nuclear}} Collaboration Matrix Diagram.}
The bubble size represents the number of coauthorship relationships for the top 50 institutions in terms of work production. 
If the number is equal to or greater than 10, it is displayed inside the bubble.}
\label{fig:halfmat_nuclear}
\end{figure}
}
\afterpage{\clearpage%
\begin{figure}[!tp]\ContinuedFloat
\centering
\vspace{-1em}
{\large \textbf{\textrm{Interinstitutional \textcolor{violet}{\textit{\nuclear}} Collaboration}}~|~2001--2020\quad {\footnotesize \emph{(continued to next page)}}}\\
\cdend{nuclear}{2010}{2000}{2011--2020}{2001--2010}\\[-1.5em]
\caption[]{\textbf{(d)~|~The Interinstitutional \textcolor{violet}{\textit{\nuclear}} Collaboration Dendrogram.}
The top 50 institutions in terms of work production, indicated by the circularised bar graphs, are displayed.
}
\label{fig:cdend1_nuclear}
\end{figure}
}
\afterpage{\clearpage%
\begin{figure}[!tp]\ContinuedFloat
\centering
\vspace{-1em}
{\large \textbf{\textrm{Interinstitutional \textcolor{violet}{\textit{\nuclear}} Collaboration}}~|~1971--2000\quad {\footnotesize \emph{(continued from previous page)}}}\\
\cdend{nuclear}{1990}{1980}{1991--2000}{1971--1990}\\[-1.5em]
\caption[]{\textbf{(d)~|~The Interinstitutional \textcolor{violet}{\textit{\nuclear}} Collaboration Dendrogram.} \emph{(Cont.)}\hfill~}
\label{fig:cdend2_nuclear}
\end{figure}
}
\afterpage{\clearpage%
\begin{landscape}
\begin{table}[!t]
\vspace{-0.5em}
\caption{\textbf{The top 100 productive institutions: \textcolor{violet}{\textit{\nuclear}}.}}
\label{tab:r1_nuclear}
\vspace{2em}
\centering
{\tiny
{\renewcommand{\arraystretch}{1.2}
\begin{tabular}{rp{5cm}lr@{\hspace{4em}}p{5cm}lr@{\hspace{4em}}p{5cm}lr}\\[-5em] \toprule[1pt] \\[-1.4em]    
 & {\scriptsize \textbf{1991--2000}} & \multicolumn{2}{c}{No.~Works} & {\scriptsize \textbf{2001--2010}} & \multicolumn{2}{c}{No.~Works} & {\scriptsize \textbf{2011--2020}} & \multicolumn{2}{r}{No.~Works} \\[-0.2em] \cmidrule[0.5pt](lr{4em}){2-4} \cmidrule[0.5pt](l{-0em}r{4em}){5-7} \cmidrule[0.5pt](l{-0em}r{1em}){8-10}
\textit{1} & Los Alamos National Laboratory & US & 1,450 & Petersburg Nuclear Physics Institute & RU & 5,898 & Chinese Academy of Sciences & CN & 3,607 \\  
\textit{2} & Oak Ridge National Laboratory & US & 1,360 & Japan Atomic Energy Agency (JAEA) & JP & 2,519 & Japan Atomic Energy Agency (JAEA) & JP & 3,583 \\  
\textit{3} & Argonne National Laboratory & US & 1,252 & Oak Ridge National Laboratory & US & 2,104 & Oak Ridge National Laboratory & US & 3,173 \\  
\textit{4} & Japan Atomic Energy Agency (JAEA) & JP & 1,142 & Los Alamos National Laboratory & US & 2,040 & Los Alamos National Laboratory & US & 3,085 \\  
\textit{5} & Lawrence Livermore National Laboratory & US & 894 & Korea Atomic Energy Research Institute & KR & 1,404 & Tsinghua University & CN & 2,284 \\  
\textit{6} & The University of Tokyo & JP & 778 & Lawrence Livermore National Laboratory & US & 1,347 & Karlsruhe Institute of Technology & DE & 1,986 \\  
\textit{7} & Max Planck Institute for Plasma Physics & DE & 704 & Argonne National Laboratory & US & 1,248 & Xi'an Jiaotong University & CN & 1,935 \\  
\textit{8} & Institut Laue-Langevin & FR & 689 & Joint Institute for Nuclear Research & RU & 1,204 & Max Planck Institute for Plasma Physics & DE & 1,885 \\  
\textit{9} & Brookhaven National Laboratory & US & 686 & Tsinghua University & CN & 1,144 & Korea Atomic Energy Research Institute & KR & 1,829 \\  
\textit{10} & Joint Institute for Nuclear Research & RU & 659 & The University of Tokyo & JP & 1,097 & Idaho National Laboratory & US & 1,781 \\  
\hdashline          
\textit{11} & Tohoku University & JP & 614 & Max Planck Institute for Plasma Physics & DE & 1,002 & Lawrence Livermore National Laboratory & US & 1,692 \\  
\textit{12} & Rutherford Appleton Laboratory & GB & 596 & Kurchatov Institute & RU & 973 & \scalebox{0.7}[1]{National Agency for New Technologies, Energy and Sustainable Economic Development} & IT & 1,580 \\  
\textit{13} & Lawrence Berkeley National Laboratory & US & 575 & Chinese Academy of Sciences & CN & 945 & Bhabha Atomic Research Centre & IN & 1,535 \\  
\textit{14} & Kurchatov Institute & RU & 555 & Tohoku University & JP & 945 & CEA Cadarache & FR & 1,521 \\  
\textit{15} & Sandia National Laboratories & US & 547 & CEA Cadarache & FR & 930 & China General Nuclear Power Corporation (China) & CN & 1,489 \\  
\textit{16} & Pacific Northwest National Laboratory & US & 525 & Paul Scherrer Institute & CH & 916 & Joint Institute for Nuclear Research & RU & 1,468 \\  
\textit{17} & Kyoto University & JP & 519 & Institut Laue-Langevin & FR & 910 & Kurchatov Institute & RU & 1,433 \\  
\textit{18} & Forschungszentrum J\"{u}lich & DE & 510 & \scalebox{0.7}[1]{National Agency for New Technologies, Energy and Sustainable Economic Development} & IT & 907 & Argonne National Laboratory & US & 1,430 \\  
\textit{19} & Research Center for Information Technology & DE & 496 & Lawrence Berkeley National Laboratory & US & 899 & The University of Tokyo & JP & 1,413 \\  
\textit{20} & National Institute of Standards and Technology & US & 476 & European Organization for Nuclear Research & CH & 898 & European Organization for Nuclear Research & CH & 1,389 \\  
\hdashline          
\textit{21} & Massachusetts Institute of Technology & US & 474 & Idaho National Laboratory & US & 872 & Paul Scherrer Institute & CH & 1,389 \\  
\textit{22} & Westinghouse Electric (United States) & US & 453 & CEA Saclay & FR & 862 & Moscow Engineering Physics Institute & RU & 1,360 \\  
\textit{23} & European Organization for Nuclear Research & CH & 446 & Tokyo Institute of Technology & JP & 844 & University of Science and Technology of China & CN & 1,283 \\  
\textit{24} & Hitachi (Japan) & JP & 435 & Research Center for Information Technology & DE & 833 & Forschungszentrum J\"{u}lich & DE & 1,219 \\  
\textit{25} & Tokyo Institute of Technology & JP & 432 & Kyoto University & JP & 832 & Kyoto University & JP & 1,212 \\  
\textit{26} & \scalebox{0.7}[1]{National Agency for New Technologies, Energy and Sustainable Economic Development} & IT & 431 & Forschungszentrum J\"{u}lich & DE & 784 & Institute of Plasma Physics & CN & 1,208 \\  
\textit{27} & Osaka University & JP & 431 & Bhabha Atomic Research Centre & IN & 780 & North China Electric Power University & CN & 1,198 \\  
\textit{28} & Idaho National Laboratory & US & 428 & Osaka University & JP & 779 & Shanghai Jiao Tong University & CN & 1,192 \\  
\textit{29} & Toshiba (Japan) & JP & 427 & High Energy Accelerator Research Organization (KEK) & JP & 749 & ITER & FR & 1,172 \\  
\textit{30} & Glenn Research Center & US & 409 & Rutherford Appleton Laboratory & GB & 728 & University of Chinese Academy of Sciences & CN & 1,095 \\  
\hdashline          
\textit{31} & Advanced Science Research Center & JP & 408 & \scalebox{0.88}[1]{Institut National de Physique Nucl\'{e}aire et de Physique des Particules} & FR & 656 & Tohoku University & JP & 1,010 \\  
\textit{32} & Laboratoire L\'{e}on Brillouin & FR & 396 & Kyushu University & JP & 616 & \scalebox{0.88}[1]{Institut National de Physique Nucl\'{e}aire et de Physique des Particules} & FR & 1,008 \\  
\textit{33} & Technical University of Munich & DE & 393 & China Institute of Atomic Energy & CN & 610 & Massachusetts Institute of Technology & US & 1,000 \\  
\textit{34} & Bhabha Atomic Research Centre & IN & 388 & Brookhaven National Laboratory & US & 610 & Institut Laue-Langevin & FR & 974 \\  
\textit{35} & Kyushu University & JP & 375 & Technical University of Munich & DE & 608 & University of Tennessee at Knoxville & US & 948 \\  
\textit{36} & \scalebox{0.88}[1]{Institut National de Physique Nucl\'{e}aire et de Physique des Particules} & FR & 373 & Massachusetts Institute of Technology & US & 592 & Direction de L'\'{E}nergie Nucl\'{e}aire & FR & 945 \\  
\textit{37} & University of California, Berkeley & US & 356 & Max Planck Society & DE & 590 & Osaka University & JP & 934 \\  
\textit{38} & Nagoya University & JP & 353 & National Institute of Standards and Technology & US & 589 & Pacific Northwest National Laboratory & US & 923 \\  
\textit{39} & Paul Scherrer Institute & CH & 340 & Atomic Energy and Alternative Energies Commission & FR & 582 & Technical University of Munich & DE & 899 \\  
\textit{40} & Atomic Energy (Canada) & CA & 337 & Sandia National Laboratories & US & 582 & Russian Academy of Sciences & RU & 861 \\  
\hdashline          
\textit{41} & Korea Atomic Energy Research Institute & KR & 336 & Advanced Science Research Center & JP & 571 & Kyushu University & JP & 830 \\  
\textit{42} & Princeton Plasma Physics Laboratory & US & 332 & General Atomics (United States) & US & 552 & University of Michigan{\textendash}Ann Arbor & US & 822 \\  
\textit{43} & Delft University of Technology & NL & 328 & Russian Academy of Sciences & RU & 546 & Tokyo Institute of Technology & JP & 818 \\  
\textit{44} & Atomic Energy and Alternative Energies Commission & FR & 319 & Pacific Northwest National Laboratory & US & 536 & Lawrence Berkeley National Laboratory & US & 804 \\  
\textit{45} & High Energy Accelerator Research Organization (KEK) & JP & 314 & Shanghai Jiao Tong University & CN & 518 & Rutherford Appleton Laboratory & GB & 801 \\  
\textit{46} & General Atomics (United States) & US & 310 & Culham Science Centre & GB & 509 & China Institute of Atomic Energy & CN & 796 \\  
\textit{47} & Joint European Torus & GB & 308 & Princeton Plasma Physics Laboratory & US & 508 & Culham Science Centre & GB & 792 \\  
\textit{48} & University of Wisconsin{\textendash}Madison & US & 308 & Michigan State University & US & 506 & High Energy Accelerator Research Organization (KEK) & JP & 792 \\  
\textit{49} & CEA Cadarache & FR & 304 & Hitachi (Japan) & JP & 502 & GSI Helmholtz Centre for Heavy Ion Research & DE & 739 \\  
\textit{50} & Princeton University & US & 302 & Nagoya University & JP & 491 & University of California, Berkeley & US & 728 \\  
          
 \\[-1.4em]
\hdashline \\[-1em]
\multicolumn{10}{r}{\scriptsize \emph{(continued to next page)}}
\end{tabular}}
}
\end{table}
\end{landscape}
}
\afterpage{\clearpage%
\begin{landscape}
\begin{table}[!t]\ContinuedFloat
\vspace{-3.3em}
\caption{\textbf{The top 100 productive institutions: \textcolor{violet}{\textit{\nuclear}}.} \emph{(Cont.)}}
\label{tab:r2_nuclear}
\vspace{2em}
{\tiny
{\renewcommand{\arraystretch}{1.2}
\begin{tabular}{rp{5cm}lr@{\hspace{4em}}p{5cm}lr@{\hspace{4em}}p{5cm}lr}\\[-5em] \toprule[1pt] \\[-1.4em]    
 & {\scriptsize \textbf{1991--2000}} & \multicolumn{2}{c}{No.\ Works} & {\scriptsize \textbf{2001--2010}} & \multicolumn{2}{c}{No.\ Works} & {\scriptsize \textbf{2011--2020}} & \multicolumn{2}{r}{No.\ Works} \\[-0.2em] \cmidrule[0.5pt](lr{4em}){2-4} \cmidrule[0.5pt](l{-0em}r{4em}){5-7} \cmidrule[0.5pt](l{-0em}r{1em}){8-10}
\textit{51} & CEA Saclay & FR & 301 & Pennsylvania State University & US & 488 & Michigan State University & US & 724 \\  
\textit{52} & China Institute of Atomic Energy & CN & 277 & University of California, Berkeley & US & 472 & Texas A\&M University & US & 707 \\  
\textit{53} & Pennsylvania State University & US & 276 & National Institute for Fusion Science & JP & 467 & Nagoya University & JP & 672 \\  
\textit{54} & Max Planck Society & DE & 275 & Toshiba (Japan) & JP & 465 & Brookhaven National Laboratory & US & 664 \\  
\textit{55} & National Institute for Fusion Science & JP & 263 & Culham Centre for Fusion Energy & GB & 460 & Atomic Energy and Alternative Energies Commission & FR & 654 \\  
\textit{56} & University of Illinois Urbana-Champaign & US & 253 & RIKEN & JP & 453 & Tianjin University & CN & 651 \\  
\textit{57} & University of California, Los Angeles & US & 252 & Hokkaido University & JP & 448 & Harbin Institute of Technology & CN & 646 \\  
\textit{58} & Michigan State University & US & 251 & Xi'an Jiaotong University & CN & 443 & Huazhong University of Science and Technology & CN & 639 \\  
\textit{59} & Texas A\&M University & US & 246 & China General Nuclear Power Corporation (China) & CN & 440 & Politecnico di Milano & IT & 631 \\  
\textit{60} & Petersburg Nuclear Physics Institute & RU & 243 & Glenn Research Center & US & 440 & University of Wisconsin{\textendash}Madison & US & 630 \\  
\hdashline          
\textit{61} & RIKEN & JP & 242 & Institute of Plasma Physics & CN & 436 & China Academy of Engineering Physics & CN & 615 \\  
\textit{62} & University of Maryland, College Park & US & 241 & Fermilab & US & 423 & Max Planck Society & DE & 614 \\  
\textit{63} & \scalebox{0.9}[1]{Fraunhofer Institute for Telecommunications, Heinrich Hertz Institute} & DE & 236 & Institute for Physics and Power Engineering & RU & 419 & \scalebox{0.8}[1]{Centro de Investigaciones Energ\'{e}ticas, Medioambientales y Tecnol\'{o}gicas} & ES & 612 \\  
\textit{64} & University of Tennessee at Knoxville & US & 232 & Delft University of Technology & NL & 415 & CEA Saclay & FR & 610 \\  
\textit{65} & Fusion (United States) & US & 231 & Royal Institute of Technology & SE & 415 & University of Paris-Saclay & FR & 610 \\  
\textit{66} & Institute for Physics and Power Engineering & RU & 228 & University of Wisconsin{\textendash}Madison & US & 415 & Hokkaido University & JP & 609 \\  
\textit{67} & Tsinghua University & CN & 224 & Direction de L'\'{E}nergie Nucl\'{e}aire & FR & 410 & Korea Advanced Institute of Science and Technology & KR & 605 \\  
\textit{68} & Chalmers University of Technology & SE & 222 & University of Michigan{\textendash}Ann Arbor & US & 407 & Sandia National Laboratories California & US & 605 \\  
\textit{69} & Johannes Gutenberg University Mainz & DE & 218 & GSI Helmholtz Centre for Heavy Ion Research & DE & 402 & \'{E}cole Polytechnique F\'{e}d\'{e}rale de Lausanne & CH & 603 \\  
\textit{70} & Kyoto University Research Reactor Institute & JP & 212 & University of Tennessee at Knoxville & US & 401 & National Institute of Standards and Technology & US & 601 \\  
\hdashline          
\textit{71} & North Carolina State University & US & 211 & Princeton University & US & 400 & Beihang University & CN & 600 \\  
\textit{72} & University of Warsaw & PL & 208 & Central Research Institute of Electric Power Industry & JP & 389 & University of Manchester & GB & 598 \\  
\textit{73} & CEA Grenoble & FR & 206 & University of Maryland, College Park & US & 378 & Sandia National Laboratories & US & 596 \\  
\textit{74} & Ricardo AEA (United Kingdom) & GB & 206 & Belgian Nuclear Research Centre & BE & 363 & Peking University & CN & 593 \\  
\textit{75} & Royal Institute of Technology & SE & 205 & National Superconducting Cyclotron Laboratory & US & 363 & Harbin Engineering University & CN & 592 \\  
\textit{76} & Uppsala University & SE & 204 & Seoul National University & KR & 352 & Institut de Radioprotection et de S\^{u}ret\'{e} Nucl\'{e}aire & FR & 587 \\  
\textit{77} & Siemens (Germany) & DE & 197 & Fusion (United States) & US & 349 & National Nuclear Energy Agency of Indonesia & ID & 581 \\  
\textit{78} & The University of Texas at Austin & US & 196 & University of Manchester & GB & 348 & National Institute for Fusion Science & JP & 580 \\  
\textit{79} & Central Research Institute of Electric Power Industry & JP & 193 & Texas A\&M University & US & 343 & Seoul National University & KR & 577 \\  
\textit{80} & Fusion Academy & US & 190 & TEPCO (Japan) & JP & 341 & Belgian Nuclear Research Centre & BE & 568 \\  
\hdashline          
\textit{81} & University of Oxford & GB & 187 & North China Electric Power University & CN & 339 & Princeton Plasma Physics Laboratory & US & 567 \\  
\textit{82} & Fermilab & US & 187 & Uppsala University & SE & 337 & Royal Institute of Technology & SE & 564 \\  
\textit{83} & Institut de Physique & FR & 186 & North Carolina State University & US & 336 & Fermilab & US & 564 \\  
\textit{84} & TEPCO (Japan) & JP & 185 & Kyoto University Research Reactor Institute & JP & 333 & Pennsylvania State University & US & 557 \\  
\textit{85} & Russian Academy of Sciences & RU & 182 & China Academy of Engineering Physics & CN & 332 & General Atomics (United States) & US & 554 \\  
\textit{86} & Joint Research Centre & IT & 178 & University of Pisa & IT & 330 & Technical University of Darmstadt & DE & 551 \\  
\textit{87} & University of Manchester & GB & 177 & Marshall Space Flight Center & US & 322 & Institute for Physics and Power Engineering & RU & 550 \\  
\textit{88} & University of Michigan{\textendash}Ann Arbor & US & 176 & Fusion Academy & US & 320 & University of Padua & IT & 549 \\  
\textit{89} & United States Nuclear Regulatory Commission & US & 175 & Westinghouse Electric (United States) & US & 316 & Indira Gandhi Centre for Atomic Research & IN & 543 \\  
\textit{90} & University of Missouri & US & 175 & Helmholtz-Zentrum Dresden-Rossendorf & DE & 315 & University of Tsukuba & JP & 543 \\  
\hdashline          
\textit{91} & National Superconducting Cyclotron Laboratory & US & 174 & Institut de Radioprotection et de S\^{u}ret\'{e} Nucl\'{e}aire & FR & 315 & Institute of Modern Physics & CN & 541 \\  
\textit{92} & University of Birmingham & GB & 168 & Georgia Institute of Technology & US & 315 & Savannah River National Laboratory & US & 540 \\  
\textit{93} & \scalebox{0.9}[1]{Science Applications International Corporation (United States)} & US & 168 & Savannah River National Laboratory & US & 304 & Tomsk Polytechnic University & RU & 538 \\  
\textit{94} & Hokkaido University & JP & 167 & Peking University & CN & 295 & Imperial College London & GB & 535 \\  
\textit{95} & Lockheed Martin (United States) & US & 165 & \scalebox{0.82}[1]{National Institute of Advanced Industrial Science and Technology (AIST)} & JP & 292 & Czech Technical University in Prague & CZ & 532 \\  
\textit{96} & Nordic Laboratory for Luminescence Dating & DK & 163 & GANIL & FR & 290 & Fusion for Energy & ES & 528 \\  
\textit{97} & Korea Advanced Institute of Science and Technology & KR & 162 & International Atomic Energy Agency & AT & 289 & University of Ontario Institute of Technology & CA & 527 \\  
\textit{98} & Imperial College London & GB & 160 & Chalmers University of Technology & SE & 289 & Georgia Institute of Technology & US & 526 \\  
\textit{99} & Hungarian Academy of Sciences & HU & 158 & Institut de Physique & FR & 286 & North Carolina State University & US & 518 \\  
\textit{100} & University of Liverpool & GB & 155 & Korea Advanced Institute of Science and Technology & KR & 282 & University of Maryland, College Park & US & 509 \\  
          
 \\[-1.4em]
\bottomrule
\end{tabular}}
}
\end{table}
\end{landscape}
}
%

\titleformat{\section}{\sc\centering\LARGE\bfseries}{\textsc{\thesection}.\!\!}{1em}{}

\afterpage{\clearpage%
\markboth{\textbf \textsc{\marine}}{}
\thispagestyle{empty}
\quad
\vspace{2cm}
\begin{center}
\pgfornament[width=0.5*\textwidth,symmetry=h]{89}\\[2em]
\section{\marine}
\vspace{1em}
\pgfornament[width=0.5*\textwidth]{89}
\end{center}
}

\afterpage{\clearpage%

\begin{figure}[!tp]
\centering
\vspace{-1em}
{\large \textbf{\textrm{{World Map of \textcolor{violet}{\textit{\marine}} Collaboration}}}~|~1971--2020}\\
\vspace{0.3cm}
\includegraphics[align=c, scale=0.054, trim={9.5cm 0 9.5cm 0},clip]{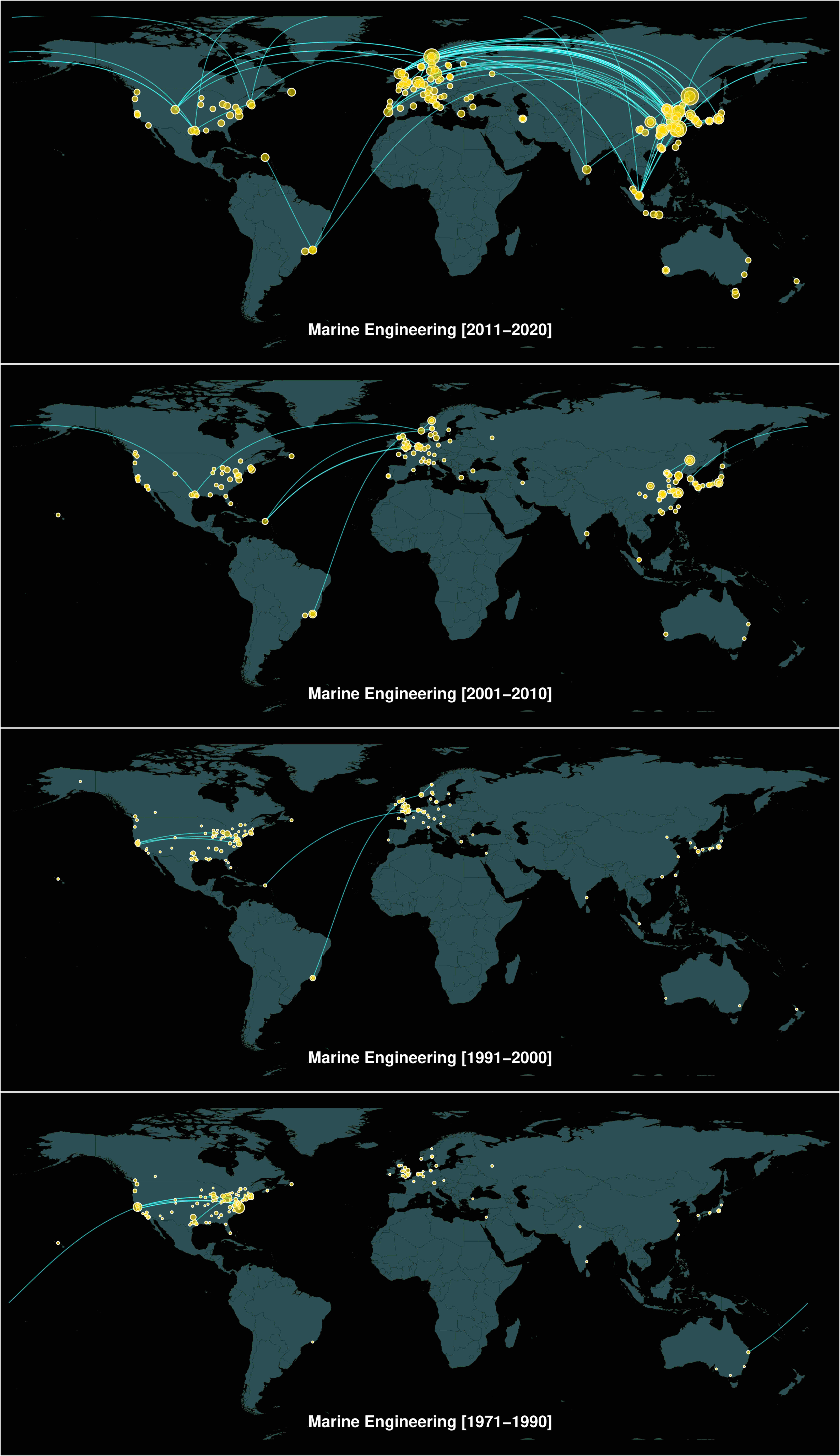}
\caption[{\marine}]{\textbf{(a)~|~The World Map of \textcolor{violet}{\textit{\marine}} Collaboration.}
The bubbles represent the top 199 institutions in terms of work production, with their sizes proportional to the work volume. 
The connecting lines depict coauthorship relationships among the top 50 institutions.}
\label{fig:wmap_marine}
\end{figure}
}
\afterpage{\clearpage%
\begin{figure}[!tp]\ContinuedFloat
\centering
\vspace{-1em}
{\large \textbf{Top 30 Productive Institutions on the World Map: \textcolor{violet}{\textit{\marine}}}~|~1991--2020}\\
\vspace{-0em}
\hspace*{-3em}                                                           
\includegraphics[align=c, scale=0.83]{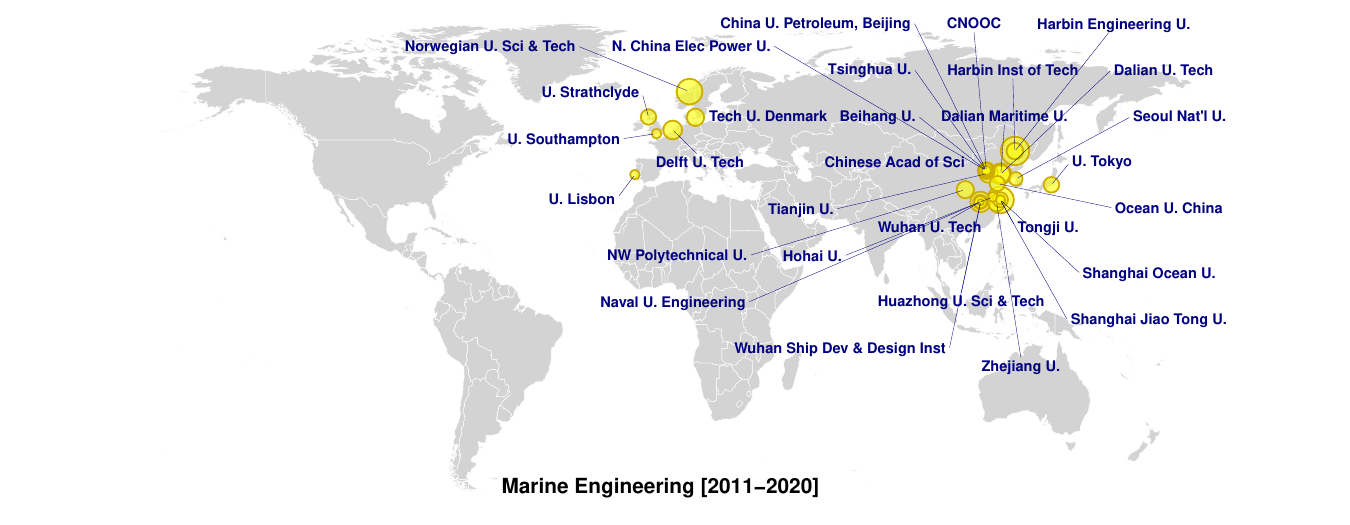}\\[-0.5em]
\quad\\[-1em]
\dotfill 
\quad\\[-0em]
\hspace*{-3em}                                                           
\includegraphics[align=c, scale=0.83]{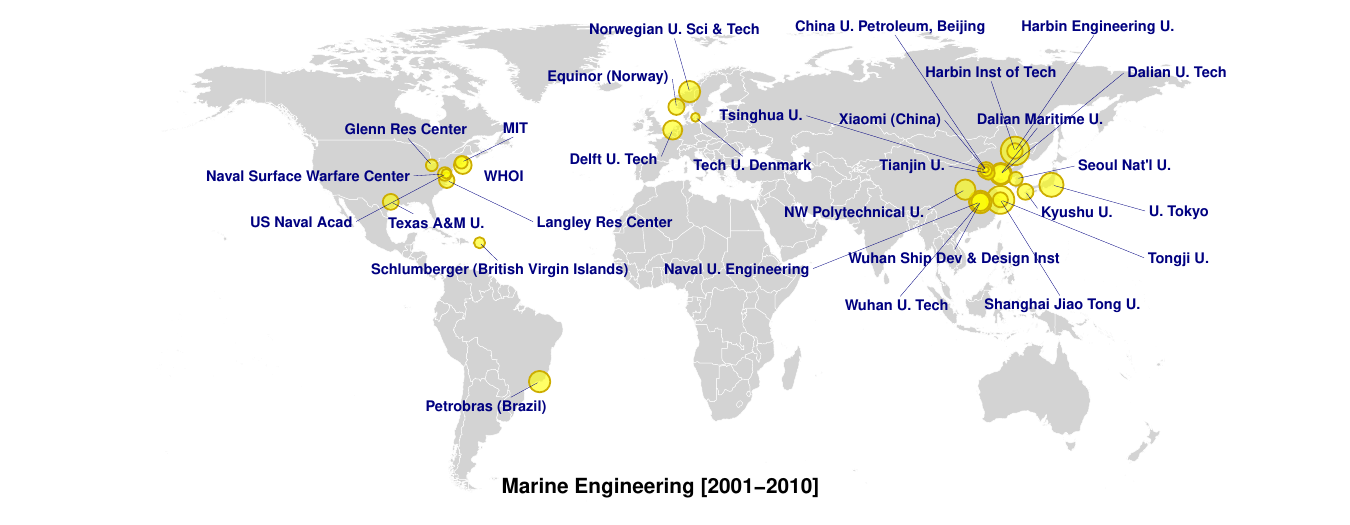}\\[-0.5em]
\quad\\[-1em]
\dotfill 
\quad\\[-0em]
\hspace*{-3em}                                                           
\includegraphics[align=c, scale=0.83]{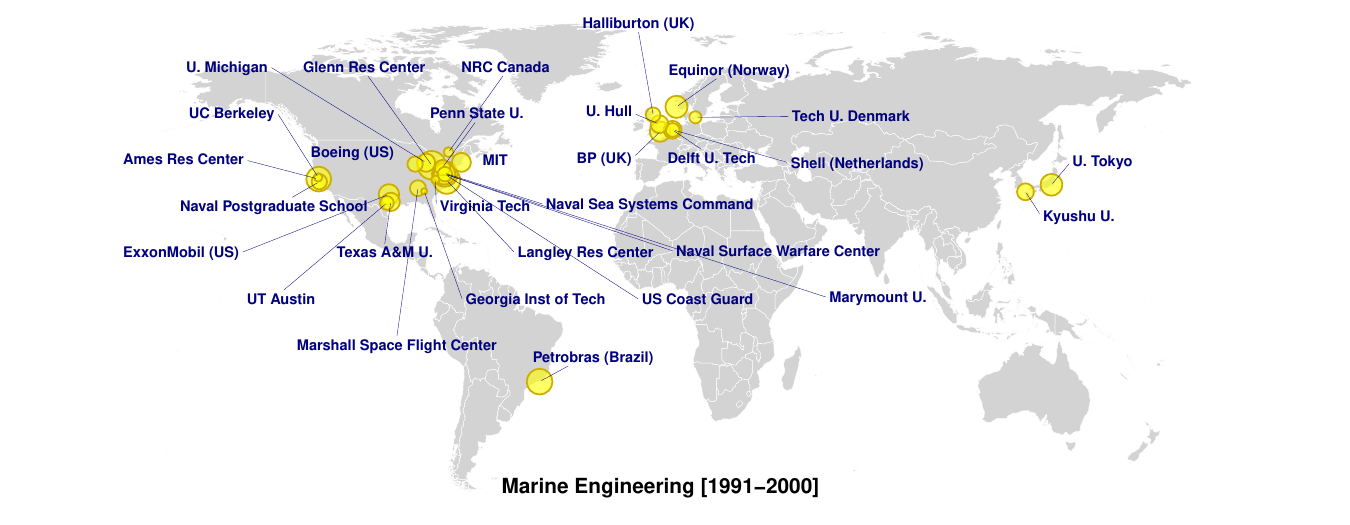}\\[-0.8em]
\caption[{\marine}]{\textbf{(b)~|~The top 30 productive institutions on the World Map: \textcolor{violet}{\textit{\marine}}.}
The bubbles represent the top 30 institutions in terms of work production, with their sizes proportional to the work volume.}
\label{fig:wmap_topinst_marine}
\end{figure}
}
\afterpage{\clearpage%
\begin{figure}[!tp]\ContinuedFloat
\centering
\vspace{-1em}
{\large \textbf{\textrm{{Interregional \textcolor{violet}{\textit{\marine}} Collaboration}}}~|~1991--2020}\\
\vspace{0.5em}
\hspace{-5em}\includegraphics[align=c, scale=1.7, vmargin=0mm]{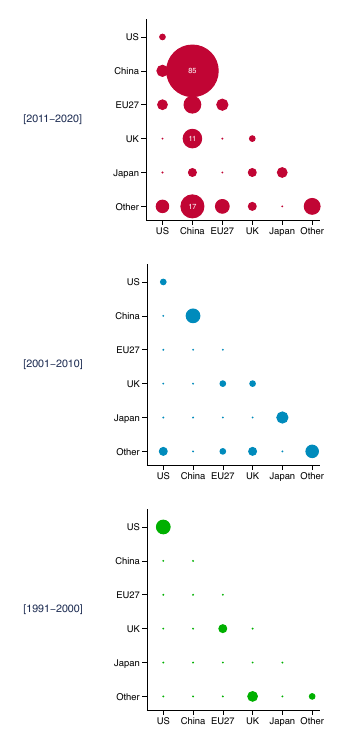}
\vspace{-1em}
\caption[]{\textbf{(c)~|~The Interregional \textcolor{violet}{\textit{\marine}} Collaboration Matrix Diagram.}
The bubble size represents the number of coauthorship relationships for the top 50 institutions in terms of work production. 
If the number is equal to or greater than 10, it is displayed inside the bubble.}
\label{fig:halfmat_marine}
\end{figure}
}
\afterpage{\clearpage%
\begin{figure}[!tp]\ContinuedFloat
\centering
\vspace{-1em}
{\large \textbf{\textrm{Interinstitutional \textcolor{violet}{\textit{\marine}} Collaboration}}~|~2001--2020\quad {\footnotesize \emph{(continued to next page)}}}\\
\cdend{marine}{2010}{2000}{2011--2020}{2001--2010}\\[-1.5em]
\caption[]{\textbf{(d)~|~The Interinstitutional \textcolor{violet}{\textit{\marine}} Collaboration Dendrogram.}
The top 50 institutions in terms of work production, indicated by the circularised bar graphs, are displayed.
}
\label{fig:cdend1_marine}
\end{figure}
}
\afterpage{\clearpage%
\begin{figure}[!tp]\ContinuedFloat
\centering
\vspace{-1em}
{\large \textbf{\textrm{Interinstitutional \textcolor{violet}{\textit{\marine}} Collaboration}}~|~1971--2000\quad {\footnotesize \emph{(continued from previous page)}}}\\
\cdend{marine}{1990}{1980}{1991--2000}{1971--1990}\\[-1.5em]
\caption[]{\textbf{(d)~|~The Interinstitutional \textcolor{violet}{\textit{\marine}} Collaboration Dendrogram.} \emph{(Cont.)}\hfill~}
\label{fig:cdend2_marine}
\end{figure}
}
\afterpage{\clearpage%
\begin{landscape}
\begin{table}[!t]
\vspace{-0.5em}
\caption{\textbf{The top 100 productive institutions: \textcolor{violet}{\textit{\marine}}.}}
\label{tab:r1_marine}
\vspace{2em}
\centering
{\tiny
{\renewcommand{\arraystretch}{1.2}
\begin{tabular}{rp{5cm}lr@{\hspace{4em}}p{5cm}lr@{\hspace{4em}}p{5cm}lr}\\[-5em] \toprule[1pt] \\[-1.4em]    
 & {\scriptsize \textbf{1991--2000}} & \multicolumn{2}{c}{No.~Works} & {\scriptsize \textbf{2001--2010}} & \multicolumn{2}{c}{No.~Works} & {\scriptsize \textbf{2011--2020}} & \multicolumn{2}{r}{No.~Works} \\[-0.2em] \cmidrule[0.5pt](lr{4em}){2-4} \cmidrule[0.5pt](l{-0em}r{4em}){5-7} \cmidrule[0.5pt](l{-0em}r{1em}){8-10}
\textit{1} & Glenn Research Center & US & 213 & Harbin Engineering University & CN & 600 & Harbin Engineering University & CN & 1,842 \\  
\textit{2} & Langley Research Center & US & 190 & Shanghai Jiao Tong University & CN & 578 & Shanghai Jiao Tong University & CN & 1,602 \\  
\textit{3} & Petrobras (Brazil) & BR & 171 & The University of Tokyo & JP & 433 & Norwegian University of Science and Technology & NO & 1,457 \\  
\textit{4} & Naval Surface Warfare Center & US & 169 & Naval University of Engineering & CN & 386 & Dalian University of Technology & CN & 978 \\  
\textit{5} & Ames Research Center & US & 159 & Dalian University of Technology & CN & 371 & Wuhan University of Technology & CN & 921 \\  
\textit{6} & Equinor (Norway) & NO & 129 & Wuhan Ship Development \& Design Institute & CN & 368 & Tianjin University & CN & 802 \\  
\textit{7} & The University of Tokyo & JP & 126 & Petrobras (Brazil) & BR & 353 & Delft University of Technology & NL & 801 \\  
\textit{8} & Naval Sea Systems Command & US & 126 & Norwegian University of Science and Technology & NO & 350 & Dalian Maritime University & CN & 778 \\  
\textit{9} & Marymount University & US & 124 & Northwestern Polytechnical University & CN & 328 & Technical University of Denmark & DK & 721 \\  
\textit{10} & ExxonMobil (United States) & US & 114 & Dalian Maritime University & CN & 317 & Northwestern Polytechnical University & CN & 714 \\  
\hdashline          
\textit{11} & BP (United Kingdom) & GB & 113 & Delft University of Technology & NL & 302 & Chinese Academy of Sciences & CN & 699 \\  
\textit{12} & Massachusetts Institute of Technology & US & 103 & Wuhan University of Technology & CN & 287 & Zhejiang University & CN & 690 \\  
\textit{13} & Texas A\&M University & US & 98 & Woods Hole Oceanographic Institution & US & 278 & Harbin Institute of Technology & CN & 682 \\  
\textit{14} & University of Michigan{\textendash}Ann Arbor & US & 98 & Xiaomi (China) & CN & 255 & The University of Tokyo & JP & 638 \\  
\textit{15} & University of Hull & GB & 95 & Equinor (Norway) & NO & 255 & China University of Petroleum, Beijing & CN & 637 \\  
\textit{16} & Shell (Netherlands) & NL & 95 & Kyushu University & JP & 251 & University of Strathclyde & GB & 626 \\  
\textit{17} & Kyushu University & JP & 90 & Texas A\&M University & US & 250 & Ocean University of China & CN & 622 \\  
\textit{18} & Marshall Space Flight Center & US & 85 & Harbin Institute of Technology & CN & 246 & Shanghai Ocean University & CN & 618 \\  
\textit{19} & Pennsylvania State University & US & 83 & Langley Research Center & US & 244 & Tsinghua University & CN & 596 \\  
\textit{20} & Boeing (United States) & US & 82 & Tongji University & CN & 240 & Naval University of Engineering & CN & 584 \\  
\hdashline          
\textit{21} & Naval Postgraduate School & US & 82 & Seoul National University & KR & 227 & Beihang University & CN & 573 \\  
\textit{22} & Halliburton (United Kingdom) & GB & 79 & Naval Surface Warfare Center & US & 224 & Seoul National University & KR & 564 \\  
\textit{23} & United States Coast Guard & US & 79 & Tianjin University & CN & 216 & North China Electric Power University & CN & 556 \\  
\textit{24} & Delft University of Technology & NL & 76 & Massachusetts Institute of Technology & US & 215 & Wuhan Ship Development \& Design Institute & CN & 527 \\  
\textit{25} & The University of Texas at Austin & US & 76 & Tsinghua University & CN & 210 & Hohai University & CN & 482 \\  
\textit{26} & Technical University of Denmark & DK & 71 & Glenn Research Center & US & 209 & University of Southampton & GB & 480 \\  
\textit{27} & National Research Council Canada & CA & 66 & Schlumberger (British Virgin Islands) & VG & 205 & Tongji University & CN & 479 \\  
\textit{28} & University of California, Berkeley & US & 64 & United States Naval Academy & US & 198 & China National Offshore Oil Corporation (China) & CN & 478 \\  
\textit{29} & Virginia Tech & US & 64 & China University of Petroleum, Beijing & CN & 194 & University of Lisbon & PT & 478 \\  
\textit{30} & University of British Columbia & CA & 63 & Technical University of Denmark & DK & 194 & Huazhong University of Science and Technology & CN & 458 \\  
\hdashline          
\textit{31} & Georgia Institute of Technology & US & 63 & Shell (Netherlands) & NL & 189 & Indian Institute of Technology Madras & IN & 450 \\  
\textit{32} & DNV GL (Norway) & NO & 60 & University of Southampton & GB & 185 & SINTEF & NO & 439 \\  
\textit{33} & Woods Hole Oceanographic Institution & US & 60 & Hohai University & CN & 183 & Jiangsu University of Science and Technology & CN & 435 \\  
\textit{34} & Loughborough University & GB & 59 & Pennsylvania State University & US & 177 & National University of Singapore & SG & 427 \\  
\textit{35} & University of Strathclyde & GB & 55 & Osaka University & JP & 175 & Aalborg University & DK & 424 \\  
\textit{36} & Norwegian University of Science and Technology & NO & 55 & Federal University of Rio de Janeiro & BR & 160 & National Renewable Energy Laboratory & US & 411 \\  
\textit{37} & Schlumberger (British Virgin Islands) & VG & 54 & China National Offshore Oil Corporation (China) & CN & 160 & Pusan National University & KR & 408 \\  
\textit{38} & Stanford University & US & 53 & Jiangsu University of Science and Technology & CN & 160 & Politecnico di Milano & IT & 379 \\  
\textit{39} & University of Glasgow & GB & 52 & Zhejiang University & CN & 160 & Federal University of Rio de Janeiro & BR & 377 \\  
\textit{40} & Norsk Hydro (Germany) & DE & 51 & Tohoku University & JP & 160 & Shanghai Maritime University & CN & 377 \\  
\hdashline          
\textit{41} & University of Cambridge & GB & 50 & Pusan National University & KR & 160 & Massachusetts Institute of Technology & US & 376 \\  
\textit{42} & University of Southampton & GB & 50 & Halliburton (United Kingdom) & GB & 158 & Sepuluh Nopember Institute of Technology & ID & 374 \\  
\textit{43} & Memorial University of Newfoundland & CA & 48 & University of Michigan{\textendash}Ann Arbor & US & 157 & Texas A\&M University & US & 374 \\  
\textit{44} & Shell (United Kingdom) & GB & 47 & BP (United Kingdom) & GB & 156 & Schlumberger (British Virgin Islands) & VG & 368 \\  
\textit{45} & University College London & GB & 46 & The University of Texas at Austin & US & 156 & Petrobras (Brazil) & BR & 366 \\  
\textit{46} & SINTEF & NO & 45 & \scalebox{0.85}[1]{Japan Agency for Marine-Earth Science and Technology (JAMSTEC)} & JP & 155 & University of Technology Malaysia & MY & 360 \\  
\textit{47} & ConocoPhillips (United States) & US & 45 & Wuhan University & CN & 154 & Nanjing University of Aeronautics and Astronautics & CN & 359 \\  
\textit{48} & University of Maryland, College Park & US & 45 & Beihang University & CN & 152 & Memorial University of Newfoundland & CA & 358 \\  
\textit{49} & Hiroshima University & JP & 44 & Chinese Academy of Sciences & CN & 150 & Kyushu University & JP & 356 \\  
\textit{50} & University of Illinois Urbana-Champaign & US & 44 & Huazhong University of Science and Technology & CN & 148 & Langley Research Center & US & 353 \\  
          
 \\[-1.4em]
\hdashline \\[-1em]
\multicolumn{10}{r}{\scriptsize \emph{(continued to next page)}}
\end{tabular}}
}
\end{table}
\end{landscape}
}
\afterpage{\clearpage%
\begin{landscape}
\begin{table}[!t]\ContinuedFloat
\vspace{-3.3em}
\caption{\textbf{The top 100 productive institutions: \textcolor{violet}{\textit{\marine}}.} \emph{(Cont.)}}
\label{tab:r2_marine}
\vspace{2em}
{\tiny
{\renewcommand{\arraystretch}{1.2}
\begin{tabular}{rp{5cm}lr@{\hspace{4em}}p{5cm}lr@{\hspace{4em}}p{5cm}lr}\\[-5em] \toprule[1pt] \\[-1.4em]    
 & {\scriptsize \textbf{1991--2000}} & \multicolumn{2}{c}{No.\ Works} & {\scriptsize \textbf{2001--2010}} & \multicolumn{2}{c}{No.\ Works} & {\scriptsize \textbf{2011--2020}} & \multicolumn{2}{r}{No.\ Works} \\[-0.2em] \cmidrule[0.5pt](lr{4em}){2-4} \cmidrule[0.5pt](l{-0em}r{4em}){5-7} \cmidrule[0.5pt](l{-0em}r{1em}){8-10}
\textit{51} & National Technical University of Athens & GR & 43 & DNV GL (Norway) & NO & 148 & Osaka University & JP & 347 \\  
\textit{52} & Naval Undersea Warfare Center & US & 43 & Hong Kong Polytechnic University & CN & 147 & Zhejiang Ocean University & CN & 340 \\  
\textit{53} & University of New Orleans & US & 43 & Ocean University of China & CN & 145 & Equinor (Norway) & NO & 333 \\  
\textit{54} & Hong Kong Polytechnic University & CN & 42 & Georgia Institute of Technology & US & 145 & Chalmers University of Technology & SE & 332 \\  
\textit{55} & Yokohama National University & JP & 42 & Virginia Tech & US & 144 & \scalebox{0.85}[1]{Japan Agency for Marine-Earth Science and Technology (JAMSTEC)} & JP & 324 \\  
\textit{56} & Federal University of Rio de Janeiro & BR & 41 & Memorial University of Newfoundland & CA & 141 & Amirkabir University of Technology & IR & 323 \\  
\textit{57} & Newcastle University & GB & 41 & Shanghai Maritime University & CN & 141 & University of Genoa & IT & 322 \\  
\textit{58} & University of Nottingham & GB & 40 & Universidade de S\~{a}o Paulo & BR & 140 & University of Western Australia & AU & 314 \\  
\textit{59} & Lockheed Martin (United States) & US & 40 & Central South University & CN & 139 & DNV GL (Norway) & NO & 314 \\  
\textit{60} & University of Washington & US & 40 & Tokyo Institute of Technology & JP & 139 & University of Tasmania & AU & 313 \\  
\hdashline          
\textit{61} & J{\o}tul (Norway) & NO & 39 & German Aerospace Center & DE & 131 & Hong Kong Polytechnic University & CN & 308 \\  
\textit{62} & United States Naval Academy & US & 39 & University of Strathclyde & GB & 131 & German Aerospace Center & DE & 307 \\  
\textit{63} & Ifremer & FR & 38 & Aalborg University & DK & 130 & Engineering (Italy) & IT & 302 \\  
\textit{64} & Nihon University & JP & 38 & National Maritime Research Institute & JP & 129 & The University of Texas at Austin & US & 301 \\  
\textit{65} & Baker Hughes (United States) & US & 38 & American Bureau of Shipping & US & 128 & Korea Maritime and Ocean University & KR & 299 \\  
\textit{66} & Heriot-Watt University & GB & 37 & Kobe University & JP & 127 & Southwest Jiaotong University & CN & 298 \\  
\textit{67} & United States Geological Survey & US & 37 & National Technical University of Athens & GR & 126 & Central South University & CN & 296 \\  
\textit{68} & Mitsubishi Heavy Industries (Germany) & DE & 36 & North China Electric Power University & CN & 124 & National Technical University of Athens & GR & 291 \\  
\textit{69} & Aalborg University & DK & 36 & Tokyo University of Marine Science and Technology & JP & 124 & Newcastle University & GB & 288 \\  
\textit{70} & United States Army & US & 36 & Total (France) & FR & 122 & Technical University of Munich & DE & 286 \\  
\hdashline          
\textit{71} & United States Naval Research Laboratory & US & 36 & University of Cambridge & GB & 119 & Cranfield University & GB & 285 \\  
\textit{72} & University of Iowa & US & 36 & Indian Institute of Technology Madras & IN & 119 & University of Stuttgart & DE & 283 \\  
\textit{73} & Weatherford College & US & 36 & Hokkaido University & JP & 118 & Universidade de S\~{a}o Paulo & BR & 282 \\  
\textit{74} & Indian Institute of Technology Madras & IN & 35 & Kyushu Institute of Technology & JP & 117 & Diponegoro University & ID & 281 \\  
\textit{75} & Saga University & JP & 35 & Shanghai University & CN & 116 & Virginia Tech & US & 278 \\  
\textit{76} & American Bureau of Shipping & US & 35 & University of Lisbon & PT & 115 & Chongqing University & CN & 276 \\  
\textit{77} & Johnson Space Center & US & 35 & United States Department of the Navy & US & 115 & Nanyang Technological University & SG & 275 \\  
\textit{78} & North Carolina State University & US & 35 & Ifremer & FR & 114 & University of Michigan{\textendash}Ann Arbor & US & 273 \\  
\textit{79} & National Cheng Kung University & TW & 34 & Nanyang Technological University & SG & 114 & Beijing Institute of Technology & CN & 269 \\  
\textit{80} & McDermott International (United States) & US & 34 & Tianjin Research Institute of Water Transport Engineering & CN & 113 & University of Edinburgh & GB & 269 \\  
\hdashline          
\textit{81} & OT Energy Services (Czechia) & CZ & 33 & Ames Research Center & US & 113 & Wuhan University & CN & 268 \\  
\textit{82} & Queen's University Belfast & GB & 33 & TechnipFMC (United States) & US & 113 & National Maritime Research Institute & JP & 266 \\  
\textit{83} & National University of Singapore & SG & 33 & California Maritime Academy & US & 112 & Glenn Research Center & US & 265 \\  
\textit{84} & Oregon State University & US & 33 & Sinopec (China) & CN & 111 & Georgia Institute of Technology & US & 260 \\  
\textit{85} & University of Hawaii System & US & 33 & Tokai University & JP & 111 & Shandong University & CN & 257 \\  
\textit{86} & Whitney Museum of American Art & US & 33 & National University of Singapore & SG & 110 & South China University of Technology & CN & 253 \\  
\textit{87} & BP (Germany) & DE & 32 & National Research Council Canada & CA & 109 & Xi'an Jiaotong University & CN & 251 \\  
\textit{88} & General Electric (United States) & US & 32 & National Renewable Energy Laboratory & US & 109 & \scalebox{0.8}[1]{Collaborative Innovation Centre for Advanced Ship and Deep-Sea Exploration} & CN & 250 \\  
\textit{89} & Shanghai University & CN & 31 & Tokyo University of Science & JP & 108 & Woods Hole Oceanographic Institution & US & 250 \\  
\textit{90} & Technical University of Berlin & DE & 31 & University of Washington & US & 107 & Gda\'{n}sk University of Technology & PL & 248 \\  
\hdashline          
\textit{91} & Norwegian Marine Technology Research Institute & NO & 31 & Yokohama National University & JP & 106 & National Cheng Kung University & TW & 248 \\  
\textit{92} & California Maritime Academy & US & 31 & Stennis Space Center & US & 106 & United States Naval Academy & US & 248 \\  
\textit{93} & Louisiana State University & US & 31 & National Cheng Kung University & TW & 104 & University of Stavanger & NO & 245 \\  
\textit{94} & National Renewable Energy Laboratory & US & 31 & Shanghai Harbour Engineering Design \& Research Institute & CN & 103 & Universitat Polit\`{e}cnica de Catalunya & ES & 244 \\  
\textit{95} & Rockwell Automation (United States) & US & 31 & South China University of Technology & CN & 103 & Tohoku University & JP & 242 \\  
\textit{96} & United States Department of the Navy & US & 31 & Hyundai Heavy Industries (South Korea) & KR & 103 & Maritime Research Institute Netherlands & NL & 241 \\  
\textit{97} & Seoul National University & KR & 30 & University of Western Australia & AU & 101 & University of Chinese Academy of Sciences & CN & 238 \\  
\textit{98} & University of Gda\'{n}sk & PL & 30 & Paderborn University & DE & 101 & Pennsylvania State University & US & 235 \\  
\textit{99} & Job Performance Systems (United States) & US & 30 & Job Performance Systems (United States) & US & 101 & Korea Advanced Institute of Science and Technology & KR & 229 \\  
\textit{100} & Imperial College London & GB & 29 & Naval Aeronautical and Astronautical University & CN & 99 & Universidad Politd\'{e}cnica de Madrid & ES & 228 \\  
          
 \\[-1.4em]
\bottomrule
\end{tabular}}
}
\end{table}
\end{landscape}
}
%

\titleformat{\section}{\sc\centering\LARGE\bfseries}{\textsc{\thesection}.\!\!}{1em}{}

\afterpage{\clearpage%
\markboth{\textbf \textsc{\neuro}}{}
\thispagestyle{empty}
\quad
\vspace{2cm}
\begin{center}
\pgfornament[width=0.5*\textwidth,symmetry=h]{89}\\[2em]
\section{\neuro}
\vspace{1em}
\pgfornament[width=0.5*\textwidth]{89}
\end{center}
}

\afterpage{\clearpage%

\begin{figure}[!tp]
\centering
\vspace{-1em}
{\large \textbf{\textrm{{World Map of \textcolor{violet}{\textit{\neuro}} Collaboration}}}~|~1971--2020}\\
\vspace{0.3cm}
\includegraphics[align=c, scale=0.054, trim={9.5cm 0 9.5cm 0},clip]{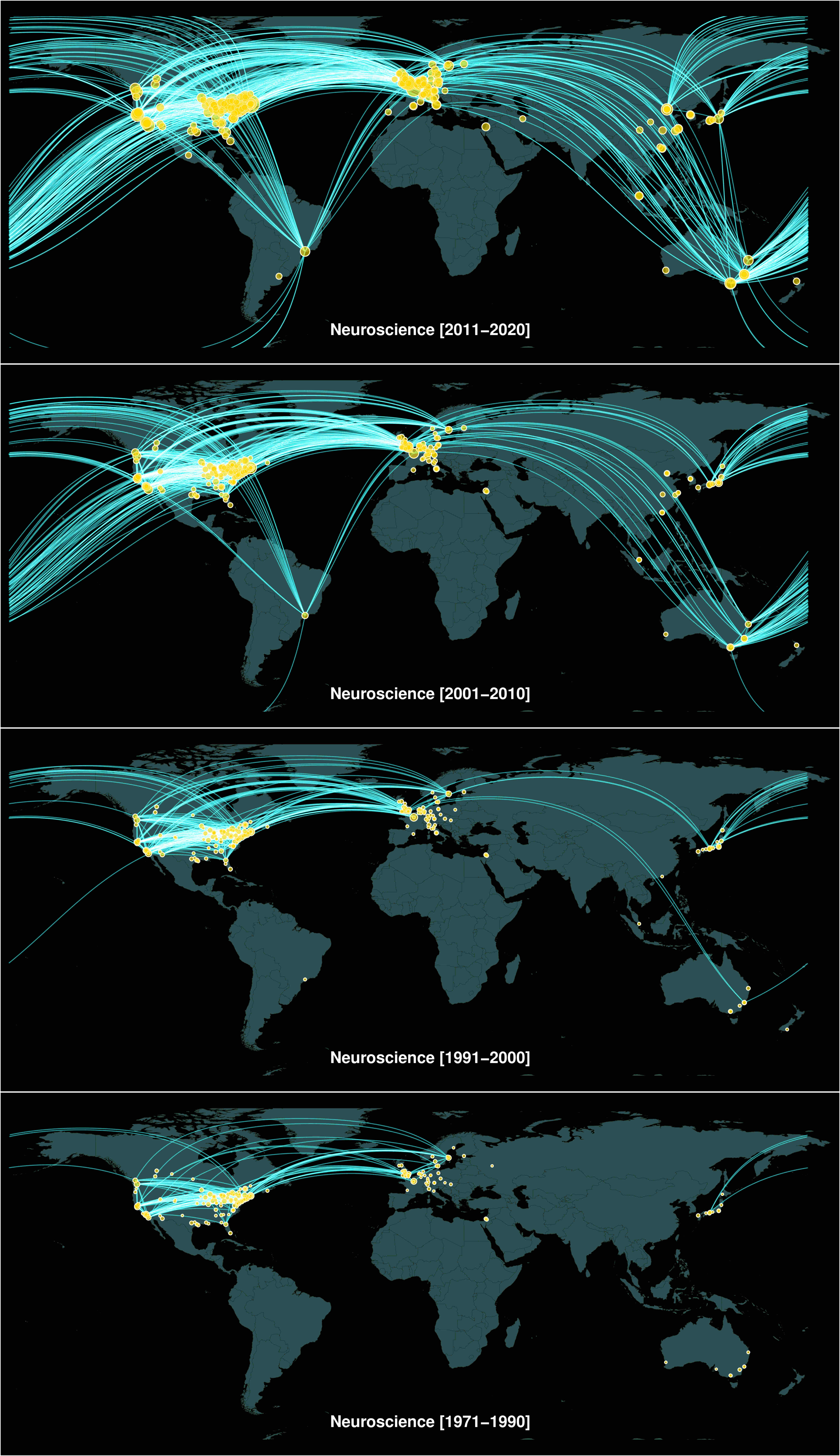}
\caption[{\neuro}]{\textbf{(a)~|~The World Map of \textcolor{violet}{\textit{\neuro}} Collaboration.}
The bubbles represent the top 199 institutions in terms of work production, with their sizes proportional to the work volume. 
The connecting lines depict coauthorship relationships among the top 50 institutions.}
\label{fig:wmap_neuro}
\end{figure}
}
\afterpage{\clearpage%
\begin{figure}[!tp]\ContinuedFloat
\centering
\vspace{-1em}
{\large \textbf{Top 30 Productive Institutions on the World Map: \textcolor{violet}{\textit{\neuro}}}~|~1991--2020}\\
\vspace{-0em}
\hspace*{-3em}                                                           
\includegraphics[align=c, scale=0.83]{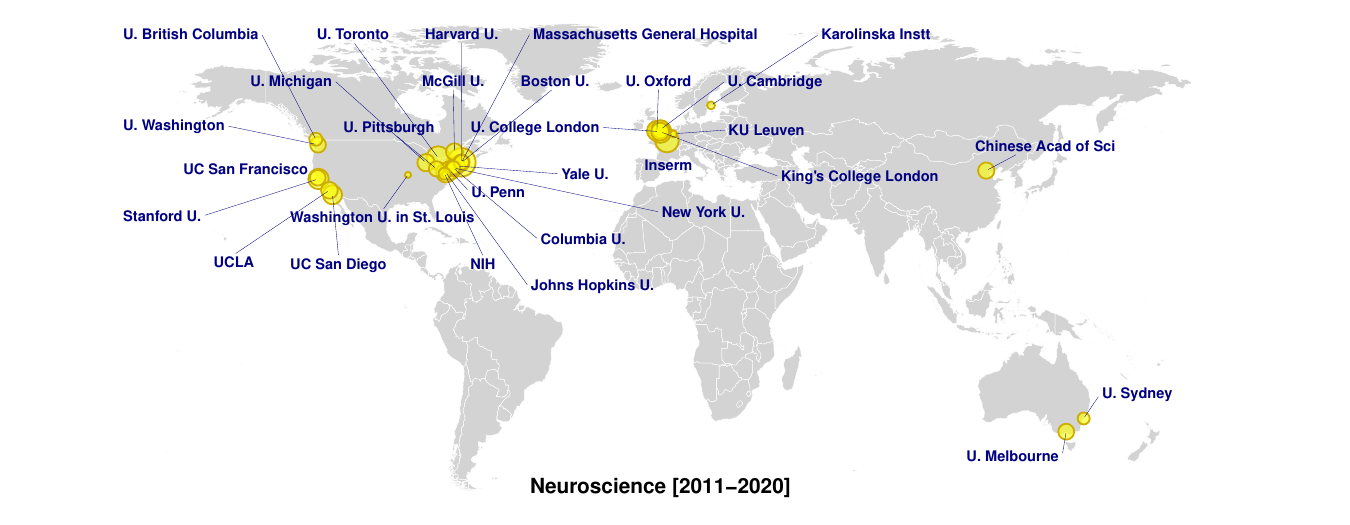}\\[-0.5em]
\quad\\[-1em]
\dotfill 
\quad\\[-0em]
\hspace*{-3em}                                                           
\includegraphics[align=c, scale=0.83]{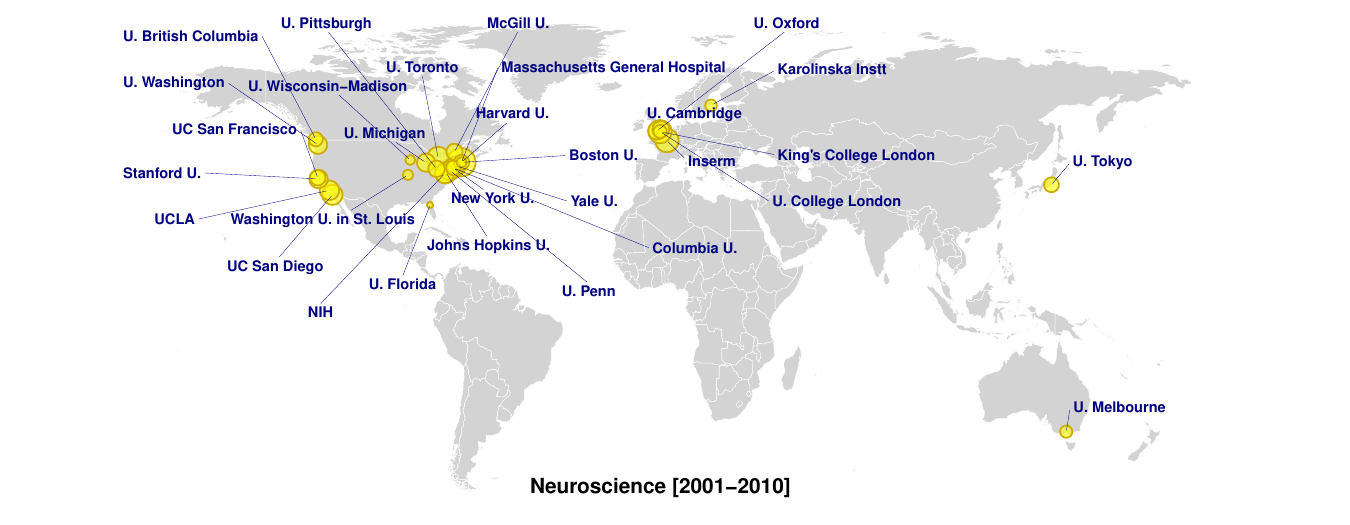}\\[-0.5em]
\quad\\[-1em]
\dotfill 
\quad\\[-0em]
\hspace*{-3em}                                                           
\includegraphics[align=c, scale=0.83]{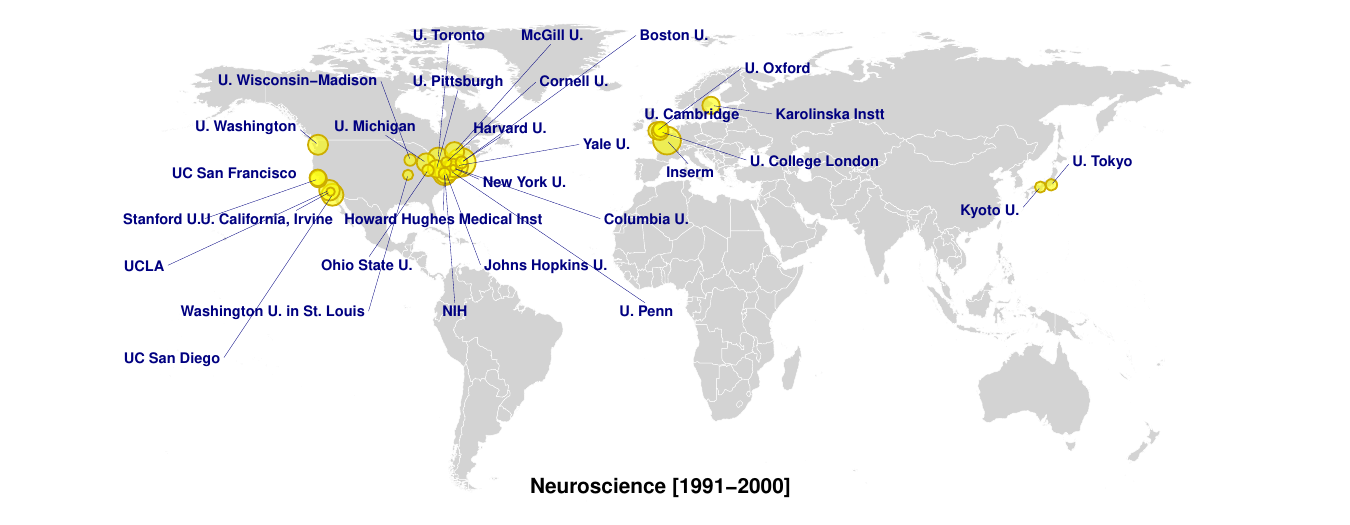}\\[-0.8em]
\caption[{\neuro}]{\textbf{(b)~|~The top 30 productive institutions on the World Map: \textcolor{violet}{\textit{\neuro}}.}
The bubbles represent the top 30 institutions in terms of work production, with their sizes proportional to the work volume.}
\label{fig:wmap_topinst_neuro}
\end{figure}
}
\afterpage{\clearpage%
\begin{figure}[!tp]\ContinuedFloat
\centering
\vspace{-1em}
{\large \textbf{\textrm{{Interregional \textcolor{violet}{\textit{\neuro}} Collaboration}}}~|~1991--2020}\\
\vspace{0.5em}
\hspace{-5em}\includegraphics[align=c, scale=1.7, vmargin=0mm]{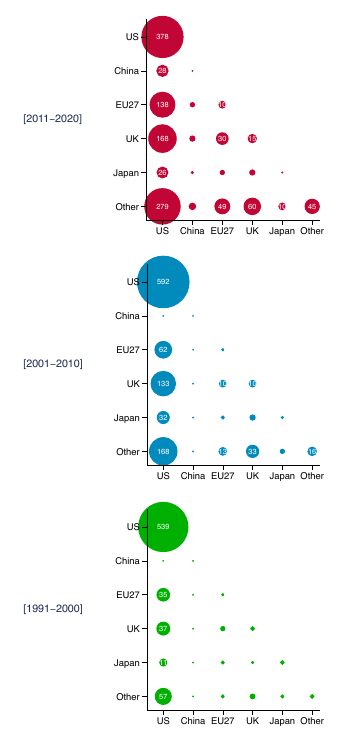}
\vspace{-1em}
\caption[]{\textbf{(c)~|~The Interregional \textcolor{violet}{\textit{\neuro}} Collaboration Matrix Diagram.}
The bubble size represents the number of coauthorship relationships for the top 50 institutions in terms of work production. 
If the number is equal to or greater than 10, it is displayed inside the bubble.}
\label{fig:halfmat_neuro}
\end{figure}
}
\afterpage{\clearpage%
\begin{figure}[!tp]\ContinuedFloat
\centering
\vspace{-1em}
{\large \textbf{\textrm{Interinstitutional \textcolor{violet}{\textit{\neuro}} Collaboration}}~|~2001--2020\quad {\footnotesize \emph{(continued to next page)}}}\\
\cdend{neuro}{2010}{2000}{2011--2020}{2001--2010}\\[-1.5em]
\caption[]{\textbf{(d)~|~The Interinstitutional \textcolor{violet}{\textit{\neuro}} Collaboration Dendrogram.}
The top 50 institutions in terms of work production, indicated by the circularised bar graphs, are displayed.
}
\label{fig:cdend1_neuro}
\end{figure}
}
\afterpage{\clearpage%
\begin{figure}[!tp]\ContinuedFloat
\centering
\vspace{-1em}
{\large \textbf{\textrm{Interinstitutional \textcolor{violet}{\textit{\neuro}} Collaboration}}~|~1971--2000\quad {\footnotesize \emph{(continued from previous page)}}}\\
\cdend{neuro}{1990}{1980}{1991--2000}{1971--1990}\\[-1.5em]
\caption[]{\textbf{(d)~|~The Interinstitutional \textcolor{violet}{\textit{\neuro}} Collaboration Dendrogram.} \emph{(Cont.)}\hfill~}
\label{fig:cdend2_neuro}
\end{figure}
}
\afterpage{\clearpage%
\begin{landscape}
\begin{table}[!t]
\vspace{-0.5em}
\caption{\textbf{The top 100 productive institutions: \textcolor{violet}{\textit{\neuro}}.}}
\label{tab:r1_neuro}
\vspace{2em}
\centering
{\tiny
{\renewcommand{\arraystretch}{1.2}
\begin{tabular}{rp{5cm}lr@{\hspace{4em}}p{5cm}lr@{\hspace{4em}}p{5cm}lr}\\[-5em] \toprule[1pt] \\[-1.4em]    
 & {\scriptsize \textbf{1991--2000}} & \multicolumn{2}{c}{No.~Works} & {\scriptsize \textbf{2001--2010}} & \multicolumn{2}{c}{No.~Works} & {\scriptsize \textbf{2011--2020}} & \multicolumn{2}{r}{No.~Works} \\[-0.2em] \cmidrule[0.5pt](lr{4em}){2-4} \cmidrule[0.5pt](l{-0em}r{4em}){5-7} \cmidrule[0.5pt](l{-0em}r{1em}){8-10}
\textit{1} & Harvard University & US & 4,585 & Harvard University & US & 10,292 & Harvard University & US & 23,906 \\  
\textit{2} & Inserm & FR & 4,379 & Inserm & FR & 7,149 & Inserm & FR & 16,215 \\  
\textit{3} & Yale University & US & 3,378 & University College London & GB & 7,075 & University of Toronto & CA & 15,897 \\  
\textit{4} & National Institutes of Health & US & 3,218 & University of Toronto & CA & 6,596 & University College London & GB & 15,006 \\  
\textit{5} & Johns Hopkins University & US & 3,046 & Johns Hopkins University & US & 6,081 & Stanford University & US & 12,900 \\  
\textit{6} & University of California, San Diego & US & 3,042 & University of California, San Diego & US & 5,744 & University of Oxford & GB & 12,388 \\  
\textit{7} & University of California, Los Angeles & US & 2,825 & Columbia University & US & 5,655 & Johns Hopkins University & US & 12,226 \\  
\textit{8} & University of Toronto & CA & 2,824 & University of Pennsylvania & US & 5,583 & University of California, San Diego & US & 11,810 \\  
\textit{9} & University of Washington & US & 2,741 & University of California, Los Angeles & US & 5,489 & University of Pennsylvania & US & 11,672 \\  
\textit{10} & University of Pennsylvania & US & 2,703 & National Institutes of Health & US & 5,362 & University of Cambridge & GB & 11,084 \\  
\hdashline          
\textit{11} & McGill University & CA & 2,645 & Yale University & US & 5,352 & King's College London & GB & 10,910 \\  
\textit{12} & University of Michigan{\textendash}Ann Arbor & US & 2,536 & Stanford University & US & 5,172 & University of Michigan{\textendash}Ann Arbor & US & 10,542 \\  
\textit{13} & Columbia University & US & 2,498 & University of Washington & US & 5,116 & Yale University & US & 10,497 \\  
\textit{14} & University of Pittsburgh & US & 2,407 & University of Michigan{\textendash}Ann Arbor & US & 5,073 & McGill University & CA & 10,270 \\  
\textit{15} & Karolinska Institutet & SE & 2,392 & University of Oxford & GB & 4,913 & University of California, Los Angeles & US & 10,128 \\  
\textit{16} & University of Oxford & GB & 2,378 & University of Cambridge & GB & 4,685 & Columbia University & US & 10,047 \\  
\textit{17} & Stanford University & US & 2,332 & University of Pittsburgh & US & 4,598 & Chinese Academy of Sciences & CN & 9,922 \\  
\textit{18} & University College London & GB & 2,314 & McGill University & CA & 4,554 & Massachusetts General Hospital & US & 9,785 \\  
\textit{19} & University of California, San Francisco & US & 2,295 & University of California, San Francisco & US & 4,362 & University of Melbourne & AU & 9,684 \\  
\textit{20} & University of Cambridge & GB & 2,182 & Boston University & US & 4,325 & University of Washington & US & 9,592 \\  
\hdashline          
\textit{21} & Boston University & US & 1,942 & The University of Tokyo & JP & 4,260 & University of California, San Francisco & US & 9,553 \\  
\textit{22} & University of Wisconsin{\textendash}Madison & US & 1,937 & New York University & US & 4,224 & University of Pittsburgh & US & 9,387 \\  
\textit{23} & The Ohio State University & US & 1,932 & University of British Columbia & CA & 4,100 & Boston University & US & 9,089 \\  
\textit{24} & Howard Hughes Medical Institute & US & 1,930 & King's College London & GB & 3,979 & New York University & US & 8,621 \\  
\textit{25} & The University of Tokyo & JP & 1,924 & University of Melbourne & AU & 3,822 & University of British Columbia & CA & 8,441 \\  
\textit{26} & Kyoto University & JP & 1,914 & Karolinska Institutet & SE & 3,755 & National Institutes of Health & US & 8,128 \\  
\textit{27} & Washington University in St. Louis & US & 1,883 & Washington University in St. Louis & US & 3,643 & University of Sydney & AU & 8,085 \\  
\textit{28} & Cornell University & US & 1,837 & Massachusetts General Hospital & US & 3,620 & Karolinska Institutet & SE & 7,320 \\  
\textit{29} & University of California, Irvine & US & 1,827 & University of Wisconsin{\textendash}Madison & US & 3,580 & KU Leuven & BE & 7,271 \\  
\textit{30} & New York University & US & 1,804 & University of Florida & US & 3,438 & Washington University in St. Louis & US & 7,252 \\  
\hdashline          
\textit{31} & University of Southern California & US & 1,791 & University of California, Berkeley & US & 3,321 & University of Southern California & US & 7,217 \\  
\textit{32} & University of British Columbia & CA & 1,723 & University of Minnesota & US & 3,298 & University of Queensland & AU & 7,161 \\  
\textit{33} & State University of New York & US & 1,707 & Emory University & US & 3,266 & UNSW Sydney & AU & 7,011 \\  
\textit{34} & Massachusetts General Hospital & US & 1,674 & University of Sydney & AU & 3,257 & Universidade de S\~{a}o Paulo & BR & 7,009 \\  
\textit{35} & University of California, Berkeley & US & 1,665 & Kyoto University & JP & 3,248 & Emory University & US & 6,956 \\  
\textit{36} & University of Minnesota & US & 1,660 & University of California, Irvine & US & 3,234 & University of Florida & US & 6,951 \\  
\textit{37} & Case Western Reserve University & US & 1,653 & Northwestern University & US & 3,194 & Duke University & US & 6,941 \\  
\textit{38} & University of North Carolina at Chapel Hill & US & 1,591 & Cornell University & US & 3,191 & University of Wisconsin{\textendash}Madison & US & 6,930 \\  
\textit{39} & Emory University & US & 1,572 & Howard Hughes Medical Institute & US & 3,176 & The University of Tokyo & JP & 6,902 \\  
\textit{40} & Osaka University & JP & 1,571 & The Ohio State University & US & 3,154 & Monash University & AU & 6,818 \\  
\hdashline          
\textit{41} & University of Iowa & US & 1,555 & University of California, Davis & US & 3,148 & University of California, Berkeley & US & 6,816 \\  
\textit{42} & Rutgers, The State University of New Jersey & US & 1,546 & University of Southern California & US & 3,078 & Northwestern University & US & 6,713 \\  
\textit{43} & University of Illinois Urbana-Champaign & US & 1,522 & Pennsylvania State University & US & 3,039 & Radboud University Nijmegen & NL & 6,710 \\  
\textit{44} & University of Florida & US & 1,504 & Vanderbilt University & US & 2,983 & University of Edinburgh & GB & 6,677 \\  
\textit{45} & Duke Medical Center & US & 1,503 & Universidade de S\~{a}o Paulo & BR & 2,975 & University of North Carolina at Chapel Hill & US & 6,666 \\  
\textit{46} & University of Arizona & US & 1,501 & University of Manchester & GB & 2,973 & University of Minnesota & US & 6,647 \\  
\textit{47} & Johns Hopkins Medicine & US & 1,480 & Duke University & US & 2,958 & University of California, Davis & US & 6,546 \\  
\textit{48} & University of Chicago & US & 1,471 & University of North Carolina at Chapel Hill & US & 2,889 & Massachusetts Institute of Technology & US & 6,350 \\  
\textit{49} & University of Sydney & AU & 1,422 & Massachusetts Institute of Technology & US & 2,806 & University of Zurich & CH & 6,264 \\  
\textit{50} & Northwestern University & US & 1,404 & Case Western Reserve University & US & 2,805 & Imperial College London & GB & 6,225 \\  
          
 \\[-1.4em]
\hdashline \\[-1em]
\multicolumn{10}{r}{\scriptsize \emph{(continued to next page)}}
\end{tabular}}
}
\end{table}
\end{landscape}
}
\afterpage{\clearpage%
\begin{landscape}
\begin{table}[!t]\ContinuedFloat
\vspace{-3.3em}
\caption{\textbf{The top 100 productive institutions: \textcolor{violet}{\textit{\neuro}}.} \emph{(Cont.)}}
\label{tab:r2_neuro}
\vspace{2em}
{\tiny
{\renewcommand{\arraystretch}{1.2}
\begin{tabular}{rp{5cm}lr@{\hspace{4em}}p{5cm}lr@{\hspace{4em}}p{5cm}lr}\\[-5em] \toprule[1pt] \\[-1.4em]    
 & {\scriptsize \textbf{1991--2000}} & \multicolumn{2}{c}{No.\ Works} & {\scriptsize \textbf{2001--2010}} & \multicolumn{2}{c}{No.\ Works} & {\scriptsize \textbf{2011--2020}} & \multicolumn{2}{r}{No.\ Works} \\[-0.2em] \cmidrule[0.5pt](lr{4em}){2-4} \cmidrule[0.5pt](l{-0em}r{4em}){5-7} \cmidrule[0.5pt](l{-0em}r{1em}){8-10}
\textit{51} & Pennsylvania State University & US & 1,393 & University of Queensland & AU & 2,796 & Sapienza University of Rome & IT & 6,164 \\  
\textit{52} & Duke University & US & 1,352 & University of Illinois Urbana-Champaign & US & 2,769 & University of Amsterdam & NL & 6,145 \\  
\textit{53} & University of California, Davis & US & 1,334 & UNSW Sydney & AU & 2,767 & Johns Hopkins Medicine & US & 6,120 \\  
\textit{54} & Massachusetts Institute of Technology & US & 1,333 & Universit\'{e} de Montr\'{e}al & CA & 2,762 & Universit\'{e} de Montr\'{e}al & CA & 5,964 \\  
\textit{55} & Baylor College of Medicine & US & 1,321 & University of Iowa & US & 2,741 & Charit\'{e} {\textendash} Universit\"{a}tsmedizin Berlin & DE & 5,855 \\  
\textit{56} & Sapienza University of Rome & IT & 1,317 & Johns Hopkins Medicine & US & 2,708 & Shanghai Jiao Tong University & CN & 5,852 \\  
\textit{57} & University of Virginia & US & 1,304 & University of Edinburgh & GB & 2,698 & The Ohio State University & US & 5,831 \\  
\textit{58} & Universit\'{e} de Montr\'{e}al & CA & 1,281 & KU Leuven & BE & 2,634 & Pennsylvania State University & US & 5,796 \\  
\textit{59} & University of Milan & IT & 1,261 & Rutgers, The State University of New Jersey & US & 2,631 & Indiana University & US & 5,686 \\  
\textit{60} & National Institute of Neurological Disorders and Stroke & US & 1,249 & Osaka University & JP & 2,627 & University of California, Irvine & US & 5,638 \\  
\hdashline          
\textit{61} & National Hospital for Neurology and Neurosurgery & GB & 1,233 & Sapienza University of Rome & IT & 2,544 & University of T\"{u}bingen & DE & 5,592 \\  
\textit{62} & National Institute of Mental Health & US & 1,227 & University of Chicago & US & 2,542 & Vanderbilt University & US & 5,590 \\  
\textit{63} & University of Melbourne & AU & 1,225 & University of Amsterdam & NL & 2,488 & Icahn School of Medicine at Mount Sinai & US & 5,545 \\  
\textit{64} & Albert Einstein College of Medicine & US & 1,224 & University of Alberta & CA & 2,451 & University of Copenhagen & DK & 5,514 \\  
\textit{65} & University of Amsterdam & NL & 1,213 & University of Zurich & CH & 2,440 & University of Manchester & GB & 5,456 \\  
\textit{66} & Tel Aviv University & IL & 1,201 & Max Planck Society & DE & 2,436 & Howard Hughes Medical Institute & US & 5,322 \\  
\textit{67} & Duke University Hospital & US & 1,189 & Indiana University & US & 2,435 & University of Calgary & CA & 5,304 \\  
\textit{68} & University of Edinburgh & GB & 1,160 & The University of Texas at Austin & US & 2,421 & University of Illinois Urbana-Champaign & US & 5,248 \\  
\textit{69} & The University of Texas at Austin & US & 1,147 & Ludwig-Maximilians-Universit\"{a}t M\"{u}nchen & DE & 2,389 & Rutgers, The State University of New Jersey & US & 5,205 \\  
\textit{70} & Kyushu University & JP & 1,145 & Oregon Health \& Science University & US & 2,389 & The University of Texas at Austin & US & 5,173 \\  
\hdashline          
\textit{71} & University of T\"{u}bingen & DE & 1,137 & Chinese Academy of Sciences & CN & 2,387 & Tsinghua University & CN & 5,096 \\  
\textit{72} & University of Miami & US & 1,134 & University of Calgary & CA & 2,371 & Istituti di Ricovero e Cura a Carattere Scientifico & IT & 5,072 \\  
\textit{73} & Lund University & SE & 1,132 & Charit\'{e} {\textendash} Universit\"{a}tsmedizin Berlin & DE & 2,345 & National University of Singapore & SG & 5,071 \\  
\textit{74} & Neurological Surgery & US & 1,132 & University of Utah & US & 2,311 & Cornell University & US & 5,062 \\  
\textit{75} & University of Manchester & GB & 1,121 & Tel Aviv University & IL & 2,310 & Peking University & CN & 4,929 \\  
\textit{76} & Western University & CA & 1,112 & University of Arizona & US & 2,304 & Seoul National University & KR & 4,906 \\  
\textit{77} & Tohoku University & JP & 1,108 & Monash University & AU & 2,272 & Maastricht University & NL & 4,896 \\  
\textit{78} & University of Utah & US & 1,098 & Imperial College London & GB & 2,268 & Brigham and Women's Hospital & US & 4,889 \\  
\textit{79} & University of Queensland & AU & 1,092 & University of T\"{u}bingen & DE & 2,263 & Kyoto University & JP & 4,867 \\  
\textit{80} & University of Kentucky & US & 1,082 & Duke Medical Center & US & 2,255 & Western University & CA & 4,858 \\  
\hdashline          
\textit{81} & University of Alberta & CA & 1,057 & Baylor College of Medicine & US & 2,220 & Zhejiang University & CN & 4,845 \\  
\textit{82} & Vanderbilt University & US & 1,056 & Tohoku University & JP & 2,147 & Ludwig-Maximilians-Universit\"{a}t M\"{u}nchen & DE & 4,805 \\  
\textit{83} & Wayne State University & US & 1,055 & Icahn School of Medicine at Mount Sinai & US & 2,145 & Tel Aviv University & IL & 4,799 \\  
\textit{84} & University of Bristol & GB & 1,033 & University of Bristol & GB & 2,134 & University of Utah & US & 4,751 \\  
\textit{85} & Oregon Health \& Science University & US & 1,024 & National Hospital for Neurology and Neurosurgery & GB & 2,133 & Baylor College of Medicine & US & 4,597 \\  
\textit{86} & University of Calgary & CA & 1,020 & University of Illinois at Chicago & US & 2,105 & University of Alberta & CA & 4,591 \\  
\textit{87} & Indiana University & US & 996 & Western University & CA & 2,099 & ETH Zurich & CH & 4,566 \\  
\textit{88} & Hokkaido University & JP & 994 & University of Helsinki & FI & 2,076 & University of Groningen & NL & 4,560 \\  
\textit{89} & Nagoya University & JP & 981 & Zhejiang University & CN & 2,073 & Utrecht University & NL & 4,533 \\  
\textit{90} & Radboud University Nijmegen & NL & 981 & Michigan State University & US & 2,026 & University of Padua & IT & 4,479 \\  
\hdashline          
\textit{91} & Stony Brook University & US & 978 & Utrecht University & NL & 2,012 & University of Chicago & US & 4,461 \\  
\textit{92} & University of Alabama at Birmingham & US & 978 & Nagoya University & JP & 1,966 & Chinese University of Hong Kong & CN & 4,437 \\  
\textit{93} & Ludwig-Maximilians-Universit\"{a}t M\"{u}nchen & DE & 977 & University of Kentucky & US & 1,952 & University of Barcelona & ES & 4,406 \\  
\textit{94} & University of Maryland, College Park & US & 969 & University of Virginia & US & 1,949 & University of Chinese Academy of Sciences & CN & 4,334 \\  
\textit{95} & Max Planck Society & DE & 968 & Wayne State University & US & 1,940 & University of Iowa & US & 4,316 \\  
\textit{96} & University of Helsinki & FI & 960 & Lund University & SE & 1,929 & University of Milan & IT & 4,255 \\  
\textit{97} & University of Wales & GB & 958 & National University of Singapore & SG & 1,925 & Heidelberg University & DE & 4,251 \\  
\textit{98} & Hebrew University of Jerusalem & IL & 949 & University of Birmingham & GB & 1,918 & Huazhong University of Science and Technology & CN & 4,249 \\  
\textit{99} & California Institute of Technology & US & 949 & Chinese University of Hong Kong & CN & 1,902 & Michigan State University & US & 4,208 \\  
\textit{100} & University of Zurich & CH & 944 & Maastricht University & NL & 1,900 & University of Helsinki & FI & 4,202 \\  
          
 \\[-1.4em]
\bottomrule
\end{tabular}}
}
\end{table}
\end{landscape}
}
%

\titleformat{\section}{\sc\centering\LARGE\bfseries}{\textsc{\thesection}.\!\!}{1em}{}

\afterpage{\clearpage%
\markboth{\textbf \textsc{\condensed}}{}
\thispagestyle{empty}
\quad
\vspace{2cm}
\begin{center}
\pgfornament[width=0.5*\textwidth,symmetry=h]{89}\\[2em]
\section{\condensed}
\vspace{1em}
\pgfornament[width=0.5*\textwidth]{89}
\end{center}
}

\afterpage{\clearpage%

\begin{figure}[!tp]
\centering
\vspace{-1em}
{\large \textbf{\textrm{{World Map of \textcolor{violet}{\textit{\condensed}} Collaboration}}}~|~1971--2020}\\
\vspace{0.3cm}
\includegraphics[align=c, scale=0.054, trim={9.5cm 0 9.5cm 0},clip]{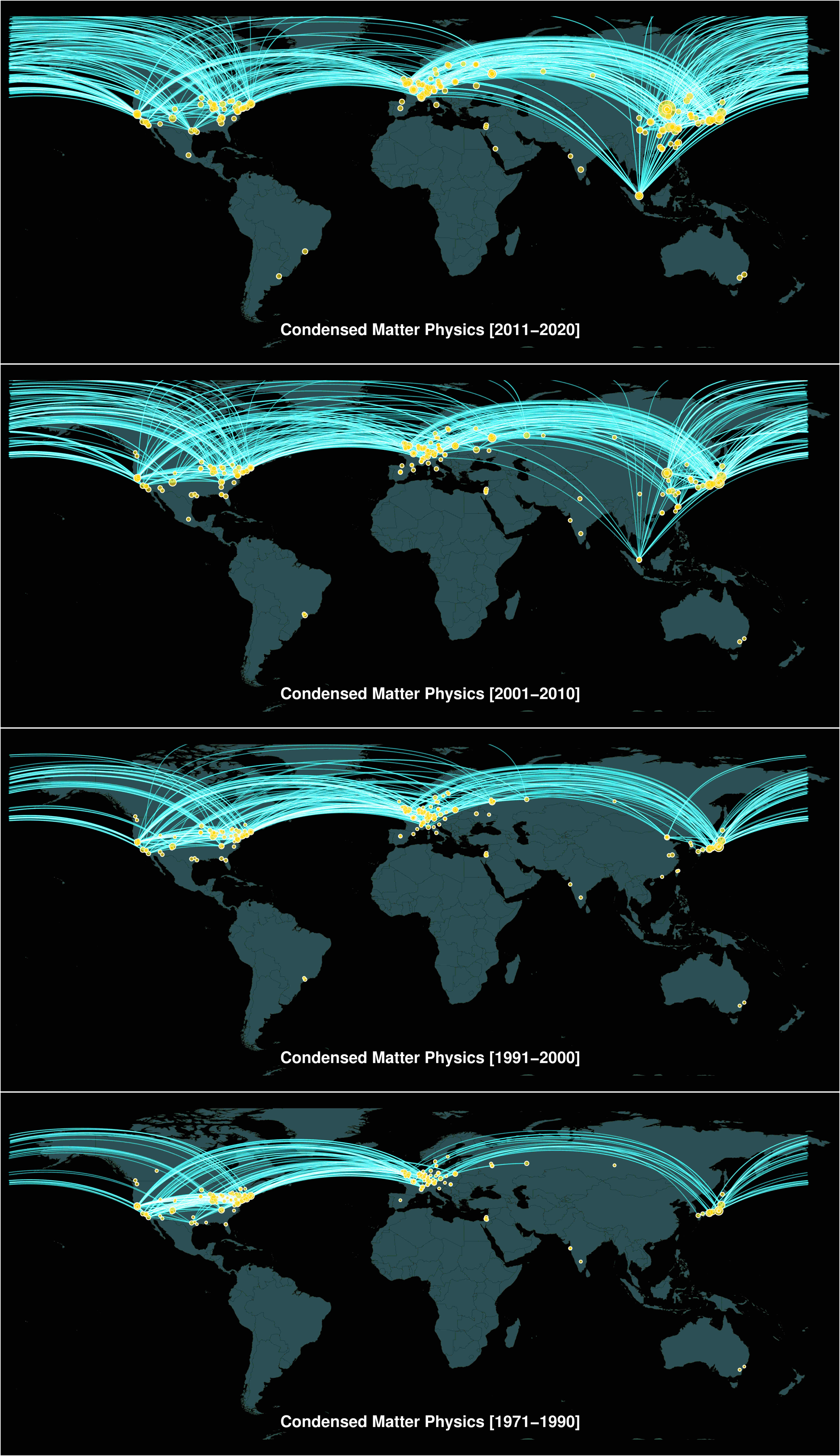}
\caption[{\condensed}]{\textbf{(a)~|~The World Map of \textcolor{violet}{\textit{\condensed}} Collaboration.}
The bubbles represent the top 199 institutions in terms of work production, with their sizes proportional to the work volume. 
The connecting lines depict coauthorship relationships among the top 50 institutions.}
\label{fig:wmap_condensed}
\end{figure}
}
\afterpage{\clearpage%
\begin{figure}[!tp]\ContinuedFloat
\centering
\vspace{-1em}
{\large \textbf{Top 30 Productive Institutions on the World Map: \textcolor{violet}{\textit{\condensed}}}~|~1991--2020}\\
\vspace{-0em}
\hspace*{-3em}                                                           
\includegraphics[align=c, scale=0.83]{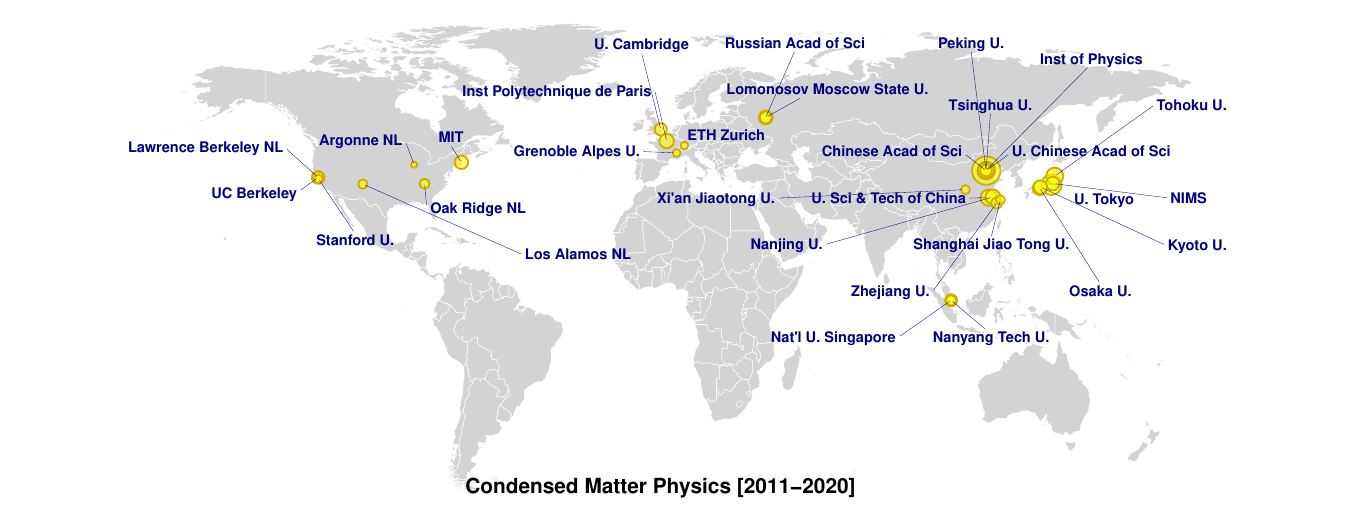}\\[-0.5em]
\quad\\[-1em]
\dotfill 
\quad\\[-0em]
\hspace*{-3em}                                                           
\includegraphics[align=c, scale=0.83]{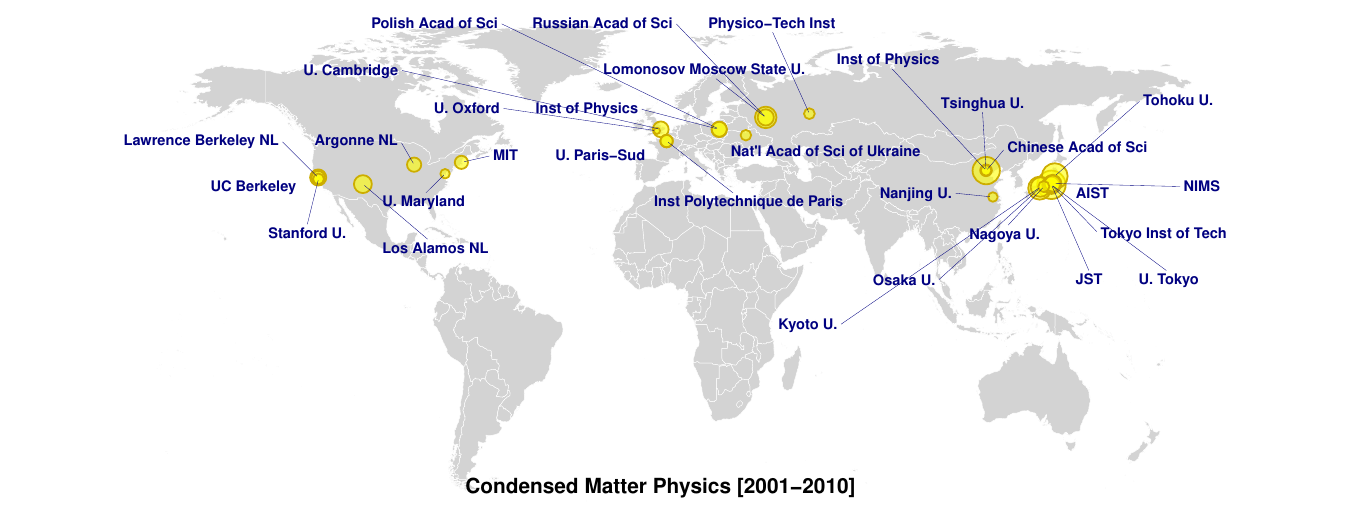}\\[-0.5em]
\quad\\[-1em]
\dotfill 
\quad\\[-0em]
\hspace*{-3em}                                                           
\includegraphics[align=c, scale=0.83]{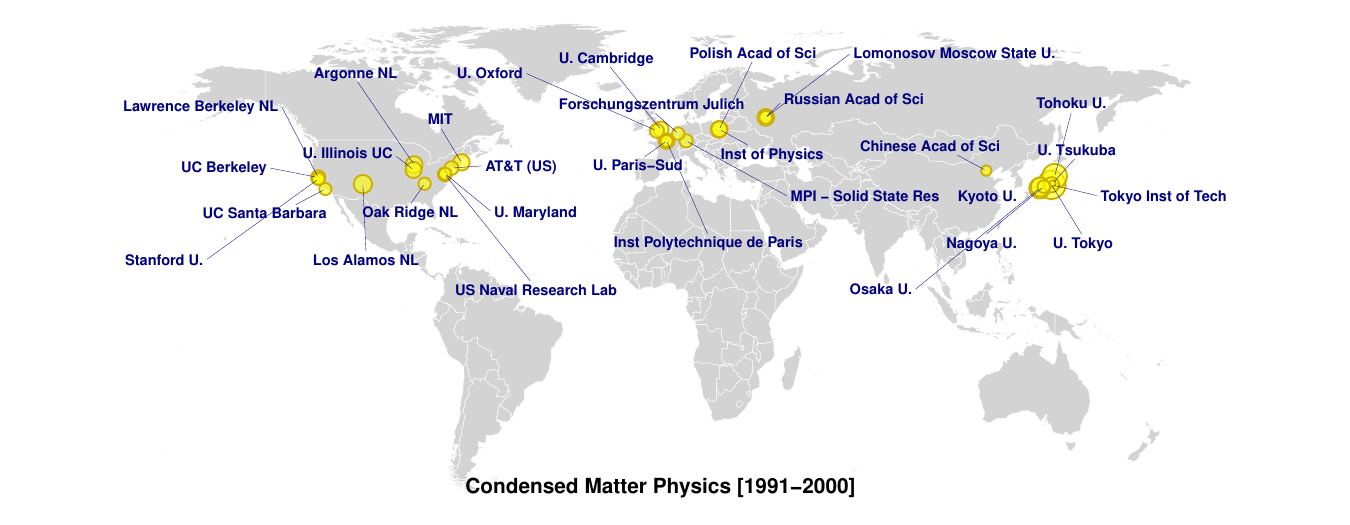}\\[-0.8em]
\caption[{\condensed}]{\textbf{(b)~|~The top 30 productive institutions on the World Map: \textcolor{violet}{\textit{\condensed}}.}
The bubbles represent the top 30 institutions in terms of work production, with their sizes proportional to the work volume.}
\label{fig:wmap_topinst_condensed}
\end{figure}
}
\afterpage{\clearpage%
\begin{figure}[!tp]\ContinuedFloat
\centering
\vspace{-1em}
{\large \textbf{\textrm{{Interregional \textcolor{violet}{\textit{\condensed}} Collaboration}}}~|~1991--2020}\\
\vspace{0.5em}
\hspace{-5em}\includegraphics[align=c, scale=1.7, vmargin=0mm]{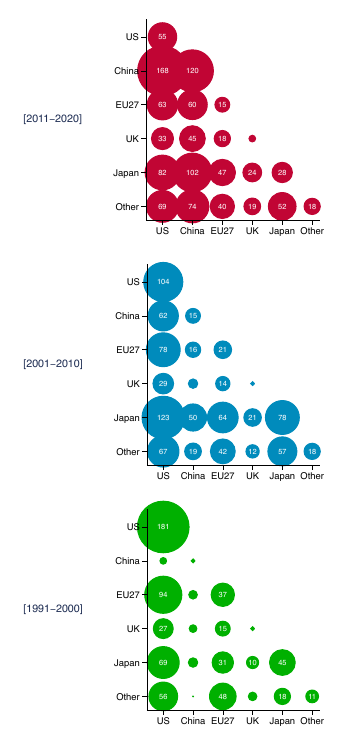}
\vspace{-1em}
\caption[]{\textbf{(c)~|~The Interregional \textcolor{violet}{\textit{\condensed}} Collaboration Matrix Diagram.}
The bubble size represents the number of coauthorship relationships for the top 50 institutions in terms of work production. 
If the number is equal to or greater than 10, it is displayed inside the bubble.}
\label{fig:halfmat_condensed}
\end{figure}
}
\afterpage{\clearpage%
\begin{figure}[!tp]\ContinuedFloat
\centering
\vspace{-1em}
{\large \textbf{\textrm{Interinstitutional \textcolor{violet}{\textit{\condensed}} Collaboration}}~|~2001--2020\quad {\footnotesize \emph{(continued to next page)}}}\\
\cdend{condensed}{2010}{2000}{2011--2020}{2001--2010}\\[-1.5em]
\caption[]{\textbf{(d)~|~The Interinstitutional \textcolor{violet}{\textit{\condensed}} Collaboration Dendrogram.}
The top 50 institutions in terms of work production, indicated by the circularised bar graphs, are displayed.
}
\label{fig:cdend1_condensed}
\end{figure}
}
\afterpage{\clearpage%
\begin{figure}[!tp]\ContinuedFloat
\centering
\vspace{-1em}
{\large \textbf{\textrm{Interinstitutional \textcolor{violet}{\textit{\condensed}} Collaboration}}~|~1971--2000\quad {\footnotesize \emph{(continued from previous page)}}}\\
\cdend{condensed}{1990}{1980}{1991--2000}{1971--1990}\\[-1.5em]
\caption[]{\textbf{(d)~|~The Interinstitutional \textcolor{violet}{\textit{\condensed}} Collaboration Dendrogram.} \emph{(Cont.)}\hfill~}
\label{fig:cdend2_condensed}
\end{figure}
}
\afterpage{\clearpage%
\begin{landscape}
\begin{table}[!t]
\vspace{-0.5em}
\caption{\textbf{The top 100 productive institutions: \textcolor{violet}{\textit{\condensed}}.}}
\label{tab:r1_condensed}
\vspace{2em}
\centering
{\tiny
{\renewcommand{\arraystretch}{1.2}
\begin{tabular}{rp{5cm}lr@{\hspace{4em}}p{5cm}lr@{\hspace{4em}}p{5cm}lr}\\[-5em] \toprule[1pt] \\[-1.4em]    
 & {\scriptsize \textbf{1991--2000}} & \multicolumn{2}{c}{No.~Works} & {\scriptsize \textbf{2001--2010}} & \multicolumn{2}{c}{No.~Works} & {\scriptsize \textbf{2011--2020}} & \multicolumn{2}{r}{No.~Works} \\[-0.2em] \cmidrule[0.5pt](lr{4em}){2-4} \cmidrule[0.5pt](l{-0em}r{4em}){5-7} \cmidrule[0.5pt](l{-0em}r{1em}){8-10}
\textit{1} & The University of Tokyo & JP & 8,443 & The University of Tokyo & JP & 11,371 & Chinese Academy of Sciences & CN & 29,323 \\  
\textit{2} & Tohoku University & JP & 6,108 & Chinese Academy of Sciences & CN & 10,374 & The University of Tokyo & JP & 12,228 \\  
\textit{3} & Osaka University & JP & 4,446 & Tohoku University & JP & 9,116 & University of Chinese Academy of Sciences & CN & 9,092 \\  
\textit{4} & Kyoto University & JP & 3,872 & Osaka University & JP & 7,277 & Tohoku University & JP & 8,637 \\  
\textit{5} & Los Alamos National Laboratory & US & 3,635 & Russian Academy of Sciences & RU & 6,508 & Tsinghua University & CN & 8,475 \\  
\textit{6} & Argonne National Laboratory & US & 3,298 & Kyoto University & JP & 5,608 & University of Science and Technology of China & CN & 7,962 \\  
\textit{7} & Institute of Physics & PL & 3,285 & Tokyo Institute of Technology & JP & 5,118 & Peking University & CN & 7,708 \\  
\textit{8} & Russian Academy of Sciences & RU & 3,258 & Los Alamos National Laboratory & US & 4,997 & Nanjing University & CN & 7,054 \\  
\textit{9} & Massachusetts Institute of Technology & US & 3,201 & \scalebox{0.82}[1]{National Institute of Advanced Industrial Science and Technology (AIST)} & JP & 4,839 & Osaka University & JP & 6,592 \\  
\textit{10} & University of Illinois Urbana-Champaign & US & 3,100 & Japan Science and Technology Agency (JST) & JP & 4,552 & Russian Academy of Sciences & RU & 6,513 \\  
\hdashline          
\textit{11} & University of Cambridge & GB & 3,030 & Lomonosov Moscow State University & RU & 4,492 & National Institute for Materials Science (NIMS) & JP & 6,338 \\  
\textit{12} & Polish Academy of Sciences & PL & 3,026 & Lawrence Berkeley National Laboratory & US & 4,478 & Kyoto University & JP & 6,336 \\  
\textit{13} & University of Paris-Sud & FR & 2,989 & National Institute for Materials Science (NIMS) & JP & 4,465 & Massachusetts Institute of Technology & US & 6,272 \\  
\textit{14} & Tokyo Institute of Technology & JP & 2,960 & University of Cambridge & GB & 4,372 & Lawrence Berkeley National Laboratory & US & 5,949 \\  
\textit{15} & Lawrence Berkeley National Laboratory & US & 2,757 & Institute of Physics & PL & 4,264 & University of Cambridge & GB & 5,753 \\  
\textit{16} & United States Naval Research Laboratory & US & 2,757 & Polish Academy of Sciences & PL & 4,200 & Nanyang Technological University & SG & 5,694 \\  
\textit{17} & University of California, Berkeley & US & 2,753 & Argonne National Laboratory & US & 4,023 & Lomonosov Moscow State University & RU & 5,634 \\  
\textit{18} & University of Oxford & GB & 2,747 & Massachusetts Institute of Technology & US & 3,825 & Institute of Physics & CN & 5,576 \\  
\textit{19} & AT\&T (United States) & US & 2,706 & University of California, Berkeley & US & 3,821 & University of California, Berkeley & US & 5,507 \\  
\textit{20} & Max Planck Institute for Solid State Research & DE & 2,634 & University of Paris-Sud & FR & 3,638 & Zhejiang University & CN & 5,435 \\  
\hdashline          
\textit{21} & Lomonosov Moscow State University & RU & 2,604 & Tsinghua University & CN & 3,596 & National University of Singapore & SG & 5,183 \\  
\textit{22} & Nagoya University & JP & 2,565 & Nagoya University & JP & 3,527 & Oak Ridge National Laboratory & US & 5,084 \\  
\textit{23} & Forschungszentrum J\"{u}lich & DE & 2,540 & Stanford University & US & 3,507 & Los Alamos National Laboratory & US & 4,956 \\  
\textit{24} & Oak Ridge National Laboratory & US & 2,536 & Physico-Technical Institute & RU & 3,492 & Shanghai Jiao Tong University & CN & 4,932 \\  
\textit{25} & University of California, Santa Barbara & US & 2,511 & Institute of Physics & CN & 3,475 & Xi'an Jiaotong University & CN & 4,910 \\  
\textit{26} & Stanford University & US & 2,510 & National Academy of Sciences of Ukraine & UA & 3,469 & Stanford University & US & 4,851 \\  
\textit{27} & University of Maryland, College Park & US & 2,492 & Nanjing University & CN & 3,371 & ETH Zurich & CH & 4,728 \\  
\textit{28} & Chinese Academy of Sciences & CN & 2,369 & University of Maryland, College Park & US & 3,369 & Argonne National Laboratory & US & 4,664 \\  
\textit{29} & University of Tsukuba & JP & 2,188 & University of Oxford & GB & 3,244 & Huazhong University of Science and Technology & CN & 4,663 \\  
\textit{30} & National Institute of Standards and Technology & US & 2,161 & Oak Ridge National Laboratory & US & 3,210 & University of Oxford & GB & 4,617 \\  
\hdashline          
\textit{31} & Princeton University & US & 2,076 & National Institute of Standards and Technology & US & 2,953 & Harbin Institute of Technology & CN & 4,492 \\  
\textit{32} & Tokyo University of Science & JP & 2,047 & ETH Zurich & CH & 2,905 & Tokyo Institute of Technology & JP & 4,441 \\  
\textit{33} & Brookhaven National Laboratory & US & 2,035 & Pennsylvania State University & US & 2,889 & Jilin University & CN & 4,257 \\  
\textit{34} & European Organization for Nuclear Research & CH & 1,971 & University of Illinois Urbana-Champaign & US & 2,849 & \scalebox{0.82}[1]{National Institute of Advanced Industrial Science and Technology (AIST)} & JP & 4,245 \\  
\textit{35} & Cornell University & US & 1,949 & Tokyo University of Science & JP & 2,841 & Fudan University & CN & 4,215 \\  
\textit{36} & Laboratory of Solid State Physics & FR & 1,939 & Princeton University & US & 2,834 & \'{E}cole Polytechnique F\'{e}d\'{e}rale de Lausanne & CH & 4,204 \\  
\textit{37} & Institute of Physics & CN & 1,889 & Max Planck Society & DE & 2,829 & Karlsruhe Institute of Technology & DE & 4,064 \\  
\textit{38} & Technical University of Munich & DE & 1,882 & University of California, Santa Barbara & US & 2,790 & \scalebox{0.95}[1]{Collaborative Innovation Center of Advanced Microstructures} & CN & 4,058 \\  
\textit{39} & Lawrence Livermore National Laboratory & US & 1,863 & Peking University & CN & 2,775 & University of Maryland, College Park & US & 3,898 \\  
\textit{40} & Max Planck Society & DE & 1,848 & University of Science and Technology of China & CN & 2,754 & Nagoya University & JP & 3,896 \\  
\hdashline          
\textit{41} & \'{E}cole Polytechnique F\'{e}d\'{e}rale de Lausanne & CH & 1,819 & Czech Academy of Sciences, Institute of Physics & CZ & 2,744 & Imperial College London & GB & 3,867 \\  
\textit{42} & Istituto Nazionale per la Fisica della Materia & IT & 1,806 & University of Michigan{\textendash}Ann Arbor & US & 2,704 & Seoul National University & KR & 3,814 \\  
\textit{43} & Pennsylvania State University & US & 1,768 & \'{E}cole Polytechnique F\'{e}d\'{e}rale de Lausanne & CH & 2,659 & University of Paris-Sud & FR & 3,776 \\  
\textit{44} & Kyushu University & JP & 1,763 & United States Naval Research Laboratory & US & 2,649 & University of Electronic Science and Technology of China & CN & 3,767 \\  
\textit{45} & University of California, San Diego & US & 1,746 & University of Tsukuba & JP & 2,621 & University of Paris-Saclay & FR & 3,719 \\  
\textit{46} & Chalmers University of Technology & SE & 1,744 & National Taiwan University & TW & 2,570 & Polish Academy of Sciences & PL & 3,719 \\  
\textit{47} & Physico-Technical Institute & RU & 1,735 & Hokkaido University & JP & 2,562 & The University of Texas at Austin & US & 3,656 \\  
\textit{48} & Hitachi (Japan) & JP & 1,726 & Kyushu University & JP & 2,554 & University of California, Los Angeles & US & 3,625 \\  
\textit{49} & ETH Zurich & CH & 1,720 & Forschungszentrum J\"{u}lich & DE & 2,515 & University of Michigan{\textendash}Ann Arbor & US & 3,611 \\  
\textit{50} & Czech Academy of Sciences, Institute of Physics & CZ & 1,704 & National University of Singapore & SG & 2,510 & National Taiwan University & TW & 3,588 \\  
          
 \\[-1.4em]
\hdashline \\[-1em]
\multicolumn{10}{r}{\scriptsize \emph{(continued to next page)}}
\end{tabular}}
}
\end{table}
\end{landscape}
}
\afterpage{\clearpage%
\begin{landscape}
\begin{table}[!t]\ContinuedFloat
\vspace{-3.3em}
\caption{\textbf{The top 100 productive institutions: \textcolor{violet}{\textit{\condensed}}.} \emph{(Cont.)}}
\label{tab:r2_condensed}
\vspace{2em}
{\tiny
{\renewcommand{\arraystretch}{1.2}
\begin{tabular}{rp{5cm}lr@{\hspace{4em}}p{5cm}lr@{\hspace{4em}}p{5cm}lr}\\[-5em] \toprule[1pt] \\[-1.4em]    
 & {\scriptsize \textbf{1991--2000}} & \multicolumn{2}{c}{No.\ Works} & {\scriptsize \textbf{2001--2010}} & \multicolumn{2}{c}{No.\ Works} & {\scriptsize \textbf{2011--2020}} & \multicolumn{2}{r}{No.\ Works} \\[-0.2em] \cmidrule[0.5pt](lr{4em}){2-4} \cmidrule[0.5pt](l{-0em}r{4em}){5-7} \cmidrule[0.5pt](l{-0em}r{1em}){8-10}
\textit{51} & University of Minnesota & US & 1,704 & Brookhaven National Laboratory & US & 2,494 & Pennsylvania State University & US & 3,557 \\  
\textit{52} & Hokkaido University & JP & 1,691 & Seoul National University & KR & 2,450 & Japan Science and Technology Agency (JST) & JP & 3,538 \\  
\textit{53} & University of Michigan{\textendash}Ann Arbor & US & 1,689 & RIKEN & JP & 2,402 & Tianjin University & CN & 3,513 \\  
\textit{54} & University of California, Los Angeles & US & 1,687 & Max Planck Institute for Solid State Research & DE & 2,389 & Ioffe Institute & RU & 3,480 \\  
\textit{55} & Iowa State University & US & 1,656 & Cornell University & US & 2,338 & Shandong University & CN & 3,447 \\  
\textit{56} & Nanjing University & CN & 1,588 & University of California, San Diego & US & 2,338 & National Institute of Standards and Technology & US & 3,445 \\  
\textit{57} & Imperial College London & GB & 1,566 & Technical University of Munich & DE & 2,206 & Princeton University & US & 3,430 \\  
\textit{58} & International Superconductivity Technology Center & JP & 1,542 & University of California, Los Angeles & US & 2,198 & Brookhaven National Laboratory & US & 3,401 \\  
\textit{59} & IBM Research - Thomas J. Watson Research Center & US & 1,523 & The University of Texas at Austin & US & 2,178 & National Academy of Sciences of Ukraine & UA & 3,330 \\  
\textit{60} & Hiroshima University & JP & 1,511 & Imperial College London & GB & 2,168 & Beihang University & CN & 3,300 \\  
\hdashline          
\textit{61} & French National Centre for Scientific Research & FR & 1,501 & Iowa State University & US & 2,152 & University of Illinois Urbana-Champaign & US & 3,295 \\  
\textit{62} & University of Wisconsin{\textendash}Madison & US & 1,495 & Hiroshima University & JP & 2,141 & Institute of Physics & PL & 3,268 \\  
\textit{63} & Technical University of Berlin & DE & 1,474 & Lawrence Livermore National Laboratory & US & 2,082 & TU Dresden & DE & 3,265 \\  
\textit{64} & China Center of Advanced Science and Technology & CN & 1,463 & Harvard University & US & 2,034 & Harvard University & US & 3,259 \\  
\textit{65} & California Institute of Technology & US & 1,437 & California Institute of Technology & US & 2,027 & Forschungszentrum J\"{u}lich & DE & 3,244 \\  
\textit{66} & University of Stuttgart & DE & 1,430 & European Organization for Nuclear Research & CH & 2,001 & Southeast University & CN & 3,170 \\  
\textit{67} & RIKEN & JP & 1,419 & University of Florida & US & 1,981 & Sichuan University & CN & 3,169 \\  
\textit{68} & The University of Texas at Austin & US & 1,414 & Karlsruhe Institute of Technology & DE & 1,974 & University of California, Santa Barbara & US & 3,100 \\  
\textit{69} & Johannes Gutenberg University Mainz & DE & 1,409 & Northwestern University & US & 1,974 & Czech Academy of Sciences, Institute of Physics & CZ & 3,097 \\  
\textit{70} & National Academy of Sciences of Ukraine & UA & 1,393 & High Energy Accelerator Research Organization (KEK) & JP & 1,947 & Northwestern University & US & 3,093 \\  
\hdashline          
\textit{71} & Universit\'{e} Paris Cit\'{e} & FR & 1,391 & Universidade de S\~{a}o Paulo & BR & 1,934 & University of Tsukuba & JP & 3,021 \\  
\textit{72} & The Ohio State University & US & 1,359 & Uppsala University & SE & 1,918 & Georgia Institute of Technology & US & 3,000 \\  
\textit{73} & University of Florida & US & 1,345 & KU Leuven & BE & 1,877 & Sungkyunkwan University & KR & 2,994 \\  
\textit{74} & Florida State University & US & 1,344 & TU Wien & AT & 1,870 & Royal Institute of Technology & SE & 2,978 \\  
\textit{75} & Rutgers, The State University of New Jersey & US & 1,342 & CEA Saclay & FR & 1,856 & Korea Advanced Institute of Science and Technology & KR & 2,971 \\  
\textit{76} & Northwestern University & US & 1,329 & National Yang Ming Chiao Tung University & TW & 1,855 & Paul Scherrer Institute & CH & 2,964 \\  
\textit{77} & Kurchatov Institute & RU & 1,324 & Georgia Institute of Technology & US & 1,852 & Moscow Institute of Physics and Technology & RU & 2,938 \\  
\textit{78} & University of Amsterdam & NL & 1,323 & Ruhr University Bochum & DE & 1,850 & Kyushu University & JP & 2,929 \\  
\textit{79} & Weizmann Institute of Science & IL & 1,316 & Rutgers, The State University of New Jersey & US & 1,837 & Technical University of Munich & DE & 2,922 \\  
\textit{80} & Institut Laue-Langevin & FR & 1,305 & Delft University of Technology & NL & 1,835 & Iowa State University & US & 2,881 \\  
\hdashline          
\textit{81} & High Energy Accelerator Research Organization (KEK) & JP & 1,304 & National Tsing Hua University & TW & 1,785 & University of Science and Technology Beijing & CN & 2,857 \\  
\textit{82} & Ames Laboratory & US & 1,296 & Nanyang Technological University & SG & 1,773 & University of Manchester & GB & 2,815 \\  
\textit{83} & Arizona State University & US & 1,293 & Japan Atomic Energy Agency (JAEA) & JP & 1,769 & Shanghai University & CN & 2,806 \\  
\textit{84} & CEA Grenoble & FR & 1,291 & Arizona State University & US & 1,769 & University of California, San Diego & US & 2,800 \\  
\textit{85} & Freie Universit\"{a}t Berlin & DE & 1,290 & University of Wisconsin{\textendash}Madison & US & 1,765 & Soochow University & CN & 2,792 \\  
\textit{86} & TU Wien & AT & 1,270 & Royal Institute of Technology & SE & 1,762 & Indian Institute of Science Bangalore & IN & 2,787 \\  
\textit{87} & Tel Aviv University & IL & 1,261 & Leibniz Institute for Solid State and Materials Research & DE & 1,757 & Uppsala University & SE & 2,785 \\  
\textit{88} & Sandia National Laboratories & US & 1,261 & Florida State University & US & 1,757 & Pohang University of Science and Technology & KR & 2,779 \\  
\textit{89} & \scalebox{0.9}[1]{P.N.\ Lebedev Physical Institute of the Russian Academy of Sciences} & RU & 1,258 & Zhejiang University & CN & 1,712 & St Petersburg University & RU & 2,778 \\  
\textit{90} & Karlsruhe Institute of Technology & DE & 1,247 & Pohang University of Science and Technology & KR & 1,712 & Purdue University West Lafayette & US & 2,765 \\  
\hdashline          
\textit{91} & Joint Institute for Nuclear Research & RU & 1,243 & Sapienza University of Rome & IT & 1,700 & Hokkaido University & JP & 2,740 \\  
\textit{92} & University of Rochester & US & 1,237 & Rutherford Appleton Laboratory & GB & 1,690 & Wuhan University & CN & 2,714 \\  
\textit{93} & Delft University of Technology & NL & 1,228 & Ames Laboratory & US & 1,688 & University College London & GB & 2,694 \\  
\textit{94} & Uppsala University & SE & 1,222 & Chalmers University of Technology & SE & 1,684 & Ural Federal University & RU & 2,656 \\  
\textit{95} & Harvard University & US & 1,217 & University of Toronto & CA & 1,675 & Sun Yat-sen University & CN & 2,633 \\  
\textit{96} & University of Pennsylvania & US & 1,214 & Kurchatov Institute & RU & 1,669 & Grenoble Alpes University & FR & 2,623 \\  
\textit{97} & KU Leuven & BE & 1,207 & University of Minnesota & US & 1,664 & TU Wien & AT & 2,615 \\  
\textit{98} & Stony Brook University & US & 1,205 & University of Stuttgart & DE & 1,642 & Moscow Engineering Physics Institute & RU & 2,610 \\  
\textit{99} & State University of New York & US & 1,201 & University of California, Davis & US & 1,639 & European Organization for Nuclear Research & CH & 2,605 \\  
\textit{100} & University of Arizona & US & 1,196 & Joint Institute for Nuclear Research & RU & 1,637 & SLAC National Accelerator Laboratory & US & 2,593 \\  
          
 \\[-1.4em]
\bottomrule
\end{tabular}}
}
\end{table}
\end{landscape}
}
%

\titleformat{\section}{\sc\centering\LARGE\bfseries}{\textsc{\thesection}.\!\!}{1em}{}

\afterpage{\clearpage%
\markboth{\textbf \textsc{\envi}}{}
\thispagestyle{empty}
\quad
\vspace{2cm}
\begin{center}
\pgfornament[width=0.5*\textwidth,symmetry=h]{89}\\[2em]
\section{\envi}
\vspace{1em}
\pgfornament[width=0.5*\textwidth]{89}
\end{center}
}

\afterpage{\clearpage%

\begin{figure}[!tp]
\centering
\vspace{-1em}
{\large \textbf{\textrm{{World Map of \textcolor{violet}{\textit{\envi}} Collaboration}}}~|~1971--2020}\\
\vspace{0.3cm}
\includegraphics[align=c, scale=0.054, trim={9.5cm 0 9.5cm 0},clip]{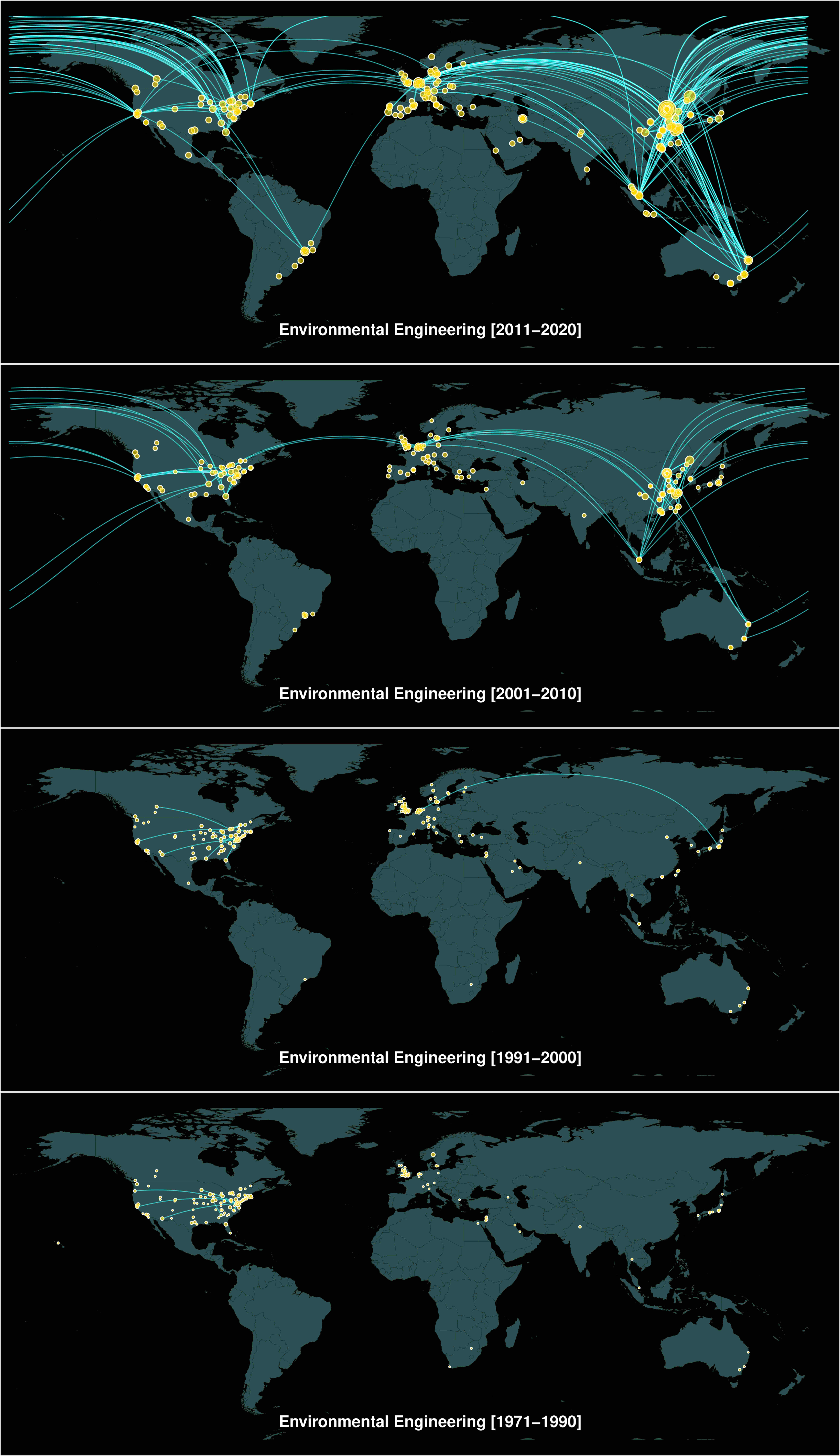}
\caption[{\envi}]{\textbf{(a)~|~The World Map of \textcolor{violet}{\textit{\envi}} Collaboration.}
The bubbles represent the top 199 institutions in terms of work production, with their sizes proportional to the work volume. 
The connecting lines depict coauthorship relationships among the top 50 institutions.}
\label{fig:wmap_envi}
\end{figure}
}
\afterpage{\clearpage%
\begin{figure}[!tp]\ContinuedFloat
\centering
\vspace{-1em}
{\large \textbf{Top 30 Productive Institutions on the World Map: \textcolor{violet}{\textit{\envi}}}~|~1991--2020}\\
\vspace{-0em}
\hspace*{-3em}                                                           
\includegraphics[align=c, scale=0.83]{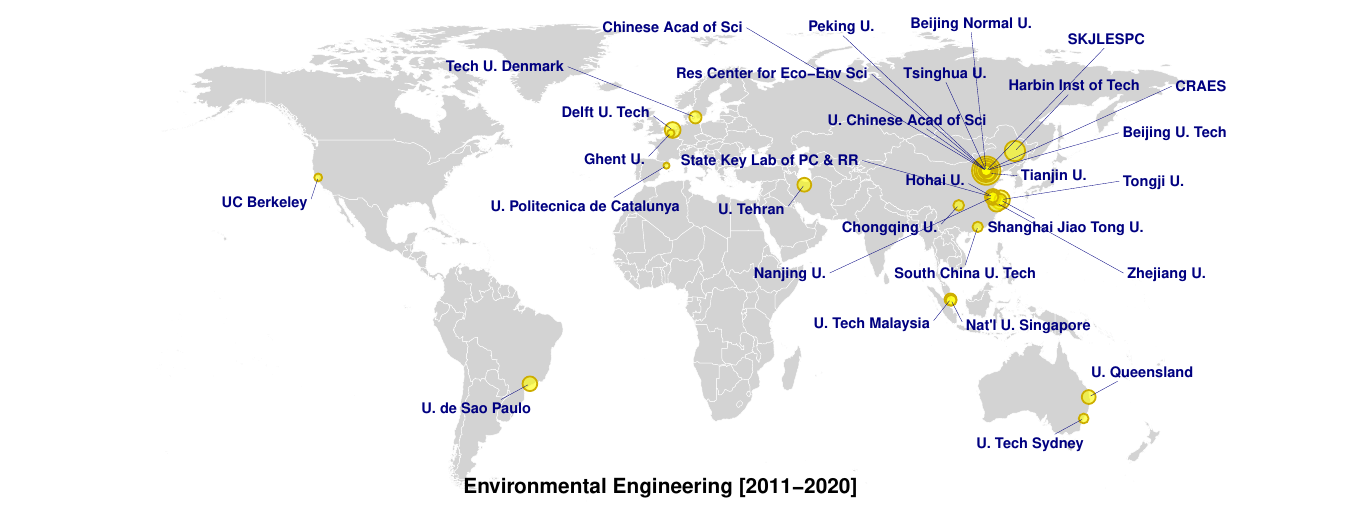}\\[-0.5em]
\quad\\[-1em]
\dotfill 
\quad\\[-0em]
\hspace*{-3em}                                                           
\includegraphics[align=c, scale=0.83]{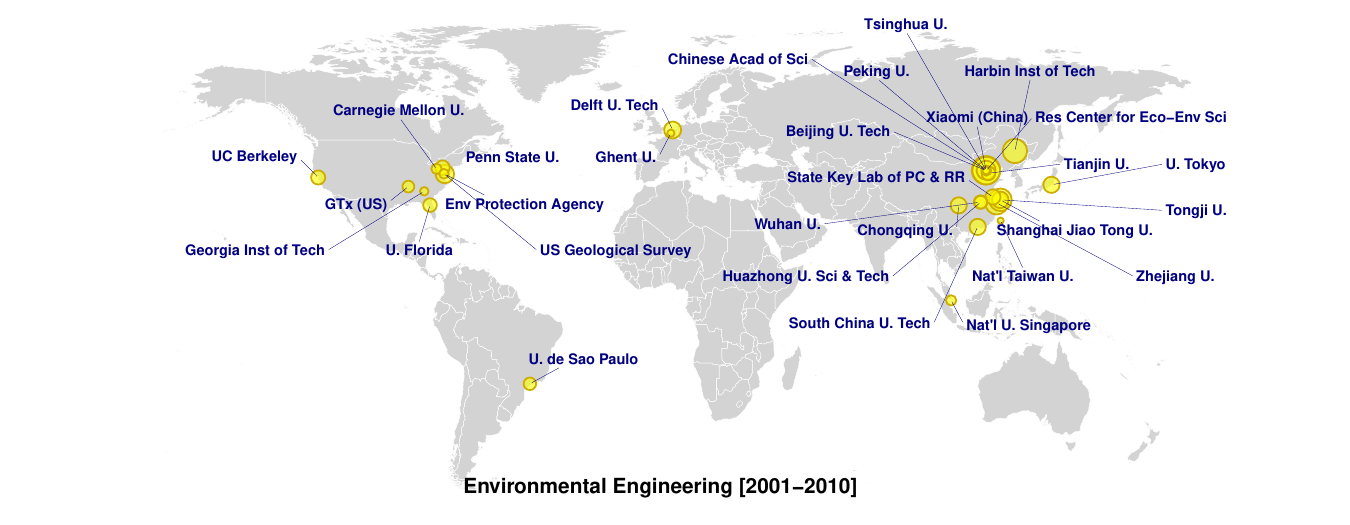}\\[-0.5em]
\quad\\[-1em]
\dotfill 
\quad\\[-0em]
\hspace*{-3em}                                                           
\includegraphics[align=c, scale=0.83]{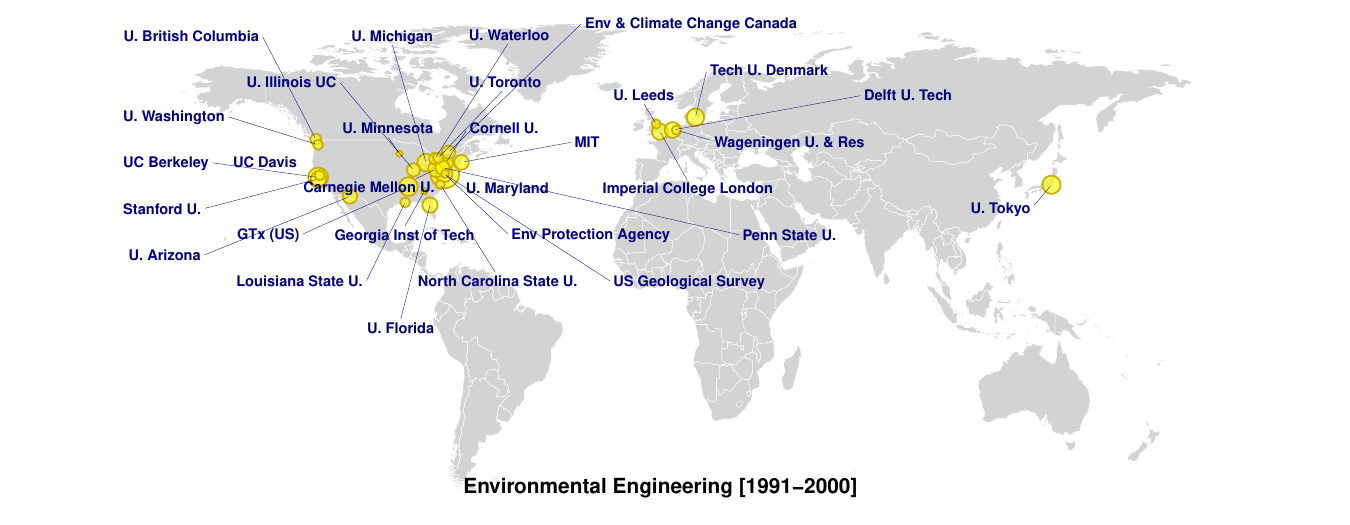}\\[-0.8em]
\caption[{\envi}]{\textbf{(b)~|~The top 30 productive institutions on the World Map: \textcolor{violet}{\textit{\envi}}.}
The bubbles represent the top 30 institutions in terms of work production, with their sizes proportional to the work volume.}
\label{fig:wmap_topinst_envi}
\end{figure}
}
\afterpage{\clearpage%
\begin{figure}[!tp]\ContinuedFloat
\centering
\vspace{-1em}
{\large \textbf{\textrm{{Interregional \textcolor{violet}{\textit{\envi}} Collaboration}}}~|~1991--2020}\\
\vspace{0.5em}
\hspace{-5em}\includegraphics[align=c, scale=1.7, vmargin=0mm]{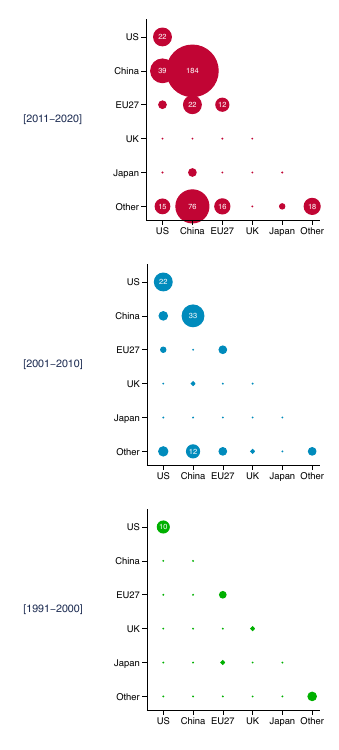}
\vspace{-1em}
\caption[]{\textbf{(c)~|~The Interregional \textcolor{violet}{\textit{\envi}} Collaboration Matrix Diagram.}
The bubble size represents the number of coauthorship relationships for the top 50 institutions in terms of work production. 
If the number is equal to or greater than 10, it is displayed inside the bubble.}
\label{fig:halfmat_envi}
\end{figure}
}
\afterpage{\clearpage%
\begin{figure}[!tp]\ContinuedFloat
\centering
\vspace{-1em}
{\large \textbf{\textrm{Interinstitutional \textcolor{violet}{\textit{\envi}} Collaboration}}~|~2001--2020\quad {\footnotesize \emph{(continued to next page)}}}\\
\cdend{envi}{2010}{2000}{2011--2020}{2001--2010}\\[-1.5em]
\caption[]{\textbf{(d)~|~The Interinstitutional \textcolor{violet}{\textit{\envi}} Collaboration Dendrogram.}
The top 50 institutions in terms of work production, indicated by the circularised bar graphs, are displayed.
}
\label{fig:cdend1_envi}
\end{figure}
}
\afterpage{\clearpage%
\begin{figure}[!tp]\ContinuedFloat
\centering
\vspace{-1em}
{\large \textbf{\textrm{Interinstitutional \textcolor{violet}{\textit{\envi}} Collaboration}}~|~1971--2000\quad {\footnotesize \emph{(continued from previous page)}}}\\
\cdend{envi}{1990}{1980}{1991--2000}{1971--1990}\\[-1.5em]
\caption[]{\textbf{(d)~|~The Interinstitutional \textcolor{violet}{\textit{\envi}} Collaboration Dendrogram.} \emph{(Cont.)}\hfill~}
\label{fig:cdend2_envi}
\end{figure}
}
\afterpage{\clearpage%
\begin{landscape}
\begin{table}[!t]
\vspace{-0.5em}
\caption{\textbf{The top 100 productive institutions: \textcolor{violet}{\textit{\envi}}.}}
\label{tab:r1_envi}
\vspace{2em}
\centering
{\tiny
{\renewcommand{\arraystretch}{1.2}
\begin{tabular}{rp{5cm}lr@{\hspace{4em}}p{5cm}lr@{\hspace{4em}}p{5cm}lr}\\[-5em] \toprule[1pt] \\[-1.4em]    
 & {\scriptsize \textbf{1991--2000}} & \multicolumn{2}{c}{No.~Works} & {\scriptsize \textbf{2001--2010}} & \multicolumn{2}{c}{No.~Works} & {\scriptsize \textbf{2011--2020}} & \multicolumn{2}{r}{No.~Works} \\[-0.2em] \cmidrule[0.5pt](lr{4em}){2-4} \cmidrule[0.5pt](l{-0em}r{4em}){5-7} \cmidrule[0.5pt](l{-0em}r{1em}){8-10}
\textit{1} & Environmental Protection Agency & US & 672 & Tsinghua University & CN & 1,805 & Chinese Academy of Sciences & CN & 5,705 \\  
\textit{2} & University of California, Berkeley & US & 331 & Chinese Academy of Sciences & CN & 1,694 & Tsinghua University & CN & 3,362 \\  
\textit{3} & GTx (United States) & US & 303 & Harbin Institute of Technology & CN & 1,306 & Harbin Institute of Technology & CN & 2,447 \\  
\textit{4} & The University of Tokyo & JP & 292 & Zhejiang University & CN & 1,168 & University of Chinese Academy of Sciences & CN & 2,244 \\  
\textit{5} & Technical University of Denmark & DK & 279 & Tongji University & CN & 1,139 & Tongji University & CN & 2,096 \\  
\textit{6} & University of Michigan{\textendash}Ann Arbor & US & 279 & Xiaomi (China) & CN & 882 & Zhejiang University & CN & 1,750 \\  
\textit{7} & United States Geological Survey & US & 267 & Environmental Protection Agency & US & 847 & \scalebox{0.9}[1]{State Key Laboratory of Pollution Control and Resource Reuse} & CN & 1,555 \\  
\textit{8} & Carnegie Mellon University & US & 264 & Delft University of Technology & NL & 775 & Delft University of Technology & NL & 1,542 \\  
\textit{9} & Imperial College London & GB & 254 & Shanghai Jiao Tong University & CN & 753 & Research Center for Eco-Environmental Sciences & CN & 1,404 \\  
\textit{10} & Delft University of Technology & NL & 250 & South China University of Technology & CN & 742 & Universidade de S\~{a}o Paulo & BR & 1,401 \\  
\hdashline          
\textit{11} & Stanford University & US & 244 & The University of Tokyo & JP & 741 & Hohai University & CN & 1,352 \\  
\textit{12} & University of Florida & US & 242 & Chongqing University & CN & 738 & Tianjin University & CN & 1,317 \\  
\textit{13} & Massachusetts Institute of Technology & US & 235 & \scalebox{0.9}[1]{State Key Laboratory of Pollution Control and Resource Reuse} & CN & 703 & University of Queensland & AU & 1,314 \\  
\textit{14} & University of Arizona & US & 232 & Pennsylvania State University & US & 675 & University of Tehran & IR & 1,313 \\  
\textit{15} & Pennsylvania State University & US & 227 & University of California, Berkeley & US & 663 & Beijing Normal University & CN & 1,295 \\  
\textit{16} & University of Illinois Urbana-Champaign & US & 217 & Tianjin University & CN & 655 & Peking University & CN & 1,184 \\  
\textit{17} & Environment and Climate Change Canada & CA & 216 & University of Florida & US & 650 & Technical University of Denmark & DK & 1,171 \\  
\textit{18} & Cornell University & US & 203 & Wuhan University & CN & 643 & Shanghai Jiao Tong University & CN & 1,147 \\  
\textit{19} & University of Waterloo & CA & 201 & Huazhong University of Science and Technology & CN & 625 & University of Technology Malaysia & MY & 1,147 \\  
\textit{20} & University of British Columbia & CA & 200 & Universidade de S\~{a}o Paulo & BR & 618 & Beijing University of Technology & CN & 1,112 \\  
\hdashline          
\textit{21} & University of Toronto & CA & 194 & GTx (United States) & US & 599 & Chinese Research Academy of Environmental Sciences & CN & 1,108 \\  
\textit{22} & University of California, Davis & US & 191 & National University of Singapore & SG & 575 & \scalebox{0.8}[1]{State Key Joint Laboratory of Environment Simulation and Pollution Control} & CN & 1,101 \\  
\textit{23} & University of Washington & US & 191 & Carnegie Mellon University & US & 575 & Nanjing University & CN & 1,100 \\  
\textit{24} & Louisiana State University & US & 190 & Beijing University of Technology & CN & 569 & National University of Singapore & SG & 1,088 \\  
\textit{25} & University of Leeds & GB & 187 & United States Geological Survey & US & 567 & Chongqing University & CN & 1,076 \\  
\textit{26} & Wageningen University \& Research & NL & 187 & Georgia Institute of Technology & US & 553 & South China University of Technology & CN & 1,061 \\  
\textit{27} & North Carolina State University & US & 186 & Peking University & CN & 548 & University of Technology Sydney & AU & 1,031 \\  
\textit{28} & University of Maryland, College Park & US & 184 & Research Center for Eco-Environmental Sciences & CN & 547 & University of California, Berkeley & US & 974 \\  
\textit{29} & University of Minnesota & US & 182 & Ghent University & BE & 545 & Ghent University & BE & 967 \\  
\textit{30} & Georgia Institute of Technology & US & 181 & National Taiwan University & TW & 543 & Universitat Polit\`{e}cnica de Catalunya & ES & 958 \\  
\hdashline          
\textit{31} & Virginia Tech & US & 181 & University of California, Davis & US & 533 & Nanyang Technological University & SG & 941 \\  
\textit{32} & University of Cincinnati & US & 178 & Hong Kong Polytechnic University & CN & 528 & Politecnico di Milano & IT & 934 \\  
\textit{33} & The University of Texas at Austin & US & 177 & University of Illinois Urbana-Champaign & US & 516 & China University of Mining and Technology & CN & 933 \\  
\textit{34} & Oak Ridge National Laboratory & US & 173 & University of Queensland & AU & 504 & Environmental Protection Agency & US & 923 \\  
\textit{35} & Texas A\&M University & US & 169 & Wageningen University \& Research & NL & 503 & UNSW Sydney & AU & 915 \\  
\textit{36} & University of Alberta & CA & 168 & Nanyang Technological University & SG & 492 & University of Lisbon & PT & 913 \\  
\textit{37} & \'{E}cole Polytechnique F\'{e}d\'{e}rale de Lausanne & CH & 164 & University of Cincinnati & US & 492 & Massachusetts Institute of Technology & US & 908 \\  
\textit{38} & National Taiwan University & TW & 164 & Hohai University & CN & 491 & China University of Petroleum, Beijing & CN & 891 \\  
\textit{39} & National University of Singapore & SG & 160 & University of British Columbia & CA & 479 & University of Florida & US & 878 \\  
\textit{40} & University of Southern California & US & 159 & Imperial College London & GB & 479 & University of Alberta & CA & 875 \\  
\hdashline          
\textit{41} & University of Queensland & AU & 158 & University of Michigan{\textendash}Ann Arbor & US & 478 & ETH Zurich & CH & 871 \\  
\textit{42} & Ghent University & BE & 158 & Virginia Tech & US & 477 & Wageningen University \& Research & NL & 871 \\  
\textit{43} & McGill University & CA & 158 & Technical University of Denmark & DK & 476 & North China Electric Power University & CN & 864 \\  
\textit{44} & University of Sheffield & GB & 158 & UNSW Sydney & AU & 461 & Wuhan University & CN & 862 \\  
\textit{45} & Lund University & SE & 158 & Stanford University & US & 461 & Dalian University of Technology & CN & 860 \\  
\textit{46} & University of Wisconsin{\textendash}Madison & US & 153 & Massachusetts Institute of Technology & US & 457 & United States Geological Survey & US & 860 \\  
\textit{47} & Newcastle University & GB & 152 & University of Toronto & CA & 448 & Shandong University & CN & 858 \\  
\textit{48} & Kyoto University & JP & 152 & National University of Defense Technology & CN & 448 & University of Illinois Urbana-Champaign & US & 857 \\  
\textit{49} & Lancaster University & GB & 151 & Seoul National University & KR & 444 & Virginia Tech & US & 856 \\  
\textit{50} & Making View (Norway) & NO & 151 & Dalian University of Technology & CN & 443 & Pennsylvania State University & US & 854 \\  
          
 \\[-1.4em]
\hdashline \\[-1em]
\multicolumn{10}{r}{\scriptsize \emph{(continued to next page)}}
\end{tabular}}
}
\end{table}
\end{landscape}
}
\afterpage{\clearpage%
\begin{landscape}
\begin{table}[!t]\ContinuedFloat
\vspace{-3.3em}
\caption{\textbf{The top 100 productive institutions: \textcolor{violet}{\textit{\envi}}.} \emph{(Cont.)}}
\label{tab:r2_envi}
\vspace{2em}
{\tiny
{\renewcommand{\arraystretch}{1.2}
\begin{tabular}{rp{5cm}lr@{\hspace{4em}}p{5cm}lr@{\hspace{4em}}p{5cm}lr}\\[-5em] \toprule[1pt] \\[-1.4em]    
 & {\scriptsize \textbf{1991--2000}} & \multicolumn{2}{c}{No.\ Works} & {\scriptsize \textbf{2001--2010}} & \multicolumn{2}{c}{No.\ Works} & {\scriptsize \textbf{2011--2020}} & \multicolumn{2}{r}{No.\ Works} \\[-0.2em] \cmidrule[0.5pt](lr{4em}){2-4} \cmidrule[0.5pt](l{-0em}r{4em}){5-7} \cmidrule[0.5pt](l{-0em}r{1em}){8-10}
\textit{51} & The Ohio State University & US & 149 & KU Leuven & BE & 434 & The University of Tokyo & JP & 850 \\  
\textit{52} & Rutgers, The State University of New Jersey & US & 146 & Arizona State University & US & 433 & Xi'an University of Architecture and Technology & CN & 847 \\  
\textit{53} & Technion {\textendash} Israel Institute of Technology & IL & 145 & Universitat Polit\`{e}cnica de Catalunya & ES & 428 & University of Toronto & CA & 825 \\  
\textit{54} & University of California, Irvine & US & 144 & The University of Texas at Austin & US & 428 & University of British Columbia & CA & 799 \\  
\textit{55} & Hong Kong Polytechnic University & CN & 142 & University of Waterloo & CA & 424 & Stanford University & US & 797 \\  
\textit{56} & Iowa State University & US & 141 & Wuhan University of Technology & CN & 424 & University of Melbourne & AU & 794 \\  
\textit{57} & University College London & GB & 140 & University of Arizona & US & 421 & Universitat Polit\`{e}cnica de Val\`{e}ncia & ES & 793 \\  
\textit{58} & Loughborough University & GB & 139 & University of Minnesota & US & 416 & Imperial College London & GB & 789 \\  
\textit{59} & National Technical University of Athens & GR & 135 & Nanjing University & CN & 405 & Huazhong University of Science and Technology & CN & 788 \\  
\textit{60} & Swiss Federal Institute of Aquatic Science and Technology & CH & 134 & Texas A\&M University & US & 402 & Texas A\&M University & US & 782 \\  
\hdashline          
\textit{61} & Purdue University West Lafayette & US & 134 & North Carolina State University & US & 395 & University of Michigan{\textendash}Ann Arbor & US & 780 \\  
\textit{62} & Research Triangle Park Foundation & US & 134 & Beijing Normal University & CN & 390 & KU Leuven & BE & 779 \\  
\textit{63} & University of California, Los Angeles & US & 134 & Swiss Federal Institute of Aquatic Science and Technology & CH & 389 & University of California, Davis & US & 771 \\  
\textit{64} & University of Manchester & GB & 133 & Cranfield University & GB & 385 & University of Porto & PT & 758 \\  
\textit{65} & University of Birmingham & GB & 132 & Kyoto University & JP & 385 & Seoul National University & KR & 745 \\  
\textit{66} & \scalebox{0.9}[1]{Commonwealth Scientific and Industrial Research Organisation} & AU & 129 & University of Southern California & US & 384 & Georgia Institute of Technology & US & 743 \\  
\textit{67} & University of Cambridge & GB & 128 & TU Wien & AT & 382 & Sapienza University of Rome & IT & 736 \\  
\textit{68} & Technical University of Berlin & DE & 127 & National Technical University of Athens & GR & 382 & Technical University of Munich & DE & 734 \\  
\textit{69} & Technical University of Munich & DE & 127 & Environment and Climate Change Canada & CA & 374 & University of Waterloo & CA & 730 \\  
\textit{70} & Cranfield University & GB & 127 & University of Maryland, College Park & US & 373 & Polytechnic University of Turin & IT & 727 \\  
\hdashline          
\textit{71} & WRc (United Kingdom) & GB & 127 & University of California, Irvine & US & 370 & National Taiwan University & TW & 721 \\  
\textit{72} & UNSW Sydney & AU & 126 & Istanbul Technical University & TR & 369 & Aarhus University & DK & 718 \\  
\textit{73} & KU Leuven & BE & 126 & China University of Mining and Technology & CN & 362 & China Agricultural University & CN & 717 \\  
\textit{74} & Istanbul Technical University & TR & 126 & Beijing University of Posts and Telecommunications & CN & 361 & Hong Kong Polytechnic University & CN & 712 \\  
\textit{75} & University of Colorado Boulder & US & 126 & University of Technology Sydney & AU & 359 & National University of Malaysia & MY & 710 \\  
\textit{76} & Tsinghua University & CN & 124 & Jilin University & CN & 358 & Carnegie Mellon University & US & 710 \\  
\textit{77} & Tohoku University & JP & 124 & Hunan University & CN & 357 & The University of Texas at Austin & US & 709 \\  
\textit{78} & Columbia University & US & 124 & Cornell University & US & 357 & Arizona State University & US & 694 \\  
\textit{79} & ExxonMobil (United States) & US & 122 & University of Washington & US & 355 & Royal Institute of Technology & SE & 693 \\  
\textit{80} & Hong Kong University of Science and Technology & CN & 119 & University of Edinburgh & GB & 354 & University of Bologna & IT & 689 \\  
\hdashline          
\textit{81} & University of Liverpool & GB & 119 & Universitat Polit\`{e}cnica de Val\`{e}ncia & ES & 351 & S\~{a}o Paulo State University & BR & 688 \\  
\textit{82} & Aalborg University & DK & 118 & Central South University & CN & 350 & University College London & GB & 684 \\  
\textit{83} & University of Amsterdam & NL & 117 & Sinopec (China) & CN & 350 & Ocean University of China & CN & 679 \\  
\textit{84} & Kyushu University & JP & 116 & \'{E}cole Polytechnique F\'{e}d\'{e}rale de Lausanne & CH & 349 & Jilin University & CN & 673 \\  
\textit{85} & AT\&T (United States) & US & 116 & Southeast University & CN & 349 & Norwegian University of Science and Technology & NO & 673 \\  
\textit{86} & Hokkaido University & JP & 113 & Columbia University & US & 349 & University of Malaya & MY & 667 \\  
\textit{87} & National Institute for Environmental Studies & JP & 113 & University of Melbourne & AU & 348 & Universidad Politd\'{e}cnica de Madrid & ES & 663 \\  
\textit{88} & Arizona State University & US & 113 & Universidad Politd\'{e}cnica de Madrid & ES & 347 & North Carolina State University & US & 663 \\  
\textit{89} & Colorado State University & US & 111 & Monash University & AU & 346 & Sun Yat-sen University & CN & 662 \\  
\textit{90} & Harvard University & US & 110 & ETH Zurich & CH & 345 & Universidade Estadual de Campinas & BR & 660 \\  
\hdashline          
\textit{91} & Royal Institute of Technology & SE & 109 & China University of Petroleum, Beijing & CN & 344 & Hunan University & CN & 653 \\  
\textit{92} & Michigan State University & US & 109 & Technical University of Munich & DE & 344 & Queensland University of Technology & AU & 648 \\  
\textit{93} & Politecnico di Milano & IT & 108 & Technical University of Berlin & DE & 342 & Universiti Putra Malaysia & MY & 646 \\  
\textit{94} & Norwegian University of Science and Technology & NO & 107 & National Cheng Kung University & TW & 342 & Swiss Federal Institute of Aquatic Science and Technology & CH & 642 \\  
\textit{95} & Agricultural Research Service & US & 107 & Politecnico di Milano & IT & 341 & University of Science and Technology Beijing & CN & 638 \\  
\textit{96} & Sapienza University of Rome & IT & 106 & The Ohio State University & US & 336 & Southeast University & CN & 634 \\  
\textit{97} & Tokyo Institute of Technology & JP & 106 & Xi'an Jiaotong University & CN & 333 & China University of Geosciences & CN & 628 \\  
\textit{98} & University of Pittsburgh & US & 106 & University of Wisconsin{\textendash}Madison & US & 331 & Cornell University & US & 628 \\  
\textit{99} & Aristotle University of Thessaloniki & GR & 104 & University of Calgary & CA & 328 & Monash University & AU & 626 \\  
\textit{100} & Universidad Nacional Aut\'{o}noma de M\'{e}xico & MX & 104 & University of Alberta & CA & 327 & University of Minnesota & US & 626 \\  
          
 \\[-1.4em]
\bottomrule
\end{tabular}}
}
\end{table}
\end{landscape}
}
%

\titleformat{\section}{\sc\centering\LARGE\bfseries}{\textsc{\thesection}.\!\!}{1em}{}

\afterpage{\clearpage%
\markboth{\textbf \textsc{\earth}}{}
\thispagestyle{empty}
\quad
\vspace{2cm}
\begin{center}
\pgfornament[width=0.5*\textwidth,symmetry=h]{89}\\[2em]
\section{\earth}
\vspace{1em}
\pgfornament[width=0.5*\textwidth]{89}
\end{center}
}

\afterpage{\clearpage%

\begin{figure}[!tp]
\centering
\vspace{-1em}
{\large \textbf{\textrm{{World Map of \textcolor{violet}{\textit{\earth}} Collaboration}}}~|~1971--2020}\\
\vspace{0.3cm}
\includegraphics[align=c, scale=0.054, trim={9.5cm 0 9.5cm 0},clip]{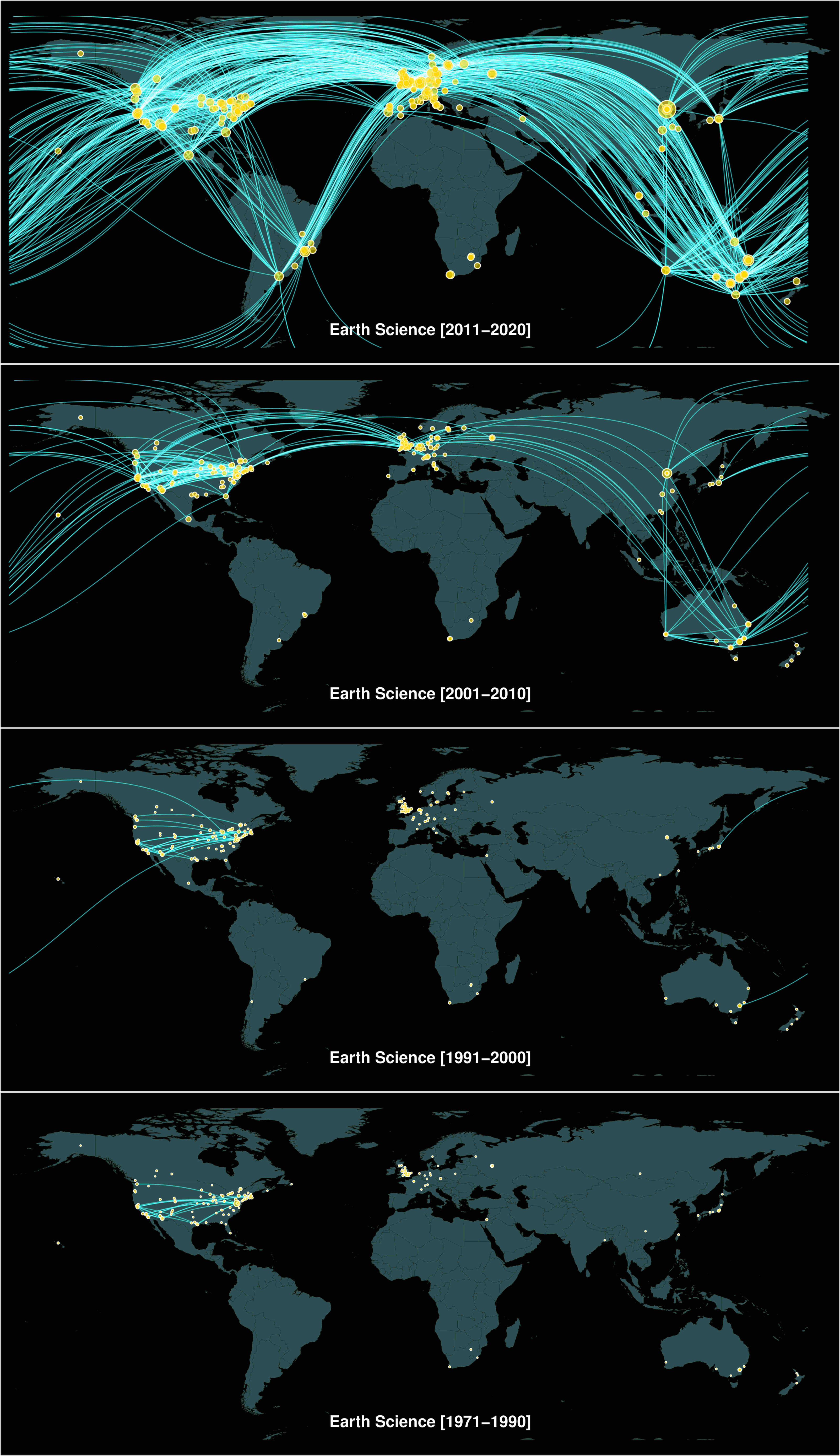}
\caption[{\earth}]{\textbf{(a)~|~The World Map of \textcolor{violet}{\textit{\earth}} Collaboration.}
The bubbles represent the top 199 institutions in terms of work production, with their sizes proportional to the work volume. 
The connecting lines depict coauthorship relationships among the top 50 institutions.}
\label{fig:wmap_earth}
\end{figure}
}
\afterpage{\clearpage%
\begin{figure}[!tp]\ContinuedFloat
\centering
\vspace{-1em}
{\large \textbf{Top 30 Productive Institutions on the World Map: \textcolor{violet}{\textit{\earth}}}~|~1991--2020}\\
\vspace{-0em}
\hspace*{-3em}                                                           
\includegraphics[align=c, scale=0.83]{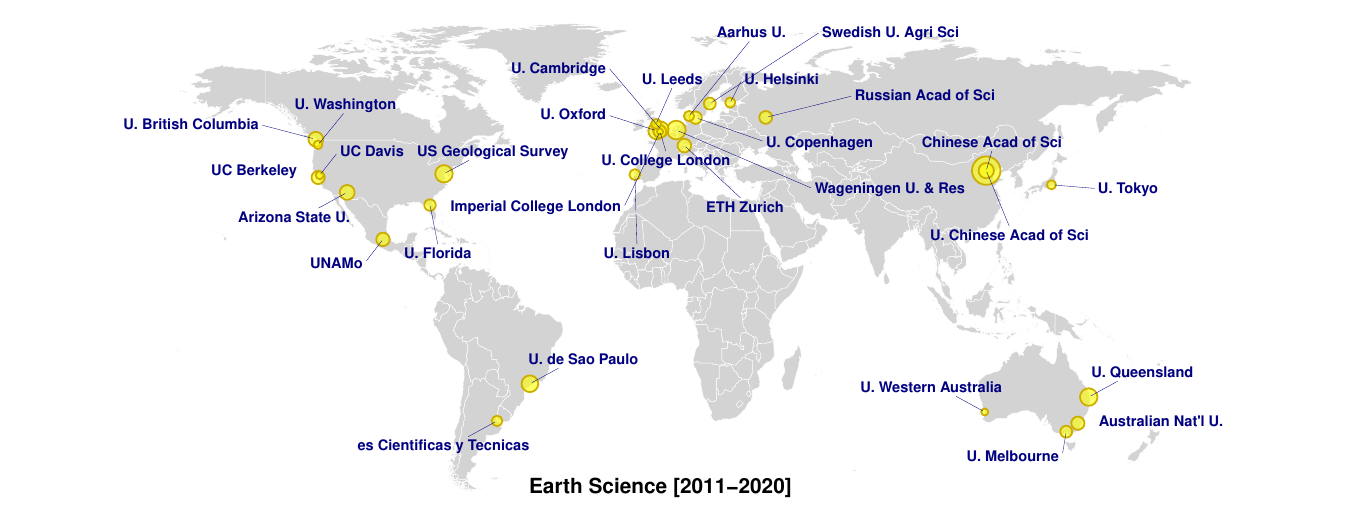}\\[-0.5em]
\quad\\[-1em]
\dotfill 
\quad\\[-0em]
\hspace*{-3em}                                                           
\includegraphics[align=c, scale=0.83]{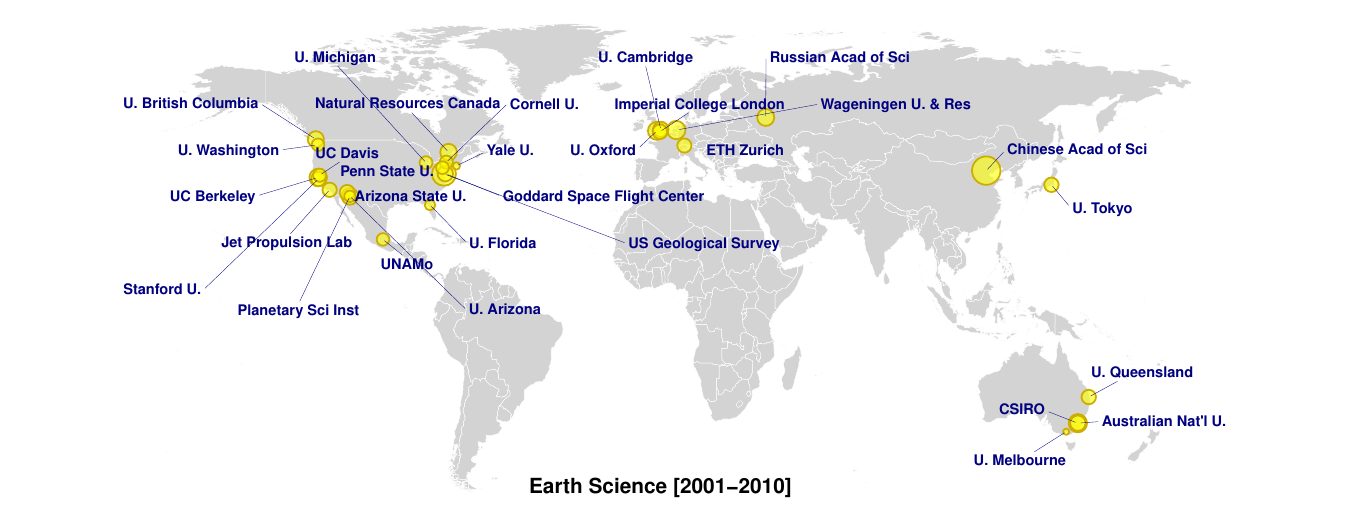}\\[-0.5em]
\quad\\[-1em]
\dotfill 
\quad\\[-0em]
\hspace*{-3em}                                                           
\includegraphics[align=c, scale=0.83]{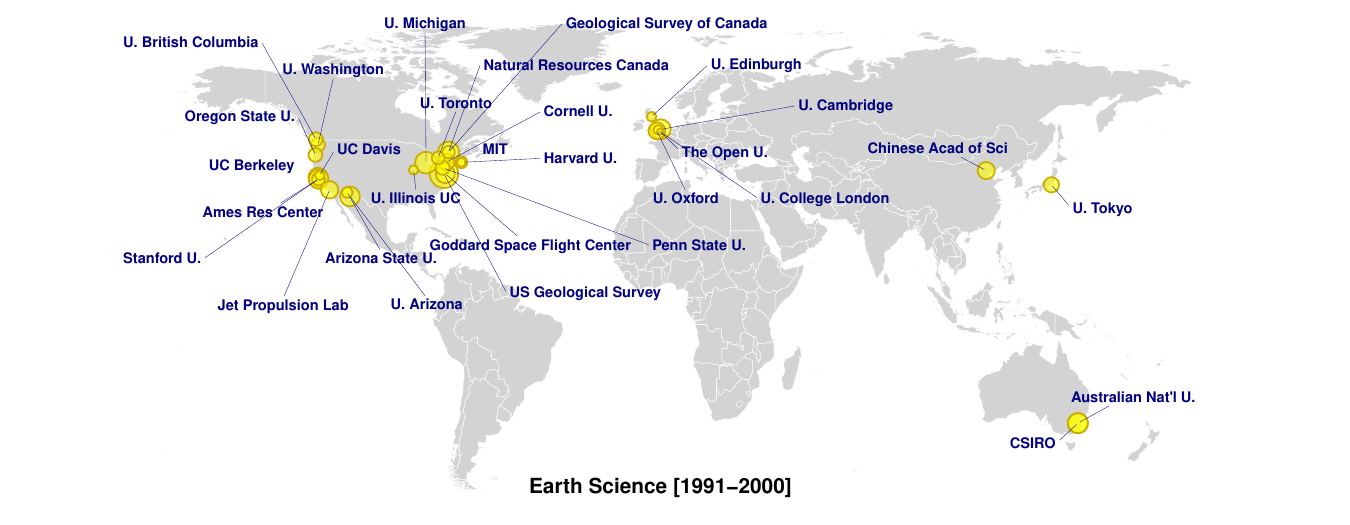}\\[-0.8em]
\caption[{\earth}]{\textbf{(b)~|~The top 30 productive institutions on the World Map: \textcolor{violet}{\textit{\earth}}.}
The bubbles represent the top 30 institutions in terms of work production, with their sizes proportional to the work volume.}
\label{fig:wmap_topinst_earth}
\end{figure}
}
\afterpage{\clearpage%
\begin{figure}[!tp]\ContinuedFloat
\centering
\vspace{-1em}
{\large \textbf{\textrm{{Interregional \textcolor{violet}{\textit{\earth}} Collaboration}}}~|~1991--2020}\\
\vspace{0.5em}
\hspace{-5em}\includegraphics[align=c, scale=1.7, vmargin=0mm]{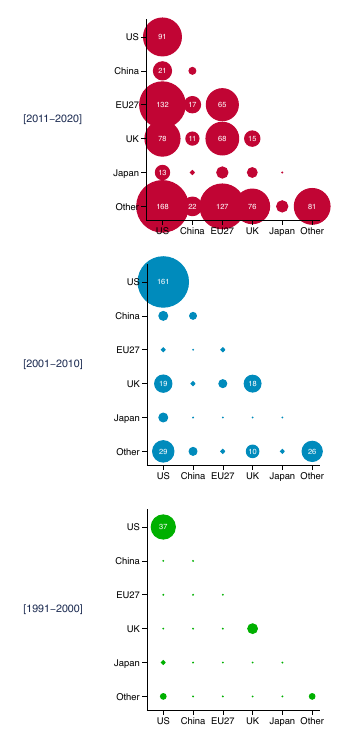}
\vspace{-1em}
\caption[]{\textbf{(c)~|~The Interregional \textcolor{violet}{\textit{\earth}} Collaboration Matrix Diagram.}
The bubble size represents the number of coauthorship relationships for the top 50 institutions in terms of work production. 
If the number is equal to or greater than 10, it is displayed inside the bubble.}
\label{fig:halfmat_earth}
\end{figure}
}
\afterpage{\clearpage%
\begin{figure}[!tp]\ContinuedFloat
\centering
\vspace{-1em}
{\large \textbf{\textrm{Interinstitutional \textcolor{violet}{\textit{\earth}} Collaboration}}~|~2001--2020\quad {\footnotesize \emph{(continued to next page)}}}\\
\cdend{earth}{2010}{2000}{2011--2020}{2001--2010}\\[-1.5em]
\caption[]{\textbf{(d)~|~The Interinstitutional \textcolor{violet}{\textit{\earth}} Collaboration Dendrogram.}
The top 50 institutions in terms of work production, indicated by the circularised bar graphs, are displayed.
}
\label{fig:cdend1_earth}
\end{figure}
}
\afterpage{\clearpage%
\begin{figure}[!tp]\ContinuedFloat
\centering
\vspace{-1em}
{\large \textbf{\textrm{Interinstitutional \textcolor{violet}{\textit{\earth}} Collaboration}}~|~1971--2000\quad {\footnotesize \emph{(continued from previous page)}}}\\
\cdend{earth}{1990}{1980}{1991--2000}{1971--1990}\\[-1.5em]
\caption[]{\textbf{(d)~|~The Interinstitutional \textcolor{violet}{\textit{\earth}} Collaboration Dendrogram.} \emph{(Cont.)}\hfill~}
\label{fig:cdend2_earth}
\end{figure}
}
\afterpage{\clearpage%
\begin{landscape}
\begin{table}[!t]
\vspace{-0.5em}
\caption{\textbf{The top 100 productive institutions: \textcolor{violet}{\textit{\earth}}.}}
\label{tab:r1_earth}
\vspace{2em}
\centering
{\tiny
{\renewcommand{\arraystretch}{1.2}
\begin{tabular}{rp{5cm}lr@{\hspace{4em}}p{5cm}lr@{\hspace{4em}}p{5cm}lr}\\[-5em] \toprule[1pt] \\[-1.4em]    
 & {\scriptsize \textbf{1991--2000}} & \multicolumn{2}{c}{No.~Works} & {\scriptsize \textbf{2001--2010}} & \multicolumn{2}{c}{No.~Works} & {\scriptsize \textbf{2011--2020}} & \multicolumn{2}{r}{No.~Works} \\[-0.2em] \cmidrule[0.5pt](lr{4em}){2-4} \cmidrule[0.5pt](l{-0em}r{4em}){5-7} \cmidrule[0.5pt](l{-0em}r{1em}){8-10}
\textit{1} & United States Geological Survey & US & 394 & Chinese Academy of Sciences & CN & 1,299 & Chinese Academy of Sciences & CN & 4,436 \\  
\textit{2} & University of Michigan{\textendash}Ann Arbor & US & 225 & United States Geological Survey & US & 878 & Wageningen University \& Research & NL & 1,839 \\  
\textit{3} & Natural Resources Canada & CA & 214 & Wageningen University \& Research & NL & 552 & University of Queensland & AU & 1,671 \\  
\textit{4} & Australian National University & AU & 199 & Australian National University & AU & 551 & United States Geological Survey & US & 1,662 \\  
\textit{5} & \scalebox{0.9}[1]{Commonwealth Scientific and Industrial Research Organisation} & AU & 194 & University of Oxford & GB & 537 & University of Oxford & GB & 1,630 \\  
\textit{6} & University of Cambridge & GB & 194 & University of California, Berkeley & US & 521 & Universidade de S\~{a}o Paulo & BR & 1,578 \\  
\textit{7} & University of Arizona & US & 192 & Russian Academy of Sciences & RU & 505 & University of Cambridge & GB & 1,463 \\  
\textit{8} & Ames Research Center & US & 187 & Natural Resources Canada & CA & 489 & University of Chinese Academy of Sciences & CN & 1,433 \\  
\textit{9} & University of California, Berkeley & US & 187 & University of Arizona & US & 483 & Arizona State University & US & 1,388 \\  
\textit{10} & Goddard Space Flight Center & US & 182 & University of British Columbia & CA & 472 & University of British Columbia & CA & 1,385 \\  
\hdashline          
\textit{11} & Jet Propulsion Laboratory & US & 162 & Goddard Space Flight Center & US & 453 & ETH Zurich & CH & 1,290 \\  
\textit{12} & Chinese Academy of Sciences & CN & 159 & \scalebox{0.9}[1]{Commonwealth Scientific and Industrial Research Organisation} & AU & 451 & Universidad Nacional Aut\'{o}noma de M\'{e}xico & MX & 1,271 \\  
\textit{13} & University of Oxford & GB & 155 & University of Cambridge & GB & 451 & University of California, Berkeley & US & 1,261 \\  
\textit{14} & Cornell University & US & 144 & Arizona State University & US & 441 & Australian National University & AU & 1,259 \\  
\textit{15} & The University of Tokyo & JP & 140 & Imperial College London & GB & 434 & Russian Academy of Sciences & RU & 1,234 \\  
\textit{16} & Pennsylvania State University & US & 133 & Stanford University & US & 430 & Imperial College London & GB & 1,220 \\  
\textit{17} & University of British Columbia & CA & 130 & The University of Tokyo & JP & 426 & University of Copenhagen & DK & 1,209 \\  
\textit{18} & University of Washington & US & 130 & University of Queensland & AU & 425 & Swedish University of Agricultural Sciences & SE & 1,179 \\  
\textit{19} & Oregon State University & US & 129 & Jet Propulsion Laboratory & US & 424 & University of Melbourne & AU & 1,176 \\  
\textit{20} & Stanford University & US & 124 & ETH Zurich & CH & 419 & University of Florida & US & 1,155 \\  
\hdashline          
\textit{21} & University of Toronto & CA & 123 & University of California, Davis & US & 396 & University of Lisbon & PT & 1,131 \\  
\textit{22} & Geological Survey of Canada & CA & 121 & University of Michigan{\textendash}Ann Arbor & US & 392 & Consejo Nacional de Investigaciones Cient\'{i}ficas y T\'{e}cnicas & AR & 1,117 \\  
\textit{23} & Massachusetts Institute of Technology & US & 118 & Pennsylvania State University & US & 389 & University of Helsinki & FI & 1,101 \\  
\textit{24} & Arizona State University & US & 115 & Universidad Nacional Aut\'{o}noma de M\'{e}xico & MX & 385 & Aarhus University & DK & 1,100 \\  
\textit{25} & The Open University & GB & 110 & University of Washington & US & 385 & University of Leeds & GB & 1,095 \\  
\textit{26} & University of Edinburgh & GB & 110 & Cornell University & US & 384 & The University of Tokyo & JP & 1,067 \\  
\textit{27} & University of California, Davis & US & 109 & Planetary Science Institute & US & 366 & University of Washington & US & 1,054 \\  
\textit{28} & University of Illinois Urbana-Champaign & US & 109 & University of Florida & US & 356 & University of California, Davis & US & 1,040 \\  
\textit{29} & Harvard University & US & 108 & Yale University & US & 332 & University of Western Australia & AU & 1,029 \\  
\textit{30} & University College London & GB & 105 & University of Melbourne & AU & 330 & University College London & GB & 1,027 \\  
\hdashline          
\textit{31} & University of Leeds & GB & 102 & Swedish University of Agricultural Sciences & SE & 330 & Utrecht University & NL & 1,016 \\  
\textit{32} & University of Minnesota & US & 102 & Ames Research Center & US & 330 & Pennsylvania State University & US & 1,002 \\  
\textit{33} & British Geological Survey & GB & 101 & China University of Geosciences & CN & 326 & Stockholm University & SE & 1,001 \\  
\textit{34} & University of Wisconsin{\textendash}Madison & US & 99 & Massachusetts Institute of Technology & US & 326 & James Cook University & AU & 999 \\  
\textit{35} & University of Sheffield & GB & 96 & Oregon State University & US & 326 & University of Arizona & US & 996 \\  
\textit{36} & Utrecht University & NL & 95 & University of Toronto & CA & 320 & University of Michigan{\textendash}Ann Arbor & US & 994 \\  
\textit{37} & University of East Anglia & GB & 94 & Macquarie University & AU & 317 & China University of Geosciences & CN & 956 \\  
\textit{38} & Scripps Institution of Oceanography & US & 93 & Smithsonian Institution & US & 317 & Helmholtz Centre for Environmental Research & DE & 942 \\  
\textit{39} & University of Tennessee at Knoxville & US & 93 & University of Minnesota & US & 315 & University of G\"{o}ttingen & DE & 929 \\  
\textit{40} & University of Cape Town & ZA & 91 & University of Wisconsin{\textendash}Madison & US & 309 & Curtin University & AU & 925 \\  
\hdashline          
\textit{41} & Imperial College London & GB & 90 & University of Leeds & GB & 307 & Stanford University & US & 924 \\  
\textit{42} & Planetary Science Institute & US & 90 & University of Sheffield & GB & 305 & Ghent University & BE & 918 \\  
\textit{43} & University of Maryland, College Park & US & 90 & Monash University & AU & 304 & University of Minnesota & US & 914 \\  
\textit{44} & Johnson Space Center & US & 89 & University of Western Australia & AU & 300 & University of Tasmania & AU & 911 \\  
\textit{45} & Texas A\&M University & US & 89 & Institute of Geology and Geophysics & CN & 300 & Lund University & SE & 910 \\  
\textit{46} & Natural History Museum & GB & 88 & The Ohio State University & US & 299 & UNSW Sydney & AU & 905 \\  
\textit{47} & The Ohio State University & US & 88 & University of Bristol & GB & 296 & Cornell University & US & 900 \\  
\textit{48} & The University of Texas at Austin & US & 86 & University of Colorado Boulder & US & 296 & Oregon State University & US & 896 \\  
\textit{49} & University of Colorado Boulder & US & 86 & University of Edinburgh & GB & 291 & \scalebox{0.9}[1]{Commonwealth Scientific and Industrial Research Organisation} & AU & 892 \\  
\textit{50} & California Institute of Technology & US & 85 & Columbia University & US & 291 & University of Wisconsin{\textendash}Madison & US & 885 \\  
          
 \\[-1.4em]
\hdashline \\[-1em]
\multicolumn{10}{r}{\scriptsize \emph{(continued to next page)}}
\end{tabular}}
}
\end{table}
\end{landscape}
}
\afterpage{\clearpage%
\begin{landscape}
\begin{table}[!t]\ContinuedFloat
\vspace{-3.3em}
\caption{\textbf{The top 100 productive institutions: \textcolor{violet}{\textit{\earth}}.} \emph{(Cont.)}}
\label{tab:r2_earth}
\vspace{2em}
{\tiny
{\renewcommand{\arraystretch}{1.2}
\begin{tabular}{rp{5cm}lr@{\hspace{4em}}p{5cm}lr@{\hspace{4em}}p{5cm}lr}\\[-5em] \toprule[1pt] \\[-1.4em]    
 & {\scriptsize \textbf{1991--2000}} & \multicolumn{2}{c}{No.\ Works} & {\scriptsize \textbf{2001--2010}} & \multicolumn{2}{c}{No.\ Works} & {\scriptsize \textbf{2011--2020}} & \multicolumn{2}{r}{No.\ Works} \\[-0.2em] \cmidrule[0.5pt](lr{4em}){2-4} \cmidrule[0.5pt](l{-0em}r{4em}){5-7} \cmidrule[0.5pt](l{-0em}r{1em}){8-10}
\textit{51} & Environmental Protection Agency & US & 85 & University of California, Santa Barbara & US & 287 & University of Exeter & GB & 878 \\  
\textit{52} & University of California, Los Angeles & US & 84 & University of Alberta & CA & 283 & Stellenbosch University & ZA & 876 \\  
\textit{53} & Woods Hole Oceanographic Institution & US & 84 & University of Copenhagen & DK & 281 & University of Alberta & CA & 871 \\  
\textit{54} & University of Queensland & AU & 83 & The Open University & GB & 275 & Sapienza University of Rome & IT & 865 \\  
\textit{55} & University of Hawaii System & US & 83 & University of East Anglia & GB & 274 & Michigan State University & US & 857 \\  
\textit{56} & University of Wales & GB & 82 & University of Helsinki & FI & 271 & University of Colorado Boulder & US & 852 \\  
\textit{57} & University of Florida & US & 82 & Lund University & SE & 269 & McGill University & CA & 843 \\  
\textit{58} & University of Amsterdam & NL & 81 & University of Maryland, College Park & US & 269 & Griffith University & AU & 841 \\  
\textit{59} & Columbia University & US & 81 & University of Cape Town & ZA & 269 & University of Sydney & AU & 835 \\  
\textit{60} & Oak Ridge National Laboratory & US & 81 & University of G\"{o}ttingen & DE & 264 & Yale University & US & 831 \\  
\hdashline          
\textit{61} & University of Adelaide & AU & 80 & Stockholm University & SE & 261 & University of Adelaide & AU & 828 \\  
\textit{62} & University of Alberta & CA & 80 & Colorado State University & US & 260 & China University of Geosciences (Beijing) & CN & 828 \\  
\textit{63} & Virginia Tech & US & 80 & Michigan State University & US & 260 & Columbia University & US & 825 \\  
\textit{64} & University of Liverpool & GB & 79 & University of Sydney & AU & 258 & Monash University & AU & 821 \\  
\textit{65} & Los Alamos National Laboratory & US & 79 & Delft University of Technology & NL & 258 & University of Toronto & CA & 821 \\  
\textit{66} & University of California, Santa Barbara & US & 77 & Lomonosov Moscow State University & RU & 257 & Lomonosov Moscow State University & RU & 789 \\  
\textit{67} & University of Bristol & GB & 76 & Natural History Museum & GB & 255 & University of Manchester & GB & 782 \\  
\textit{68} & University of Reading & GB & 76 & University College London & GB & 253 & University of Bristol & GB & 781 \\  
\textit{69} & University of California, San Diego & US & 76 & Utrecht University & NL & 249 & Delft University of Technology & NL & 774 \\  
\textit{70} & University of Western Australia & AU & 75 & Peking University & CN & 244 & Planetary Science Institute & US & 773 \\  
\hdashline          
\textit{71} & Russian Academy of Sciences & RU & 75 & Curtin University & AU & 241 & University of Edinburgh & GB & 772 \\  
\textit{72} & Vrije Universiteit Amsterdam & NL & 74 & University of Reading & GB & 241 & Macquarie University & AU & 771 \\  
\textit{73} & Royal Holloway University of London & GB & 73 & Harvard University & US & 241 & University of Maryland, College Park & US & 769 \\  
\textit{74} & Universidad Nacional Aut\'{o}noma de M\'{e}xico & MX & 73 & Cardiff University & GB & 240 & Colorado State University & US & 761 \\  
\textit{75} & University of Copenhagen & DK & 72 & California Institute of Technology & US & 240 & University of Southampton & GB & 757 \\  
\textit{76} & Colorado State University & US & 72 & Universidade de S\~{a}o Paulo & BR & 239 & University of Bern & CH & 755 \\  
\textit{77} & Michigan State University & US & 72 & James Cook University & AU & 238 & Goddard Space Flight Center & US & 746 \\  
\textit{78} & University of Bern & CH & 70 & University of Li\`{e}ge & BE & 237 & University of Zurich & CH & 743 \\  
\textit{79} & Lamont-Doherty Earth Observatory & US & 70 & Ghent University & BE & 235 & National University of Singapore & SG & 739 \\  
\textit{80} & Rutgers, The State University of New Jersey & US & 70 & The University of Texas at Austin & US & 235 & Beijing Normal University & CN & 731 \\  
\hdashline          
\textit{81} & Kiel University & DE & 69 & Johnson Space Center & US & 233 & Jet Propulsion Laboratory & US & 728 \\  
\textit{82} & Yale University & US & 69 & Griffith University & AU & 232 & University of Sheffield & GB & 723 \\  
\textit{83} & Macquarie University & AU & 67 & University of California, Los Angeles & US & 232 & The University of Texas at Austin & US & 720 \\  
\textit{84} & Newcastle University & GB & 67 & Helmholtz Centre for Environmental Research & DE & 230 & KU Leuven & BE & 717 \\  
\textit{85} & Wageningen University \& Research & NL & 67 & Duke University & US & 230 & University of Li\`{e}ge & BE & 713 \\  
\textit{86} & Stockholm University & SE & 67 & University of Manchester & GB & 229 & University of Illinois Urbana-Champaign & US & 707 \\  
\textit{87} & Max Planck Institute for Chemistry & DE & 66 & Texas A\&M University & US & 229 & Texas A\&M University & US & 706 \\  
\textit{88} & Swedish University of Agricultural Sciences & SE & 66 & University of Hawaii System & US & 229 & S\~{a}o Paulo State University & BR & 700 \\  
\textit{89} & Max Planck Society & DE & 65 & McGill University & CA & 228 & University of Cape Town & ZA & 700 \\  
\textit{90} & University of Manchester & GB & 64 & Princeton University & US & 225 & Harvard University & US & 699 \\  
\hdashline          
\textit{91} & Lund University & SE & 64 & University of Illinois Urbana-Champaign & US & 224 & University of Porto & PT & 694 \\  
\textit{92} & Brown University & US & 64 & China University of Geosciences (Beijing) & CN & 223 & Massachusetts Institute of Technology & US & 689 \\  
\textit{93} & US Forest Service & US & 64 & University of Adelaide & AU & 222 & University of Bologna & IT & 686 \\  
\textit{94} & Lunar and Planetary Institute & US & 63 & Woods Hole Oceanographic Institution & US & 221 & University of Oslo & NO & 671 \\  
\textit{95} & Princeton University & US & 63 & Nanjing University & CN & 219 & Peking University & CN & 667 \\  
\textit{96} & Lomonosov Moscow State University & RU & 61 & University of Oslo & NO & 218 & Natural History Museum & GB & 667 \\  
\textit{97} & University of New Mexico & US & 61 & University of Tasmania & AU & 217 & University of Florence & IT & 665 \\  
\textit{98} & University of Sydney & AU & 60 & Vrije Universiteit Amsterdam & NL & 216 & University of Auckland & NZ & 663 \\  
\textit{99} & University of Tasmania & AU & 60 & University of Tennessee at Knoxville & US & 216 & The Ohio State University & US & 652 \\  
\textit{100} & ETH Zurich & CH & 60 & Sapienza University of Rome & IT & 215 & Norwegian University of Science and Technology & NO & 651 \\  
          
 \\[-1.4em]
\bottomrule
\end{tabular}}
}
\end{table}
\end{landscape}
}
%

\titleformat{\section}{\sc\centering\LARGE\bfseries}{\textsc{\thesection}.\!\!}{1em}{}

\afterpage{\clearpage%
\markboth{\textbf \textsc{\astro}}{}
\thispagestyle{empty}
\quad
\vspace{2cm}
\begin{center}
\pgfornament[width=0.5*\textwidth,symmetry=h]{89}\\[2em]
\section{\astro}
\vspace{1em}
\pgfornament[width=0.5*\textwidth]{89}
\end{center}
}

\afterpage{\clearpage%

\begin{figure}[!tp]
\centering
\vspace{-1em}
{\large \textbf{\textrm{{World Map of \textcolor{violet}{\textit{\astro}} Collaboration}}}~|~1971--2020}\\
\vspace{0.3cm}
\includegraphics[align=c, scale=0.054, trim={9.5cm 0 9.5cm 0},clip]{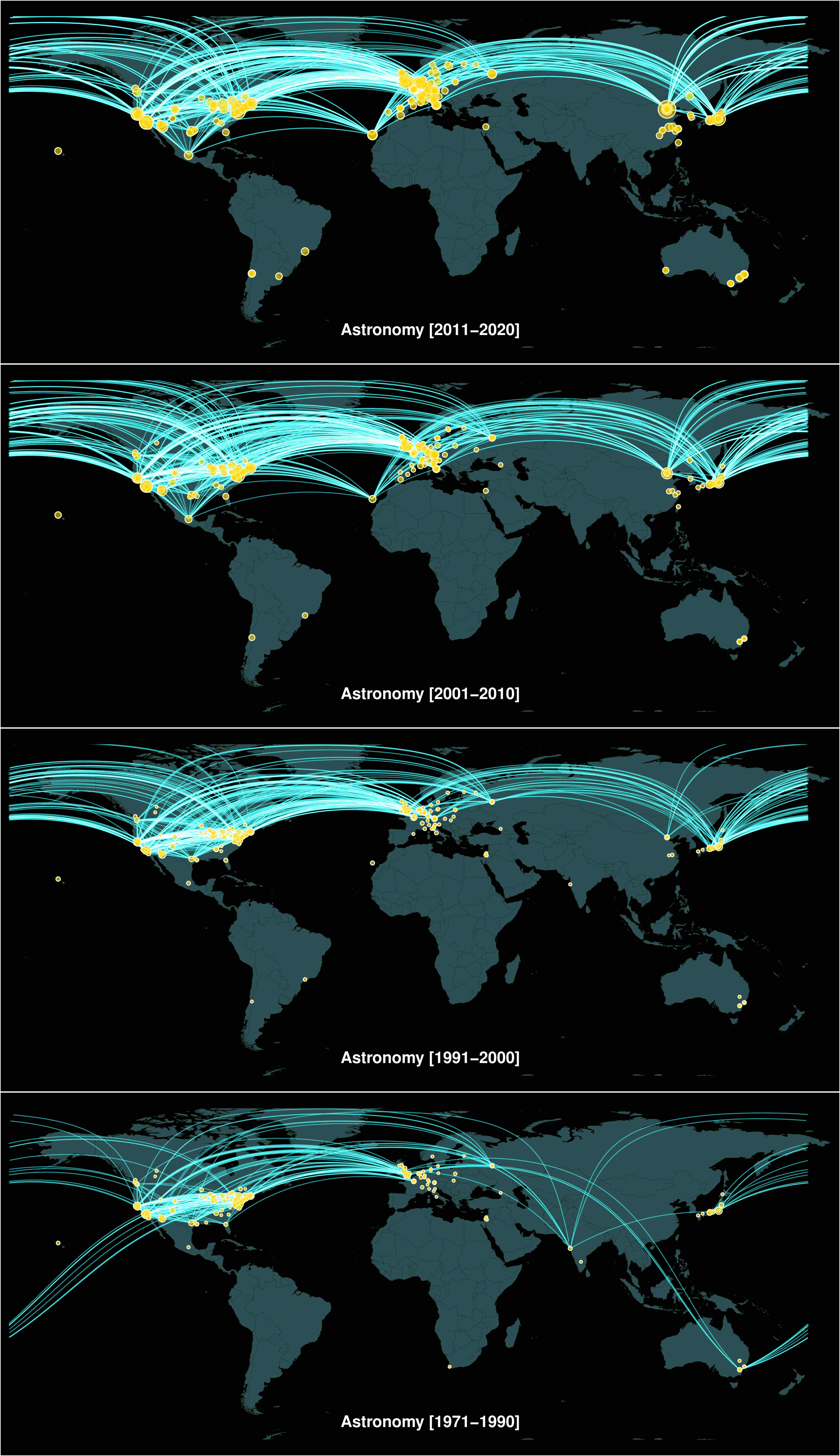}
\caption[{\astro}]{\textbf{(a)~|~The World Map of \textcolor{violet}{\textit{\astro}} Collaboration.}
The bubbles represent the top 199 institutions in terms of work production, with their sizes proportional to the work volume. 
The connecting lines depict coauthorship relationships among the top 50 institutions.}
\label{fig:wmap_astro}
\end{figure}
}
\afterpage{\clearpage%
\begin{figure}[!tp]\ContinuedFloat
\centering
\vspace{-1em}
{\large \textbf{Top 30 Productive Institutions on the World Map: \textcolor{violet}{\textit{\astro}}}~|~1991--2020}\\
\vspace{-0em}
\hspace*{-3em}                                                           
\includegraphics[align=c, scale=0.83]{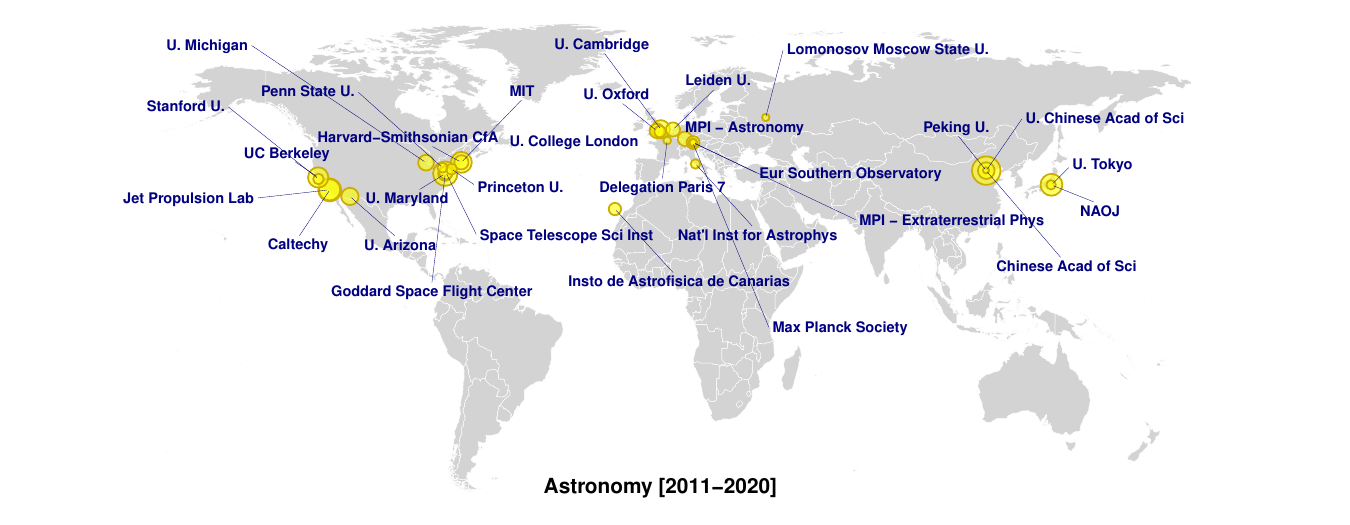}\\[-0.5em]
\quad\\[-1em]
\dotfill 
\quad\\[-0em]
\hspace*{-3em}                                                           
\includegraphics[align=c, scale=0.83]{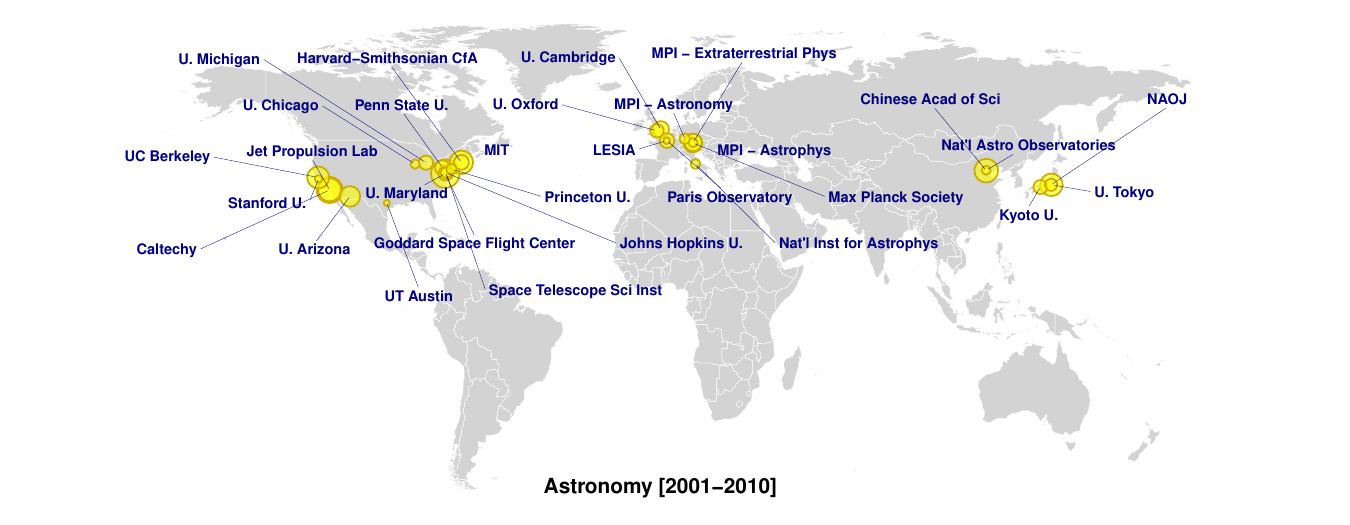}\\[-0.5em]
\quad\\[-1em]
\dotfill 
\quad\\[-0em]
\hspace*{-3em}                                                           
\includegraphics[align=c, scale=0.83]{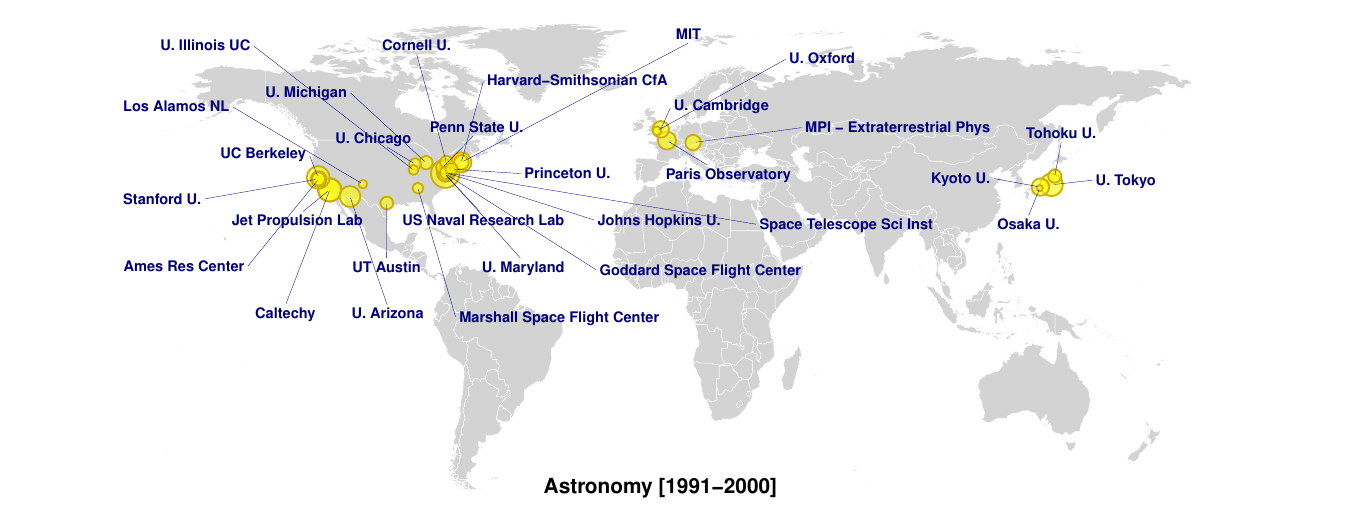}\\[-0.8em]
\caption[{\astro}]{\textbf{(b)~|~The top 30 productive institutions on the World Map: \textcolor{violet}{\textit{\astro}}.}
The bubbles represent the top 30 institutions in terms of work production, with their sizes proportional to the work volume.}
\label{fig:wmap_topinst_astro}
\end{figure}
}
\afterpage{\clearpage%
\begin{figure}[!tp]\ContinuedFloat
\centering
\vspace{-1em}
{\large \textbf{\textrm{{Interregional \textcolor{violet}{\textit{\astro}} Collaboration}}}~|~1991--2020}\\
\vspace{0.5em}
\hspace{-5em}\includegraphics[align=c, scale=1.7, vmargin=0mm]{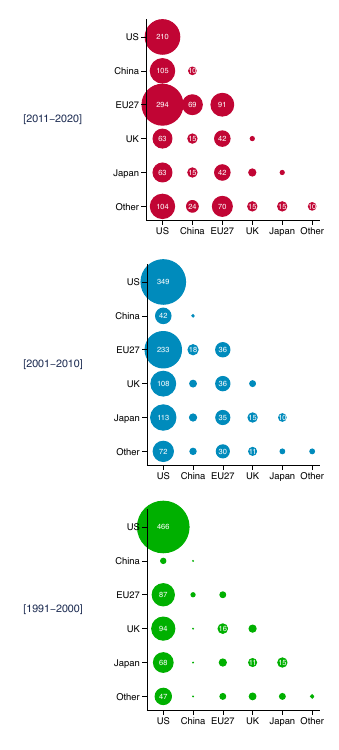}
\vspace{-1em}
\caption[]{\textbf{(c)~|~The Interregional \textcolor{violet}{\textit{\astro}} Collaboration Matrix Diagram.}
The bubble size represents the number of coauthorship relationships for the top 50 institutions in terms of work production. 
If the number is equal to or greater than 10, it is displayed inside the bubble.}
\label{fig:halfmat_astro}
\end{figure}
}
\afterpage{\clearpage%
\begin{figure}[!tp]\ContinuedFloat
\centering
\vspace{-1em}
{\large \textbf{\textrm{Interinstitutional \textcolor{violet}{\textit{\astro}} Collaboration}}~|~2001--2020\quad {\footnotesize \emph{(continued to next page)}}}\\
\cdend{astro}{2010}{2000}{2011--2020}{2001--2010}\\[-1.5em]
\caption[]{\textbf{(d)~|~The Interinstitutional \textcolor{violet}{\textit{\astro}} Collaboration Dendrogram.}
The top 50 institutions in terms of work production, indicated by the circularised bar graphs, are displayed.
}
\label{fig:cdend1_astro}
\end{figure}
}
\afterpage{\clearpage%
\begin{figure}[!tp]\ContinuedFloat
\centering
\vspace{-1em}
{\large \textbf{\textrm{Interinstitutional \textcolor{violet}{\textit{\astro}} Collaboration}}~|~1971--2000\quad {\footnotesize \emph{(continued from previous page)}}}\\
\cdend{astro}{1990}{1980}{1991--2000}{1971--1990}\\[-1.5em]
\caption[]{\textbf{(d)~|~The Interinstitutional \textcolor{violet}{\textit{\astro}} Collaboration Dendrogram.} \emph{(Cont.)}\hfill~}
\label{fig:cdend2_astro}
\end{figure}
}
\afterpage{\clearpage%
\begin{landscape}
\begin{table}[!t]
\vspace{-0.5em}
\caption{\textbf{The top 100 productive institutions: \textcolor{violet}{\textit{\astro}}.}}
\label{tab:r1_astro}
\vspace{2em}
\centering
{\tiny
{\renewcommand{\arraystretch}{1.2}
\begin{tabular}{rp{5cm}lr@{\hspace{4em}}p{5cm}lr@{\hspace{4em}}p{5cm}lr}\\[-5em] \toprule[1pt] \\[-1.4em]    
 & {\scriptsize \textbf{1991--2000}} & \multicolumn{2}{c}{No.~Works} & {\scriptsize \textbf{2001--2010}} & \multicolumn{2}{c}{No.~Works} & {\scriptsize \textbf{2011--2020}} & \multicolumn{2}{r}{No.~Works} \\[-0.2em] \cmidrule[0.5pt](lr{4em}){2-4} \cmidrule[0.5pt](l{-0em}r{4em}){5-7} \cmidrule[0.5pt](l{-0em}r{1em}){8-10}
\textit{1} & Goddard Space Flight Center & US & 4,783 & Goddard Space Flight Center & US & 8,442 & Chinese Academy of Sciences & CN & 14,968 \\  
\textit{2} & Jet Propulsion Laboratory & US & 3,162 & California Institute of Technology & US & 7,064 & Goddard Space Flight Center & US & 10,596 \\  
\textit{3} & California Institute of Technology & US & 2,942 & Chinese Academy of Sciences & CN & 5,800 & California Institute of Technology & US & 9,298 \\  
\textit{4} & University of California, Berkeley & US & 2,906 & Center for Astrophysics Harvard \& Smithsonian & US & 5,686 & The University of Tokyo & JP & 8,570 \\  
\textit{5} & The University of Tokyo & JP & 2,876 & Jet Propulsion Laboratory & US & 5,536 & Jet Propulsion Laboratory & US & 8,249 \\  
\textit{6} & University of Arizona & US & 2,458 & The University of Tokyo & JP & 5,253 & Center for Astrophysics Harvard \& Smithsonian & US & 7,948 \\  
\textit{7} & Center for Astrophysics Harvard \& Smithsonian & US & 2,318 & University of California, Berkeley & US & 5,002 & University of California, Berkeley & US & 7,567 \\  
\textit{8} & Space Telescope Science Institute & US & 1,998 & University of Arizona & US & 4,498 & University of Cambridge & GB & 6,539 \\  
\textit{9} & Paris Observatory & FR & 1,970 & Max Planck Society & DE & 4,051 & University of Arizona & US & 6,070 \\  
\textit{10} & Kyoto University & JP & 1,886 & Max Planck Institute for Extraterrestrial Physics & DE & 3,637 & University of Chinese Academy of Sciences & CN & 5,824 \\  
\hdashline          
\textit{11} & University of Cambridge & GB & 1,782 & University of Cambridge & GB & 3,428 & University of Michigan{\textendash}Ann Arbor & US & 5,490 \\  
\textit{12} & Ames Research Center & US & 1,683 & Space Telescope Science Institute & US & 3,221 & University of Oxford & GB & 5,120 \\  
\textit{13} & Max Planck Institute for Extraterrestrial Physics & DE & 1,636 & Pennsylvania State University & US & 3,174 & Max Planck Institute for Astronomy & DE & 5,049 \\  
\textit{14} & University of Maryland, College Park & US & 1,593 & Massachusetts Institute of Technology & US & 3,139 & Massachusetts Institute of Technology & US & 5,029 \\  
\textit{15} & Massachusetts Institute of Technology & US & 1,553 & Paris Observatory & FR & 3,134 & Leiden University & NL & 4,993 \\  
\textit{16} & University of Michigan{\textendash}Ann Arbor & US & 1,450 & University of Michigan{\textendash}Ann Arbor & US & 3,033 & University of Maryland, College Park & US & 4,916 \\  
\textit{17} & Tohoku University & JP & 1,440 & Kyoto University & JP & 2,998 & Space Telescope Science Institute & US & 4,774 \\  
\textit{18} & Stanford University & US & 1,436 & University of Oxford & GB & 2,955 & Max Planck Institute for Extraterrestrial Physics & DE & 4,707 \\  
\textit{19} & United States Naval Research Laboratory & US & 1,426 & University of Maryland, College Park & US & 2,888 & Instituto de Astrof\'{\i}sica de Canarias & ES & 4,572 \\  
\textit{20} & Pennsylvania State University & US & 1,414 & National Astronomical Observatory of Japan & JP & 2,780 & University College London & GB & 4,400 \\  
\hdashline          
\textit{21} & The University of Texas at Austin & US & 1,406 & Johns Hopkins University & US & 2,661 & Princeton University & US & 4,350 \\  
\textit{22} & Cornell University & US & 1,372 & Max Planck Institute for Astronomy & DE & 2,641 & Stanford University & US & 4,286 \\  
\textit{23} & Johns Hopkins University & US & 1,368 & Princeton University & US & 2,626 & Pennsylvania State University & US & 4,199 \\  
\textit{24} & Princeton University & US & 1,363 & National Institute for Astrophysics & IT & 2,572 & National Institute for Astrophysics & IT & 4,149 \\  
\textit{25} & University of Chicago & US & 1,338 & University of Chicago & US & 2,547 & Max Planck Society & DE & 4,138 \\  
\textit{26} & Marshall Space Flight Center & US & 1,301 & Max Planck Institute for Astrophysics & DE & 2,533 & National Astronomical Observatory of Japan & JP & 4,131 \\  
\textit{27} & University of Illinois Urbana-Champaign & US & 1,259 & National Astronomical Observatories & CN & 2,523 & D\'{e}l\'{e}gation Paris 7 & FR & 4,019 \\  
\textit{28} & University of Oxford & GB & 1,226 & Stanford University & US & 2,478 & Lomonosov Moscow State University & RU & 4,005 \\  
\textit{29} & Los Alamos National Laboratory & US & 1,218 & The University of Texas at Austin & US & 2,457 & European Southern Observatory & DE & 3,994 \\  
\textit{30} & Osaka University & JP & 1,193 & University of California, Los Angeles & US & 2,447 & Peking University & CN & 3,983 \\  
\hdashline          
\textit{31} & Max Planck Society & DE & 1,191 & \scalebox{0.9}[1]{Laboratory of Space Studies and Instrumentation in Astrophysics} & FR & 2,430 & Kyoto University & JP & 3,979 \\  
\textit{32} & Chinese Academy of Sciences & CN & 1,181 & Tohoku University & JP & 2,409 & University of Toronto & CA & 3,912 \\  
\textit{33} & Nagoya University & JP & 1,160 & Instituto de Astrof\'{\i}sica de Canarias & ES & 2,402 & National Astronomical Observatories & CN & 3,881 \\  
\textit{34} & University of Colorado Boulder & US & 1,158 & Universidad Nacional Aut\'{o}noma de M\'{e}xico & MX & 2,394 & The University of Texas at Austin & US & 3,879 \\  
\textit{35} & Institute of Astronomy & RU & 1,155 & European Southern Observatory & DE & 2,384 & Johns Hopkins University & US & 3,846 \\  
\textit{36} & University of Washington & US & 1,150 & Cornell University & US & 2,234 & University of La Laguna & ES & 3,705 \\  
\textit{37} & Lawrence Livermore National Laboratory & US & 1,131 & University College London & GB & 2,209 & University of California, Santa Cruz & US & 3,654 \\  
\textit{38} & University of Wisconsin{\textendash}Madison & US & 1,121 & University of Toronto & CA & 2,193 & University of California, Los Angeles & US & 3,627 \\  
\textit{39} & The Ohio State University & US & 1,098 & University of California, Santa Cruz & US & 2,179 & Ames Research Center & US & 3,616 \\  
\textit{40} & University of California, Los Angeles & US & 1,055 & Los Alamos National Laboratory & US & 2,172 & University of Colorado Boulder & US & 3,568 \\  
\hdashline          
\textit{41} & Columbia University & US & 1,030 & Imperial College London & GB & 2,159 & University of Geneva & CH & 3,539 \\  
\textit{42} & University of Toronto & CA & 1,024 & University of Wisconsin{\textendash}Madison & US & 2,154 & Max Planck Institute for Radio Astronomy & DE & 3,486 \\  
\textit{43} & National Astronomical Observatory of Japan & JP & 1,014 & The Ohio State University & US & 2,140 & ETH Zurich & CH & 3,453 \\  
\textit{44} & University of Manchester & GB & 1,010 & University of Washington & US & 2,120 & Universidad Nacional Aut\'{o}noma de M\'{e}xico & MX & 3,449 \\  
\textit{45} & University of California, San Diego & US & 1,005 & University of Colorado Boulder & US & 2,100 & Tsinghua University & CN & 3,422 \\  
\textit{46} & Max Planck Institute for Astrophysics & DE & 967 & United States Naval Research Laboratory & US & 2,096 & Heidelberg University & DE & 3,418 \\  
\textit{47} & Lawrence Berkeley National Laboratory & US & 952 & Ames Research Center & US & 2,095 & The Ohio State University & US & 3,378 \\  
\textit{48} & University College London & GB & 950 & Lomonosov Moscow State University & RU & 2,087 & University of Washington & US & 3,353 \\  
\textit{49} & University of California, Santa Cruz & US & 940 & Russian Academy of Sciences & RU & 2,086 & Institut d'Astrophysique de Paris & FR & 3,351 \\  
\textit{50} & Imperial College London & GB & 928 & Columbia University & US & 2,056 & Russian Academy of Sciences & RU & 3,345 \\  
          
 \\[-1.4em]
\hdashline \\[-1em]
\multicolumn{10}{r}{\scriptsize \emph{(continued to next page)}}
\end{tabular}}
}
\end{table}
\end{landscape}
}
\afterpage{\clearpage%
\begin{landscape}
\begin{table}[!t]\ContinuedFloat
\vspace{-3.3em}
\caption{\textbf{The top 100 productive institutions: \textcolor{violet}{\textit{\astro}}.} \emph{(Cont.)}}
\label{tab:r2_astro}
\vspace{2em}
{\tiny
{\renewcommand{\arraystretch}{1.2}
\begin{tabular}{rp{5cm}lr@{\hspace{4em}}p{5cm}lr@{\hspace{4em}}p{5cm}lr}\\[-5em] \toprule[1pt] \\[-1.4em]    
 & {\scriptsize \textbf{1991--2000}} & \multicolumn{2}{c}{No.\ Works} & {\scriptsize \textbf{2001--2010}} & \multicolumn{2}{c}{No.\ Works} & {\scriptsize \textbf{2011--2020}} & \multicolumn{2}{r}{No.\ Works} \\[-0.2em] \cmidrule[0.5pt](lr{4em}){2-4} \cmidrule[0.5pt](l{-0em}r{4em}){5-7} \cmidrule[0.5pt](l{-0em}r{1em}){8-10}
\textit{51} & Lomonosov Moscow State University & RU & 918 & Osaka University & JP & 2,014 & Nagoya University & JP & 3,338 \\  
\textit{52} & National Radio Astronomy Observatory & US & 912 & University of Hawaii System & US & 1,963 & Australian National University & AU & 3,290 \\  
\textit{53} & Max Planck Institute for Astronomy & DE & 886 & Nagoya University & JP & 1,959 & University of Chicago & US & 3,263 \\  
\textit{54} & University of Hawaii System & US & 883 & Institut d'Astrophysique de Paris & FR & 1,956 & Osservatorio Astronomico di Padova & IT & 3,261 \\  
\textit{55} & European Space Research and Technology Centre & NL & 882 & Max Planck Institute for Radio Astronomy & DE & 1,868 & University of Science and Technology of China & CN & 3,232 \\  
\textit{56} & Institute of Space and Astronautical Science (ISAS) & JP & 865 & University of Manchester & GB & 1,866 & Columbia University & US & 3,208 \\  
\textit{57} & Durham University & GB & 846 & Osservatorio Astronomico di Padova & IT & 1,854 & Durham University & GB & 3,151 \\  
\textit{58} & Fermilab & US & 821 & Leiden University & NL & 1,843 & University of Edinburgh & GB & 3,112 \\  
\textit{59} & University of Florida & US & 804 & University of Illinois Urbana-Champaign & US & 1,836 & University of Manchester & GB & 3,103 \\  
\textit{60} & Tokyo Institute of Technology & JP & 786 & National Radio Astronomy Observatory & US & 1,835 & Imperial College London & GB & 3,044 \\  
\hdashline          
\textit{61} & European Southern Observatory & DE & 777 & Institute of Astronomy & RU & 1,818 & Aix-Marseille University & FR & 3,039 \\  
\textit{62} & Instituto de Astrof\'{\i}sica de Canarias & ES & 773 & University of Leicester & GB & 1,791 & University of Sydney & AU & 3,009 \\  
\textit{63} & University of California, Santa Barbara & US & 773 & European Space Research and Technology Centre & NL & 1,784 & Lawrence Berkeley National Laboratory & US & 2,972 \\  
\textit{64} & University of Amsterdam & NL & 771 & University of Padua & IT & 1,761 & University of Amsterdam & NL & 2,956 \\  
\textit{65} & Tel Aviv University & IL & 759 & Lawrence Berkeley National Laboratory & US & 1,760 & Planetary Science Institute & US & 2,946 \\  
\textit{66} & Australian National University & AU & 757 & European Southern Observatory & CL & 1,739 & University of Wisconsin{\textendash}Madison & US & 2,926 \\  
\textit{67} & University of Leicester & GB & 756 & University of Florida & US & 1,681 & Sapienza University of Rome & IT & 2,853 \\  
\textit{68} & University of Paris-Sud & FR & 749 & Durham University & GB & 1,679 & University of Bologna & IT & 2,826 \\  
\textit{69} & University of Alabama in Huntsville & US & 748 & Tsinghua University & CN & 1,656 & Cornell University & US & 2,823 \\  
\textit{70} & National Institute of Standards and Technology & US & 738 & Tokyo Institute of Technology & JP & 1,646 & Max Planck Institute for Astrophysics & DE & 2,798 \\  
\hdashline          
\textit{71} & Yale University & US & 735 & Lawrence Livermore National Laboratory & US & 1,634 & University of Copenhagen & DK & 2,794 \\  
\textit{72} & University of Virginia & US & 734 & Laboratoire d'Astrophysique de Marseille & FR & 1,630 & Tohoku University & JP & 2,786 \\  
\textit{73} & European Organization for Nuclear Research & CH & 730 & University of Bologna & IT & 1,590 & Harbin Institute of Technology & CN & 2,756 \\  
\textit{74} & University of Sydney & AU & 729 & University of California, San Diego & US & 1,574 & University of Li\`{e}ge & BE & 2,754 \\  
\textit{75} & Polish Academy of Sciences & PL & 722 & Arcetri Astrophysical Observatory & IT & 1,573 & University of Padua & IT & 2,747 \\  
\textit{76} & Russian Academy of Sciences & RU & 716 & University of Virginia & US & 1,540 & University of Southampton & GB & 2,720 \\  
\textit{77} & Max Planck Institute for Radio Astronomy & DE & 708 & University of Amsterdam & NL & 1,512 & Stockholm University & SE & 2,716 \\  
\textit{78} & Institut d'Astrophysique de Paris & FR & 706 & Institute of Space and Astronautical Science (ISAS) & JP & 1,505 & Yale University & US & 2,694 \\  
\textit{79} & University of Minnesota & US & 703 & Marshall Space Flight Center & US & 1,503 & Instituto de Astrof\'{\i}sica de Andaluc\'{\i}a & ES & 2,683 \\  
\textit{80} & Universidad Nacional Aut\'{o}noma de M\'{e}xico & MX & 698 & CEA Saclay & FR & 1,492 & Universit\'{e} de Toulouse & FR & 2,676 \\  
\hdashline          
\textit{81} & Rutherford Appleton Laboratory & GB & 695 & ETH Zurich & CH & 1,485 & Laboratoire d'Astrophysique de Marseille & FR & 2,667 \\  
\textit{82} & Langley Research Center & US & 693 & Australian National University & AU & 1,483 & Beihang University & CN & 2,659 \\  
\textit{83} & Sapienza University of Rome & IT & 684 & Fermilab & US & 1,476 & Harvard University & US & 2,651 \\  
\textit{84} & Universities Space Research Association & US & 682 & University of Sydney & AU & 1,474 & Universidade de S\~{a}o Paulo & BR & 2,643 \\  
\textit{85} & Arizona State University & US & 679 & University of Southampton & GB & 1,450 & Nanjing University & CN & 2,641 \\  
\textit{86} & Boston University & US & 678 & Tokyo University of Science & JP & 1,450 & National Radio Astronomy Observatory & US & 2,639 \\  
\textit{87} & RIKEN & JP & 676 & Universidade de S\~{a}o Paulo & BR & 1,436 & University of California, San Diego & US & 2,624 \\  
\textit{88} & University of Padua & IT & 663 & Osservatorio astronomico di Bologna & IT & 1,432 & European Southern Observatory & CL & 2,602 \\  
\textit{89} & University of Li\`{e}ge & BE & 655 & University of Li\`{e}ge & BE & 1,421 & \scalebox{0.9}[1]{Laboratory of Space Studies and Instrumentation in Astrophysics} & FR & 2,597 \\  
\textit{90} & University of Bonn & DE & 653 & Leibniz Institute for Astrophysics Potsdam & DE & 1,416 & German Aerospace Center & DE & 2,568 \\  
\hdashline          
\textit{91} & Johnson Space Center & US & 648 & Instituto de Astrof\'{\i}sica de Andaluc\'{\i}a & ES & 1,416 & \scalebox{0.87}[1]{Kavli Institute for the Physics and Mathematics of the Universe (IPMU)} & JP & 2,563 \\  
\textit{92} & University of Pennsylvania & US & 647 & University of California, Santa Barbara & US & 1,406 & University of Illinois Urbana-Champaign & US & 2,561 \\  
\textit{93} & Queen Mary University of London & GB & 644 & Yale University & US & 1,405 & Kavli Institute for Particle Astrophysics and Cosmology & US & 2,499 \\  
\textit{94} & National Astronomical Observatories & CN & 642 & Sapienza University of Rome & IT & 1,379 & Max Planck Institute for Solar System Research & DE & 2,483 \\  
\textit{95} & Tokyo University of Science & JP & 639 & Japan Aerospace Exploration Agency (JAXA) & JP & 1,373 & KU Leuven & BE & 2,462 \\  
\textit{96} & Leiden University & NL & 627 & University of Paris-Sud & FR & 1,363 & Arizona State University & US & 2,448 \\  
\textit{97} & University of British Columbia & CA & 616 & Brera Astronomical Observatory & IT & 1,348 & Arcetri Astrophysical Observatory & IT & 2,432 \\  
\textit{98} & Kitt Peak National Observatory & US & 597 & University of British Columbia & CA & 1,345 & University of Paris-Saclay & FR & 2,423 \\  
\textit{99} & Universidade de S\~{a}o Paulo & BR & 588 & University of Edinburgh & GB & 1,344 & Paris Observatory & FR & 2,411 \\  
\textit{100} & University of Sussex & GB & 586 & University of Bonn & DE & 1,335 & Leibniz Institute for Astrophysics Potsdam & DE & 2,367 \\  
          
 \\[-1.4em]
\bottomrule
\end{tabular}}
}
\end{table}
\end{landscape}
}
%

\titleformat{\section}{\sc\centering\LARGE\bfseries}{\textsc{\thesection}.\!\!}{1em}{}

\afterpage{\clearpage%
\markboth{\textbf \textsc{\math}}{}
\thispagestyle{empty}
\quad
\vspace{2cm}
\begin{center}
\pgfornament[width=0.5*\textwidth,symmetry=h]{89}\\[2em]
\section{\math}
\vspace{1em}
\pgfornament[width=0.5*\textwidth]{89}
\end{center}
}

\afterpage{\clearpage%

\begin{figure}[!tp]
\centering
\vspace{-1em}
{\large \textbf{\textrm{{World Map of \textcolor{violet}{\textit{\math}} Collaboration}}}~|~1971--2020}\\
\vspace{0.3cm}
\includegraphics[align=c, scale=0.054, trim={9.5cm 0 9.5cm 0},clip]{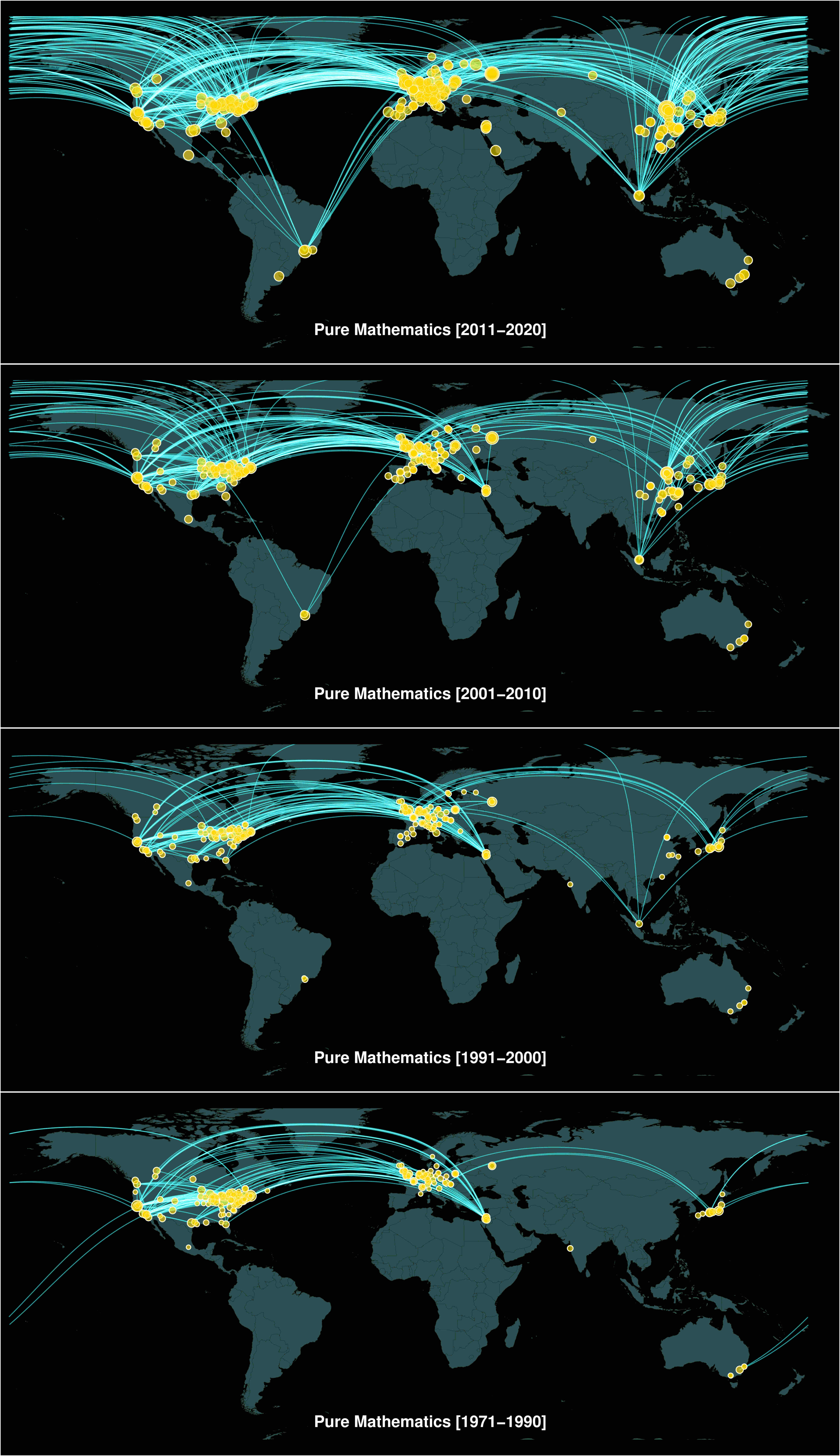}
\caption[{\math}]{\textbf{(a)~|~The World Map of \textcolor{violet}{\textit{\math}} Collaboration.}
The bubbles represent the top 199 institutions in terms of work production, with their sizes proportional to the work volume. 
The connecting lines depict coauthorship relationships among the top 50 institutions.}
\label{fig:wmap_math}
\end{figure}
}
\afterpage{\clearpage%
\begin{figure}[!tp]\ContinuedFloat
\centering
\vspace{-1em}
{\large \textbf{Top 30 Productive Institutions on the World Map: \textcolor{violet}{\textit{\math}}}~|~1991--2020}\\
\vspace{-0em}
\hspace*{-3em}                                                           
\includegraphics[align=c, scale=0.83]{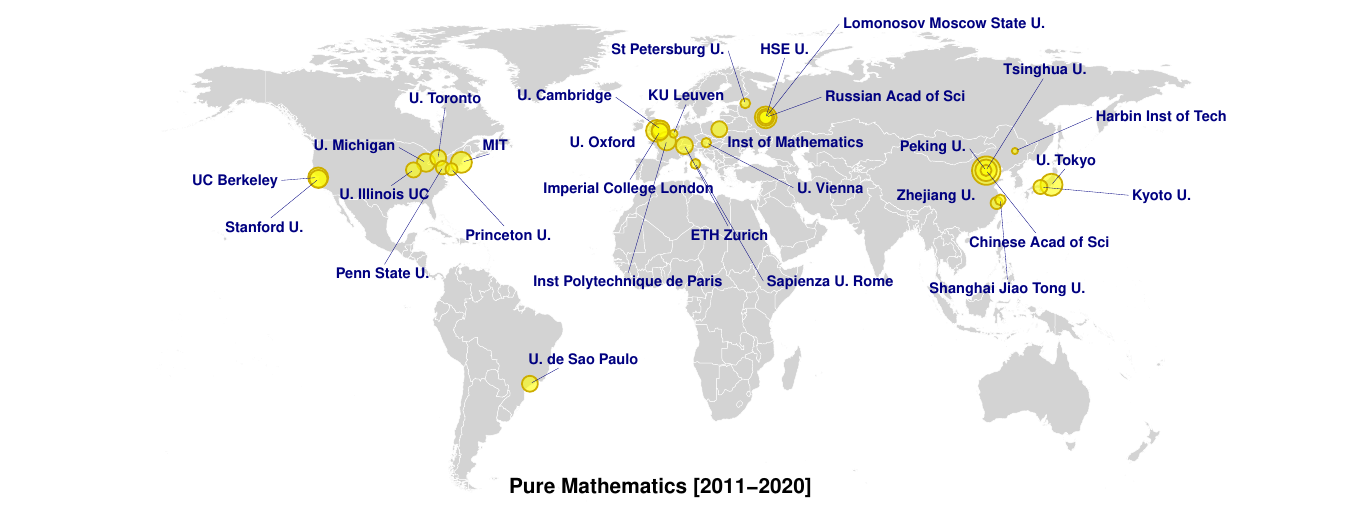}\\[-0.5em]
\quad\\[-1em]
\dotfill 
\quad\\[-0em]
\hspace*{-3em}                                                           
\includegraphics[align=c, scale=0.83]{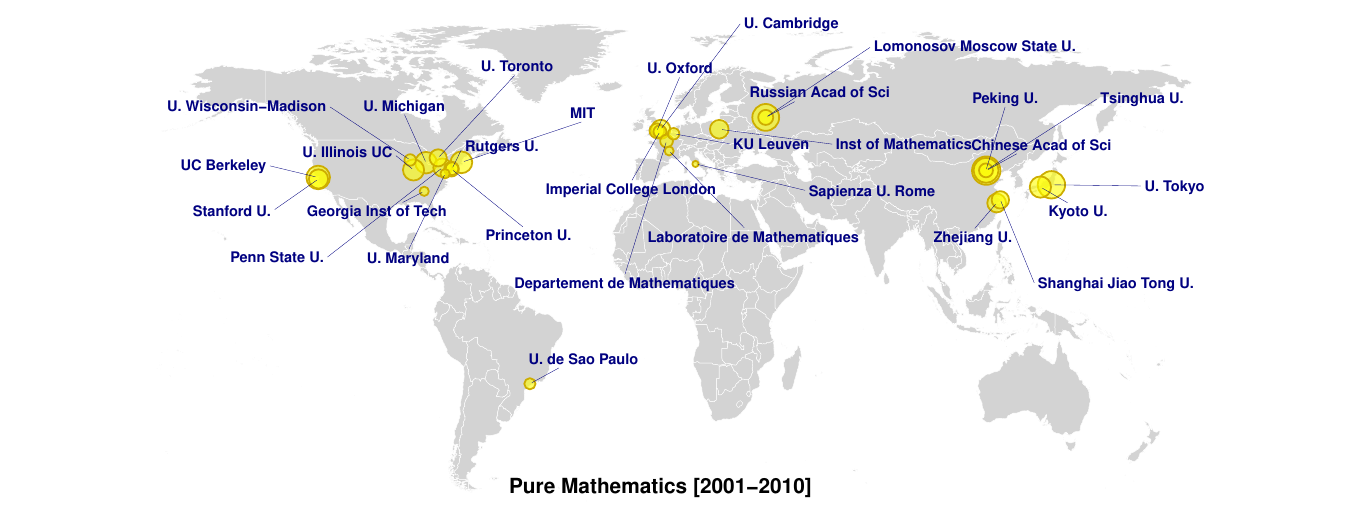}\\[-0.5em]
\quad\\[-1em]
\dotfill 
\quad\\[-0em]
\hspace*{-3em}                                                           
\includegraphics[align=c, scale=0.83]{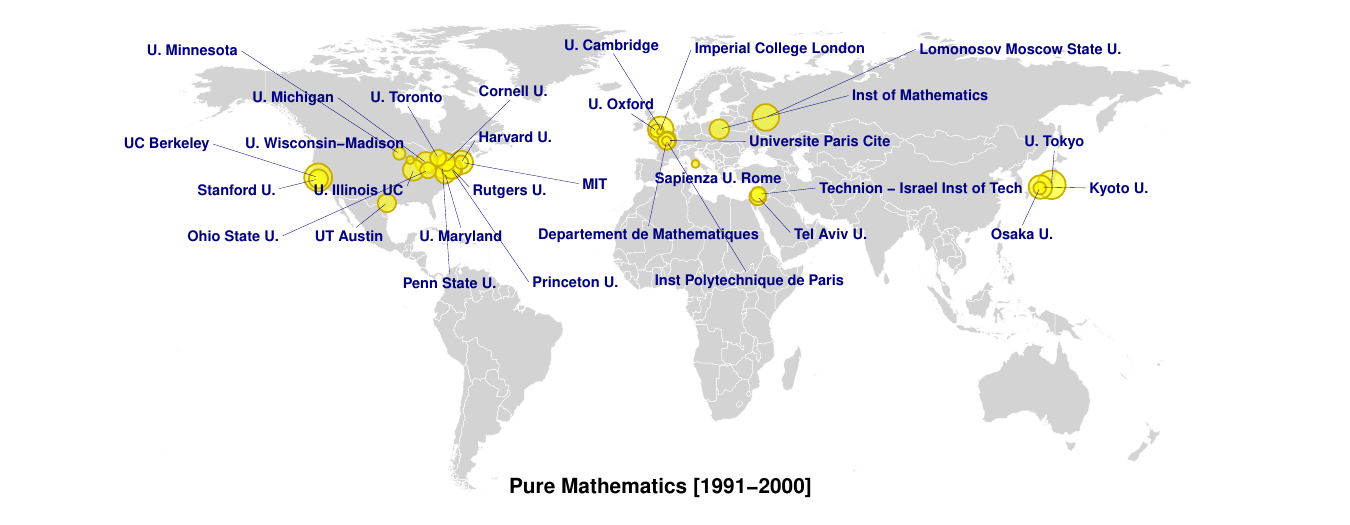}\\[-0.8em]
\caption[{\math}]{\textbf{(b)~|~The top 30 productive institutions on the World Map: \textcolor{violet}{\textit{\math}}.}
The bubbles represent the top 30 institutions in terms of work production, with their sizes proportional to the work volume.}
\label{fig:wmap_topinst_math}
\end{figure}
}
\afterpage{\clearpage%
\begin{figure}[!tp]\ContinuedFloat
\centering
\vspace{-1em}
{\large \textbf{\textrm{{Interregional \textcolor{violet}{\textit{\math}} Collaboration}}}~|~1991--2020}\\
\vspace{0.5em}
\hspace{-5em}\includegraphics[align=c, scale=1.7, vmargin=0mm]{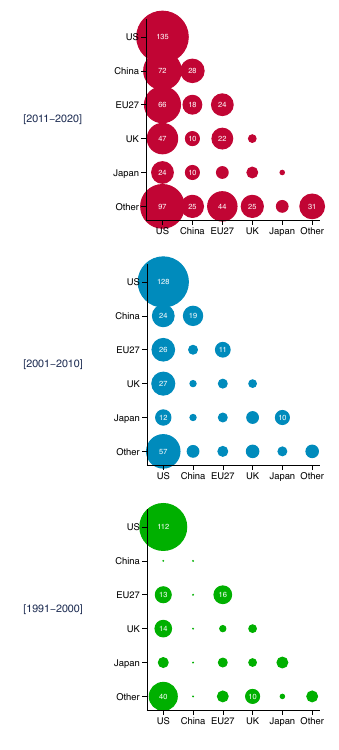}
\vspace{-1em}
\caption[]{\textbf{(c)~|~The Interregional \textcolor{violet}{\textit{\math}} Collaboration Matrix Diagram.}
The bubble size represents the number of coauthorship relationships for the top 50 institutions in terms of work production. 
If the number is equal to or greater than 10, it is displayed inside the bubble.}
\label{fig:halfmat_math}
\end{figure}
}
\afterpage{\clearpage%
\begin{figure}[!tp]\ContinuedFloat
\centering
\vspace{-1em}
{\large \textbf{\textrm{Interinstitutional \textcolor{violet}{\textit{\math}} Collaboration}}~|~2001--2020\quad {\footnotesize \emph{(continued to next page)}}}\\
\cdend{math}{2010}{2000}{2011--2020}{2001--2010}\\[-1.5em]
\caption[]{\textbf{(d)~|~The Interinstitutional \textcolor{violet}{\textit{\math}} Collaboration Dendrogram.}
The top 50 institutions in terms of work production, indicated by the circularised bar graphs, are displayed.
}
\label{fig:cdend1_math}
\end{figure}
}
\afterpage{\clearpage%
\begin{figure}[!tp]\ContinuedFloat
\centering
\vspace{-1em}
{\large \textbf{\textrm{Interinstitutional \textcolor{violet}{\textit{\math}} Collaboration}}~|~1971--2000\quad {\footnotesize \emph{(continued from previous page)}}}\\
\cdend{math}{1990}{1980}{1991--2000}{1971--1990}\\[-1.5em]
\caption[]{\textbf{(d)~|~The Interinstitutional \textcolor{violet}{\textit{\math}} Collaboration Dendrogram.} \emph{(Cont.)}\hfill~}
\label{fig:cdend2_math}
\end{figure}
}
\afterpage{\clearpage%
\begin{landscape}
\begin{table}[!t]
\vspace{-0.5em}
\caption{\textbf{The top 100 productive institutions: \textcolor{violet}{\textit{\math}}.}}
\label{tab:r1_math}
\vspace{2em}
\centering
{\tiny
{\renewcommand{\arraystretch}{1.2}
\begin{tabular}{rp{5cm}lr@{\hspace{4em}}p{5cm}lr@{\hspace{4em}}p{5cm}lr}\\[-5em] \toprule[1pt] \\[-1.4em]    
 & {\scriptsize \textbf{1991--2000}} & \multicolumn{2}{c}{No.~Works} & {\scriptsize \textbf{2001--2010}} & \multicolumn{2}{c}{No.~Works} & {\scriptsize \textbf{2011--2020}} & \multicolumn{2}{r}{No.~Works} \\[-0.2em] \cmidrule[0.5pt](lr{4em}){2-4} \cmidrule[0.5pt](l{-0em}r{4em}){5-7} \cmidrule[0.5pt](l{-0em}r{1em}){8-10}
\textit{1} & The University of Tokyo & JP & 1,604 & Chinese Academy of Sciences & CN & 2,881 & Chinese Academy of Sciences & CN & 5,153 \\  
\textit{2} & University of California, Berkeley & US & 1,557 & The University of Tokyo & JP & 2,716 & The University of Tokyo & JP & 3,541 \\  
\textit{3} & Lomonosov Moscow State University & RU & 1,458 & Lomonosov Moscow State University & RU & 2,642 & Lomonosov Moscow State University & RU & 3,490 \\  
\textit{4} & University of Cambridge & GB & 1,399 & Tsinghua University & CN & 2,401 & University of Oxford & GB & 3,476 \\  
\textit{5} & Kyoto University & JP & 1,262 & University of California, Berkeley & US & 2,243 & Tsinghua University & CN & 3,440 \\  
\textit{6} & Massachusetts Institute of Technology & US & 1,248 & Massachusetts Institute of Technology & US & 2,051 & Massachusetts Institute of Technology & US & 3,333 \\  
\textit{7} & Pennsylvania State University & US & 1,171 & University of Michigan{\textendash}Ann Arbor & US & 2,038 & University of California, Berkeley & US & 3,160 \\  
\textit{8} & University of Michigan{\textendash}Ann Arbor & US & 1,144 & Kyoto University & JP & 2,012 & University of Michigan{\textendash}Ann Arbor & US & 2,914 \\  
\textit{9} & University of Illinois Urbana-Champaign & US & 1,141 & University of Illinois Urbana-Champaign & US & 2,000 & Stanford University & US & 2,883 \\  
\textit{10} & Rutgers, The State University of New Jersey & US & 1,108 & Stanford University & US & 1,844 & University of Cambridge & GB & 2,819 \\  
\hdashline          
\textit{11} & Institute of Mathematics & PL & 1,071 & University of Cambridge & GB & 1,830 & ETH Zurich & CH & 2,777 \\  
\textit{12} & University of Maryland, College Park & US & 1,055 & Zhejiang University & CN & 1,813 & Russian Academy of Sciences & RU & 2,744 \\  
\textit{13} & Princeton University & US & 1,045 & Institute of Mathematics & PL & 1,803 & Imperial College London & GB & 2,733 \\  
\textit{14} & Stanford University & US & 1,043 & Pennsylvania State University & US & 1,785 & University of Toronto & CA & 2,655 \\  
\textit{15} & The University of Texas at Austin & US & 1,012 & Shanghai Jiao Tong University & CN & 1,724 & Universidade de S\~{a}o Paulo & BR & 2,642 \\  
\textit{16} & Cornell University & US & 974 & University of Toronto & CA & 1,704 & Institute of Mathematics & PL & 2,640 \\  
\textit{17} & D\'{e}partement de Math\'{e}matiques & FR & 968 & Russian Academy of Sciences & RU & 1,597 & University of Illinois Urbana-Champaign & US & 2,568 \\  
\textit{18} & Tel Aviv University & IL & 965 & University of Oxford & GB & 1,580 & Kyoto University & JP & 2,468 \\  
\textit{19} & University of Toronto & CA & 942 & Princeton University & US & 1,580 & Pennsylvania State University & US & 2,416 \\  
\textit{20} & The Ohio State University & US & 936 & Peking University & CN & 1,544 & National Research University Higher School of Economics & RU & 2,415 \\  
\hdashline          
\textit{21} & Technion {\textendash} Israel Institute of Technology & IL & 912 & Imperial College London & GB & 1,505 & Zhejiang University & CN & 2,339 \\  
\textit{22} & Universit\'{e} Paris Cit\'{e} & FR & 911 & Rutgers, The State University of New Jersey & US & 1,503 & Princeton University & US & 2,330 \\  
\textit{23} & Harvard University & US & 873 & D\'{e}partement de Math\'{e}matiques & FR & 1,494 & Peking University & CN & 2,290 \\  
\textit{24} & University of Oxford & GB & 870 & KU Leuven & BE & 1,464 & Shanghai Jiao Tong University & CN & 2,266 \\  
\textit{25} & Osaka University & JP & 857 & University of Wisconsin{\textendash}Madison & US & 1,463 & St Petersburg University & RU & 2,236 \\  
\textit{26} & University of Minnesota & US & 849 & Universidade de S\~{a}o Paulo & BR & 1,459 & Sapienza University of Rome & IT & 2,231 \\  
\textit{27} & Sapienza University of Rome & IT & 804 & University of Maryland, College Park & US & 1,420 & University of Vienna & AT & 2,221 \\  
\textit{28} & University of Wisconsin{\textendash}Madison & US & 804 & Georgia Institute of Technology & US & 1,416 & KU Leuven & BE & 2,183 \\  
\textit{29} & Imperial College London & GB & 799 & Laboratoire de Math\'{e}matiques & FR & 1,410 & Harbin Institute of Technology & CN & 2,167 \\  
\textit{30} & KU Leuven & BE & 798 & Sapienza University of Rome & IT & 1,395 & The University of Texas at Austin & US & 2,113 \\  
\hdashline          
\textit{31} & Hebrew University of Jerusalem & IL & 798 & The University of Texas at Austin & US & 1,392 & Texas A\&M University & US & 2,083 \\  
\textit{32} & University of California, Los Angeles & US & 764 & \scalebox{0.9}[1]{French Institute for Research in Computer Science and Automation} & FR & 1,383 & Laboratoire de Math\'{e}matiques & FR & 2,082 \\  
\textit{33} & Czech Academy of Sciences, Institute of Mathematics & CZ & 755 & Tohoku University & JP & 1,382 & University of Science and Technology of China & CN & 2,029 \\  
\textit{34} & University of Chicago & US & 754 & Osaka University & JP & 1,364 & Columbia University & US & 2,029 \\  
\textit{35} & University of Amsterdam & NL & 751 & Texas A\&M University & US & 1,335 & Harvard University & US & 2,001 \\  
\textit{36} & European Organization for Nuclear Research & CH & 748 & National University of Singapore & SG & 1,322 & University of Waterloo & CA & 1,958 \\  
\textit{37} & University of California, Santa Barbara & US & 744 & Harbin Institute of Technology & CN & 1,311 & Georgia Institute of Technology & US & 1,941 \\  
\textit{38} & Texas A\&M University & US & 741 & Wuhan University & CN & 1,311 & Rutgers, The State University of New Jersey & US & 1,930 \\  
\textit{39} & National University of Singapore & SG & 726 & The Ohio State University & US & 1,311 & University of British Columbia & CA & 1,903 \\  
\textit{40} & University of Paris-Sud & FR & 709 & University of Waterloo & CA & 1,279 & Technical University of Munich & DE & 1,869 \\  
\hdashline          
\textit{41} & Purdue University West Lafayette & US & 699 & Tokyo Institute of Technology & JP & 1,274 & University of Washington & US & 1,863 \\  
\textit{42} & University of California, San Diego & US & 691 & University of California, Los Angeles & US & 1,265 & Max Planck Institute for Mathematics & DE & 1,861 \\  
\textit{43} & California Institute of Technology & US & 690 & Max Planck Institute for Mathematics & DE & 1,255 & University of Wisconsin{\textendash}Madison & US & 1,861 \\  
\textit{44} & University of Warsaw & PL & 689 & University of Washington & US & 1,213 & The Ohio State University & US & 1,860 \\  
\textit{45} & University of Waterloo & CA & 685 & Tel Aviv University & IL & 1,208 & Politecnico di Milano & IT & 1,844 \\  
\textit{46} & Yale University & US & 683 & Technion {\textendash} Israel Institute of Technology & IL & 1,202 & National University of Singapore & SG & 1,843 \\  
\textit{47} & Russian Academy of Sciences & RU & 681 & ETH Zurich & CH & 1,201 & Beihang University & CN & 1,841 \\  
\textit{48} & Tohoku University & JP & 675 & Beijing Normal University & CN & 1,195 & University of California, San Diego & US & 1,826 \\  
\textit{49} & Polish Academy of Sciences & PL & 665 & University of Minnesota & US & 1,193 & University of Bologna & IT & 1,817 \\  
\textit{50} & University of Florida & US & 659 & Cornell University & US & 1,191 & University of Padua & IT & 1,815 \\  
          
 \\[-1.4em]
\hdashline \\[-1em]
\multicolumn{10}{r}{\scriptsize \emph{(continued to next page)}}
\end{tabular}}
}
\end{table}
\end{landscape}
}
\afterpage{\clearpage%
\begin{landscape}
\begin{table}[!t]\ContinuedFloat
\vspace{-3.3em}
\caption{\textbf{The top 100 productive institutions: \textcolor{violet}{\textit{\math}}.} \emph{(Cont.)}}
\label{tab:r2_math}
\vspace{2em}
{\tiny
{\renewcommand{\arraystretch}{1.2}
\begin{tabular}{rp{5cm}lr@{\hspace{4em}}p{5cm}lr@{\hspace{4em}}p{5cm}lr}\\[-5em] \toprule[1pt] \\[-1.4em]    
 & {\scriptsize \textbf{1991--2000}} & \multicolumn{2}{c}{No.\ Works} & {\scriptsize \textbf{2001--2010}} & \multicolumn{2}{c}{No.\ Works} & {\scriptsize \textbf{2011--2020}} & \multicolumn{2}{r}{No.\ Works} \\[-0.2em] \cmidrule[0.5pt](lr{4em}){2-4} \cmidrule[0.5pt](l{-0em}r{4em}){5-7} \cmidrule[0.5pt](l{-0em}r{1em}){8-10}
\textit{51} & Stony Brook University & US & 656 & Steklov Mathematical Institute & RU & 1,183 & Institut de Math\'{e}matiques de Jussieu & FR & 1,795 \\  
\textit{52} & McGill University & CA & 641 & University of Science and Technology of China & CN & 1,179 & Steklov Mathematical Institute & RU & 1,785 \\  
\textit{53} & University of Washington & US & 636 & Fudan University & CN & 1,178 & University of California, Los Angeles & US & 1,785 \\  
\textit{54} & University of Manchester & GB & 617 & California Institute of Technology & US & 1,175 & Wuhan University & CN & 1,783 \\  
\textit{55} & University of Alberta & CA & 615 & Columbia University & US & 1,165 & King Abdulaziz University & SA & 1,766 \\  
\textit{56} & University of Pennsylvania & US & 611 & University of Paris-Sud & FR & 1,163 & \'{E}cole Polytechnique F\'{e}d\'{e}rale de Lausanne & CH & 1,763 \\  
\textit{57} & Nagoya University & JP & 609 & Harvard University & US & 1,152 & University of Chinese Academy of Sciences & CN & 1,761 \\  
\textit{58} & Columbia University & US & 605 & University of California, San Diego & US & 1,148 & University of Minnesota & US & 1,743 \\  
\textit{59} & ETH Zurich & CH & 602 & University of Chicago & US & 1,148 & University College London & GB & 1,739 \\  
\textit{60} & Kyushu University & JP & 601 & University of Bologna & IT & 1,142 & University of Lisbon & PT & 1,729 \\  
\hdashline          
\textit{61} & Tokyo Institute of Technology & JP & 598 & Kyushu University & JP & 1,139 & University of Maryland, College Park & US & 1,721 \\  
\textit{62} & Delft University of Technology & NL & 596 & University of Pisa & IT & 1,099 & Beijing Institute of Technology & CN & 1,711 \\  
\textit{63} & Utrecht University & NL & 596 & Nanjing University & CN & 1,085 & Universidad Nacional Aut\'{o}noma de M\'{e}xico & MX & 1,711 \\  
\textit{64} & Los Alamos National Laboratory & US & 591 & Delft University of Technology & NL & 1,080 & Fudan University & CN & 1,707 \\  
\textit{65} & Chinese Academy of Sciences & CN & 586 & University of Florida & US & 1,077 & Tel Aviv University & IL & 1,686 \\  
\textit{66} & University of Pisa & IT & 575 & University of Alberta & CA & 1,071 & University of Warsaw & PL & 1,677 \\  
\textit{67} & University of Edinburgh & GB & 574 & St Petersburg University & RU & 1,055 & University of Chicago & US & 1,677 \\  
\textit{68} & University of Southern California & US & 570 & Ghent University & BE & 1,054 & University of Warwick & GB & 1,674 \\  
\textit{69} & University of Iowa & US & 564 & University of Manchester & GB & 1,053 & Polish Academy of Sciences & PL & 1,670 \\  
\textit{70} & E\"{o}tv\"{o}s Lor\'{a}nd University & HU & 562 & University of Bonn & DE & 1,045 & UNSW Sydney & AU & 1,653 \\  
\hdashline          
\textit{71} & University of British Columbia & CA & 561 & \'{E}cole Polytechnique F\'{e}d\'{e}rale de Lausanne & CH & 1,042 & Delft University of Technology & NL & 1,640 \\  
\textit{72} & Steklov Mathematical Institute & RU & 561 & Dalian University of Technology & CN & 1,040 & University of Granada & ES & 1,629 \\  
\textit{73} & University of Arizona & US & 560 & University of Granada & ES & 1,038 & D\'{e}partement de Math\'{e}matiques & FR & 1,617 \\  
\textit{74} & University of Bonn & DE & 559 & University of California, Santa Barbara & US & 1,036 & Tohoku University & JP & 1,615 \\  
\textit{75} & Michigan State University & US & 558 & Carnegie Mellon University & US & 1,015 & Carnegie Mellon University & US & 1,603 \\  
\textit{76} & Australian National University & AU & 551 & Universidad Nacional Aut\'{o}noma de M\'{e}xico & MX & 1,011 & \scalebox{0.9}[1]{French Institute for Research in Computer Science and Automation} & FR & 1,601 \\  
\textit{77} & State University of New York & US & 550 & Michigan State University & US & 1,010 & Technion {\textendash} Israel Institute of Technology & IL & 1,600 \\  
\textit{78} & Universidade de S\~{a}o Paulo & BR & 548 & University of British Columbia & CA & 1,009 & Ghent University & BE & 1,592 \\  
\textit{79} & Georgia Institute of Technology & US & 544 & Universidad Complutense de Madrid & ES & 1,000 & Tongji University & CN & 1,590 \\  
\textit{80} & Carnegie Mellon University & US & 542 & Hebrew University of Jerusalem & IL & 1,000 & University of Pennsylvania & US & 1,587 \\  
\hdashline          
\textit{81} & University of Warwick & GB & 537 & University of Warsaw & PL & 1,000 & University of Amsterdam & NL & 1,569 \\  
\textit{82} & Bielefeld University & DE & 536 & University of Edinburgh & GB & 986 & Cornell University & US & 1,557 \\  
\textit{83} & Weizmann Institute of Science & IL & 526 & Polish Academy of Sciences & PL & 984 & Dalian University of Technology & CN & 1,556 \\  
\textit{84} & China Center of Advanced Science and Technology & CN & 524 & Huazhong University of Science and Technology & CN & 982 & University of Bonn & DE & 1,553 \\  
\textit{85} & Ruhr University Bochum & DE & 524 & University of Amsterdam & NL & 982 & Osaka University & JP & 1,548 \\  
\textit{86} & Hokkaido University & JP & 523 & Xi'an Jiaotong University & CN & 972 & University of Sydney & AU & 1,543 \\  
\textit{87} & Universidad Complutense de Madrid & ES & 521 & Purdue University West Lafayette & US & 968 & Nanyang Technological University & SG & 1,539 \\  
\textit{88} & Institute for Advanced Study & US & 519 & Institut de Math\'{e}matiques de Jussieu & FR & 966 & California Institute of Technology & US & 1,536 \\  
\textit{89} & University of Milan & IT & 515 & TU Wien & AT & 962 & Beijing Normal University & CN & 1,533 \\  
\textit{90} & TU Wien & AT & 514 & Xiaomi (China) & CN & 958 & Michigan State University & US & 1,530 \\  
\hdashline          
\textit{91} & University of Stuttgart & DE & 514 & University of Pennsylvania & US & 958 & Consejo Nacional de Investigaciones Cient\'{i}ficas y T\'{e}cnicas & AR & 1,527 \\  
\textit{92} & \'{E}cole Polytechnique & FR & 511 & Seoul National University & KR & 954 & Shandong University & CN & 1,521 \\  
\textit{93} & Technical University of Munich & DE & 510 & Nagoya University & JP & 949 & University of Edinburgh & GB & 1,512 \\  
\textit{94} & Ben-Gurion University of the Negev & IL & 509 & Max Planck Society & DE & 948 & TU Wien & AT & 1,508 \\  
\textit{95} & Boston University & US & 507 & Sichuan University & CN & 945 & University of Alberta & CA & 1,504 \\  
\textit{96} & Max Planck Institute for Mathematics & DE & 504 & Ben-Gurion University of the Negev & IL & 936 & Huazhong University of Science and Technology & CN & 1,504 \\  
\textit{97} & Iowa State University & US & 501 & Beijing Institute of Technology & CN & 927 & University of Electronic Science and Technology of China & CN & 1,499 \\  
\textit{98} & Peking University & CN & 500 & University of Vienna & AT & 926 & University of Manchester & GB & 1,497 \\  
\textit{99} & Applied Mathematics (United States) & US & 500 & University of Padua & IT & 911 & University of Paris-Sud & FR & 1,487 \\  
\textit{100} & \scalebox{0.9}[1]{French Institute for Research in Computer Science and Automation} & FR & 499 & Yale University & US & 908 & McGill University & CA & 1,485 \\  
          
 \\[-1.4em]
\bottomrule
\end{tabular}}
}
\end{table}
\end{landscape}
}

\afterpage{\clearpage%
{\small
\setlength{\baselineskip}{13pt}

\vspace{6mm}
\paragraph{\textbf{Acknowledgements.}}
The contents of this booklet solely reflect the views of the author and should not be interpreted as necessarily representing the official policies or endorsements, either expressed or implied, of any of the organisations with which the author is currently or has been affiliated in the past.

\vspace{-1em}
\paragraph{\textbf{Disclaimer.}}
The author does not assume any legal responsibility for errors, omissions, or claims, nor does the author offer any warranty, express or implied, concerning the information present herein and in the associated data repository. Additionally, the author is not liable should this report be utilised for a purpose other than its intended use.

\vspace{-1em}
\paragraph{\textbf{Author Contributions.}}
Keisuke Okamura: 
Conceptualisation, Methodology, Software, Validation, Formal analysis, Investigation, Data curation, Writing (Original Draft, Review \& Editing), Visualisation, Project administration.

\vspace{-1em}
\paragraph{\textbf{Competing Interests.}}
The author has no competing interests.

\vspace{-1em}
\paragraph{\textbf{Funding Information.}}
The author did not receive any funding for this research.

\vspace{-1em}
\paragraph{\textbf{Data Availability.}}
The datasets and figures generated and/or analysed during this study can be found in the Zenodo repository at \url{https://doi.org/10.5281/zenodo.8266166}.
}

\vspace{3em}
\titleformat{\section}{\centering\large\bfseries}{}{1em}{}

\bibliographystyle{unsrt}

\markboth{\textsc{References}}{}
\addcontentsline{toc}{section}{References}

\setlength{\bibsep}{0\baselineskip plus 0.2\baselineskip}
\setlength{\bibsep}{6pt plus 0.3ex}
\renewcommand*{\bibfont}{\small}

}

\end{document}